\documentclass[aps,prd,twocolumn,amsmath,amssymb,showpacs]{revtex4}

\usepackage{graphicx}
\usepackage{dcolumn}
\usepackage{bm}

\begin{document}

\title{Off-diagonal coefficients of \\
the DeWitt-Schwinger and Hadamard representations of the Feynman
propagator}

\author{Yves D\'ecanini}
\email{decanini@univ-corse.fr}
\author{Antoine Folacci}
\email{folacci@univ-corse.fr}
\affiliation{UMR CNRS 6134 SPE,
Equipe Physique Semi-Classique (et) de la Mati\`ere Condens\'ee, \\
Universit\'e de Corse, Facult\'e des Sciences, BP 52, 20250 Corte,
France}


\date{February 22, 2006}

\begin{abstract}

Having in mind applications to gravitational wave theory (in
connection with the radiation reaction problem), stochastic
semiclassical gravity (in connection with the regularization of the
noise kernel) and quantum field theory in higher-dimensional curved
spacetime (in connection with the Hadamard regularization of the
stress-energy tensor), we improve the DeWitt-Schwinger and Hadamard
representations of the Feynman propagator of a massive scalar field
theory defined on an arbitrary gravitational background by deriving
higher-order terms for the covariant Taylor series expansions of the
geometrical coefficients -- i.e., the DeWitt and Hadamard
coefficients -- that define them.

\end{abstract}


\maketitle

\section{Introduction}

The short-distance behavior of the Green functions of a field theory
defined on a curved spacetime is of fundamental importance at both
the classical and the quantum level. This has been emphasized at the
beginning of the 1960s by Bryce DeWitt in his pioneering works
dealing with i) the radiation emission of a particle moving in a
gravitational background and the radiation reaction force or
self-force felt by this particle \cite{DeWittBrehme} and ii) the
general problem of the quantization of fields in curved spacetime
\cite{DeWitt65}. In order to describe this short-distance behavior,
DeWitt, extending some ideas developed in
Refs.~\cite{Fock1937,Schwinger1951,Hadamard}, introduced the
so-called DeWitt-Schwinger and Hadamard representations of the Green
functions. These two tools have since been extensively and
successfully used to analyze and understand various aspects of
gravitational physics, from gravitational wave theory to
renormalization in quantum gravity. We refer to the monographs of
Birrell and Davies \cite{BirrellDavies}, Fulling \cite{Fulling89},
Wald \cite{Wald94}, Avramidi \cite{AvramidiLNP2000} and to the
recent review articles by Vassilevich \cite{VassilevichPR2003} and
Poisson \cite{PoissonLR2004} as well as to references therein for a
non-exhaustive state of affairs of the literature concerning the
status and the use of these two representations.

On a curved spacetime, the DeWitt-Schwinger representation is
constructed from the sequence of the DeWitt coefficients (also
called heat-kernel coefficients in the Riemannian framework)
$A_n(x,x')$ with $n \in \mathbb{N}$ which are purely geometrical
two-point objects formally independent of the dimension $d$ of
spacetime and defined by a recursion relation. The DeWitt
coefficients $A_n(x,x')$ of lowest orders encode the short-distance
singular behavior of the Green functions and, as a consequence,
their determination is an important problem. Unfortunately, in
general, these coefficients can not be determined exactly. It is
however possible to look for them in the form of a covariant Taylor
series expansion for $x'$ in the neighborhood of $x$
\begin{eqnarray}
&  & A_n(x,x')=a_n(x) -  a_{n \, \mu_1} (x) \sigma^{; \mu_1}(x,x')
\nonumber \\ & &  \qquad \qquad + \frac{1}{2!} a_{n \,\mu_1 \mu_2}
(x) \sigma^{;\mu_1}(x,x')\sigma^{;\mu_2}(x,x') + \dots \nonumber
\end{eqnarray}
(here $\sigma (x,x')$ denotes the geodetic interval between $x$ and
$x'$ \cite{DeWittBrehme,DeWitt65}) and to try then to ``solve" the
recursion relation defining them. This is not an easy task and
computational complications increase very rapidly with the orders
$n$ and $p$ of the coefficient $a_{n \,\mu_1 \dots \mu_p} (x)$ which
is a scalar of order $R^{n+p}$ constructed from the Riemann tensor
$R$ and its derivatives. In fact, during the last forty years, it is
mainly the determination of the so-called diagonal DeWitt
coefficients $a_n(x)=A_n(x,x)$ with $n \ge 1$ (we have the trivial
result $a_0(x)=1$) which has attracted the attention of theoretical
physicists in connection with renormalization in the effective
action for quantum field theories and quantum gravity and with
gravitational anomalies. In addition, mathematicians have calculated
these coefficients in connection with spectral geometry, topology of
manifolds and the Atiyah-Singer index theorem \cite{Gilkey84}. Among
the numerous important results obtained by very different technical
approaches, it is worthwhile pointing out the derivation of the
following:

\qquad i) $a_1(x)$ and $a_2(x)$ by DeWitt \cite{DeWitt65} for a
scalar field (in the presence of a Yang-Mills background) and for
the Dirac spinor field both propagating on an arbitrary curved
spacetime;

\qquad ii) $a_3(x)$ by Sakai \cite{Sakai71} for an ordinary scalar
field theory defined on an arbitrary curved space and by Gilkey
\cite{Gilkey75} for the general case, i.e. for tensorial field
theories defined on Riemannian manifolds in the presence of external
gauge fields;

\qquad iii) $a_4(x)$ by Amsterdamski, Berkin and O'Connors
\cite{Amsterdamski89} for an ordinary scalar field and by Avramidi,
in Ref.~\cite{Avramidi91} (see also the corresponding erratum
\cite{Avramidi98}), for the general case;

\qquad iv) $a_5(x)$ by Van de Ven \cite{vandeVen98} for the
general case.

The importance, from the physical point of view, of the off-diagonal
DeWitt coefficients has been clearly realized in the mid-1970s when
interest in the regularization and renormalization of the
stress-energy tensor associated with a quantum field propagating on
a curved spacetime began to grow
\cite{DeWitt75,Christensen1,Christensen2}. Indeed, it appeared that,
in this context, the knowledge of the first terms of the covariant
Taylor series expansions of the DeWitt coefficients of lowest orders
was crucial. Christensen then derived the covariant Taylor series
expansions of the DeWitt coefficients $A_0(x,x')$, $A_1(x,x')$, and
$A_2(x,x')$ up to orders $\sigma^{2}$, $\sigma^1$ and $\sigma^{0}$
respectively for an ordinary scalar field theory in 1976
\cite{Christensen1} and for the spin $1/2$- and spin $1$- theories
in 1978 \cite{Christensen2}. Christensen's work has had a great
impact on quantum field theory in curved spacetime: in connection
with the point-splitting prescription \cite{DeWitt65}, it has
provided a general technique for the regularization and
renormalization of the stress-energy tensor. However, it is
important to note that Christensen's results have a limited domain
of applicability: they have been used to regularize the
stress-energy tensor in a four-dimensional curved spacetime and they
could also permit us to develop the regularization process in a
three-dimensional curved spacetime but, to our knowledge, this has
never been explicitly realized. Nowadays, supergravity theories,
string theories and M-theory predict that spacetime has more
dimensions than the four we observe. In this context, it is
therefore necessary to extend the Christensen's method taking into
account the possible extra dimensions: in order to be able to work
in five dimensions, it is necessary to derive the DeWitt
coefficients $A_0(x,x')$, $A_1(x,x')$ and $A_2(x,x')$ up to orders
$\sigma^{5/2}$, $\sigma^{3/2}$ and $\sigma^{1/2}$; in order to be
able to work in six dimensions, it is necessary to derive the DeWitt
coefficients $A_0(x,x')$, $A_1(x,x')$, $A_2(x,x')$ and $A_3(x,x')$
up to orders $\sigma^3$, $\sigma^2$, $\sigma^1$ and $\sigma^{0}$;
... in order to be able to work in ten dimensions, it is necessary
to derive the DeWitt coefficients $A_0(x,x')$, $A_1(x,x')$,
$A_2(x,x')$, $A_3(x,x')$, $A_4(x,x')$ and $A_5(x,x')$ up to orders
$\sigma^5$, $\sigma^4$, $\sigma^3$, $\sigma^2$ , $\sigma^1$ and
$\sigma^{0}$; and in order to be able to work in eleven dimensions,
it is necessary to derive the DeWitt coefficients $A_0(x,x')$,
$A_1(x,x')$, $A_2(x,x')$, $A_3(x,x')$, $A_4(x,x')$ and $A_5(x,x')$
up to orders $\sigma^{11/2}$, $\sigma^{9/2}$, $\sigma^{7/2}$,
$\sigma^{5/2}$, $\sigma^{3/2}$ and $\sigma^{1/2}$.

In fact, we do not need to appeal to supergravity theories, string
theories and M-theory as well as the possible extra-dimensions of
spacetime to justify the necessity to go beyond Christensen's
results. In recent works dealing with four-dimensional gravitational
physics, such a necessity has clearly appeared in two different
contexts: in the quantum domain of stochastic semiclassical gravity,
in connection with the regularization of the noise kernel, but also
in the classical domain of gravitational wave theory, in connection
with the radiation reaction force. As far as the noise kernel is
concerned, it should be recalled that it is a measure of the
fluctuations of the stress-energy tensor associated with a quantum
field theory defined on a curved spacetime. It is defined as the
vacuum expectation value of a bitensor constructed by taking the
product of the stress-energy-tensor operator with itself
\cite{PhillipsHu01,PhillipsHu03}. It plays a central role in
stochastic semiclassical gravity (see Ref.~\cite{HuVerdaguerLR2004}
for a review on this topic) permitting us to define the stochastic
part of the source in the Einstein-Langevin equations. Its
regularization, in the coincidence limit, necessitates the knowledge
of the divergent part of the Feynman propagator up to order
$\sigma^2$ \cite{PhillipsHu03} and therefore the knowledge of the
DeWitt coefficients $A_0(x,x')$, $A_1(x,x')$, $A_2(x,x')$ and
$A_3(x,x')$ up to orders $\sigma^3$, $\sigma^2$, $\sigma^1$ and
$\sigma^{0}$. As far as the radiation reaction force is concerned,
it should be recalled that its computation in Schwarzschild and Kerr
spacetimes for arbitrary orbits is now an urgent problem of
gravitational wave theory (see Ref.~\cite{PoissonLR2004} for a
review as well as Refs.~\cite{Barack_et_al_2002,BarackOri2003} for
recent important progress). In particular, the computation of the
nonlocal part of this force (it is an integral of the retarded Green
function over the past trajectory of the particle moving in the
gravitational background and which is the source of the
gravitational radiation) has been considered in recent works
\cite{AndersonHu04,AndersonFlanaganOttewill05} and the necessity, in
this context, to go beyond Christensen's expansions of the DeWitt
coefficients has been pointed out.

On a $d$-dimensional curved spacetime, the Hadamard representation
is constructed from a set of two-point coefficients, the so-called
Hadamard coefficients, which are also defined by recursion
relations. For $d$ even, there exists 3 families of Hadamard
coefficients noted $U_n(x,x')$ with $n=0,1, \dots, d/2-2$ and
$V_n(x,x')$ and $W_n(x,x')$ with $n \in \mathbb{N}$, while for $d$
odd,  there exists 2 families of Hadamard coefficients noted
$U_n(x,x')$ and $W_n(x,x')$ with $n \in \mathbb{N}$. The Hadamard
coefficients $U_n(x,x')$ and $V_n(x,x')$ are, like the DeWitt
coefficients $A_n(x,x')$, purely geometrical objects and here again
those of lowest orders encode the short-distance singular behavior
of the Green functions. In fact, the Hadamard coefficients
$U_n(x,x')$ and $V_n(x,x')$ can be constructed from the DeWitt
coefficients $A_n(x,x')$. Thus, the knowledge of the covariant
Taylor series expansions for $x'$ in the neighborhood of $x$ of the
DeWitt coefficients permits us to construct immediately the
corresponding expansions of the geometrical Hadamard coefficients.
As far as the coefficients $W_n(x,x')$ are concerned, it is
important to note that they correspond to a finite part of the Green
functions and that they are neither determined in terms of the local
geometry nor uniquely defined by a recursion relation. As a
consequence, they can be used to encode supplementary physical
information concerning the studied field (boundary conditions,
quantum state dependence, ...). Because of that property, the
Hadamard representation is in our opinion more interesting than the
DeWitt-Schwinger one. Moreover, in the context of the regularization
of the stress-energy tensor, the Christensen approach has been
replaced by a variant based on the Hadamard representation, the
so-called Hadamard method
\cite{Wald77,AdlerETAL1,AdlerETAL2,Wald78,BrownOttewill83,Castagnino84,
BrownOttewill86,BernardFolacci86,Tadaki87,AFO,Folacci91}. It is more
general than the original method and more efficient. Furthermore,
because of its axiomatic foundations \cite{Wald77,Wald78,Wald94}, it
is more rigorous. It has been developed in a four-dimensional
framework and its extension in higher dimensions necessitates the
derivation of the covariant Taylor series expansions of the Hadamard
coefficients beyond the orders reached in
Refs.~\cite{Wald77,AdlerETAL1,AdlerETAL2,Wald78,BrownOttewill83,Castagnino84,
BrownOttewill86,BernardFolacci86,Tadaki87,AFO,Folacci91}.

In the present article, we shall consider the DeWitt-Schwinger and
Hadamard representations of the Feynman propagator of a massive
scalar field theory and we shall improve these two representations
by obtaining higher-order terms for the covariant Taylor series
expansions of the coefficients -- i.e., the DeWitt and geometrical
Hadamard coefficients -- that define them. More precisely, we shall
first provide the covariant Taylor series expansions of the DeWitt
coefficients $A_0(x,x')$, $A_1(x,x')$, $A_2(x,x')$ and $A_3(x,x')$
up to orders $\sigma^{3}$, $\sigma^{2}$, $\sigma^1 $ and
$\sigma^{0}$ respectively. We shall then provide the following:

i) in three dimensions, the covariant Taylor series expansions of
the geometrical Hadamard coefficients $U_0(x,x')$,  $U_1(x,x')$,
$U_2(x,x')$ and $U_3(x,x')$ up to orders $\sigma^{3}$, $\sigma^{2}$,
$\sigma^1 $ and $\sigma^{0}$ respectively or, in other words, the
covariant Taylor series expansion of the divergent part
$U(x,x')/\sigma^{1/2}(x,x')$ of the Hadamard representation up to
order $\sigma^{5/2}$;

ii) in four dimensions, the covariant Taylor series expansions of
the geometrical Hadamard coefficients $U_0(x,x')$, $V_0(x,x')$,
$V_1(x,x')$, and $V_2 (x,x')$ up to orders $\sigma^{3}$,
$\sigma^{2}$, $\sigma^1 $ and $\sigma^{0}$ respectively or, in other
words, the covariant Taylor series expansions of the divergent parts
$U(x,x')/\sigma(x,x')$ and $V(x,x')\ln \sigma(x,x')$ of the Hadamard
representation up to order $\sigma^{2}$ and $\sigma^{2} \ln \sigma$
respectively;

iii) in five dimensions, the covariant Taylor series expansions of
the geometrical Hadamard coefficients $U_0(x,x')$,  $U_1(x,x')$,
$U_2(x,x')$ and $U_3(x,x')$ up to orders $\sigma^{3}$, $\sigma^{2}$,
$\sigma^1 $ and $\sigma^{0}$ respectively or, in other words, the
covariant Taylor series expansion of the divergent part
$U(x,x')/\sigma^{3/2}(x,x')$ of the Hadamard representation up to
order $\sigma^{3/2}$;

iv) in six dimensions, the covariant Taylor series expansions of the
geometrical Hadamard coefficients $U_0(x,x')$,  $U_1(x,x')$,
$V_0(x,x')$ and $V_1 (x,x')$ up to orders $\sigma^{3}$,
$\sigma^{2}$, $\sigma^1 $ and $\sigma^{0}$ respectively or, in other
words, the covariant Taylor series expansions of the divergent parts
$U(x,x')/\sigma^2(x,x')$ and $V(x,x')\ln \sigma(x,x')$ of the
Hadamard representation up to order $\sigma $ and $\sigma  \ln
\sigma $ respectively.

Our article is organized as follows. In Sec.~II, we establish the
framework of our study as well as our notations. In particular, we
establish the relationship linking the DeWitt and the geometrical
Hadamard coefficients and we also prove that the DeWitt-Schwinger
representation possesses the Hadamard form. In Sec.~III, by
combining the old covariant recursive method of DeWitt
\cite{DeWittBrehme,DeWitt65} with results obtained from the modern
covariant non-recursive approach of Avramidi
\cite{Avramidi_PhD,AvramidiLNP2000}, we explicitly construct the
covariant Taylor series expansions of the DeWitt coefficients
$A_0(x,x')$, $A_1(x,x')$, $A_2(x,x')$ and $A_3(x,x')$ up to orders
$\sigma^{3}$, $\sigma^{2}$, $\sigma^1 $ and $\sigma^{0}$
respectively. In Sec.~IV, we translate the results previously
obtained in the framework of the Hadamard formalism and we provide
the explicit expressions for the covariant Taylor series expansions
of the corresponding geometrical Hadamard coefficients. Finally, in
Sec.~V, we discuss possible extensions of our work as well as
immediate applications. In five appendixes, we gather some technical
details which have been used to derive our results. In these
appendixes we have also provided the covariant Taylor series
expansions of the bitensors $\Delta^{1/2}$, $ \Delta ^{-1/2}{\Delta
^{1/2}}_{;\mu} \sigma^{; \mu}$, $\sigma_{;\mu \nu}$ and $g_{\nu
 \nu'}\sigma_{;\mu}^{\phantom{;\mu} \nu'}$ beyond the orders needed
 in the present article, i.e. up to
orders $\sigma^{11/2}$, $\sigma^{9/2}$, $\sigma^{9/2}$ and
$\sigma^{9/2}$ respectively. We think that these results could be
very useful in a near future for people working in the field of
gravitational physics.

It should be noted that we shall use the geometrical conventions of
Hawking and Ellis \cite{HawkingEllis} concerning the definitions of
the scalar curvature $R$, the Ricci tensor $R_{\mu \nu}$ and the
Riemann tensor $R_{\mu \nu \rho \sigma}$ and we shall extensively
use the commutation of covariant derivatives in the form
\begin{eqnarray}\label{CD_NabNabTensor}
&  & T^{\rho \dots}_{\phantom{\rho} \sigma \dots ;\nu \mu} -
T^{\rho \dots}_{\phantom{\rho} \sigma \dots ;\mu \nu} = \nonumber \\
&   & \qquad  + R^{\rho}_{\phantom{\rho} \tau \mu \nu} T^{\tau
\dots}_{\phantom{\tau} \sigma \dots } + \dots -
R^{\tau}_{\phantom{\tau} \sigma \mu \nu} T^{\rho
\dots}_{\phantom{\rho} \tau \dots} - \dots
\end{eqnarray}

\section{DeWitt-Schwinger and Hadamard representations}

We shall consider a massive scalar field $\Phi$ propagating on a
$d$-dimensional curved spacetime $({\cal M},g)$ and obeying the
wave equation
\begin{equation}\label{WEQ}
\left( \Box -m^2 -\xi R \right) \Phi =0.
\end{equation}
Here $m$ is the mass of the scalar field, $\xi$ is a dimensionless
factor which accounts for the possible coupling between the scalar
field and the gravitational background and we shall assume that
$d>2$. We shall focus our attention on the Feynman propagator
$G^{\mathrm{F}}(x,x')$ solution of
\begin{equation}\label{WEQ_G1}
\left( \Box_x -m^2 -\xi R \right) G^{\mathrm{F}}(x,x') = -\delta^d
(x,x')
\end{equation}
with $\delta^d (x,x')= [-g(x)]^{-1/2}(x) \delta^d (x-x')$, or more
precisely on the way in which its DeWitt-Schwinger and Hadamard
representations encode its short-distance behavior. It should be
noted that our presentation does not pretend to be mathematically
rigorous. It is however possible to find precisions concerning the
mathematical status of the DeWitt-Schwinger and Hadamard
representations as well as the nature of the series defining them
in
Refs.~\cite{Hadamard,Garabedian,Friedlander,Fulling89,Wald94,MorettiCMP1999,
MorettiCMP2000}.

\subsection{DeWitt-Schwinger representation of
$G^{\mathrm{F}}(x,x')$}

We first recall that the DeWitt-Schwinger representation of the
Feynman propagator $G^{\mathrm{F}}(x,x')$ is given by (see
Refs.~\cite{DeWitt65,DeWitt03,BirrellDavies,Fulling89})
\begin{equation}\label{DSrep1}
G^{\mathrm{F}}_{\mathrm{DS}} (x,x') = i \int_0^{+\infty} H(s;x,x')
\, ds
\end{equation}
where $H(s;x,x')$ is a function which satisfies
\begin{subequations}\label{DSrep2}
\begin{equation}\label{DSrep2a}
\left( i \frac{\partial}{\partial s} + \Box_x -m^2 -\xi R \right)
H(s;x,x') = 0 \quad \mathrm{for} \quad s>0
\end{equation}
with the boundary condition
\begin{equation}\label{DSrep2b}
H(s;x,x') \to  \delta^d (x,x') \quad \mathrm{as} \quad s \to 0,
\end{equation}
\end{subequations}
and which can be formally written, for $s \to 0$ and $x'$ near $x$,
on the form
\begin{eqnarray}\label{DSrep2c}
&  & H(s;x,x') = i (4\pi is)^{-d/2} \, e^{(i/2s)[
\sigma(x,x')+i\epsilon] -i m^2 s} \nonumber \\
&  &  \qquad \qquad \qquad \qquad \times \sum_{n=0}^{+\infty}
A_n(x,x') (is)^n.
\end{eqnarray}
Here the factor $i\epsilon$ with $\epsilon \to 0_+$ is introduced to
give to $G^{\mathrm{F}}_{\mathrm{DS}} (x,x')$ a singularity
structure that is consistent with the definition of the Feynman
propagator as a time-ordered product. Furthermore, the DeWitt
coefficients $A_n(x,x')$ labelled by $n \in \mathbb{N}$ are a
sequence of biscalar functions, symmetric in the exchange of $x$ and
$x'$, regular for $x' \to x$, and defined by the recursion relations
\begin{subequations}\label{DSrep3}
\begin{eqnarray}\label{DSrep3a}
& & (n+1) A_{n+1} + A_{n+1 ; \mu} \sigma ^{; \mu} - A_{n+1} \Delta
^{-1/2}{\Delta ^{1/2}}_{;\mu} \sigma^{; \mu}  \nonumber \\
&  &  \qquad \qquad \qquad \qquad  = \left( \Box_x -\xi R \right)
A_n \quad \mathrm{for}~n \in \mathbb{N}
\end{eqnarray}
and the boundary condition
\begin{equation}\label{DSrep3b}
A_0= \Delta ^{1/2}.
\end{equation}
\end{subequations}
In Eqs.~(\ref{DSrep2c}) and (\ref{DSrep3a}), $\sigma(x,x')$ is the
geodetic interval -- i.e., $2\sigma(x,x')$ is the square of the
geodesic distance between $x$ and $x'$ -- and we have $\sigma(x,x')
< 0$ if $x$ and $x'$ are timelike related, $\sigma(x,x') = 0$ if $x$
and $x'$ are null related and $\sigma(x,x') > 0$ if $x$ and $x'$ are
spacelike related. It is a biscalar function that satisfies
\begin{equation}\label{DSrep4}
2 \sigma= \sigma^{; \mu}\sigma_{; \mu}.
\end{equation}
In Eqs.~(\ref{DSrep3a}) and (\ref{DSrep3b}), $\Delta (x,x')$ is
the biscalar form of the Van Vleck-Morette determinant
\cite{DeWitt65}. It is defined by
\begin{equation}\label{DSrep5}
\Delta (x,x')= -[-g(x)]^{-1/2} \mathrm{det} (- \sigma_{; \mu
\nu'}(x,x')) [-g(x')]^{-1/2}
\end{equation}
and it satisfies the partial differential equation
\begin{subequations}\label{DSrep6}
\begin{equation}\label{DSrep6a}
\Box_x \sigma = d - 2 \Delta ^{-1/2}{\Delta ^{1/2}}_{;\mu}
\sigma^{; \mu}
\end{equation}
and the boundary condition
\begin{equation}\label{DSrep6b}
\lim_{x' \to x} \Delta (x,x')=1.
\end{equation}
\end{subequations}
The recursion relations (\ref{DSrep3a}), the boundary condition
(\ref{DSrep3b}) and the relations (\ref{DSrep4}) and
(\ref{DSrep6}) insure that the function $H(s;x,x')$ given by
(\ref{DSrep2c}) is a solution of (\ref{DSrep2a}) and
(\ref{DSrep2b}) and therefore that (\ref{DSrep1}) solves the wave
equation (\ref{WEQ_G1}). The DeWitt coefficients $A_n(x,x')$ can
be formally obtained by solving the recursion relations
(\ref{DSrep3a}) taking into account the boundary condition
(\ref{DSrep3b}). This can be realized by integrating along the
geodesic joining $x$ to $x'$ (it is unique for $x'$ near $x$ or
more generally for $x'$ in a convex normal neighborhood of $x$).
As a consequence, the DeWitt coefficients are determined uniquely
and are purely geometrical objects, i.e. they only depend on the
geometry along this geodesic.

\subsection{Hadamard representation of
$G^{\mathrm{F}}(x,x')$}

As far as the structure of the Hadamard representation of the
Feynman propagator $G^{\mathrm{F}}(x,x')$ is concerned, we recall
that it depends on whether the dimension $d$ of spacetime is even or
odd. For $d$ even, it is given by (here we extend considerations
developed in Refs.~\cite{Hadamard,Garabedian,Friedlander})
\begin{eqnarray}\label{HevRep1}
& & G^{\mathrm{F}}_{\mathrm{H}} (x,x') = i \, \frac{(d/2-2)!}{2
(2\pi)^{d/2}}
\left[\frac{U(x,x')}{[\sigma(x,x')+i\epsilon]^{d/2-1}} \right. \nonumber \\
& &  \left. \phantom{\frac{U}{\sigma^{d}}} + V(x,x') \ln
[\sigma(x,x')+i\epsilon] + W(x,x') \right]
\end{eqnarray}
where $U(x,x')$, $V(x,x')$ and $W(x,x')$ are symmetric biscalars,
regular for $x' \to x$ and which possess expansions of the form
\begin{subequations}\label{HevRep2}
\begin{eqnarray}
& & U(x,x')= \sum_{n=0}^{d/2-2} U_n(x,x')\sigma^n(x,x'), \label{HevRep2a} \\
& & V(x,x')= \sum_{n=0}^{+\infty} V_n(x,x')\sigma^n(x,x'), \label{HevRep2b} \\
& & W(x,x')= \sum_{n=0}^{+\infty} W_n(x,x')\sigma^n(x,x').
\label{HevRep2c}
\end{eqnarray}
\end{subequations}
For $d$ odd, it is given by (see
Refs.~\cite{Hadamard,Garabedian,Friedlander})
\begin{eqnarray}\label{HodRep1}
& & G^{\mathrm{F}}_{\mathrm{H}} (x,x') = i \, \frac{\Gamma(d/2-1)}{2
(2\pi)^{d/2}}
\left[\frac{U(x,x')}{[\sigma(x,x')+i\epsilon]^{d/2-1}} \right. \nonumber \\
& &  \left. \phantom{\frac{U }{\sigma(x,x')^{\frac{d-2}{2}}}} \qquad
\qquad + W(x,x') \right]
\end{eqnarray}
where $U(x,x')$ and $W(x,x')$ are again symmetric and regular
biscalar functions which now possess expansions of the form
\begin{subequations}\label{HodRep2}
\begin{eqnarray}
& & U(x,x')= \sum_{n=0}^{+\infty} U_n(x,x')\sigma^n(x,x'), \label{Hodrep2a} \\
& & W(x,x')= \sum_{n=0}^{+\infty} W_n(x,x')\sigma^n(x,x').
\label{Hodrep2b}
\end{eqnarray}
\end{subequations}
In Eqs.~(\ref{HevRep1}) and (\ref{HodRep1}), the factor $i\epsilon$
with $\epsilon \to 0_+$ is again introduced to give to
$G^{\mathrm{F}}_{\mathrm{H}} (x,x')$ a singularity structure that is
consistent with the definition of the Feynman propagator as a
time-ordered product.

For $d$ even, the Hadamard coefficients $U_n(x,x')$, $V_n(x,x')$ and
$W_n(x,x')$ are symmetric and regular biscalar functions. The
coefficients $U_n(x,x')$ satisfy the recursion relations
\begin{subequations}\label{HevRep5A}
\begin{eqnarray}\label{HevRep5a}
& & (n+1)(2n+4-d) U_{n+1} + (2n+4-d) U_{n+1 ; \mu} \sigma ^{; \mu}
\nonumber \\
&  &  \quad  - (2n+4-d) U_{n+1} \Delta ^{-1/2}{\Delta
^{1/2}}_{;\mu} \sigma^{; \mu}   \nonumber \\
&  &  \quad  + \left( \Box_x -m^2 -\xi R \right) U_n =0 \nonumber \\
& & \qquad \qquad\qquad\qquad \mathrm{for}~ n=0,1, \dots , d/2-3
\end{eqnarray}
with the boundary condition
\begin{equation}\label{HevRep5b}
U_0= \Delta ^{1/2}.
\end{equation}
\end{subequations}
The coefficients $V_n(x,x')$ satisfy the recursion relations
\begin{subequations}\label{HevRep5B}
\begin{eqnarray}\label{HevRep5c}
& & (n+1)(2n+d) V_{n+1} + 2(n+1) V_{n+1 ; \mu} \sigma ^{; \mu}
\nonumber \\
&  &  \quad  - 2(n+1) V_{n+1} \Delta ^{-1/2}{\Delta
^{1/2}}_{;\mu} \sigma^{; \mu}   \nonumber \\
&  &  \quad  + \left( \Box_x -m^2 -\xi R \right) V_n =0  \quad
\mathrm{for}~n \in \mathbb{N}
\end{eqnarray}
with the boundary condition
\begin{eqnarray}\label{HevRep5d}
& & (d-2) V_0 + 2 V_{0 ; \mu} \sigma ^{; \mu} - 2 V_0 \Delta
^{-1/2}{\Delta
^{1/2}}_{;\mu} \sigma^{; \mu}   \nonumber \\
&  &  \quad  + \left( \Box_x -m^2 -\xi R \right) U_{d/2-2} =0.
\end{eqnarray}
\end{subequations}
The coefficients $W_n(x,x')$  satisfy the recursion relations
\begin{eqnarray}\label{HevRep6}
& & (n+1)(2n+d) W_{n+1} + 2(n+1) W_{n+1 ; \mu} \sigma ^{; \mu}
\nonumber \\
&  &  \quad  - 2(n+1) W_{n+1} \Delta ^{-1/2}{\Delta
^{1/2}}_{;\mu} \sigma^{; \mu}   \nonumber \\
& &  \quad + (4n+2+d) V_{n+1} + 2 V_{n+1 ; \mu} \sigma ^{; \mu}
\nonumber \\
&  &  \quad  - 2 V_{n+1} \Delta ^{-1/2}{\Delta
^{1/2}}_{;\mu} \sigma^{; \mu}   \nonumber \\
&  &  \quad  + \left( \Box_x -m^2 -\xi R \right) W_n =0 \quad
\mathrm{for}~n \in \mathbb{N}.
\end{eqnarray}
From the recursion relations (\ref{HevRep5a}), (\ref{HevRep5c})
and (\ref{HevRep6}), the boundary conditions (\ref{HevRep5b}) and
(\ref{HevRep5d}) and the relations (\ref{DSrep4}) and
(\ref{DSrep6}) it is possible to prove that the Hadamard
representation (\ref{HevRep1})-(\ref{HevRep2}) solves the wave
equation (\ref{WEQ_G1}). This can be done easily by noting that we
have
\begin{equation}\label{HevRep_EQ_V}
\left( \Box_x -m^2 -\xi R \right) V =0
\end{equation}
as a consequence of (\ref{HevRep5c}) and
\begin{eqnarray}\label{HevRep_EQ_W}
& & \sigma \left( \Box_x -m^2 -\xi R \right) W = - \left( \Box_x
-m^2 -\xi R \right) U_{d/2-2} \nonumber \\
& & \quad -(d-2) V - 2 V_{; \mu} \sigma ^{; \mu} + 2 V \Delta
^{-1/2}{\Delta ^{1/2}}_{;\mu} \sigma^{; \mu}
\end{eqnarray}
as a consequence of (\ref{HevRep5d}) and (\ref{HevRep6}). The
Hadamard coefficients $U_n(x,x')$ can be formally obtained by
integrating the recursion relations (\ref{HevRep5a}) along the
geodesic joining $x$ to $x'$. {\it Mutatis mutandis}, the Hadamard
coefficients $V_n(x,x')$ can be obtained by solving the recursion
relations (\ref{HevRep5c}). As a consequence, the Hadamard
coefficients $U_n(x,x')$ and $V_n(x,x')$ are purely geometric
biscalars. As far the Hadamard coefficients $W_n(x,x')$ are
concerned, it should be noted that the biscalar $W_0(x,x')$ is
unrestrained by the recursion relations (\ref{HevRep6}). These
relations only determine the $W_n(x,x')$ with $n \ge 1$ once
$W_0(x,x')$ is specified.

For $d$ odd, the Hadamard coefficients $U_n(x,x')$ and $W_n(x,x')$
are symmetric and regular biscalar functions. The coefficients
$U_n(x,x')$ satisfy the recursion relations
\begin{subequations}\label{HodRep3}
\begin{eqnarray}\label{HodRep3a}
& & (n+1)(2n+4-d) U_{n+1} + (2n+4-d) U_{n+1 ; \mu} \sigma ^{; \mu}
\nonumber \\
&  &  \quad  - (2n+4-d) U_{n+1} \Delta ^{-1/2}{\Delta
^{1/2}}_{;\mu} \sigma^{; \mu}   \nonumber \\
&  &  \quad  + \left( \Box_x -m^2 -\xi R \right) U_n =0 \quad
\mathrm{for}~ n \in \mathbb{N}
\end{eqnarray}
with the boundary condition
\begin{equation}\label{HodRep3b}
U_0= \Delta ^{1/2}.
\end{equation}
\end{subequations}
The coefficients $W_n(x,x')$ satisfy the recursion relations
\begin{eqnarray}\label{HodRep4}
& & (n+1)(2n+d) W_{n+1} + 2(n+1) W_{n+1 ; \mu} \sigma ^{; \mu}
\nonumber \\
&  &  \quad  - 2(n+1) W_{n+1} \Delta ^{-1/2}{\Delta ^{1/2}}_{;\mu}
\sigma^{; \mu}
\nonumber \\
&  &  \quad  + \left( \Box_x -m^2 -\xi R \right) W_n =0 \quad
\mathrm{for}~ n \in \mathbb{N}.
\end{eqnarray}
From the recursion relations (\ref{HodRep3a}) and (\ref{HodRep4}),
the boundary conditions (\ref{HodRep3b}) and the relations
(\ref{DSrep4}) and (\ref{DSrep6}) it is possible to prove that the
Hadamard representation (\ref{HodRep1})-(\ref{HodRep2}) solves the
wave equation (\ref{WEQ_G1}). This can be done easily from
\begin{equation}\label{HodRep_EQ_W}
\left( \Box_x -m^2 -\xi R \right) W = 0
\end{equation}
which is a consequence of (\ref{HodRep4}). Here again, it should be
noted that the Hadamard coefficients $U_n(x,x')$ are purely
geometric biscalars which can be formally obtained by integrating
the recursion relations (\ref{HodRep3a}) along the geodesic joining
$x$ to $x'$. Here again the biscalar $W_0(x,x')$ is unrestrained by
the recursion relations (\ref{HodRep4}).

\subsection{From the DeWitt coefficients $A_n(x,x')$ to the geometrical
Hadamard coefficients $U_n(x,x')$ and $V_n(x,x')$}

It is possible to establish the relationship linking the DeWitt
coefficients $A_n(x,x')$ and the geometrical Hadamard coefficients
$U_n(x,x')$ and $V_n(x,x')$. In order to do this, we first introduce
a new sequence ${\tilde{A}}_n(m^2;x,x')$ with $n \in \mathbb{N}$ of
geometrical coefficients which we shall call the mass-dependent
DeWitt coefficients. They are defined as the sequence of biscalar
functions, symmetric in the exchange of $x$ and $x'$, regular for
$x' \to x$, which satisfy the recursion relations
\begin{subequations}\label{MDWCoeff1}
\begin{eqnarray}\label{MDWCoeff1a}
& & (n+1) {\tilde{A}}_{n+1}  + {\tilde{A}}_{n+1 ; \mu}  \sigma ^{;
\mu} - {\tilde{A}}_{n+1} \Delta
^{-1/2}{\Delta ^{1/2}}_{;\mu} \sigma^{; \mu}  \nonumber \\
&  &  \qquad \qquad   = \left( \Box_x -m^2 -\xi R \right)
{\tilde{A}}_n \quad \mathrm{for}~ n \in \mathbb{N}
\end{eqnarray}
and the boundary condition
\begin{equation}\label{MDWCoeff1b}
{\tilde{A}}_0 = \Delta ^{1/2}.
\end{equation}
\end{subequations}
Of course, they are linked to the (ordinary) DeWitt coefficients
$A_n(x,x')$. We have
\begin{equation}\label{RelAetA1}
A_n(x,x')={\tilde{A}}_n(m^2=0;x,x')
\end{equation}
and a direct comparison of Eqs.~(\ref{MDWCoeff1}) and (\ref{DSrep3})
permits us to obtain easily
\begin{equation}\label{RelAetA2}
{\tilde{A}}_n(m^2;x,x') = \sum_{k=0}^{n} \frac{(-1)^k}{k!}(m^2)^k
A_{n-k}(x,x').
\end{equation}
Now, by comparing the equations (\ref{HevRep5A}) and
(\ref{HevRep5B}) defining the Hadamard coefficients for $d$ even
with the equations (\ref{MDWCoeff1}) defining the mass-dependent
DeWitt coefficients, we can obtain
\begin{subequations}\label{RelUetAp1}
\begin{eqnarray}
& & U_n(x,x') = \frac{(d/2-2-n)!}{2^n (d/2-2)!}
{\tilde{A}}_n(m^2;x,x') \nonumber \\
& & \qquad \qquad\qquad\qquad\quad \mathrm{for}~ n=0,1, \dots ,
d/2-2,  \label{RelUetAp1a} \\
& & V_n(x,x') = \frac{
(-1)^{n+1}}{2^{n+d/2-1} n! (d/2-2)!}
{\tilde{A}}_{n+d/2-1}(m^2;x,x') \nonumber \\
& & \qquad \qquad\qquad\qquad\quad \mathrm{for}~ n \in \mathbb{N},
\label{RelUetAp1b}
\end{eqnarray}
\end{subequations}
and therefore we establish from (\ref{RelAetA2}) the relations
\begin{subequations}\label{RelUetAp2}
\begin{eqnarray}
& & U_n(x,x') = \frac{(d/2-2-n)!}{2^n (d/2-2)!} \sum_{k=0}^{n}
\frac{(-1)^k}{k!}(m^2)^k A_{n-k}(x,x') \nonumber \\& & \qquad
\qquad\qquad\qquad\quad \mathrm{for}~ n=0,1, \dots , d/2-2,
\label{RelUetAp2a}  \\
& & V_n(x,x') = \frac{ (-1)^{n+1}}{2^{n+d/2-1} n! (d/2-2)!} \nonumber \\
& & \qquad \quad \times  \sum_{k=0}^{n+d/2-1}
\frac{(-1)^k}{k!}(m^2)^k A_{n+d/2-1-k}(x,x') \nonumber \\
& & \qquad \qquad\qquad\qquad\quad \mathrm{for}~ n \in \mathbb{N}.
\label{RelUetAp2b}
\end{eqnarray}
\end{subequations}

Similarly, by comparing the equations (\ref{HodRep3}) defining the
Hadamard coefficients for $d$ odd with the equations
(\ref{MDWCoeff1}) defining the mass-dependent DeWitt coefficients,
we can obtain
\begin{eqnarray}
& & U_n(x,x') = \frac{\Gamma(d/2-1-n)}{2^n \Gamma(d/2-1)}
{\tilde{A}}_n(m^2;x,x')   \quad   \mathrm{for}~ n \in \mathbb{N}  \nonumber \\
\label{RelUetAo1}
\end{eqnarray}
and therefore we establish from (\ref{RelAetA2}) the relation
\begin{eqnarray}
& & U_n(x,x') = \frac{\Gamma(d/2-1-n)}{2^n \Gamma(d/2-1)} \nonumber \\
& & \qquad \quad \times \sum_{k=0}^{n} \frac{(-1)^k}{k!}(m^2)^k
A_{n-k}(x,x')    \nonumber \\
& & \qquad \qquad\qquad\qquad\quad   \mathrm{for}~ n \in \mathbb{N}.
\label{RelUetAo2}
\end{eqnarray}

\subsection{Hadamard form of the DeWitt-Schwinger representation}

The short-distance behavior of the DeWitt-Schwinger representation
$G^{\mathrm{F}}_{\mathrm{DS}}(x,x')$ of the Feynman propagator does
not explicitly appear in its expression given by
Eqs.~(\ref{DSrep1})-(\ref{DSrep2c}). In fact, this behavior is of
the same form as that of the Hadamard representation
$G^{\mathrm{F}}_{\mathrm{H}}(x,x')$. Indeed, it is possible to prove
that the DeWitt-Schwinger representation is a particular case of the
Hadamard one (see Appendix A for details). It corresponds to the
Hadamard representation constructed from the biscalar $W_0(x,x')$
given by
\begin{widetext}
\begin{eqnarray}\label{DSformH_sec_ev}
&  & W_0(x,x') = \left[ \ln (m^2/2) + \gamma -
\psi(d/2) \right]V_0(x,x')  \nonumber \\
& &  \qquad -\frac{1}{2^{d/2-1} (d/2-2)!} \left[\sum_{k=0}^{d/2-2}
\frac{(-1)^k {(m^2)}^k}{k!} \left( \sum_{\ell=k+1}^{d/2-1}
\frac{1}{\ell} \right) A_{d/2-1-k}(x,x')-\sum_{k=0}^{+\infty}
\frac{k!}{{(m^2)}^{k+1}} A_{d/2+k}(x,x') \right]
\end{eqnarray}
for $d$ even and by
\begin{eqnarray}\label{DSform_sec_od}
& & W_0(x,x') =   -\frac{1}{2^{d/2-1} \Gamma(d/2-1)}
\left[\sum_{k=0}^{d/2-3/2} \frac{(-1)^k
{(m^2)}^{k+1/2}}{\Gamma(k+3/2)} \, \pi \, A_{d/2-3/2-k}(x,x') -
\sum_{k=0}^{+\infty} \frac{\Gamma(k+1/2)}{{(m^2)}^{k+1/2}}
A_{d/2-1/2+k}(x,x') \right] \nonumber \\
\end{eqnarray}
for $d$ odd.
\end{widetext}
In Eq.~(\ref{DSformH_sec_ev}), $\psi$ denotes the logarithm
derivative of the gamma function and $\gamma$ is the Euler constant.
The pathological behavior for $m^2 \to 0$ (infra-red divergence) of
this Hadamard coefficient must be noted. Of course, such a behavior
also exists for the DeWitt-Schwinger representation
$G^{\mathrm{F}}_{\mathrm{DS}}(x,x')$ of the Feynman propagator.
Furthermore, it should be also noted that for $d=4$ we recover the
result derived by Brown and Ottewill in Ref.~\cite{BrownOttewill83}.

\section{Covariant Taylor
series expansions of the DeWitt coefficients}

In this section, we shall solve the recursion relations
(\ref{DSrep3}) by looking for their solutions $A_n(x,x')$ with
$n=0,1,2$ and $3$ as covariant Taylor series expansions for $x'$
near $x$ of the form
\begin{equation}\label{CTSExpMasslessA1}
A_n(x,x')=a_n(x) +\sum_{p=1}^{+\infty} \frac{(-1)^p}{p!} a_{n \,
(p)}(x,x')
\end{equation}
where the $a_{n \, (p)}(x,x')$ with $p=1,2, \dots $ are biscalars in
$x$ and $x'$ which are of the form
\begin{eqnarray}
&  & a_{n \, (p)}(x,x')=a_{n \,\, a_1 \dots a_p}(x)
\sigma^{;a_1}(x,x') \dots \sigma^{;a_p}(x,x'). \nonumber \\
\label{CTSExpMasslessA2}
\end{eqnarray}
In fact, we shall first construct the covariant Taylor series
expansions of the mass-dependent DeWitt coefficients
${\tilde{A}}_n(m^2;x,x')$ with $n=0,1,2$ and $3$ defined by
(\ref{MDWCoeff1}). Indeed, from these results, we shall then
immediately obtain the expansions of the DeWitt coefficients
$A_n(x,x')$ with $n=0,1,2$ and $3$ by using (\ref{RelAetA1}) and, in
the next section, we shall be able to easily obtain the expansions
of the corresponding geometrical Hadamard coefficients by using
(\ref{RelUetAp1}) and (\ref{RelUetAo1}). We shall write the
covariant Taylor series expansions of the mass-dependent DeWitt
coefficients in the form
\begin{equation}\label{CTSExpMassdepA1}
{\tilde{A}}_n(m^2;x,x')={\tilde{a}}_n(m^2;x) +\sum_{p=1}^{+\infty}
\frac{(-1)^p}{p!} {\tilde{a}}_{n \, (p)}(m^2;x,x')
\end{equation}
where the ${\tilde{a}}_{n \, (p)}(m^2;x,x')$ with $p=1,2, \dots $
are biscalars in $x$ and $x'$ which are of the form
\begin{eqnarray}
&  & {\tilde{a}}_{n \, (p)}(m^2;x,x')= {\tilde{a}}_{n \,\, a_1 \dots
a_p}(m^2;x)  \nonumber \\
& & \qquad \qquad \qquad \times \sigma^{;a_1}(x,x') \dots
\sigma^{;a_p}(x,x'). \label{CTSExpMassdepA2}
\end{eqnarray}

We shall use the covariant recursive method invented by DeWitt
\cite{DeWittBrehme,DeWitt65} and developed by many others (see
Refs.~\cite{Christensen1,Christensen2,
PhillipsHu03,AndersonFlanaganOttewill05} and references therein).
This method requires preliminarily the knowledge of the covariant
Taylor series expansions of various bitensors such as $\sigma_{;\mu
\nu}$, $g_{\nu \nu'}\sigma_{;\mu}^{\phantom{;\mu} \nu'}$,
$\Delta^{1/2}$, $ \Delta ^{-1/2}{\Delta ^{1/2}}_{;\mu} \sigma^{;
\mu}$, $\Box \Delta^{1/2}$... Here $g_{\mu \nu'}$ denotes the
bivector of parallel transport from $x$ to $x'$ (see
Refs.~\cite{DeWittBrehme,DeWitt65}) which is defined by the partial
differential equation
\begin{subequations}
\begin{equation}\label{Geodetic1a} g_{\mu \nu';
\rho}\sigma^{;\rho}=0
\end{equation}
and the boundary condition
\begin{equation}\label{Geodetic1b}
\lim_{x' \to x} g_{\mu \nu'} = g_{\mu \nu}.
\end{equation}
\end{subequations}
The construction of all the previously mentioned expansions is a
rather hard task. DeWitt has shown that it necessitates the
knowledge of the coincidence limits
\begin{equation}\label{CoincLimits}
   \lim_{x' \to x} \sigma_{;a_1 \dots a_p}.
\end{equation}
They can be obtained by repeatedly differentiating the relation
(\ref{DSrep4}) and can be expressed as complicated sums of terms
involving products of derivatives of the Riemann tensor.
Unfortunately, obtaining the coincidence limits (\ref{CoincLimits})
becomes more and more difficult as the order $p$ increases (see the
discussion on pp.~180-183 of Ref.~\cite{Fulling89}) and is even a
formidable computational challenge (see the ``recent" analysis by
Christensen in Ref.~\cite{DMA4}). In the 1960s, DeWitt derived the
coincidence limits (\ref{CoincLimits}) up to order $p=4$ and the
covariant Taylor series expansion of $\Delta^{1/2}$ up to order
$\sigma$ \cite{DeWittBrehme,DeWitt65} . In the mid-1970s,
Christensen was able to obtain them up to orders $p=6$ and
$\sigma^2$ respectively \cite{Christensen1,Christensen2} and in the
mid-1980s, Brown and Ottewill have slightly improved Christensen's
results by reaching the orders $p=7$ (the corresponding results do
not appear in their article but they appear in a recent article by
Anderson, Flanagan and Ottewill \cite{AndersonFlanaganOttewill05})
and $\sigma^{5/2}$ respectively. It should be also noted that
Phillips and Hu in Ref.~\cite{PhillipsHu03} claim to have reached
the order $p=8$ for the coincidence limits (\ref{CoincLimits}) but
we think that their results are not correct because they lead to
covariant Taylor series expansions of $\sigma_{;\mu \nu}$ and
$\Delta^{1/2}$ up to order $\sigma^3$ which are wrong (see
Appendixes B and C).

Happily, as early as 1986, Avramidi introduced in his PhD thesis
(see Ref.~\cite{Avramidi_PhD} for the English translation and
Ref.~\cite{AvramidiLNP2000} for a revised and expanded version) a
set of new and powerful non-recursive techniques permitting the
construction of the covariant Taylor series expansions of various
bitensors needed in quantum gravity which avoid the preliminary
construction of the coincidence limits (\ref{CoincLimits}). By using
Avramidi's techniques, we have explicitly obtained all the covariant
Taylor series expansions of the bitensors we need in order to solve
the recursion relations (\ref{MDWCoeff1}) up to the orders announced
in Sec.~I. All our results are displayed in Appendixes B, C and E.
In these appendixes, we have also provided the covariant Taylor
series expansions of $\Delta^{1/2}$, $ \Delta ^{-1/2}{\Delta
^{1/2}}_{;\mu} \sigma^{; \mu}$, $\sigma_{;\mu \nu}$ and $g_{\nu
 \nu'}\sigma_{;\mu}^{\phantom{;\mu} \nu'}$ beyond the orders needed here, i.e. up to
orders $\sigma^{11/2}$, $\sigma^{9/2}$, $\sigma^{9/2}$ and
$\sigma^{9/2}$ respectively. Such results show the power of
Avramidi's techniques. In fact, even if we do not need them in the
present article, we think that they could be very useful in a near
future for other people working in the field of gravitational
physics. Furthermore, we shall use them in our next article
\cite{DecaniniFolacci2005b} where we intend to develop the Hadamard
regularization of the stress-energy tensor for a quantized scalar
field in a general spacetime of arbitrary dimension.

In summary, we shall now solve the recursion relations
(\ref{MDWCoeff1}) (and therefore the recursion relations
(\ref{DSrep3})) by combining the old covariant recursive method of
DeWitt with results obtained from the modern covariant non-recursive
techniques developed by Avramidi. In order to simplify our
calculations, we shall in addition use the symmetry of the
mass-dependent DeWitt coefficients ${\tilde{A}}_n(m^2;x,x')$ with $n
\in \mathbb{N}$ in the exchange of $x$ and $x'$. This property
induces constraints on the coefficients ${\tilde{a}}_{n \,
(p)}(m^2;x,x')$ with $p$ odd and, in Appendix D, we have obtained
and displayed various associated results which will be very useful
in this section. In the same appendix, we have also collected
important results concerning the covariant Taylor series expansions
of the covariant derivative, the second covariant derivative and the
d'Alembertian of an arbitrary biscalar.

\subsection{Covariant Taylor series expansion of ${\tilde{A}}_0(m^2;x,x')$}

The  mass-dependent DeWitt coefficient  ${\tilde{A}}_0(m^2;x,x')$ is
equal to $\Delta^{1/2}(x,x')$ (see Eq.~(\ref{MDWCoeff1b})). Its
covariant Taylor series expansion is then given by (see Appendix C
and Eqs.~(\ref{AppSCU01}), (\ref{AppSCU02}) and
(\ref{AppSCU04_2})-(\ref{AppSCU04_6}))
\begin{eqnarray}\label{CTSexpA0_1}
&  & {\tilde{A}}_0 ={\tilde{a}}_0 - {\tilde{a}}_{0 \,\, a}
\sigma^{;a}+\frac{1}{2!} {\tilde{a}}_{0 \,\, a b}
\sigma^{;a}\sigma^{;b}  -\frac{1}{3!} {\tilde{a}}_{0 \,\, a b c}
\sigma^{;a}\sigma^{;b}\sigma^{;c}  \nonumber \\
& & \quad + \frac{1}{4!} {\tilde{a}}_{0 \,\, a b c d}
\sigma^{;a}\sigma^{;b}\sigma^{;c}\sigma^{;d}
  -\frac{1}{5!} {\tilde{a}}_{0 \,\, a b c d
e}\sigma^{;a}\sigma^{;b}\sigma^{;c}\sigma^{;d}\sigma^{;e} \nonumber \\
& & \quad + \frac{1}{6!} {\tilde{a}}_{0 \,\, a b c d e
f}\sigma^{;a}\sigma^{;b}\sigma^{;c}\sigma^{;d}\sigma^{;e}\sigma^{;f}+
O \left(\sigma^{7/2} \right)
\end{eqnarray}
with
\begin{widetext}
\begin{subequations} \label{CTSexpA0_2}
\begin{eqnarray}
& & {\tilde{a}}_0=1 \\
& & {\tilde{a}}_{0 \,\, a}=0 \\
& & {\tilde{a}}_{0 \,\, a b} =
(1/6)  \,  R_{a b}  \\
& & {\tilde{a}}_{0 \,\, a b c} =
(1/4)  \,   R_{(a b; c)}  \\
& & {\tilde{a}}_{0 \,\, a b c d} = (3/10)  \, R_{(a b; c d)} +
(1/15) \, R^{\rho}_{\phantom{\rho}(a |\tau| b}
R^{\tau}_{\phantom{\tau} c |\rho |  d)}
+ (1/12)  R_{(a b} R_{c d)} \\
& &{\tilde{a}}_{0 \,\, a b c d e} =  (1/3) \, R_{(a b; c d e)} +
(1/3) \, R^{\rho}_{\phantom{\rho}(a |\tau| b}
R^{\tau}_{\phantom{\tau}
c |\rho |  d;e)} + (5/12)  R_{(a b} R_{c d; e)}\\
& & {\tilde{a}}_{0 \,\, a b c d e f} = (5/14) \, R_{(a b; c d e f)}
+(4/7) \, R^{\rho}_{\phantom{\rho}(a |\tau| b}
R^{\tau}_{\phantom{\tau} c |\rho |  d;e f)}  + (15/28) \,
R^{\rho}_{\phantom{\rho}(a |\tau| b; c} R^{\tau}_{\phantom{\tau} d
|\rho |  e; f)} + (3/4) \, R_{(a b }R_{c d; e f)} \nonumber
\\
&  & \quad  + (5/8) \, R_{(a b ; c}R_{ d e ; f)}  + (8/63) \,
R^{\rho}_{\phantom{\rho}(a |\tau | b }R^{\tau}_{\phantom{\tau} c
|\sigma | d}R^{\sigma}_{\phantom{\sigma} e |\rho | f)} + (1/6) \,
R_{(a b} R^{\rho}_{\phantom{\rho} c |\tau| d}
R^{\tau}_{\phantom{\tau} e |\rho |  f)} + (5/72) \, R_{(a b}R_{c
d}R_{e f)}
\end{eqnarray}
\end{subequations}
and we can also write
\begin{eqnarray}\label{CTSexpA0_3}
& & {\tilde{A}}_0 = 1 + \frac{1}{12}    R_{a b}
\sigma^{;a}\sigma^{;b}
-\frac{1}{24}     R_{a b; c}  \sigma^{;a}\sigma^{;b} \sigma^{;c} \nonumber \\
& & \quad + \left[ \frac{1}{80}  R_{a b; c d} + \frac{1}{360}
R^{\rho}_{\phantom{\rho} a \tau b} R^{\tau}_{\phantom{\tau} c \rho
d} + \frac{1}{288} R_{a b} R_{c d} \right]
\sigma^{;a}\sigma^{;b} \sigma^{;c} \sigma^{;d}    \nonumber \\
& & \quad - \left[ \frac{1}{360}  R_{a b; c d e} + \frac{1}{360}
R^{\rho}_{\phantom{\rho} a \tau b} R^{\tau}_{\phantom{\tau} c \rho
d;e } + \frac{1}{288} R_{a b} R_{c d; e} \right]
\sigma^{;a}\sigma^{;b} \sigma^{;c} \sigma^{;d} \sigma^{;e} \nonumber \\
& & \quad + \left[ \frac{1}{2016}  R_{a b; c d e f} + \frac{1}{1260}
\, R^{\rho}_{\phantom{\rho} a \tau b} R^{\tau}_{\phantom{\tau} c
\rho  d;e f } + \frac{1}{1344}  R^{\rho}_{\phantom{\rho} a \tau b;
c} R^{\tau}_{\phantom{\tau} d \rho  e; f }  + \frac{1}{960}  R_{a b
}R_{c d; e f}  + \frac{1}{1152}  R_{a b ; c}R_{ d e ; f} \right.
\nonumber
\\
&  & \qquad  \left. + \frac{1}{5670} R^{\rho}_{\phantom{\rho}a \tau
b }R^{\tau}_{\phantom{\tau} c \sigma d}R^{\sigma}_{\phantom{\sigma}
e \rho  f}  + \frac{1}{4320}  R_{a b} R^{\rho}_{\phantom{\rho} c
\tau d} R^{\tau}_{\phantom{\tau} e \rho f }
 + \frac{1}{10368}  R_{a b}R_{c d}R_{e f} \right]
 \sigma^{;a}\sigma^{;b} \sigma^{;c} \sigma^{;d}
 \sigma^{;e}\sigma^{;f}
+ O \left(\sigma^{7/2} \right).
\end{eqnarray}
\end{widetext}

\subsection{Covariant Taylor series expansion of ${\tilde{A}}_1(m^2;x,x')$}

The  mass-dependent DeWitt coefficient  ${\tilde{A}}_1(m^2;x,x')$ is
the solution of Eq.~(\ref{MDWCoeff1a}) with $n=0$, i.e. it satisfies
\begin{eqnarray}\label{CTSexpA1_1}
& & {\tilde{A}}_1 + {\tilde{A}}_{1 ; \mu} \sigma ^{; \mu} -
{\tilde{A}}_1 \Delta ^{-1/2}{\Delta ^{1/2}}_{;\mu} \sigma^{; \mu}
\nonumber \\
& & \qquad \qquad \qquad \qquad= \left( \Box_x - m^2 - \xi R \right)
{\tilde{A}}_0.
\end{eqnarray}
In this equation, we replace ${\tilde{A}}_1$ by its covariant Taylor
series expansion for $x'$ in the neighborhood of $x$ given by
\begin{eqnarray}\label{CTSexpA1_2}
& & {\tilde{A}}_1 ={\tilde{a}}_1 - {\tilde{a}}_{1 \,\, a}
\sigma^{;a}+\frac{1}{2!} {\tilde{a}}_{1 \,\, a b}
\sigma^{;a}\sigma^{;b} -\frac{1}{3!} {\tilde{a}}_{1 \,\, a b c}
\sigma^{;a}\sigma^{;b}\sigma^{;c} \nonumber \\
& & \qquad  \qquad  + \frac{1}{4!} {\tilde{a}}_{1 \,\, a b c d}
\sigma^{;a}\sigma^{;b}\sigma^{;c}\sigma^{;d} + O \left(\sigma^{5/2}
\right).
\end{eqnarray}
By using the covariant Taylor series expansion of $\Delta
^{-1/2}{\Delta ^{1/2}}_{;\mu} \sigma^{; \mu}$ constructed in
Appendix C (see Eq.~(\ref{AppSTU0p4})) as well as  the constraints
induced by the symmetry of ${\tilde{A}}_1(m^2;x,x')$ under the
exchange of $x$ and $x'$ (see Appendix D and
Eqs.~(\ref{AppConstrF_2a}), (\ref{AppConstrF_2b}) and
(\ref{AppFS_cse})), we can easily obtain the covariant Taylor series
expansion of the left-hand side of Eq.~(\ref{CTSexpA1_1}) up to
order $\sigma^2$. The covariant Taylor series expansion of the
right-hand side of Eq.~(\ref{CTSexpA1_1}) up to order $\sigma^2$ can
be found from the expansions of $\Box {\tilde{A}}_0=\Box
\Delta^{1/2}$ and ${\tilde{A}}_0= \Delta^{1/2}$ respectively given
by (\ref{AppVBoxVVMfinal_2})-(\ref{AppBoxVVMfinal_3}) or
(\ref{AppVBoxVVMfinal_4}) and by
(\ref{CTSexpA0_1})-(\ref{CTSexpA0_2}) or (\ref{CTSexpA0_3}). The
direct comparison of the expansions of the left- and right-hand
sides of Eq.~(\ref{CTSexpA1_1}) yields the coefficients
${\tilde{a}}_1$, ${\tilde{a}}_{1 \,\, a}$, ${\tilde{a}}_{1 \,\,
ab}$, ${\tilde{a}}_{1 \,\, abc}$ and ${\tilde{a}}_{1 \,\, abcd}$:
\begin{widetext}
\begin{subequations}\label{CTSexpA1_3}
\begin{eqnarray}
& & {\tilde{a}}_1= -m^2 -(\xi -1/6) \, R  \\
& & {\tilde{a}}_{1 \,\, a}= -(1/2)\, (\xi -1/6) \, R_{;a} \\
& & {\tilde{a}}_{1 \,\, a b} = (1/60)  \,  \Box R_{a b} -(1/3) \,
(\xi-3/20) \, R_{;ab}  -(1/6)\,m^2 \,  R_{a b} \nonumber \\
& & \qquad     -(1/6)\,(\xi -1/6) \, R R_{a b} -(1/45)
\,R^\rho_{\phantom{\rho}a} R_{\rho b} + (1/90) \, R^{\rho
\sigma}R_{\rho a \sigma b}
+ (1/90)  \, R^{\rho \sigma \tau}_{\phantom{\rho \sigma \tau} a }R_{\rho \sigma \tau b} \\
& & {\tilde{a}}_{1 \,\, a b c} = - (1/4 ) \, (\xi -2/15) \, R _{ ;
(a b c)} + (1/40) \, ( \Box R_{(a b} )_{;c)}  - (1/4 ) \, m^2 \, R_{
(a b; c)}  \nonumber \\
& & \qquad     - (1/4 ) \, (\xi -1/6) \, R R_{ (a b; c)}- (1/4 ) \,
(\xi -1/6) \, R_{;(a } R_{ bc)} -
(1/15) \, R^{ \rho }_{\phantom{ \rho} (a } R_{ |\rho| b;c) } \nonumber \\
& & \qquad   + (1/60) \, R^{ \rho  }_{\phantom{ \rho}  \sigma}
R^{\sigma}_{\phantom{\sigma}   (a |\rho | b;  c)}
  + (1/60) \, R^{ \rho  }_{\phantom{ \rho}
 \sigma ;(a } R^{\sigma}_{\phantom{\sigma} b |\rho | c)}
 + (1/30) \, R^{
\rho  \sigma \tau}_{\phantom{ \rho \sigma \tau} (a}
R_{|\rho \sigma \tau|   b;  c)}  \\
& & {\tilde{a}}_{1 \,\, a b c d} = (1/35) \, ( \Box R_{(a b})_{;c
d)} - (1/5) \, (\xi-5/42) \, R _{ ; (a b c d)} - (3/10) \, m^2 \, R
_{
 (a b; c d)} - (3/10) \, (\xi
-1/6) \, R R _{
 (a b; c d)} \nonumber \\
&  & \qquad  -(1/2) \, (\xi-1/6) \, R_{;(a}R_{bc;d)}  -(1/3) \,
(\xi-3/20) \, R_{;(ab}R_{cd)} + (1/60) \, R _{ (a b}\Box R _{ c d)}
- (1/12) \, m^2 \, R _{ (a b}R _{ c d)}\nonumber \\
&  & \qquad   -(3/35) \, R^\rho_{\phantom{ \rho}  (a } R_{ |\rho|
b;c d) }+ (1/105) \, R^{ \rho  }_{\phantom{ \rho}  (a } R_{ b c;
|\rho| d) } - (11/210) \, R^{ \rho  }_{\phantom{ \rho} (a;b } R_{
|\rho|   c;d) } - (3/70) \, R^{ \rho }_{\phantom{ \rho} (a ; b }
R_{ c d );   \rho  } \nonumber \\
&  & \qquad   + (17/840) \, R_{ (a b }^{ \phantom{(a b } ; \rho }R_{
c d ); \rho  }  + (2/105) \, R^{ \rho  }_{\phantom{ \rho}
 \sigma } R^{\sigma}_{\phantom{\sigma} (a |\rho | b; c d)}
 + (1/105) \, R^{ \rho  }_{\phantom{ \rho}
 (a ; |\sigma |} R^{\sigma}_{\phantom{\sigma} b |\rho | c; d)}
+ (1/30) \, R^{ \rho  }_{\phantom{ \rho}
 \sigma ;(a } R^{\sigma}_{\phantom{\sigma} b |\rho | c;
 d)} \nonumber \\
&  & \qquad  - (4/175) \, R^{ \rho  }_{\phantom{ \rho}
 (a ; |\sigma | b} R^{\sigma}_{\phantom{\sigma} c |\rho |
 d)}  + (11/525) \, R_{(a b \phantom{;\rho}
\sigma}^{\phantom{(a b } ; \rho  } R^{\sigma}_{\phantom{\sigma} c
|\rho | d)}    + (11/525) \, R^{ \rho  }_{\phantom{ \rho}
 \sigma ;(a b} R^{\sigma}_{\phantom{\sigma} c |\rho | d)}
 + (4/525) \, R^{\rho}_{\phantom{ \rho} (a |\sigma| b}
 \Box R^{\sigma}_{\phantom{\sigma} c |\rho | d)} \nonumber \\
&  & \qquad   - (1/15) \, m^2 \,
 R^{\rho}_{\phantom{ \rho} (a |\sigma| b}
  R^{\sigma}_{\phantom{\sigma} c |\rho | d)} + (4/105) \,
 R^{\rho \sigma \tau}_{\phantom{\rho \sigma \tau} (a} R_{|\rho \sigma \tau| b; c d)} + (1/140) \,
 R^{\rho \phantom{(a |\sigma| b}    ;\tau}_{\phantom{ \rho} (a |\sigma| b}
R^{\sigma}_{\phantom{\sigma} c |\rho | d) ; \tau}  \nonumber \\
&  & \qquad  + (1/28) \,
 R^{\rho \sigma \tau}_{\phantom{\rho \sigma \tau} (a;b} R_{|\rho \sigma \tau|  c; d)}
- (1/12) \, (\xi -1/6) \, R R _{(a b}R _{ c d)} - (1/45) \, R^{ \rho
}_{\phantom{ \rho}  (a  } R_{ |\rho| b}R_{c d) }  + (1/315) \, R^{
\rho  }_{\phantom{ \rho}
 (a } R_{|\sigma| b}
 R^{\sigma}_{\phantom{\sigma} c |\rho | d)} \nonumber \\
& & \qquad    +  (1/90) \, R^{\rho
 \sigma}R_{(a b}R_{|\rho| c
 |\sigma| d)}
- (1/15) \, (\xi-1/6) \,  R R^{\rho}_{\phantom{\rho} (a |\sigma | b}
 R^{\sigma}_{\phantom{\sigma} c |\rho | d)} + (1/90) \, R_{(a b} R^{ \rho  \sigma
\tau}_{\phantom{ \rho \sigma \tau} c} R_{|\rho \sigma \tau|   d)} \nonumber \\
&  & \qquad   + (26/1575) \, R^{ \rho }_{\phantom{ \rho}
 \sigma } R^{\sigma}_{\phantom{\sigma} (a |\tau | b}
 R^{\tau}_{\phantom{\tau} c |\rho | d)} + (2/63) \, R^{ \rho  }_{\phantom{ \rho}
 (a } R^{\sigma \phantom{b} \tau}_{\phantom{\sigma} b \phantom{ \tau } c}
 R_{|\rho \sigma \tau| d)} + (4/1575) \, R^{\rho \sigma \tau
 \kappa} R_{ \rho  (a |\tau|
 b} R_{|\sigma| c |\kappa|
 d)} \nonumber \\
&  & \qquad   + (4/525) \, R^{\rho  \kappa \tau }_{\phantom{\rho
\kappa \tau } (a} R_{|\rho \tau| \phantom{\sigma}
 b}^{\phantom{ |\rho \tau|} \sigma} R_{|\sigma| c |\kappa|
 d)}  + (16/1575) \, R^{\rho  \kappa \tau }_{\phantom{\rho
\kappa \tau } (a} R_{|\rho|  \phantom{\sigma}
 |\tau| b}^{\phantom{ |\rho| } \sigma} R_{|\sigma| c |\kappa|
 d)}  + (8/1575) \, R^{\rho  \tau \kappa
}_{\phantom{\rho
 \tau  \kappa} (a} R_{|\rho \tau| \phantom{\sigma}
 b}^{\phantom{ |\rho \tau|} \sigma} R_{|\sigma| c |\kappa|
 d)}.
\end{eqnarray}
\end{subequations}
By replacing (\ref{CTSexpA1_3}) into (\ref{CTSexpA1_2}) we can then
write
\begin{eqnarray}\label{CTSexpA1_4}
& & {\tilde{A}}_1 = -m^2 -\left(\xi -\frac{1}{6}\right) R  +
\frac{1}{2}
\left(\xi -\frac{1}{6}\right) R_{;a}\sigma^{;a} \nonumber \\
& & \quad + \left[\frac{1}{120}  \Box R_{a b} -\frac{1}{6}
\left(\xi- \frac{3}{20} \right)  R_{;ab} - \frac{1}{12}  m^2 \,
R_{a b} \right. \nonumber \\
& & \quad   \left. -\frac{1}{12} \left(\xi -\frac{1}{6}\right) R
R_{a b} -\frac{1}{90} \,R^\rho_{\phantom{\rho}a} R_{\rho b} +
\frac{1}{180} R^{\rho \sigma}R_{\rho a \sigma b} + \frac{1}{180}
R^{\rho \sigma \tau}_{\phantom{\rho \sigma \tau} a }R_{\rho \sigma
\tau b} \right]
\sigma^{;a}\sigma^{;b} \nonumber \\
& & \quad + \left[ \frac{1}{24}  \left(\xi - \frac{2}{15} \right) R
_{ ; a b c} - \frac{1}{240}  ( \Box R_{a b} )_{;c} + \frac{1}{24}
m^2 \, R_{ a b; c} + \frac{1}{24}  \left(\xi -\frac{1}{6}\right) R
R_{ a b; c} + \frac{1}{24}  \left(\xi -\frac{1}{6}\right)
R_{;a } R_{ bc} \right. \nonumber \\
& & \quad   \left. + \frac{1}{90}  R^{ \rho }_{\phantom{ \rho} a }
R_{ \rho b;c } -\frac{1}{360}  R^{ \rho }_{\phantom{ \rho} \sigma}
R^{\sigma}_{\phantom{\sigma}   a \rho  b; c}
 - \frac{1}{360}  R^{ \rho  }_{\phantom{
\rho}
 \sigma ;a } R^{\sigma}_{\phantom{\sigma} b \rho  c}
 - \frac{1}{180}  R^{
\rho  \sigma \tau}_{\phantom{ \rho \sigma \tau} a} R_{\rho \sigma
\tau   b;  c} \right] \sigma^{;a}\sigma^{;b} \sigma^{;c} \nonumber
\\
& & \quad + \left[  \frac{1}{840} ( \Box R_{a b})_{;c d} -
\frac{1}{120} \left(\xi-\frac{5}{42}\right) R _{ ; a b c d} -
\frac{1}{80} m^2 \, R _{
 a b; c d} -
\frac{1}{80} \left(\xi-\frac{1}{6}\right) R R _{
 a b; c d} - \frac{1}{48}  \left(\xi-\frac{1}{6}\right) R_{;a}R_{bc;d}  \right. \nonumber \\
& & \quad  \left. -\frac{1}{72} \left(\xi-\frac{3}{20}\right)
R_{;ab}R_{cd}  + \frac{1}{1440} R _{ a b}\Box R _{ c d} -
\frac{1}{288} m^2 \, R _{a b}R _{ c d}
 -\frac{1}{280} R^\rho_{\phantom{ \rho} a } R_{ \rho b;c d }+
\frac{1}{2520} R^{ \rho }_{\phantom{ \rho}  a } R_{ b c; \rho d } -
\frac{11}{5040} R^{ \rho }_{\phantom{ \rho} a;b } R_{ \rho   c;d }  \right. \nonumber \\
& & \quad  \left. - \frac{1}{560} R^{ \rho }_{\phantom{ \rho} a ; b
} R_{ c d ; \rho  }
 + \frac{17}{20160} R_{ a b }^{ \phantom{a
b } ; \rho }R_{ c d; \rho  }   + \frac{1}{1260}  R^{ \rho
}_{\phantom{ \rho}
 \sigma } R^{\sigma}_{\phantom{\sigma} a \rho  b; c d}
 + \frac{1}{2520} R^{ \rho }_{\phantom{
\rho}
 a ; \sigma } R^{\sigma}_{\phantom{\sigma} b \rho  c; d}
  + \frac{1}{720} R^{ \rho  }_{\phantom{ \rho}
 \sigma ;a } R^{\sigma}_{\phantom{\sigma} b \rho  c;
 d}\right. \nonumber \\
& & \quad  \left. - \frac{1}{1050} R^{ \rho  }_{\phantom{ \rho}
 a ; \sigma  b} R^{\sigma}_{\phantom{\sigma} c \rho
 d}  + \frac{11}{12600} R_{a b \phantom{;\rho}
\sigma}^{\phantom{a b } ; \rho  } R^{\sigma}_{\phantom{\sigma} c
\rho  d}   + \frac{11}{12600} R^{ \rho }_{\phantom{ \rho}
 \sigma ;a b} R^{\sigma}_{\phantom{\sigma} c \rho  d}
 + \frac{1}{3150} R^{\rho}_{\phantom{ \rho} a \sigma b}
 \Box R^{\sigma}_{\phantom{\sigma} c \rho  d}
 - \frac{1}{360} m^2 \, R^{\rho}_{\phantom{\rho} a
\sigma  b}
 R^{\sigma}_{\phantom{\sigma} c \rho  d} \right. \nonumber \\
& & \quad  \left.  + \frac{1}{630}
 R^{\rho \sigma \tau}_{\phantom{\rho \sigma \tau} a} R_{\rho \sigma \tau b; c d} + \frac{1}{3360}
 R^{\rho \phantom{a \sigma b}    ;\tau}_{\phantom{ \rho} a \sigma b}
R^{\sigma}_{\phantom{\sigma} c \rho  d ; \tau}  + \frac{1}{672}
 R^{\rho \sigma \tau}_{\phantom{\rho \sigma \tau} a;b} R_{\rho \sigma \tau  c; d}
 - \frac{1}{288} \left(\xi-\frac{1}{6}\right)  R
R _{a b}R _{ c d} - \frac{1}{1080} R^{ \rho }_{\phantom{ \rho}  a  }
R_{ \rho b}R_{c d } \right. \nonumber \\
& & \quad   \left. + \frac{1}{7560} R^{ \rho }_{\phantom{ \rho}
 a } R_{\sigma b}
 R^{\sigma}_{\phantom{\sigma} c \rho  d}  +  \frac{1}{2160} R^{\rho
 \sigma}R_{a b}R_{\rho c
 \sigma d}  - \frac{1}{360} \left(\xi-\frac{1}{6}\right)   R
R^{\rho}_{\phantom{\rho} a \sigma b}
 R^{\sigma}_{\phantom{\sigma} c \rho  d} + \frac{1}{2160} R_{a b} R^{ \rho \sigma
\tau}_{\phantom{ \rho \sigma \tau} c} R_{\rho \sigma \tau d} \right. \nonumber \\
& & \quad   \left. + \frac{13}{18900} R^{ \rho }_{\phantom{ \rho}
 \sigma } R^{\sigma}_{\phantom{\sigma} a \tau  b}
 R^{\tau}_{\phantom{\tau} c \rho  d}  + \frac{1}{756} R^{ \rho  }_{\phantom{ \rho}
 a } R^{\sigma \phantom{b} \tau}_{\phantom{\sigma} b \phantom{ \tau } c}
 R_{\rho \sigma \tau d} + \frac{1}{9450} R^{\rho \sigma \tau
 \kappa} R_{ \rho  a \tau
 b} R_{\sigma c \kappa
 d} + \frac{1}{3150}  R^{\rho  \kappa \tau }_{\phantom{\rho
\kappa \tau } a} R_{\rho \tau \phantom{\sigma}
 b}^{\phantom{ \rho \tau} \sigma} R_{\sigma c \kappa
 d} \right. \nonumber \\
 & & \quad   \left.  + \frac{2}{4725} R^{\rho  \kappa \tau }_{\phantom{\rho
\kappa \tau } a} R_{\rho  \phantom{\sigma}
 \tau b}^{\phantom{ \rho } \sigma} R_{\sigma c \kappa
 d}  + \frac{1}{4725}  R^{\rho  \tau \kappa
}_{\phantom{\rho
 \tau  \kappa} a} R_{\rho \tau \phantom{\sigma}
 b}^{\phantom{ \rho \tau} \sigma} R_{\sigma c \kappa
 d}\right]
 \sigma^{;a}\sigma^{;b} \sigma^{;c} \sigma^{;d}
+ O \left(\sigma^{5/2} \right).
\end{eqnarray}
\end{widetext}

\subsection{Covariant Taylor series expansion of ${\tilde{A}}_2(m^2;x,x')$}

The  mass-dependent DeWitt coefficient  ${\tilde{A}}_2(m^2;x,x')$ is
the solution of Eq.~(\ref{MDWCoeff1a}) with $n=1$, i.e. it satisfies
\begin{eqnarray}\label{CTSexpA2_1}
& & 2{\tilde{A}}_2 + {\tilde{A}}_{2 ; \mu} \sigma ^{; \mu} -
{\tilde{A}}_2 \Delta ^{-1/2}{\Delta
^{1/2}}_{;\mu} \sigma^{; \mu}   \nonumber \\
& & \qquad \qquad \qquad \qquad = \left( \Box_x - m^2 -\xi R \right)
{\tilde{A}}_1.
\end{eqnarray}
In this equation, we replace ${\tilde{A}}_2$ by its covariant Taylor
series expansion for $x'$ in the neighborhood of $x$ given by
\begin{equation}\label{CTSexpA2_2}
{\tilde{A}}_2 ={\tilde{a}}_2 - {\tilde{a}}_{2 \,\, a}
\sigma^{;a}+\frac{1}{2!} {\tilde{a}}_{2 \,\, a b}
\sigma^{;a}\sigma^{;b}  + O \left(\sigma^{3/2} \right).
\end{equation}
By using the covariant Taylor series expansion of $\Delta
^{-1/2}{\Delta ^{1/2}}_{;\mu} \sigma^{; \mu}$ given in
Eq.~(\ref{AppSTU0p4}) as well as  the constraints induced by the
symmetry of ${\tilde{A}}_2(x,x')$ under the exchange of $x$ and $x'$
(see Appendix D and Eqs.~(\ref{AppConstrF_2a}) and
(\ref{AppFS_cse})), we can easily obtain the covariant Taylor series
expansion of the left-hand side of Eq.~(\ref{CTSexpA2_1}) up to
order $\sigma$. The covariant Taylor series expansion of the
right-hand side of Eq.~(\ref{CTSexpA2_1}) up to order $\sigma$ can
be found from the expansion ${\tilde{A}}_1$ given by
(\ref{CTSexpA1_2})-(\ref{CTSexpA1_3}) or (\ref{CTSexpA1_4}) and from
the expansion of $\Box {\tilde{A}}_1$. The later can be constructed
by using the theory developed in Appendix D and more particularly
Eqs.~(\ref{CTSExpBoxF1}), (\ref{CTSExpBoxF2}) and
(\ref{AppBoxFsymm}). After the most tedious calculation of this
article using extensively the commutation of covariant derivatives
in the form (\ref{CD_NabNabTensor}) as well as the Bianchi
identities (\ref{AppBianchi_1}) and their consequences
(\ref{AppBianchi_2})-(\ref{AppBianchi_4}), we obtain
\begin{equation}\label{CTSexpA2_3}
\Box {\tilde{A}}_1 ={\tilde{a}}''_1 - {\tilde{a}}''_{1 \,\, a}
\sigma^{;a}+\frac{1}{2!} {\tilde{a}}''_{1 \,\, a b}
\sigma^{;a}\sigma^{;b}  + O \left(\sigma^{3/2} \right)
\end{equation}
with
\begin{widetext}
\begin{subequations}\label{CTSexpA2_4}
\begin{eqnarray}
& & {\tilde{a}}''_1= -(1/3) \, (\xi -1/5) \, \Box R - (1/6) \, m^2
\, R - (1/6) \, (\xi -1/6) \, R^2 - (1/90) \, R_{\rho \sigma}R^{\rho
\sigma}
+  (1/90) \, R_{\rho \sigma \tau \kappa}R^{\rho \sigma  \tau \kappa}\\
& & {\tilde{a}}''_{1 \,\, a}= -(1/12) \, (\xi -1/5) \, (\Box R)_{;a}
- (1/12) \, (\xi -1/6) \, RR_{;a} - (1/180) \, R_{\rho
\sigma}R^{\rho \sigma}_{\phantom{\rho \sigma};a} +  (1/180) \,
R_{\rho \sigma \tau
\kappa}R^{\rho \sigma  \tau \kappa}_{\phantom{\rho \sigma \tau \kappa};a} \\
& & {\tilde{a}}''_{1 \,\, a b} = (1/210)  \,  \Box \Box R_{a b}
-(1/30) \, (\xi-1/7) \, (\Box R)_{;ab} - (1/20) \, m^2 \, \Box  R_{a
b} + (1/60) \, m^2 \,  R_{;a b} -(7/180)\,(\xi -1/7) \, R R_{;a b}
 \nonumber \\
& & \qquad -(2/45) \, (\xi -1/7) \, R_{;\rho (a}
R^\rho_{\phantom{\rho}b)}
 -(1/18) \,(\xi -1/5) \,  (\Box R) R_{a b} +(2/15) \, (\xi
-3/14) \, R_{;\rho } R^\rho_{\phantom{\rho}(a;b)}\nonumber \\
& & \qquad  - (2/15) \, (\xi -17/84) \, R_{;\rho }
R_{ab}^{\phantom{ab}; \rho} -(1/20)\,(\xi -2/9) \,  R \Box R_{a b}
-(1/36)\,m^2 \,  R R_{a b} -(1/63) \, R_{\rho (a} \Box
R^\rho_{\phantom{\rho}b)}\nonumber \\
& & \qquad + (1/15) \, m^2 \, R_{\rho a} R^\rho_{\phantom{\rho}b}
-(1/350) \, R^{\rho \sigma} R_{\rho \sigma;(a b)} -(2/525) \,
R^{\rho \sigma} R_{\rho (a ; b)\sigma} + (8/1575) \, R^{\rho \sigma}
R_{ab; \rho \sigma}  + (1/315) \, R^\rho_{\phantom{\rho} a;\sigma}
R_{\rho b}^{\phantom{\rho b};\sigma} \nonumber \\
& & \qquad  - (1/63) \, R^\rho_{\phantom{\rho} a;\sigma}
R^\sigma_{\phantom{\sigma} b;\rho }- (4/45) \, (\xi -3/14) \,
R^{;\rho \sigma}
 R_{\rho a \sigma b} + (2/315) \, (\Box R^{\rho \sigma})R_{\rho a \sigma b} -
(1/30) \, m^2 \, R^{\rho \sigma} R_{\rho a \sigma b}  \nonumber \\
& & \qquad + (2/315) \, R^{\rho \sigma;\tau} R_{\tau \sigma \rho
(a;b)}  + (1/225) \, R^{\rho \sigma}\Box R_{\rho a \sigma b}  +
(1/105) \, R^{\rho \sigma;\tau} R_{\rho a \sigma b;\tau} - (8/1575)
\, R^{\rho \sigma;\tau}_{\phantom{\rho \sigma;\tau} (a} R_{|\tau
\sigma \rho| b)}  \nonumber \\
& &  \qquad  + (23/1575) \, R^{\rho \phantom{(a}; \sigma
\tau}_{\phantom{\rho } (a} R_{|\tau \sigma \rho| b)} - (4/225) \,
R^{\rho \phantom{(a}; \sigma \tau}_{\phantom{\rho } (a} R_{|\rho
\sigma \tau | b)} -(2/175) \, R^{\rho \sigma \tau \kappa} R_{\rho
\sigma \tau (a;b) \kappa}  + (16/1575) \, R^{\rho \sigma
\tau}_{\phantom{\rho \sigma \tau}a} \Box R_{\rho \sigma \tau b}\nonumber \\
& &  \qquad - (1/30) \, m^2 \, R^{\rho \sigma \tau}_{\phantom{\rho
\sigma \tau}a} R_{\rho \sigma \tau b}+ (23/3150) \, R^{\rho \sigma
\tau \kappa} R_{\rho \sigma \tau \kappa ; (a b) }  + (1/105) \,
R^{\rho \sigma \tau}_{\phantom{\rho \sigma \tau}a;\kappa}R_{\rho
\sigma \tau b}^{\phantom{\rho \sigma \tau b};\kappa} + (1/1260) \,
R^{\rho \sigma \tau \kappa}_{\phantom{\rho \sigma \tau \kappa} ;a}
R_{\rho \sigma \tau \kappa ; b} \nonumber \\
& &  \qquad -(1/36) \, (\xi -1/6) \, R^2 R_{ab} + (1/15) \, (\xi
-2/9) \, R R_{\rho a} R^\rho_{\phantom{\rho}b}-(1/540) \, R^{\rho
\sigma}R_{\rho \sigma}R_{ab}+ (4/945) \, R^{\rho \sigma}R_{\rho
a}R_{\sigma b} \nonumber \\
& &  \qquad - (1/30) \, (\xi -2/9) \, R R^{\rho \sigma}R_{\rho a
\sigma b} -(2/945) \, R^{\rho \tau}R^\sigma_{\phantom{\sigma}
\tau}R_{\rho a \sigma b} + (32/4725) \, R^{\rho
\sigma}R^\tau_{\phantom{\tau} (a}R_{|\tau \sigma \rho| b)}  \nonumber \\
& &  \qquad + (2/4725) \, R_{\rho \sigma}R^{\rho \kappa \sigma
\lambda}R_{\kappa a \lambda b} -(1/30) \, (\xi -2/9) \, R R^{\rho
\sigma \tau}_{\phantom{\rho \sigma \tau}a }R_{\rho \sigma \tau b} +
(1/540) \, R_{ab}R^{\rho \sigma \tau \kappa} R_{\rho \sigma \tau
\kappa } \nonumber \\
& &  \qquad  + (31/4725) \, R_{\rho \sigma}R^{\rho \kappa
\lambda}_{\phantom{\rho \kappa \lambda}a}R^\sigma_{\phantom{\sigma}
\kappa \lambda b} - (1/75) \, R_{\rho \sigma}R^{\rho \kappa
\lambda}_{\phantom{\rho \kappa \lambda}a}R^\sigma_{\phantom{\sigma}
\lambda \kappa b} + (17/4725) \, R^{\rho \sigma}R^{\kappa
\lambda}_{\phantom{\kappa \lambda} \rho a}R_{\kappa \lambda \sigma
b} \nonumber \\
& &  \qquad - (17/1890) \, R^\kappa_{\phantom{\kappa} (a}R^{\rho
\sigma \tau}_{\phantom{\rho \sigma \tau} |\kappa| }R_{|\rho \sigma
\tau| b)} -(34/4725) \, R^{\rho \sigma \tau}_{\phantom{\rho \sigma
\tau} \lambda } R_{\rho \sigma \tau \kappa}R^{\lambda \phantom{a}
\kappa}_{\phantom{\lambda} a  \phantom{\kappa} b} + (4/189) \,
R^{\rho \kappa \sigma \lambda}R^\tau_{\phantom{\tau} \rho \sigma
a}R_{\tau \kappa \lambda b} \nonumber \\
& & \qquad - (2/225) \, R^{\rho \kappa \sigma \lambda}R_{\rho \sigma
\tau a }R_{\kappa \lambda \phantom{\tau} b}^{\phantom{\kappa
\lambda} \tau} + (76/4725) \, R^{\rho \sigma \kappa \lambda}R_{\rho
\sigma \tau a }R_{\kappa \lambda \phantom{\tau} b}^{\phantom{\kappa
\lambda} \tau}.
\end{eqnarray}
\end{subequations}
\end{widetext}
The direct comparison of the expansions of the left- and right-hand
sides of Eq.~(\ref{CTSexpA2_1}) then yields the coefficients
${\tilde{a}}_2$, ${\tilde{a}}_{2 \,\, a}$ and ${\tilde{a}}_{2 \,\,
ab}$. We have
\begin{widetext}
\begin{subequations}\label{CTSexpA2_5}
\begin{eqnarray}
& & {\tilde{a}}_2= (1/2) \, m^4 -(1/6) \, (\xi -1/5) \, \Box R +
(\xi -1/6) \, m^2 \, R \nonumber \\
& & \qquad + (1/2) \, (\xi -1/6)^2 \, R^2 - (1/180) \, R_{\rho
\sigma}R^{\rho \sigma}
+  (1/180) \, R_{\rho \sigma \tau \kappa}R^{\rho \sigma  \tau \kappa}\\
& & {\tilde{a}}_{2 \,\, a}= -(1/12) \, (\xi -1/5) \, (\Box R)_{;a}
+(1/2) \, (\xi -1/6) \, m^2 \, R_{;a} \nonumber \\
& & \qquad  + (1/2) \, (\xi -1/6)^2 \, RR_{;a} - (1/180) \, R_{\rho
\sigma}R^{\rho \sigma}_{\phantom{\rho \sigma};a} +  (1/180) \,
R_{\rho \sigma \tau \kappa}R^{\rho \sigma  \tau
\kappa}_{\phantom{\rho \sigma \tau \kappa};a}
\end{eqnarray}
and
\begin{eqnarray}
& & {\tilde{a}}_{2 \,\, a b} = (1/840)  \,  \Box \Box R_{a b}
-(1/20) \, (\xi-4/21) \, (\Box R)_{;ab} - (1/60) \, m^2 \, \Box R_{a
b}
 + (1/3) \, (\xi-3/20) \, m^2 \, R_{;a b} + (1/12) \, m^4 \,  R_{a b}
  \nonumber \\
& & \quad + (1/3) \, (\xi -1/6)(\xi-3/20) \,  R R_{;a b}
   -(1/90) \, (\xi -1/7)
\, R_{;\rho (a} R^\rho_{\phantom{\rho}b)}
 -(1/36) \,(\xi -1/5) \,  (\Box R) R_{a b}  \nonumber \\
& & \quad+ (1/4) (\xi -1/6)^2 \, R_{;a} R_{;b} +(1/30) \, (\xi
-3/14) \, R_{;\rho } R^\rho_{\phantom{\rho}(a;b)} - (1/30) \, (\xi
-17/84) \, R_{;\rho } R_{ab}^{\phantom{ab}; \rho} +(1/6)\, (\xi-1/6)
\, m^2 \,  R R_{a b} \nonumber \\
& & \quad  -(1/60)\,(\xi -1/6) \,  R \Box R_{a b} -(1/252) \,
R_{\rho (a} \Box R^\rho_{\phantom{\rho}b)} + (1/45) \, m^2 R_{\rho
a} R^\rho_{\phantom{\rho}b} -(11/3150) \, R^{\rho \sigma} R_{\rho
\sigma;(a b)} - (1/360) \, R^{\rho \sigma}_{\phantom{\rho
\sigma};a}R_{\rho \sigma;b}\nonumber \\
& & \quad  -(1/1050) \, R^{\rho \sigma} R_{\rho (a ; b)\sigma} +
(2/1575) \, R^{\rho \sigma} R_{ab; \rho \sigma}  + (1/1260) \,
R^\rho_{\phantom{\rho} a;\sigma} R_{\rho b}^{\phantom{\rho
b};\sigma} - (1/252) \, R^\rho_{\phantom{\rho} a;\sigma}
R^\sigma_{\phantom{\sigma} b;\rho } \nonumber \\
& & \quad - (1/45) \, (\xi -3/14) \, R^{;\rho \sigma}R_{\rho a
\sigma b}+ (1/630) \, (\Box R^{\rho \sigma})R_{\rho a \sigma b} -
(1/90) \, m^2 \, R^{\rho \sigma} R_{\rho a \sigma b} + (1/630) \,
R^{\rho \sigma;\tau} R_{\tau \sigma \rho (a;b)}  \nonumber \\
& & \quad + (1/900) \, R^{\rho \sigma}\Box R_{\rho a \sigma b} +
(1/420) \, R^{\rho \sigma;\tau} R_{\rho a \sigma b;\tau}  - (2/1575)
\, R^{\rho \sigma;\tau}_{\phantom{\rho \sigma;\tau} (a} R_{|\tau
\sigma \rho| b)} + (23/6300) \, R^{\rho \phantom{(a}; \sigma
\tau}_{\phantom{\rho } (a} R_{|\tau \sigma \rho| b)} \nonumber \\
& & \quad - (1/225) \, R^{\rho \phantom{(a}; \sigma
\tau}_{\phantom{\rho } (a} R_{|\rho \sigma \tau | b)} -(1/350) \,
R^{\rho \sigma \tau \kappa} R_{\rho \sigma \tau (a;b) \kappa} +
(4/1575) \, R^{\rho \sigma \tau}_{\phantom{\rho \sigma \tau}a} \Box
R_{\rho \sigma \tau b} - (1/90) \, m^2 \, R^{\rho \sigma
\tau}_{\phantom{\rho \sigma \tau}a} R_{\rho \sigma \tau b}  \nonumber \\
& &  \quad + (29/6300) \, R^{\rho \sigma \tau \kappa} R_{\rho \sigma
\tau \kappa ; (a b) } + (1/420) \, R^{\rho \sigma
\tau}_{\phantom{\rho \sigma \tau}a;\kappa}R_{\rho \sigma \tau
b}^{\phantom{\rho \sigma \tau b};\kappa} + (1/336) \, R^{\rho \sigma
\tau \kappa}_{\phantom{\rho \sigma \tau \kappa};a} R_{\rho \sigma
\tau \kappa ; b } + (1/12) \, (\xi -1/6)^2 \, R^2 R_{ab}\nonumber \\
& &  \quad  + (1/45) \, (\xi -1/6) \, R R_{\rho a}
R^\rho_{\phantom{\rho}b}  -(1/1080) \, R^{\rho \sigma}R_{\rho
\sigma}R_{ab} + (1/945) \, R^{\rho \sigma}R_{\rho a}R_{\sigma b}  -
(1/90) \, (\xi -1/6) \, R R^{\rho \sigma}R_{\rho a \sigma b} \nonumber \\
& &  \quad -(1/1890) \, R^{\rho \tau}R^\sigma_{\phantom{\sigma}
\tau}R_{\rho a \sigma b}  + (8/4725) \, R^{\rho
\sigma}R^\tau_{\phantom{\tau} (a}R_{|\tau \sigma \rho| b)} +
(1/9450) \, R_{\rho \sigma}R^{\rho \kappa \sigma \lambda}R_{\kappa a
\lambda b} -(1/90) \, (\xi -1/6) \, R R^{\rho
\sigma \tau}_{\phantom{\rho \sigma \tau}a }R_{\rho \sigma \tau b} \nonumber \\
& &  \quad + (1/1080) \, R_{ab}R^{\rho \sigma \tau \kappa} R_{\rho
\sigma \tau \kappa }   + (31/18900) \, R_{\rho \sigma}R^{\rho \kappa
\lambda}_{\phantom{\rho \kappa \lambda}a}R^\sigma_{\phantom{\sigma}
\kappa \lambda b} - (1/300) \, R_{\rho \sigma}R^{\rho \kappa
\lambda}_{\phantom{\rho \kappa \lambda}a}R^\sigma_{\phantom{\sigma}
\lambda \kappa b}  \nonumber \\
& &  \quad + (17/18900) \, R^{\rho \sigma}R^{\kappa
\lambda}_{\phantom{\kappa \lambda} \rho a}R_{\kappa \lambda \sigma
b} - (17/7560) \, R^\kappa_{\phantom{\kappa} (a}R^{\rho \sigma
\tau}_{\phantom{\rho \sigma \tau} |\kappa| }R_{|\rho \sigma \tau|
b)} -(17/9450) \, R^{\rho \sigma \tau}_{\phantom{\rho \sigma \tau}
\lambda } R_{\rho \sigma \tau \kappa}R^{\lambda \phantom{a}
\kappa}_{\phantom{\lambda} a  \phantom{\kappa} b} \nonumber \\
& & \quad  + (1/189) \, R^{\rho \kappa \sigma
\lambda}R^\tau_{\phantom{\tau} \rho \sigma a}R_{\tau \kappa \lambda
b} - (1/450) \, R^{\rho \kappa \sigma \lambda}R_{\rho \sigma \tau a
}R_{\kappa \lambda \phantom{\tau} b}^{\phantom{\kappa \lambda} \tau}
+ (19/4725) \, R^{\rho \sigma \kappa \lambda}R_{\rho \sigma \tau a
}R_{\kappa \lambda \phantom{\tau} b}^{\phantom{\kappa \lambda} \tau}
.
\end{eqnarray}
\end{subequations}
By replacing (\ref{CTSexpA2_5}) into (\ref{CTSexpA2_2}) we can then
write
\begin{eqnarray}\label{CTSexpA2_6}
& & {\tilde{A}}_2 =  \frac{1}{2}  m^4  -\frac{1}{6} \left(\xi -
\frac{1}{5} \right) \Box R + \left(\xi -\frac{1}{6}\right) m^2 \, R
+ \frac{1}{2}   \left(\xi -\frac{1}{6}\right)^2  R^2 - \frac{1}{180}
R_{\rho \sigma}R^{\rho \sigma} + \frac{1}{180}
R_{\rho \sigma \tau \kappa}R^{\rho \sigma  \tau \kappa} \nonumber \\
& & \quad  +  \left[ \frac{1}{12}   \left(\xi - \frac{1}{5} \right)
 (\Box R)_{;a}  - \frac{1}{2} \left(\xi -\frac{1}{6}\right) m^2 \, R_{;a} - \frac{1}{2} \left(\xi -\frac{1}{6}\right)^2   RR_{;a}
 + \frac{1}{180}
R_{\rho \sigma}R^{\rho \sigma}_{\phantom{\rho \sigma};a} -
\frac{1}{180} R_{\rho \sigma \tau \kappa}R^{\rho \sigma  \tau
\kappa}_{\phantom{\rho \sigma \tau \kappa};a} \right]
\sigma^{;a} \nonumber \\
& & \quad + \left[ \frac{1}{1680} \Box \Box R_{a b} -\frac{1}{40}
\left(\xi-\frac{4}{21}\right)  (\Box R)_{;ab}  - \frac{1}{120} m^2
\, \Box R_{a b} + \frac{1}{6} \left(\xi-\frac{3}{20}\right) m^2 \,
R_{;a b} + \frac{1}{24} m^4 \, R_{ab} \right. \nonumber \\
& & \qquad \left.  + \frac{1}{6} \left(\xi
-\frac{1}{6}\right)\left(\xi-\frac{3}{20}\right)  R R_{;a b}
-\frac{1}{180} \left(\xi -\frac{1}{7}\right) R_{;\rho a}
R^\rho_{\phantom{\rho}b}
 -\frac{1}{72} \left(\xi -\frac{1}{5}\right)  (\Box R) R_{a b} +
\frac{1}{8} \left(\xi -\frac{1}{6}\right)^2 \, R_{;a} R_{;b} \right. \nonumber \\
& & \qquad \left. +\frac{1}{60} \left(\xi -\frac{3}{14}\right)
R_{;\rho } R^\rho_{\phantom{\rho}a;b}  - \frac{1}{60}  \left(\xi
-\frac{17}{84}\right) R_{;\rho } R_{ab}^{\phantom{ab}; \rho}
+\frac{1}{12} \left(\xi-\frac{1}{6}\right)  m^2 \,  R R_{a b}
-\frac{1}{120} \left(\xi -\frac{1}{6}\right) \,  R \Box R_{a b} \right. \nonumber \\
& & \qquad \left. -\frac{1}{504} R_{\rho a} \Box
R^\rho_{\phantom{\rho}b} + \frac{1}{90}  m^2 \, R_{\rho a}
R^\rho_{\phantom{\rho}b} -\frac{11}{6300} R^{\rho \sigma} R_{\rho
\sigma;a b}  - \frac{1}{720} R^{\rho \sigma}_{\phantom{\rho
\sigma};a}R_{\rho \sigma;b} -\frac{1}{2100} R^{\rho \sigma} R_{\rho
a ; b\sigma} + \frac{1}{1575} R^{\rho \sigma} R_{ab; \rho \sigma} \right. \nonumber \\
& & \qquad \left. + \frac{1}{2520} R^\rho_{\phantom{\rho} a;\sigma}
R_{\rho b}^{\phantom{\rho b};\sigma}  - \frac{1}{504}
R^\rho_{\phantom{\rho} a;\sigma} R^\sigma_{\phantom{\sigma} b;\rho }
- \frac{1}{90} \left(\xi -\frac{3}{14} \right) \, R^{;\rho
\sigma}R_{\rho a \sigma b}+ \frac{1}{1260} (\Box R^{\rho
\sigma})R_{\rho a \sigma b} -
\frac{1}{180} m^2 \, R^{\rho \sigma} R_{\rho a \sigma b} \right. \nonumber \\
& & \qquad \left. + \frac{1}{1260} R^{\rho \sigma;\tau} R_{\tau
\sigma \rho a;b} +\frac{1}{1800} R^{\rho \sigma}\Box R_{\rho a
\sigma b} + \frac{1}{840} R^{\rho \sigma;\tau} R_{\rho a \sigma
b;\tau} - \frac{1}{1575} R^{\rho \sigma;\tau}_{\phantom{\rho
\sigma;\tau} a} R_{\tau \sigma \rho b} + \frac{23}{12600} R^{\rho
\phantom{a}; \sigma \tau}_{\phantom{\rho }
a} R_{\tau \sigma \rho b} \right. \nonumber \\
& & \qquad \left. - \frac{1}{450} R^{\rho \phantom{a}; \sigma
\tau}_{\phantom{\rho } a} R_{\rho \sigma \tau  b} -\frac{1}{700}
R^{\rho \sigma \tau \kappa} R_{\rho \sigma \tau a;b \kappa} +
\frac{2}{1575} R^{\rho \sigma \tau}_{\phantom{\rho \sigma \tau}a}
\Box R_{\rho \sigma \tau b} - \frac{1}{180} m^2 \, R^{\rho \sigma
\tau}_{\phantom{\rho \sigma \tau}a} R_{\rho \sigma \tau b} +
\frac{29}{12600} R^{\rho \sigma \tau \kappa}
R_{\rho \sigma \tau \kappa ; a b } \right. \nonumber \\
& & \qquad \left. + \frac{1}{840} R^{\rho \sigma
\tau}_{\phantom{\rho \sigma \tau}a;\kappa}R_{\rho \sigma \tau
b}^{\phantom{\rho \sigma \tau b};\kappa} + \frac{1}{672} R^{\rho
\sigma \tau \kappa}_{\phantom{\rho \sigma \tau \kappa};a} R_{\rho
\sigma \tau \kappa ; b } + \frac{1}{24} \left(\xi
-\frac{1}{6}\right)^2 \, R^2 R_{ab} + \frac{1}{90} \left(\xi
-\frac{1}{6}\right)  R
R_{\rho a} R^\rho_{\phantom{\rho}b} \right. \nonumber \\
& & \qquad \left. -\frac{1}{2160} R^{\rho \sigma}R_{\rho
\sigma}R_{ab} + \frac{1}{1890} R^{\rho \sigma}R_{\rho a}R_{\sigma b}
- \frac{1}{180} \left(\xi -\frac{1}{6}\right) \, R R^{\rho
\sigma}R_{\rho a \sigma b} -\frac{1}{3780} R^{\rho
\tau}R^\sigma_{\phantom{\sigma} \tau}R_{\rho a \sigma b}   +
\frac{4}{4725} R^{\rho \sigma}R^\tau_{\phantom{\tau} a}R_{\tau
\sigma \rho b} \right. \nonumber \\
& & \qquad \left. + \frac{1}{18900} R_{\rho \sigma}R^{\rho \kappa
\sigma \lambda}R_{\kappa a \lambda b} -\frac{1}{180} \left(\xi
-\frac{1}{6}\right)  R R^{\rho \sigma \tau}_{\phantom{\rho \sigma
\tau}a }R_{\rho \sigma \tau b}  + \frac{1}{2160} R_{ab}R^{\rho
\sigma \tau \kappa} R_{\rho \sigma \tau \kappa }   +
\frac{31}{37800} R_{\rho \sigma}R^{\rho \kappa
\lambda}_{\phantom{\rho \kappa
\lambda}a}R^\sigma_{\phantom{\sigma} \kappa \lambda b} \right. \nonumber \\
& & \qquad \left. - \frac{1}{600} R_{\rho \sigma}R^{\rho \kappa
\lambda}_{\phantom{\rho \kappa \lambda}a}R^\sigma_{\phantom{\sigma}
\lambda \kappa b}
  + \frac{17}{37800} R^{\rho \sigma}R^{\kappa
\lambda}_{\phantom{\kappa \lambda} \rho a}R_{\kappa \lambda \sigma
b} - \frac{17}{15120} R^\kappa_{\phantom{\kappa} a}R^{\rho \sigma
\tau}_{\phantom{\rho \sigma \tau} \kappa }R_{\rho \sigma \tau b}
-\frac{17}{18900} R^{\rho \sigma \tau}_{\phantom{\rho \sigma \tau}
\lambda } R_{\rho \sigma \tau \kappa}R^{\lambda \phantom{a}
\kappa}_{\phantom{\lambda} a  \phantom{\kappa} b} \right. \nonumber \\
& & \qquad \left. + \frac{1}{378} R^{\rho \kappa \sigma
\lambda}R^\tau_{\phantom{\tau} \rho \sigma a}R_{\tau \kappa \lambda
b} - \frac{1}{900} R^{\rho \kappa \sigma \lambda}R_{\rho \sigma \tau
a }R_{\kappa \lambda \phantom{\tau} b}^{\phantom{\kappa \lambda}
\tau} + \frac{19}{9450} R^{\rho \sigma \kappa \lambda}R_{\rho \sigma
\tau a }R_{\kappa \lambda \phantom{\tau} b}^{\phantom{\kappa
\lambda} \tau}\right] \sigma^{;a}\sigma^{;b} + O
\left(\sigma^{3/2}\right).
\end{eqnarray}
\end{widetext}

\subsection{Covariant Taylor series expansion of ${\tilde{A}}_3(m^2;x,x')$}

The mass-dependent DeWitt coefficient  ${\tilde{A}}_3(m^2;x,x')$ is
the solution of Eq.~(\ref{MDWCoeff1a}) with $n=2$, i.e. it satisfies
\begin{eqnarray}\label{CTSexpA3_1}
& & 3{\tilde{A}}_3 + {\tilde{A}}_{3 ; \mu} \sigma ^{; \mu} -
{\tilde{A}}_3 \Delta ^{-1/2}{\Delta
^{1/2}}_{;\mu} \sigma^{; \mu}    \nonumber \\
& & \qquad \qquad \qquad \qquad = \left( \Box_x -m^2 - \xi R \right)
{\tilde{A}}_2.
\end{eqnarray}
In this equation, we replace ${\tilde{A}}_3$ by its covariant Taylor
series expansion for $x'$ in the neighborhood of $x$ given by
\begin{equation}\label{CTSexpA3_2}
{\tilde{A}}_3 ={\tilde{a}}_3  + O \left(\sigma^{1/2} \right).
\end{equation}
By noting that ${\tilde{A}}_{3 ; \mu} \sigma ^{; \mu} -
{\tilde{A}}_3 \Delta ^{-1/2}{\Delta ^{1/2}}_{;\mu} \sigma^{; \mu}= O
\left(\sigma^{1/2} \right) $, we can see easily  that the left-hand
side of Eq.~(\ref{CTSexpA3_1}) to order $\sigma^0$ reduces to $a_3$.
The covariant Taylor series expansion of the right-hand side of
Eq.~(\ref{CTSexpA3_1}) to order $\sigma^0$ can be found from the
expansion ${\tilde{A}}_2$ given by (\ref{CTSexpA2_2}) and
(\ref{CTSexpA2_5}) or (\ref{CTSexpA2_6}) and from the expansion of
$\Box {\tilde{A}}_2$. The later can be constructed by using the
theory developed in Appendix D and more particularly
Eqs.~(\ref{CTSExpBoxF1}), (\ref{CTSExpBoxF2}) and
(\ref{AppBoxFsymm}). We easily obtain
\begin{equation}\label{CTSexpA3_3}
\Box {\tilde{A}}_2 ={\tilde{a}}''_2  + O \left(\sigma^{1/2} \right)
\end{equation}
with
\begin{widetext}
\begin{eqnarray}\label{CTSexpA3_4}
& & {\tilde{a}}''_2=  -(1/20)  \, (\xi -3/14) \, \Box \Box R + (1/3)
\, (\xi-1/5) \, m^2 \, \Box R + (1/12) \, m^4 \, R + (1/3) \,
(\xi-1/4) \,(\xi-1/5) \, R\Box R   \nonumber \\
& & \qquad + (1/4) \, [\xi^2- (2/5) \, \xi +17/420] \,
R_{;\rho}R^{;\rho} + (1/6) \, (\xi-1/6) \, m^2 \, R^2 - (1/30) \,
(\xi -3/14) \, R_{;\rho \sigma } R^{\rho \sigma}   \nonumber \\
& & \qquad -(1/210)\, R_{\rho \sigma } \Box R^{\rho \sigma} + (1/90)
\, m^2 \, R_{\rho \sigma } R^{\rho \sigma} -(1/840)\, R_{\rho \sigma
;\tau} R^{\rho \sigma ;\tau} -(1/420)\, R_{\rho \tau ;\sigma}
R^{\sigma \tau ;\rho }  \nonumber \\
& & \qquad + (1/140) \, R_{\rho \sigma \tau \kappa} \Box R^{\rho
\sigma \tau \kappa} - (1/90) \, m^2 \, R_{\rho \sigma \tau \kappa}
R^{\rho \sigma \tau \kappa} + (3/560) \, R_{\rho \sigma \tau \kappa
;\lambda} R^{\rho \sigma \tau \kappa ;\lambda} + (1/12) \,
(\xi-1/6)^2 \, R^3 \nonumber \\
& & \qquad + (1/90)\, (\xi-1/4) \, RR_{\rho \sigma } R^{\rho \sigma}
+ (1/1890) \, R_{\rho \sigma } R^{\rho}_{\phantom{\rho}
\tau}R^{\sigma \tau} -(1/630) \, R_{\rho \sigma }R_{\kappa \lambda
}R^{\rho \kappa \sigma \lambda} -(1/90) \, (\xi-1/4) \, RR_{\rho
\sigma \tau \kappa}  R^{\rho \sigma \tau \kappa} \nonumber \\
& & \qquad -(1/315) \, R_{\kappa \lambda}R^{\kappa \rho \sigma
\tau}R^\lambda_{\phantom{\lambda} \rho \sigma \tau} + (1/189) \,
R^{\rho \kappa \sigma \lambda}R_{\rho \alpha \sigma \beta}R_{\kappa
\phantom{\alpha} \lambda \phantom{\beta}}^{\phantom{\kappa} \alpha
\phantom{\lambda} \beta} + (11/3780) \, R^{\rho \sigma \kappa
\lambda}R_{\rho \sigma \alpha \beta}R_{\kappa
\lambda}^{\phantom{\kappa \lambda} \alpha \beta}.
\end{eqnarray}
The direct comparison of the expansions of the left- and right-hand
sides of Eq.~(\ref{CTSexpA3_1}) then yields the coefficients $a_3$:
\begin{eqnarray} \label{CTSexpA3_5}
& & {\tilde{a}}_3= -(1/6) \, m^6 -(1/60)  \, (\xi -3/14) \, \Box
\Box R +  (1/6) \, (\xi-1/5) \, m^2 \, \Box R - (1/2) \, (\xi -1/6)
 \, m^4 \, R \nonumber \\
& & \qquad + (1/6) \, (\xi-1/6) \,(\xi-1/5) \, R\Box R + (1/12) \,
[\xi^2- (2/5) \, \xi
+17/420] \, R_{;\rho}R^{;\rho} -(1/2) \, (\xi-1/6)^2 \, m^2 \, R^2 \nonumber \\
& & \qquad - (1/90) \, (\xi -3/14) \, R_{;\rho \sigma } R^{\rho
\sigma}   -(1/630)\, R_{\rho \sigma } \Box R^{\rho \sigma}
+(1/180)\, m^2 \,  R_{\rho \sigma }  R^{\rho \sigma}  -(1/2520)\,
R_{\rho \sigma ;\tau} R^{\rho \sigma ;\tau} \nonumber \\
& & \qquad -(1/1260)\, R_{\rho \tau ;\sigma} R^{\sigma \tau ;\rho }
+ (1/420) \, R_{\rho \sigma \tau \kappa} \Box R^{\rho \sigma \tau
\kappa} - (1/180) \, m^2 \, R_{\rho \sigma \tau \kappa} R^{\rho
\sigma \tau \kappa} + (1/560) \, R_{\rho \sigma \tau \kappa
;\lambda} R^{\rho \sigma \tau \kappa ;\lambda} \nonumber \\
& & \qquad  - (1/6) \, (\xi-1/6)^3 \, R^3 + (1/180) \, (\xi-1/6) \,
RR_{\rho \sigma } R^{\rho \sigma} + (1/5670) \, R_{\rho \sigma }
R^{\rho}_{\phantom{\rho} \tau}R^{\sigma \tau} -(1/1890) \, R_{\rho
\sigma }R_{\kappa \lambda }R^{\rho \kappa \sigma \lambda}\nonumber \\
& & \qquad  -(1/180) \, (\xi-1/6) \, RR_{\rho \sigma \tau \kappa}
R^{\rho \sigma \tau \kappa} -(1/945) \, R_{\kappa \lambda}R^{\kappa
\rho \sigma \tau}R^\lambda_{\phantom{\lambda} \rho \sigma \tau}+
(1/567) \, R^{\rho \kappa \sigma \lambda}R_{\rho \alpha \sigma
\beta}R_{\kappa \phantom{\alpha} \lambda
\phantom{\beta}}^{\phantom{\kappa} \alpha \phantom{\lambda} \beta} \nonumber \\
& & \qquad  + (11/11340) \, R^{\rho \sigma \kappa \lambda}R_{\rho
\sigma \alpha \beta}R_{\kappa \lambda}^{\phantom{\kappa \lambda}
\alpha \beta}.
\end{eqnarray}
\end{widetext}
This result agrees with that of Gilkey \cite{Gilkey75,Gilkey84}
obtained by using totally different methods, i.e. in the framework
of the pseudo-differential operator theory. The comparison of our
result with his own is immediate in spite of our different
conventions with regard to the metric signature, the Riemann tensor
and the commutation of covariant derivatives. This agreement permits
us to believe in the validity of all the calculations previously
carried out and therefore in the validity of the covariant Taylor
series obtained for the mass-dependent DeWitt coefficients. Finally,
by replacing (\ref{CTSexpA3_5}) into (\ref{CTSexpA3_2}) we can now
write
\begin{widetext}
\begin{eqnarray}\label{CTSexpA3_6}
& & {\tilde{A}}_3 =  -\frac{1}{6} m^6 -\frac{1}{60} \left(\xi -
\frac{3}{14} \right) \, \Box \Box R +  \frac{1}{6}
\left(\xi-\frac{1}{5}\right) \, m^2 \, \Box R - \frac{1}{2}
\left(\xi -\frac{1}{6}\right)
 \, m^4 \, R   + \frac{1}{6} \left(\xi-\frac{1}{6}\right)
\left(\xi-\frac{1}{5}\right) \, R\Box R  \nonumber \\
& & \qquad  + \frac{1}{12} \left(\xi^2- \frac{2}{5}  \xi +
\frac{17}{420} \right) \, R_{;\rho}R^{;\rho} -\frac{1}{2} \left(\xi
-\frac{1}{6}\right)^2  m^2 \, R^2  - \frac{1}{90} \,  \left(\xi -
\frac{3}{14} \right) R_{;\rho \sigma } R^{\rho \sigma}
-\frac{1}{630} R_{\rho \sigma } \Box R^{\rho \sigma} \nonumber \\
& & \qquad +\frac{1}{180} m^2 \, R_{\rho \sigma } R^{\rho \sigma}
 -\frac{1}{2520} R_{\rho \sigma ;\tau} R^{\rho \sigma ;\tau}
-\frac{1}{1260} R_{\rho \tau ;\sigma} R^{\sigma \tau ;\rho } +
\frac{1}{420} \, R_{\rho \sigma \tau \kappa} \Box R^{\rho \sigma
\tau \kappa} - \frac{1}{180} m^2 \, R_{\rho \sigma \tau \kappa}
R^{\rho \sigma \tau \kappa} \nonumber \\
& & \qquad + \frac{1}{560} R_{\rho \sigma \tau \kappa ;\lambda}
R^{\rho \sigma \tau \kappa ;\lambda}   - \frac{1}{6}
 \left(\xi- \frac{1}{6} \right)^3 \, R^3  + \frac{1}{180}
 \left(\xi-  \frac{1}{6} \right) \, RR_{\rho \sigma }
R^{\rho \sigma}  + \frac{1}{5670} R_{\rho \sigma }
R^{\rho}_{\phantom{\rho} \tau}R^{\sigma \tau} \nonumber \\
& & \qquad -\frac{1}{1890} R_{\rho \sigma }R_{\kappa \lambda
}R^{\rho \kappa \sigma \lambda} -\frac{1}{180} \left(\xi-
\frac{1}{6} \right) RR_{\rho \sigma \tau \kappa} R^{\rho \sigma \tau
\kappa} -\frac{1}{945} R_{\kappa \lambda}R^{\kappa \rho \sigma
\tau}R^\lambda_{\phantom{\lambda} \rho \sigma \tau}+ \frac{1}{567}
R^{\rho \kappa \sigma \lambda}R_{\rho \alpha \sigma \beta}R_{\kappa
\phantom{\alpha} \lambda \phantom{\beta}}^{\phantom{\kappa} \alpha
\phantom{\lambda} \beta} \nonumber \\
& & \qquad  + \frac{11}{11340} R^{\rho \sigma \kappa \lambda}R_{\rho
\sigma \alpha \beta}R_{\kappa \lambda}^{\phantom{\kappa \lambda}
\alpha \beta} + O \left(\sigma^{1/2} \right).
\end{eqnarray}
\end{widetext}

\subsection{Covariant Taylor series expansion of the DeWitt
coefficients $A_n(x,x')$ with $n=0,1,2,3$}

We think it is unnecessary to write explicitly the covariant Taylor
series expansion of the DeWitt coefficients $A_n(x,x')$ with
$n=0,1,2,3$. Indeed, we know that
$A_n(x,x')={\tilde{A}}_n(m^2=0;x,x')$ (see (\ref{RelAetA1})). Then,
the coefficients $a_n(x)$ and $a_{n \,\, a_1 \dots a_p}(x)$ defining
the expansion of the DeWitt coefficients $A_n(x,x')$ (see
(\ref{CTSExpMasslessA1}) and (\ref{CTSExpMasslessA2})) and  the
coefficients ${\tilde{a}}_n(m^2;x)$ and ${\tilde{a}}_{n \,\, a_1
\dots a_p}(m^2;x)$ defining the expansion of the mass-dependent
DeWitt coefficients ${\tilde{A}}_n(m^2; x,x')$ (see
(\ref{CTSExpMassdepA1}) and (\ref{CTSExpMassdepA2})) are linked by
\begin{subequations} \label{CTSExpRelDW_MassdepDW}
\begin{eqnarray}
& & a_n(x)= {\tilde{a}}_n(m^2=0;x)  \\
& & a_{n \,\, a_1 \dots a_p}(x)={\tilde{a}}_{n \,\, a_1 \dots
a_p}(m^2=0;x).
\end{eqnarray}
\end{subequations}
As a consequence:

i) the expansion of $A_0(x,x')$ up to order $\sigma^3$ is directly
given by (\ref{CTSexpA0_1}) and (\ref{CTSexpA0_2}) or equivalently
by (\ref{CTSexpA0_3}),

ii) the expansion of $A_1(x,x')$ up to order $\sigma^2$ is obtained
from (\ref{CTSexpA1_2}) and (\ref{CTSexpA1_3}) or equivalently from
(\ref{CTSexpA1_4}) by taking $m^2=0$ in these relations,

iii) the expansion of $A_2(x,x')$ up to order $\sigma$ is given by
(\ref{CTSexpA2_2}) and (\ref{CTSexpA2_5}) or equivalently by
(\ref{CTSexpA2_6}) by taking $m^2=0$ in these relations,

iv) the expansion of $A_3(x,x')$ up to order $\sigma^0$ is given by
(\ref{CTSexpA3_2}) and (\ref{CTSexpA3_5}) or equivalently by
(\ref{CTSexpA3_6}) by taking $m^2=0$ in these relations.

\section{Covariant Taylor
series expansions of the Hadamard coefficients}

In this section, we shall provide the covariant Taylor series
expansions of the geometrical Hadamard coefficients $U_n(x,x')$ and
$V_n(x,x')$ of lowest orders for the dimensions $d=3,4,5$ and $6$ of
spacetime. For $x'$ near $x$, we shall write these expansions in the
form
\begin{eqnarray}
& & U_n(x,x')=u_n(x) +\sum_{p=1}^{+\infty} \frac{(-1)^p}{p!} u_{n \,
(p)}(x,x') \label{CTSExpU1} \\
& & V_n(x,x')=v_n(x) +\sum_{p=1}^{+\infty} \frac{(-1)^p}{p!} v_{n \,
(p)}(x,x') \label{CTSExpV1}
\end{eqnarray}
where the $u_{n \, (p)}(x,x')$ and $v_{n \, (p)}(x,x')$ with $p=1,2,
\dots $ are all biscalars in $x$ and $x'$ which are of the form
\begin{eqnarray}
&  & u_{n \, (p)}(x,x')=u_{n \,\, a_1 \dots a_p}(x)
\sigma^{;a_1}(x,x') \dots \sigma^{;a_p}(x,x') \nonumber \\
& & \label{CTSExpU2} \\
&  & v_{n \, (p)}(x,x')=v_{n \,\, a_1 \dots a_p}(x)
\sigma^{;a_1}(x,x') \dots \sigma^{;a_p}(x,x'). \nonumber
\\
& & \label{CTSExpV2}
\end{eqnarray}
These expansions are easily obtained from those of the
mass-dependent DeWitt coefficients constructed in the previous
section by using the relations (\ref{RelUetAp1}) for $d$ even and
the relation (\ref{RelUetAo1}) for $d$ odd. By using the relation
(\ref{RelUetAo1}), we obtain for $d=3$
\begin{subequations} \label{CTSExpHAD_d3_1}
\begin{eqnarray}
& & U_0(x,x')= {\tilde{A}}_0(m^2;x,x')  \\
& & U_1(x,x')= - {\tilde{A}}_1(m^2;x,x') \\
& & U_2(x,x')= (1/3) \, {\tilde{A}}_2(m^2;x,x')  \\
& & U_3(x,x')= -(1/15) \,{\tilde{A}}_3(m^2;x,x').
\end{eqnarray}
\end{subequations}
From the relation (\ref{RelUetAp1}), we obtain for $d=4$
\begin{subequations} \label{CTSExpHAD_d4_1}
\begin{eqnarray}
& & U_0(x,x')= {\tilde{A}}_0(m^2;x,x')  \\
& & V_0(x,x')= -(1/2) \, {\tilde{A}}_1(m^2;x,x') \\
& & V_1(x,x')= (1/4) \, {\tilde{A}}_2(m^2;x,x')  \\
& & V_2(x,x')= -(1/16) \,{\tilde{A}}_3(m^2;x,x').
\end{eqnarray}
\end{subequations}
From the relation (\ref{RelUetAo1}), we obtain for $d=5$
\begin{subequations} \label{CTSExpHAD_d5_1}
\begin{eqnarray}
& & U_0(x,x')= {\tilde{A}}_0(m^2;x,x')  \\
& & U_1(x,x')= {\tilde{A}}_1(m^2;x,x') \\
& & U_2(x,x')= - \, {\tilde{A}}_2(m^2;x,x')  \\
& & U_3(x,x')= (1/3) \,{\tilde{A}}_3(m^2;x,x').
\end{eqnarray}
\end{subequations}
and finally, from the relation (\ref{RelUetAp1}), we obtain for
$d=6$
\begin{subequations} \label{CTSExpHAD_d6_1}
\begin{eqnarray}
& & U_0(x,x')= {\tilde{A}}_0(m^2;x,x')  \\
& & U_1(x,x')= (1/2) \,  {\tilde{A}}_1(m^2;x,x') \\
& & V_0(x,x')= -(1/4) \, {\tilde{A}}_2(m^2;x,x')  \\
& & V_1(x,x')= (1/8) \,{\tilde{A}}_3(m^2;x,x').
\end{eqnarray}
\end{subequations}
We could stop here this section. However, we prefer to provide the
explicit expansions of all these Hadamard coefficients for the
reader who simply needs them and is not specially interested in
following the derivation of the expansions of the DeWitt
coefficients carried out in the previous section.

For $d=3,4,5$ and $6$ we have
\begin{eqnarray}\label{CTSexpHAD_gen1}
&  & U_0 =u_0 - u_{0 \,\, a} \sigma^{;a}+\frac{1}{2!} u_{0 \,\, a b}
\sigma^{;a}\sigma^{;b}  -\frac{1}{3!} u_{0 \,\, a b c}
\sigma^{;a}\sigma^{;b}\sigma^{;c}  \nonumber \\
& & \quad + \frac{1}{4!} u_{0 \,\, a b c d}
\sigma^{;a}\sigma^{;b}\sigma^{;c}\sigma^{;d}
  -\frac{1}{5!} u_{0 \,\, a b c d
e}\sigma^{;a}\sigma^{;b}\sigma^{;c}\sigma^{;d}\sigma^{;e} \nonumber \\
& & \quad + \frac{1}{6!} u_{0 \,\, a b c d e
f}\sigma^{;a}\sigma^{;b}\sigma^{;c}\sigma^{;d}\sigma^{;e}\sigma^{;f}+
O \left(\sigma^{7/2} \right)
\end{eqnarray}
with
\begin{widetext}
\begin{subequations}
\begin{eqnarray}\label{CTSexpHAD_gen2}
& & u_0=1 \\
& & u_{0 \,\, a}=0 \\
& & u_{0 \,\, a b} =
(1/6)  \,  R_{a b}  \\
& & u_{0 \,\, a b c} =
(1/4)  \,   R_{(a b; c)}  \\
& & u_{0 \,\, a b c d} = (3/10)  \, R_{(a b; c d)} + (1/15) \,
R^{\rho}_{\phantom{\rho}(a |\tau| b} R^{\tau}_{\phantom{\tau} c
|\rho |  d)}
+ (1/12)  R_{(a b} R_{c d)} \\
& & u_{0 \,\, a b c d e} =  (1/3) \, R_{(a b; c d e)} + (1/3) \,
R^{\rho}_{\phantom{\rho}(a |\tau| b} R^{\tau}_{\phantom{\tau}
c |\rho |  d;e)} + (5/12)  R_{(a b} R_{c d; e)}\\
& & u_{0 \,\, a b c d e f} = (5/14) \, R_{(a b; c d e f)} +(4/7) \,
R^{\rho}_{\phantom{\rho}(a |\tau| b} R^{\tau}_{\phantom{\tau} c
|\rho |  d;e f)}  + (15/28) \, R^{\rho}_{\phantom{\rho}(a |\tau| b;
c} R^{\tau}_{\phantom{\tau} d |\rho |  e; f)} + (3/4) \, R_{(a b
}R_{c d; e f)} \nonumber
\\
&  & \quad  + (5/8) \, R_{(a b ; c}R_{ d e ; f)}  + (8/63) \,
R^{\rho}_{\phantom{\rho}(a |\tau | b }R^{\tau}_{\phantom{\tau} c
|\sigma | d}R^{\sigma}_{\phantom{\sigma} e |\rho | f)} + (1/6) \,
R_{(a b} R^{\rho}_{\phantom{\rho} c |\tau| d}
R^{\tau}_{\phantom{\tau} e |\rho |  f)} + (5/72) \, R_{(a b}R_{c
d}R_{e f)}.
\end{eqnarray}
\end{subequations}
\end{widetext}

Furthermore, for $d=3$, we have
\begin{eqnarray}
& & U_1 =u_1 - u_{1 \,\, a} \sigma^{;a}+\frac{1}{2!} u_{1 \,\, a b}
\sigma^{;a}\sigma^{;b} -\frac{1}{3!} u_{1 \,\, a b c}
\sigma^{;a}\sigma^{;b}\sigma^{;c} \nonumber \\
& & \qquad  \qquad  + \frac{1}{4!} u_{1 \,\, a b c d}
\sigma^{;a}\sigma^{;b}\sigma^{;c}\sigma^{;d} + O \left(\sigma^{5/2}
\right)  \label{CTSExpHAD_d3_2}\\
& & U_2 =u_2 - u_{2 \,\, a} \sigma^{;a}+\frac{1}{2!} u_{2 \,\, a b}
\sigma^{;a}\sigma^{;b}  + O \left(\sigma^{3/2} \right)
\label{CTSExpHAD_d3_4}\\
& & U_3 =u_3  + O \left(\sigma^{1/2} \right) \label{CTSExpHAD_d3_6}
\end{eqnarray}
with
\begin{widetext}
\begin{subequations}\label{CTSExpHAD_d3_3}
\begin{eqnarray}
& & u_1= m^2 + (\xi -1/6) \, R  \\
& & u_{1 \,\, a}= (1/2)\, (\xi -1/6) \, R_{;a} \\
& & u_{1 \,\, a b} = -(1/60)  \,  \Box R_{a b} + (1/3) \,
(\xi-3/20) \, R_{;ab}  +(1/6)\,m^2 \,  R_{a b} \nonumber \\
& & \qquad     +(1/6)\,(\xi -1/6) \, R R_{a b} +(1/45)
\,R^\rho_{\phantom{\rho}a} R_{\rho b} - (1/90) \, R^{\rho
\sigma}R_{\rho a \sigma b}
- (1/90)  \, R^{\rho \sigma \tau}_{\phantom{\rho \sigma \tau} a }R_{\rho \sigma \tau b} \\
& & u_{1 \,\, a b c} =  (1/4 ) \, (\xi -2/15) \, R _{ ; (a b c)} -
(1/40) \, ( \Box R_{(a b} )_{;c)}  + (1/4 ) \, m^2 \, R_{
(a b; c)}  \nonumber \\
& & \qquad     + (1/4 ) \, (\xi -1/6) \, R R_{ (a b; c)} + (1/4 ) \,
(\xi -1/6) \, R_{;(a } R_{ bc)} +
(1/15) \, R^{ \rho }_{\phantom{ \rho} (a } R_{ |\rho| b;c) } \nonumber \\
& & \qquad   - (1/60) \, R^{ \rho  }_{\phantom{ \rho}  \sigma}
R^{\sigma}_{\phantom{\sigma}   (a |\rho | b;  c)}
  - (1/60) \, R^{ \rho  }_{\phantom{ \rho}
 \sigma ;(a } R^{\sigma}_{\phantom{\sigma} b |\rho | c)}
 - (1/30) \, R^{
\rho  \sigma \tau}_{\phantom{ \rho \sigma \tau} (a}
R_{|\rho \sigma \tau|   b;  c)}  \\
& & u_{1 \,\, a b c d} = -(1/35) \, ( \Box R_{(a b})_{;c d)} + (1/5)
\, (\xi-5/42) \, R _{ ; (a b c d)} + (3/10) \, m^2 \, R _{
 (a b; c d)} + (3/10) \, (\xi
-1/6) \, R R _{
 (a b; c d)} \nonumber \\
&  & \qquad  +(1/2) \, (\xi-1/6) \, R_{;(a}R_{bc;d)}  +(1/3) \,
(\xi-3/20) \, R_{;(ab}R_{cd)} - (1/60) \, R _{ (a b}\Box R _{ c d)}
+ (1/12) \, m^2 \, R _{ (a b}R _{ c d)}\nonumber \\
&  & \qquad   +(3/35) \, R^\rho_{\phantom{ \rho}  (a } R_{ |\rho|
b;c d) } - (1/105) \, R^{ \rho  }_{\phantom{ \rho}  (a } R_{ b c;
|\rho| d) } + (11/210) \, R^{ \rho  }_{\phantom{ \rho} (a;b } R_{
|\rho| c;d) } + (3/70) \, R^{ \rho }_{\phantom{ \rho} (a ; b }
R_{ c d );   \rho  } \nonumber \\
&  & \qquad   - (17/840) \, R_{ (a b }^{ \phantom{(a b } ; \rho }R_{
c d ); \rho  }  - (2/105) \, R^{ \rho  }_{\phantom{ \rho}
 \sigma } R^{\sigma}_{\phantom{\sigma} (a |\rho | b; c d)}
 - (1/105) \, R^{ \rho  }_{\phantom{ \rho}
 (a ; |\sigma |} R^{\sigma}_{\phantom{\sigma} b |\rho | c; d)}
- (1/30) \, R^{ \rho  }_{\phantom{ \rho}
 \sigma ;(a } R^{\sigma}_{\phantom{\sigma} b |\rho | c;
 d)} \nonumber \\
&  & \qquad  + (4/175) \, R^{ \rho  }_{\phantom{ \rho}
 (a ; |\sigma | b} R^{\sigma}_{\phantom{\sigma} c |\rho |
 d)}  - (11/525) \, R_{(a b \phantom{;\rho}
\sigma}^{\phantom{(a b } ; \rho  } R^{\sigma}_{\phantom{\sigma} c
|\rho | d)}    - (11/525) \, R^{ \rho  }_{\phantom{ \rho}
 \sigma ;(a b} R^{\sigma}_{\phantom{\sigma} c |\rho | d)}
 - (4/525) \, R^{\rho}_{\phantom{ \rho} (a |\sigma| b}
 \Box R^{\sigma}_{\phantom{\sigma} c |\rho | d)} \nonumber \\
&  & \qquad   + (1/15) \, m^2 \,
 R^{\rho}_{\phantom{ \rho} (a |\sigma| b}
  R^{\sigma}_{\phantom{\sigma} c |\rho | d)} - (4/105) \,
 R^{\rho \sigma \tau}_{\phantom{\rho \sigma \tau} (a} R_{|\rho \sigma \tau| b; c d)} - (1/140) \,
 R^{\rho \phantom{(a |\sigma| b}    ;\tau}_{\phantom{ \rho} (a |\sigma| b}
R^{\sigma}_{\phantom{\sigma} c |\rho | d) ; \tau}  \nonumber \\
&  & \qquad  - (1/28) \,
 R^{\rho \sigma \tau}_{\phantom{\rho \sigma \tau} (a;b} R_{|\rho \sigma \tau|  c; d)}
+ (1/12) \, (\xi -1/6) \, R R _{(a b}R _{ c d)} + (1/45) \, R^{ \rho
}_{\phantom{ \rho}  (a  } R_{ |\rho| b}R_{c d) }  - (1/315) \, R^{
\rho  }_{\phantom{ \rho}
 (a } R_{|\sigma| b}
 R^{\sigma}_{\phantom{\sigma} c |\rho | d)} \nonumber \\
& & \qquad    -  (1/90) \, R^{\rho
 \sigma}R_{(a b}R_{|\rho| c
 |\sigma| d)}
+ (1/15) \, (\xi-1/6) \,  R R^{\rho}_{\phantom{\rho} (a |\sigma | b}
 R^{\sigma}_{\phantom{\sigma} c |\rho | d)} - (1/90) \, R_{(a b} R^{ \rho  \sigma
\tau}_{\phantom{ \rho \sigma \tau} c} R_{|\rho \sigma \tau|   d)} \nonumber \\
&  & \qquad   - (26/1575) \, R^{ \rho }_{\phantom{ \rho}
 \sigma } R^{\sigma}_{\phantom{\sigma} (a |\tau | b}
 R^{\tau}_{\phantom{\tau} c |\rho | d)} - (2/63) \, R^{ \rho  }_{\phantom{ \rho}
 (a } R^{\sigma \phantom{b} \tau}_{\phantom{\sigma} b \phantom{ \tau } c}
 R_{|\rho \sigma \tau| d)} - (4/1575) \, R^{\rho \sigma \tau
 \kappa} R_{ \rho  (a |\tau|
 b} R_{|\sigma| c |\kappa|
 d)} \nonumber \\
&  & \qquad   - (4/525) \, R^{\rho  \kappa \tau }_{\phantom{\rho
\kappa \tau } (a} R_{|\rho \tau| \phantom{\sigma}
 b}^{\phantom{ |\rho \tau|} \sigma} R_{|\sigma| c |\kappa|
 d)}  - (16/1575) \, R^{\rho  \kappa \tau }_{\phantom{\rho
\kappa \tau } (a} R_{|\rho|  \phantom{\sigma}
 |\tau| b}^{\phantom{ |\rho| } \sigma} R_{|\sigma| c |\kappa|
 d)}  - (8/1575) \, R^{\rho  \tau \kappa
}_{\phantom{\rho
 \tau  \kappa} (a} R_{|\rho \tau| \phantom{\sigma}
 b}^{\phantom{ |\rho \tau|} \sigma} R_{|\sigma| c |\kappa|
 d)} \nonumber \\
\end{eqnarray}
\end{subequations}
and
\begin{subequations}\label{CTSExpHAD_d3_5}
\begin{eqnarray}
& & u_2= (1/6) \, m^4 -(1/18) \, (\xi -1/5) \, \Box R + (1/3) \,
(\xi -1/6) \, m^2 \, R \nonumber \\
& & \qquad + (1/6) \, (\xi -1/6)^2 \, R^2 - (1/540) \, R_{\rho
\sigma}R^{\rho \sigma}
+  (1/540) \, R_{\rho \sigma \tau \kappa}R^{\rho \sigma  \tau \kappa}\\
& & u_{2 \,\, a}= -(1/36) \, (\xi -1/5) \, (\Box R)_{;a}
+(1/6) \, (\xi -1/6) \, m^2 \, R_{;a} \nonumber \\
& & \qquad  + (1/6) \, (\xi -1/6)^2 \, RR_{;a} - (1/540) \, R_{\rho
\sigma}R^{\rho \sigma}_{\phantom{\rho \sigma};a} +  (1/540) \,
R_{\rho \sigma \tau \kappa}R^{\rho \sigma  \tau
\kappa}_{\phantom{\rho \sigma \tau \kappa};a}
\end{eqnarray}
and
\begin{eqnarray}
& & u_{2 \,\, a b} = (1/2520)  \,  \Box \Box R_{a b} -(1/60) \,
(\xi-4/21) \, (\Box R)_{;ab} - (1/180) \, m^2 \, \Box R_{a b}
 + (1/9) \, (\xi-3/20) \, m^2 \, R_{;a b} \nonumber \\
& & \quad + (1/36) \, m^4 \,  R_{a b}
   + (1/9) \, (\xi -1/6)(\xi-3/20) \,  R R_{;a b}
   -(1/270) \, (\xi -1/7)
\, R_{;\rho (a} R^\rho_{\phantom{\rho}b)}
 -(1/108) \,(\xi -1/5) \,  (\Box R) R_{a b}  \nonumber \\
& & \quad+ (1/12) (\xi -1/6)^2 \, R_{;a} R_{;b} +(1/90) \, (\xi
-3/14) \, R_{;\rho } R^\rho_{\phantom{\rho}(a;b)} - (1/90) \, (\xi
-17/84) \, R_{;\rho } R_{ab}^{\phantom{ab}; \rho} \nonumber \\
& & \quad +(1/18)\, (\xi-1/6) \, m^2 \,  R R_{a b}  -(1/180)\,(\xi
-1/6) \,  R \Box R_{a b} -(1/756) \, R_{\rho (a} \Box
R^\rho_{\phantom{\rho}b)} + (1/135) \, m^2 R_{\rho a}
R^\rho_{\phantom{\rho}b} \nonumber \\
& & \quad -(11/9450) \, R^{\rho \sigma} R_{\rho \sigma;(a b)} -
(1/1080) \, R^{\rho \sigma}_{\phantom{\rho \sigma};a}R_{\rho
\sigma;b}  -(1/3150) \, R^{\rho \sigma} R_{\rho (a ; b)\sigma} +
(2/4725) \, R^{\rho \sigma} R_{ab; \rho \sigma}  \nonumber \\
& & \quad + (1/3780) \, R^\rho_{\phantom{\rho} a;\sigma} R_{\rho
b}^{\phantom{\rho b};\sigma} - (1/756) \, R^\rho_{\phantom{\rho}
a;\sigma} R^\sigma_{\phantom{\sigma} b;\rho } - (1/135) \, (\xi
-3/14) \, R^{;\rho \sigma}R_{\rho a \sigma b}+ (1/1890) \, (\Box
R^{\rho \sigma})R_{\rho a \sigma b} \nonumber \\
& & \quad - (1/270) \, m^2 \, R^{\rho \sigma} R_{\rho a \sigma b} +
(1/1890) \, R^{\rho \sigma;\tau} R_{\tau \sigma \rho (a;b)}  +
(1/2700) \, R^{\rho \sigma}\Box R_{\rho a \sigma b} + (1/1260) \,
R^{\rho \sigma;\tau} R_{\rho a \sigma b;\tau} \nonumber \\
& & \quad  - (2/4725) \, R^{\rho \sigma;\tau}_{\phantom{\rho
\sigma;\tau} (a} R_{|\tau \sigma \rho| b)} + (23/18900) \, R^{\rho
\phantom{(a}; \sigma \tau}_{\phantom{\rho } (a} R_{|\tau \sigma
\rho| b)} - (1/675) \, R^{\rho \phantom{(a}; \sigma
\tau}_{\phantom{\rho } (a} R_{|\rho \sigma \tau | b)} -(1/1050) \,
R^{\rho \sigma \tau \kappa} R_{\rho \sigma \tau (a;b) \kappa}\nonumber \\
& &  \quad + (4/4725) \, R^{\rho \sigma \tau}_{\phantom{\rho \sigma
\tau}a} \Box R_{\rho \sigma \tau b} - (1/270) \, m^2 \, R^{\rho
\sigma \tau}_{\phantom{\rho \sigma \tau}a} R_{\rho \sigma \tau b} +
(29/18900) \, R^{\rho \sigma \tau \kappa} R_{\rho \sigma \tau \kappa
; (a b) } + (1/1260) \, R^{\rho \sigma \tau}_{\phantom{\rho \sigma
\tau}a;\kappa}R_{\rho \sigma \tau b}^{\phantom{\rho \sigma \tau
b};\kappa} \nonumber \\
& &  \quad + (1/1008) \, R^{\rho \sigma \tau \kappa}_{\phantom{\rho
\sigma \tau \kappa};a} R_{\rho \sigma \tau \kappa ; b } + (1/36) \,
(\xi -1/6)^2 \, R^2 R_{ab}  + (1/135) \, (\xi -1/6) \, R R_{\rho a}
R^\rho_{\phantom{\rho}b}  -(1/3240) \, R^{\rho \sigma}R_{\rho
\sigma}R_{ab}  \nonumber \\
& &  \quad  + (1/2835) \, R^{\rho \sigma}R_{\rho a}R_{\sigma b}  -
(1/270) \, (\xi -1/6) \, R R^{\rho \sigma}R_{\rho a \sigma
b}-(1/5670) \, R^{\rho \tau}R^\sigma_{\phantom{\sigma} \tau}R_{\rho
a \sigma b}  + (8/14175) \, R^{\rho \sigma}R^\tau_{\phantom{\tau}
(a}R_{|\tau \sigma \rho| b)} \nonumber \\
& &  \quad + (1/28350) \, R_{\rho \sigma}R^{\rho \kappa \sigma
\lambda}R_{\kappa a \lambda b} -(1/270) \, (\xi -1/6) \, R R^{\rho
\sigma \tau}_{\phantom{\rho \sigma \tau}a }R_{\rho \sigma \tau b} +
(1/3240) \, R_{ab}R^{\rho \sigma \tau \kappa} R_{\rho \sigma \tau
\kappa } \nonumber \\
& &  \quad  + (31/56700) \, R_{\rho \sigma}R^{\rho \kappa
\lambda}_{\phantom{\rho \kappa \lambda}a}R^\sigma_{\phantom{\sigma}
\kappa \lambda b} - (1/900) \, R_{\rho \sigma}R^{\rho \kappa
\lambda}_{\phantom{\rho \kappa \lambda}a}R^\sigma_{\phantom{\sigma}
\lambda \kappa b}   + (17/56700) \, R^{\rho \sigma}R^{\kappa
\lambda}_{\phantom{\kappa \lambda} \rho a}R_{\kappa \lambda \sigma
b} \nonumber \\
& & \quad - (17/22680) \, R^\kappa_{\phantom{\kappa} (a}R^{\rho
\sigma \tau}_{\phantom{\rho \sigma \tau} |\kappa| }R_{|\rho \sigma
\tau| b)} -(17/28350) \, R^{\rho \sigma \tau}_{\phantom{\rho \sigma
\tau} \lambda } R_{\rho \sigma \tau \kappa}R^{\lambda \phantom{a}
\kappa}_{\phantom{\lambda} a  \phantom{\kappa} b}  + (1/567) \,
R^{\rho \kappa \sigma \lambda}R^\tau_{\phantom{\tau} \rho \sigma
a}R_{\tau \kappa \lambda b} \nonumber \\
& &  \quad - (1/1350) \, R^{\rho \kappa \sigma \lambda}R_{\rho
\sigma \tau a }R_{\kappa \lambda \phantom{\tau} b}^{\phantom{\kappa
\lambda} \tau} + (19/14175) \, R^{\rho \sigma \kappa \lambda}R_{\rho
\sigma \tau a }R_{\kappa \lambda \phantom{\tau} b}^{\phantom{\kappa
\lambda} \tau}
\end{eqnarray}
\end{subequations}
and
\begin{eqnarray}\label{CTSExpHAD_d3_7}
& & u_3= (1/90) \, m^6 + (1/900)  \, (\xi -3/14) \, \Box \Box R -
(1/90) \, (\xi-1/5) \, m^2 \, \Box R + (1/30) \, (\xi -1/6)
 \, m^4 \, R \nonumber \\
& & \qquad - (1/90) \, (\xi-1/6) \,(\xi-1/5) \, R\Box R - (1/180) \,
[\xi^2- (2/5) \, \xi
+17/420] \, R_{;\rho}R^{;\rho} +(1/30) \, (\xi-1/6)^2 \, m^2 \, R^2 \nonumber \\
& & \qquad + (1/1350) \, (\xi -3/14) \, R_{;\rho \sigma } R^{\rho
\sigma}   +(1/9450)\, R_{\rho \sigma } \Box R^{\rho \sigma}
-(1/2700)\, m^2 \,  R_{\rho \sigma }  R^{\rho \sigma}  +(1/37800)\,
R_{\rho \sigma ;\tau} R^{\rho \sigma ;\tau} \nonumber \\
& & \qquad +(1/18900)\, R_{\rho \tau ;\sigma} R^{\sigma \tau ;\rho }
- (1/6300) \, R_{\rho \sigma \tau \kappa} \Box R^{\rho \sigma \tau
\kappa} + (1/2700) \, m^2 \, R_{\rho \sigma \tau \kappa} R^{\rho
\sigma \tau \kappa} - (1/8400) \, R_{\rho \sigma \tau \kappa
;\lambda} R^{\rho \sigma \tau \kappa ;\lambda} \nonumber \\
& & \qquad  + (1/90) \, (\xi-1/6)^3 \, R^3 - (1/2700) \, (\xi-1/6)
\, RR_{\rho \sigma } R^{\rho \sigma} - (1/85050) \, R_{\rho \sigma }
R^{\rho}_{\phantom{\rho} \tau}R^{\sigma \tau} +(1/28350) \, R_{\rho
\sigma }R_{\kappa \lambda }R^{\rho \kappa \sigma \lambda}\nonumber \\
& & \qquad  +(1/2700) \, (\xi-1/6) \, RR_{\rho \sigma \tau \kappa}
R^{\rho \sigma \tau \kappa} +(1/14175) \, R_{\kappa
\lambda}R^{\kappa \rho \sigma \tau}R^\lambda_{\phantom{\lambda} \rho
\sigma \tau} - (1/8505) \, R^{\rho \kappa \sigma \lambda}R_{\rho
\alpha \sigma \beta}R_{\kappa \phantom{\alpha} \lambda
\phantom{\beta}}^{\phantom{\kappa} \alpha \phantom{\lambda} \beta} \nonumber \\
& & \qquad  - (11/170100) \, R^{\rho \sigma \kappa \lambda}R_{\rho
\sigma \alpha \beta}R_{\kappa \lambda}^{\phantom{\kappa \lambda}
\alpha \beta}.
\end{eqnarray}
\end{widetext}

For $d=4$, we have
\begin{eqnarray}
& & V_0 =v_0 - v_{0 \,\, a} \sigma^{;a}+\frac{1}{2!} v_{0 \,\, a b}
\sigma^{;a}\sigma^{;b} -\frac{1}{3!} v_{0 \,\, a b c}
\sigma^{;a}\sigma^{;b}\sigma^{;c} \nonumber \\
& & \qquad  \qquad  + \frac{1}{4!} v_{0 \,\, a b c d}
\sigma^{;a}\sigma^{;b}\sigma^{;c}\sigma^{;d} + O \left(\sigma^{5/2}
\right)  \label{CTSExpHAD_d4_2}\\
& & V_1 =v_1 - v_{1 \,\, a} \sigma^{;a}+\frac{1}{2!} v_{1 \,\, a b}
\sigma^{;a}\sigma^{;b}  + O \left(\sigma^{3/2} \right)
\label{CTSExpHAD_d4_4}\\
& & V_2 =v_2  + O \left(\sigma^{1/2} \right) \label{CTSExpHAD_d4_6}
\end{eqnarray}
with
\begin{widetext}
\begin{subequations}\label{CTSExpHAD_d4_3}
\begin{eqnarray}
& & v_0= (1/2) \, m^2 + (1/2) \,(\xi -1/6) \, R  \\
& & v_{0 \,\, a}= (1/4)\, (\xi -1/6) \, R_{;a} \\
& & v_{0 \,\, a b} = -(1/120)  \,  \Box R_{a b} +(1/6) \,
(\xi-3/20) \, R_{;ab}  +(1/12)\,m^2 \,  R_{a b} \nonumber \\
& & \qquad     +(1/12)\,(\xi -1/6) \, R R_{a b} +(1/90)
\,R^\rho_{\phantom{\rho}a} R_{\rho b} - (1/180) \, R^{\rho
\sigma}R_{\rho a \sigma b}
- (1/180)  \, R^{\rho \sigma \tau}_{\phantom{\rho \sigma \tau} a }R_{\rho \sigma \tau b} \\
& & v_{0 \,\, a b c} =  (1/8 ) \, (\xi -2/15) \, R _{ ; (a b c)} -
(1/80) \, ( \Box R_{(a b} )_{;c)}  + (1/8) \, m^2 \, R_{
(a b; c)}  \nonumber \\
& & \qquad     + (1/8) \, (\xi -1/6) \, R R_{ (a b; c)} + (1/8) \,
(\xi -1/6) \, R_{;(a } R_{ bc)} +
(1/30) \, R^{ \rho }_{\phantom{ \rho} (a } R_{ |\rho| b;c) } \nonumber \\
& & \qquad   - (1/120) \, R^{ \rho  }_{\phantom{ \rho}  \sigma}
R^{\sigma}_{\phantom{\sigma}   (a |\rho | b;  c)}
  - (1/120) \, R^{ \rho  }_{\phantom{ \rho}
 \sigma ;(a } R^{\sigma}_{\phantom{\sigma} b |\rho | c)}
 - (1/60) \, R^{
\rho  \sigma \tau}_{\phantom{ \rho \sigma \tau} (a}
R_{|\rho \sigma \tau|   b;  c)}  \\
& & v_{0 \,\, a b c d} = -(1/70) \, ( \Box R_{(a b})_{;c d)} +
(1/10) \, (\xi-5/42) \, R _{ ; (a b c d)} + (3/20) \, m^2 \, R _{
 (a b; c d)} + (3/20) \, (\xi
-1/6) \, R R _{
 (a b; c d)} \nonumber \\
&  & \qquad  +(1/4) \, (\xi-1/6) \, R_{;(a}R_{bc;d)}  +(1/6) \,
(\xi-3/20) \, R_{;(ab}R_{cd)} - (1/120) \, R _{ (a b}\Box R _{ c d)}
+ (1/24) \, m^2 \, R _{ (a b}R _{ c d)}\nonumber \\
&  & \qquad   +(3/70) \, R^\rho_{\phantom{ \rho}  (a } R_{ |\rho|
b;c d) }- (1/210) \, R^{ \rho  }_{\phantom{ \rho}  (a } R_{ b c;
|\rho| d) } + (11/420) \, R^{ \rho  }_{\phantom{ \rho} (a;b } R_{
|\rho| c;d) } + (3/140) \, R^{ \rho }_{\phantom{ \rho} (a ; b }
R_{ c d );   \rho  } \nonumber \\
&  & \qquad   - (17/1680) \, R_{ (a b }^{ \phantom{(a b } ; \rho
}R_{ c d ); \rho  }  - (1/105) \, R^{ \rho  }_{\phantom{ \rho}
 \sigma } R^{\sigma}_{\phantom{\sigma} (a |\rho | b; c d)}
 - (1/210) \, R^{ \rho  }_{\phantom{ \rho}
 (a ; |\sigma |} R^{\sigma}_{\phantom{\sigma} b |\rho | c; d)}
- (1/60) \, R^{ \rho  }_{\phantom{ \rho}
 \sigma ;(a } R^{\sigma}_{\phantom{\sigma} b |\rho | c;
 d)} \nonumber \\
&  & \qquad  + (2/175) \, R^{ \rho  }_{\phantom{ \rho}
 (a ; |\sigma | b} R^{\sigma}_{\phantom{\sigma} c |\rho |
 d)}  - (11/1050) \, R_{(a b \phantom{;\rho}
\sigma}^{\phantom{(a b } ; \rho  } R^{\sigma}_{\phantom{\sigma} c
|\rho | d)}    - (11/1050) \, R^{ \rho  }_{\phantom{ \rho}
 \sigma ;(a b} R^{\sigma}_{\phantom{\sigma} c |\rho | d)}
 - (2/525) \, R^{\rho}_{\phantom{ \rho} (a |\sigma| b}
 \Box R^{\sigma}_{\phantom{\sigma} c |\rho | d)} \nonumber \\
&  & \qquad   +(1/30) \, m^2 \,
 R^{\rho}_{\phantom{ \rho} (a |\sigma| b}
  R^{\sigma}_{\phantom{\sigma} c |\rho | d)} - (2/105) \,
 R^{\rho \sigma \tau}_{\phantom{\rho \sigma \tau} (a} R_{|\rho \sigma \tau| b; c d)}
 - (1/280) \,
 R^{\rho \phantom{(a |\sigma| b}    ;\tau}_{\phantom{ \rho} (a |\sigma| b}
R^{\sigma}_{\phantom{\sigma} c |\rho | d) ; \tau}  \nonumber \\
&  & \qquad  - (1/56) \,
 R^{\rho \sigma \tau}_{\phantom{\rho \sigma \tau} (a;b} R_{|\rho \sigma \tau|  c; d)}
+ (1/24) \, (\xi -1/6) \, R R _{(a b}R _{ c d)} + (1/90) \, R^{ \rho
}_{\phantom{ \rho}  (a  } R_{ |\rho| b}R_{c d) }  - (1/630) \, R^{
\rho }_{\phantom{ \rho}
 (a } R_{|\sigma| b}
 R^{\sigma}_{\phantom{\sigma} c |\rho | d)} \nonumber \\
& & \qquad    -  (1/180) \, R^{\rho
 \sigma}R_{(a b}R_{|\rho| c
 |\sigma| d)}
+ (1/30) \, (\xi-1/6) \,  R R^{\rho}_{\phantom{\rho} (a |\sigma | b}
 R^{\sigma}_{\phantom{\sigma} c |\rho | d)} -(1/180) \, R_{(a b} R^{ \rho  \sigma
\tau}_{\phantom{ \rho \sigma \tau} c} R_{|\rho \sigma \tau|   d)} \nonumber \\
&  & \qquad   - (13/1575) \, R^{ \rho }_{\phantom{ \rho}
 \sigma } R^{\sigma}_{\phantom{\sigma} (a |\tau | b}
 R^{\tau}_{\phantom{\tau} c |\rho | d)} - (1/63) \, R^{ \rho  }_{\phantom{ \rho}
 (a } R^{\sigma \phantom{b} \tau}_{\phantom{\sigma} b \phantom{ \tau } c}
 R_{|\rho \sigma \tau| d)} - (2/1575) \, R^{\rho \sigma \tau
 \kappa} R_{ \rho  (a |\tau|
 b} R_{|\sigma| c |\kappa|
 d)} \nonumber \\
&  & \qquad   - (2/525) \, R^{\rho  \kappa \tau }_{\phantom{\rho
\kappa \tau } (a} R_{|\rho \tau| \phantom{\sigma}
 b}^{\phantom{ |\rho \tau|} \sigma} R_{|\sigma| c |\kappa|
 d)} - (8/1575) \, R^{\rho  \kappa \tau }_{\phantom{\rho
\kappa \tau } (a} R_{|\rho|  \phantom{\sigma}
 |\tau| b}^{\phantom{ |\rho| } \sigma} R_{|\sigma| c |\kappa|
 d)}  -(4/1575) \, R^{\rho  \tau \kappa
}_{\phantom{\rho
 \tau  \kappa} (a} R_{|\rho \tau| \phantom{\sigma}
 b}^{\phantom{ |\rho \tau|} \sigma} R_{|\sigma| c |\kappa|
 d)}
\end{eqnarray}
\end{subequations}
and
\begin{subequations}\label{CTSExpHAD_d4_5}
\begin{eqnarray}
& & v_1= (1/8) \, m^4 -(1/24) \, (\xi -1/5) \, \Box R +
(1/4) \, (\xi -1/6) \, m^2 \, R \nonumber \\
& & \qquad + (1/8) \, (\xi -1/6)^2 \, R^2 - (1/720) \, R_{\rho
\sigma}R^{\rho \sigma}
+  (1/720) \, R_{\rho \sigma \tau \kappa}R^{\rho \sigma  \tau \kappa}\\
& & v_{1 \,\, a}= -(1/48) \, (\xi -1/5) \, (\Box R)_{;a}
+(1/8) \, (\xi -1/6) \, m^2 \, R_{;a} \nonumber \\
& & \qquad  + (1/8) \, (\xi -1/6)^2 \, RR_{;a} - (1/720) \, R_{\rho
\sigma}R^{\rho \sigma}_{\phantom{\rho \sigma};a} +  (1/720) \,
R_{\rho \sigma \tau \kappa}R^{\rho \sigma  \tau
\kappa}_{\phantom{\rho \sigma \tau \kappa};a}
\end{eqnarray}
and
\begin{eqnarray}
& & v_{1 \,\, a b} = (1/3360)  \,  \Box \Box R_{a b} -(1/80) \,
(\xi-4/21) \, (\Box R)_{;ab} - (1/240) \, m^2 \, \Box R_{a b}
 + (1/12) \, (\xi-3/20) \, m^2 \, R_{;a b}   \nonumber \\
& & \quad + (1/48) \, m^4 \,  R_{a b} + (1/12) \, (\xi
-1/6)(\xi-3/20) \,  R R_{;a b}
   -(1/360) \, (\xi -1/7)
\, R_{;\rho (a} R^\rho_{\phantom{\rho}b)}
 -(1/144) \,(\xi -1/5) \,  (\Box R) R_{a b}  \nonumber \\
& & \quad+ (1/16) (\xi -1/6)^2 \, R_{;a} R_{;b} +(1/120) \, (\xi
-3/14) \, R_{;\rho } R^\rho_{\phantom{\rho}(a;b)} - (1/120) \, (\xi
-17/84) \, R_{;\rho } R_{ab}^{\phantom{ab}; \rho} \nonumber \\
& & \quad  +(1/24)\, (\xi-1/6) \, m^2 \,  R R_{a b}  -(1/240)\,(\xi
-1/6) \,  R \Box R_{a b} -(1/1008) \, R_{\rho (a} \Box
R^\rho_{\phantom{\rho}b)} + (1/180) \, m^2 R_{\rho a}
R^\rho_{\phantom{\rho}b} \nonumber \\
& & \quad -(11/12600) \, R^{\rho \sigma} R_{\rho \sigma;(a b)} -
(1/1440) \, R^{\rho \sigma}_{\phantom{\rho \sigma};a}R_{\rho
\sigma;b}  -(1/4200) \, R^{\rho \sigma} R_{\rho (a ; b)\sigma} +
(1/3150) \, R^{\rho \sigma} R_{ab; \rho \sigma} \nonumber \\
& & \quad  + (1/5040) \, R^\rho_{\phantom{\rho} a;\sigma} R_{\rho
b}^{\phantom{\rho b};\sigma} - (1/1008) \, R^\rho_{\phantom{\rho}
a;\sigma} R^\sigma_{\phantom{\sigma} b;\rho } - (1/180) \, (\xi
-3/14) \, R^{;\rho \sigma}R_{\rho a \sigma b}+ (1/2520) \, (\Box
R^{\rho \sigma})R_{\rho a \sigma b} \nonumber \\
& & \quad - (1/360) \, m^2 \, R^{\rho \sigma} R_{\rho a \sigma b} +
(1/2520) \, R^{\rho \sigma;\tau} R_{\tau \sigma \rho (a;b)}  +
(1/3600) \, R^{\rho \sigma}\Box R_{\rho a \sigma b} + (1/1680) \,
R^{\rho \sigma;\tau} R_{\rho a \sigma b;\tau} \nonumber \\
& & \quad  - (1/3150) \, R^{\rho \sigma;\tau}_{\phantom{\rho
\sigma;\tau} (a} R_{|\tau \sigma \rho| b)} + (23/25200) \, R^{\rho
\phantom{(a}; \sigma \tau}_{\phantom{\rho } (a} R_{|\tau \sigma
\rho| b)}  - (1/900) \, R^{\rho \phantom{(a}; \sigma
\tau}_{\phantom{\rho } (a} R_{|\rho \sigma \tau | b)} -(1/1400) \,
R^{\rho \sigma \tau \kappa} R_{\rho \sigma \tau (a;b) \kappa} \nonumber \\
& & \quad + (1/1575) \, R^{\rho \sigma \tau}_{\phantom{\rho \sigma
\tau}a} \Box R_{\rho \sigma \tau b} - (1/360) \, m^2 \, R^{\rho
\sigma \tau}_{\phantom{\rho \sigma \tau}a} R_{\rho \sigma \tau b}  +
(29/25200) \, R^{\rho \sigma \tau \kappa} R_{\rho \sigma \tau \kappa
; (a b) } + (1/1680) \, R^{\rho \sigma \tau}_{\phantom{\rho \sigma
\tau}a;\kappa}R_{\rho \sigma \tau b}^{\phantom{\rho \sigma \tau
b};\kappa} \nonumber \\
& &  \quad + (1/1344) \, R^{\rho \sigma \tau \kappa}_{\phantom{\rho
\sigma \tau \kappa};a} R_{\rho \sigma \tau \kappa ; b } + (1/48) \,
(\xi -1/6)^2 \, R^2 R_{ab} + (1/180) \, (\xi -1/6) \, R R_{\rho a}
R^\rho_{\phantom{\rho}b}  -(1/4320) \, R^{\rho \sigma}R_{\rho
\sigma}R_{ab} \nonumber \\
& &  \quad + (1/3780) \, R^{\rho \sigma}R_{\rho a}R_{\sigma b}  -
(1/360) \, (\xi -1/6) \, R R^{\rho \sigma}R_{\rho a \sigma b}
-(1/7560) \, R^{\rho \tau}R^\sigma_{\phantom{\sigma} \tau}R_{\rho a
\sigma b}  + (2/4725) \, R^{\rho \sigma}R^\tau_{\phantom{\tau}
(a}R_{|\tau \sigma \rho| b)}  \nonumber \\
& &  \quad + (1/37800) \, R_{\rho \sigma}R^{\rho \kappa \sigma
\lambda}R_{\kappa a \lambda b} -(1/360) \, (\xi -1/6) \, R R^{\rho
\sigma \tau}_{\phantom{\rho \sigma \tau}a }R_{\rho \sigma \tau b}  +
(1/4320) \, R_{ab}R^{\rho \sigma \tau \kappa} R_{\rho \sigma \tau
\kappa }   \nonumber \\
& &  \quad + (31/75600) \, R_{\rho \sigma}R^{\rho \kappa
\lambda}_{\phantom{\rho \kappa \lambda}a}R^\sigma_{\phantom{\sigma}
\kappa \lambda b} - (1/1200) \, R_{\rho \sigma}R^{\rho \kappa
\lambda}_{\phantom{\rho \kappa \lambda}a}R^\sigma_{\phantom{\sigma}
\lambda \kappa b}  + (17/75600) \, R^{\rho \sigma}R^{\kappa
\lambda}_{\phantom{\kappa \lambda} \rho a}R_{\kappa \lambda \sigma
b}\nonumber \\
& &  \quad  - (17/30240) \, R^\kappa_{\phantom{\kappa} (a}R^{\rho
\sigma \tau}_{\phantom{\rho \sigma \tau} |\kappa| }R_{|\rho \sigma
\tau| b)} -(17/37800) \, R^{\rho \sigma \tau}_{\phantom{\rho \sigma
\tau} \lambda } R_{\rho \sigma \tau \kappa}R^{\lambda \phantom{a}
\kappa}_{\phantom{\lambda} a  \phantom{\kappa} b}  + (1/756) \,
R^{\rho \kappa \sigma \lambda}R^\tau_{\phantom{\tau} \rho \sigma
a}R_{\tau \kappa \lambda b} \nonumber \\
& & \quad - (1/1800) \, R^{\rho \kappa \sigma \lambda}R_{\rho \sigma
\tau a }R_{\kappa \lambda \phantom{\tau} b}^{\phantom{\kappa
\lambda} \tau} + (19/18900) \, R^{\rho \sigma \kappa \lambda}R_{\rho
\sigma \tau a }R_{\kappa \lambda \phantom{\tau} b}^{\phantom{\kappa
\lambda} \tau}
\end{eqnarray}
\end{subequations}
and
\begin{eqnarray}\label{CTSExpHAD_d4_7}
& & v_2= (1/96) \, m^6 +(1/960)  \, (\xi -3/14) \, \Box \Box R -
(1/96) \, (\xi-1/5) \, m^2 \, \Box R + (1/32) \, (\xi -1/6)
 \, m^4 \, R \nonumber \\
& & \qquad - (1/96) \, (\xi-1/6) \,(\xi-1/5) \, R\Box R - (1/192) \,
[\xi^2- (2/5) \, \xi
+17/420] \, R_{;\rho}R^{;\rho} +(1/32) \, (\xi-1/6)^2 \, m^2 \, R^2 \nonumber \\
& & \qquad + (1/1440) \, (\xi -3/14) \, R_{;\rho \sigma } R^{\rho
\sigma}   +(1/10080)\, R_{\rho \sigma } \Box R^{\rho \sigma}
-(1/2880)\, m^2 \,  R_{\rho \sigma }  R^{\rho \sigma}  +(1/40320)\,
R_{\rho \sigma ;\tau} R^{\rho \sigma ;\tau} \nonumber \\
& & \qquad +(1/20160)\, R_{\rho \tau ;\sigma} R^{\sigma \tau ;\rho }
- (1/6720) \, R_{\rho \sigma \tau \kappa} \Box R^{\rho \sigma \tau
\kappa} + (1/2880) \, m^2 \, R_{\rho \sigma \tau \kappa} R^{\rho
\sigma \tau \kappa} - (1/8960) \, R_{\rho \sigma \tau \kappa
;\lambda} R^{\rho \sigma \tau \kappa ;\lambda} \nonumber \\
& & \qquad  + (1/96) \, (\xi-1/6)^3 \, R^3 - (1/2880) \, (\xi-1/6)
\, RR_{\rho \sigma } R^{\rho \sigma} - (1/90720) \, R_{\rho \sigma }
R^{\rho}_{\phantom{\rho} \tau}R^{\sigma \tau} + (1/30240) \, R_{\rho
\sigma }R_{\kappa \lambda }R^{\rho \kappa \sigma \lambda}\nonumber \\
& & \qquad  +(1/2880) \, (\xi-1/6) \, RR_{\rho \sigma \tau \kappa}
R^{\rho \sigma \tau \kappa} +(1/15120) \, R_{\kappa
\lambda}R^{\kappa \rho \sigma \tau}R^\lambda_{\phantom{\lambda} \rho
\sigma \tau}- (1/9072) \, R^{\rho \kappa \sigma \lambda}R_{\rho
\alpha \sigma \beta}R_{\kappa \phantom{\alpha} \lambda
\phantom{\beta}}^{\phantom{\kappa} \alpha \phantom{\lambda} \beta} \nonumber \\
& & \qquad  - (11/181440) \, R^{\rho \sigma \kappa \lambda}R_{\rho
\sigma \alpha \beta}R_{\kappa \lambda}^{\phantom{\kappa \lambda}
\alpha \beta}.
\end{eqnarray}
\end{widetext}

It should be noted that our expressions of the coefficients $v_{0}$,
$v_{0 \,\, a b}$ and  $v_{1}$ are in agreement with those existing
in the literature (see, for example, Ref.~\cite{BrownOttewill86}).
In contrast, our expressions of the coefficients $v_{0 \,\, a b c
d}$, $v_{1 \,\, a b}$ and $v_{2}$ disagree with the only known
results, i.e. those obtained by Phillips and Hu in
Ref.~\cite{PhillipsHu03}. The comparison of our results with theirs
is far from being obvious. Indeed, contrary to Phillips and Hu we
have systematically used the Bianchi identities (\ref{AppBianchi_1})
and their consequences (\ref{AppBianchi_2})-(\ref{AppBianchi_4}) in
order to simplify all our calculations. As a consequence, our
results are more compact while we consider a more general scalar
theory (Phillips and Hu have limited their study to the conformally
invariant theory, i.e. they have worked with $m^2=0$ and $\xi
=1/6$). For example, our expressions of $v_{0 \,\, a b c d}$ and
$v_{1 \,\, a b}$ have respectively $36$ and $54$ terms while those
of Phillips and Hu have respectively $52$ and $71$ terms. Because of
that, we have been obliged to first simplify their results and then
we have emphasized the disagreement with ours. In fact, we are sure
that the results of Phillips and Hu are wrong. Indeed, we know that
in the four-dimensional framework the coefficients $v_2$ and $a_3$
must be proportional and we have found that the result of Phillips
and Hu does not reproduce that of Gilkey. This is not really
surprising: they have constructed the covariant Taylor series
expansions of the Hadamard coefficients from the expansions of
$\sigma_{;\mu \nu}$ and $\Delta ^{1/2}$ and we have noted in
Appendixes B and C that the expansions they have obtained for these
two bitensors are incorrect.

For $d=5$, we have
\begin{eqnarray}
& & U_1 =u_1 - u_{1 \,\, a} \sigma^{;a}+\frac{1}{2!} u_{1 \,\, a b}
\sigma^{;a}\sigma^{;b} -\frac{1}{3!} u_{1 \,\, a b c}
\sigma^{;a}\sigma^{;b}\sigma^{;c} \nonumber \\
& & \qquad  \qquad  + \frac{1}{4!} u_{1 \,\, a b c d}
\sigma^{;a}\sigma^{;b}\sigma^{;c}\sigma^{;d} + O \left(\sigma^{5/2}
\right)  \label{CTSExpHAD_d5_2}\\
& & U_2 =u_2 - u_{2 \,\, a} \sigma^{;a}+\frac{1}{2!} u_{2 \,\, a b}
\sigma^{;a}\sigma^{;b}  + O \left(\sigma^{3/2} \right)
\label{CTSExpHAD_d5_4}\\
& & U_3 =u_3  + O \left(\sigma^{1/2} \right) \label{CTSExpHAD_d5_6}
\end{eqnarray}
with
\begin{widetext}
\begin{subequations}\label{CTSExpHAD_d5_3}
\begin{eqnarray}
& & u_1= -m^2 -(\xi -1/6) \, R  \\
& & u_{1 \,\, a}= -(1/2)\, (\xi -1/6) \, R_{;a} \\
& & u_{1 \,\, a b} = (1/60)  \,  \Box R_{a b} -(1/3) \,
(\xi-3/20) \, R_{;ab}  -(1/6)\,m^2 \,  R_{a b} \nonumber \\
& & \qquad     -(1/6)\,(\xi -1/6) \, R R_{a b} -(1/45)
\,R^\rho_{\phantom{\rho}a} R_{\rho b} + (1/90) \, R^{\rho
\sigma}R_{\rho a \sigma b}
+ (1/90)  \, R^{\rho \sigma \tau}_{\phantom{\rho \sigma \tau} a }R_{\rho \sigma \tau b} \\
& & u_{1 \,\, a b c} = - (1/4 ) \, (\xi -2/15) \, R _{ ; (a b c)} +
(1/40) \, ( \Box R_{(a b} )_{;c)}  - (1/4 ) \, m^2 \, R_{
(a b; c)}  \nonumber \\
& & \qquad     - (1/4 ) \, (\xi -1/6) \, R R_{ (a b; c)}- (1/4 ) \,
(\xi -1/6) \, R_{;(a } R_{ bc)} -
(1/15) \, R^{ \rho }_{\phantom{ \rho} (a } R_{ |\rho| b;c) } \nonumber \\
& & \qquad   + (1/60) \, R^{ \rho  }_{\phantom{ \rho}  \sigma}
R^{\sigma}_{\phantom{\sigma}   (a |\rho | b;  c)}
  + (1/60) \, R^{ \rho  }_{\phantom{ \rho}
 \sigma ;(a } R^{\sigma}_{\phantom{\sigma} b |\rho | c)}
 + (1/30) \, R^{
\rho  \sigma \tau}_{\phantom{ \rho \sigma \tau} (a}
R_{|\rho \sigma \tau|   b;  c)}  \\
& & u_{1 \,\, a b c d} = (1/35) \, ( \Box R_{(a b})_{;c d)} - (1/5)
\, (\xi-5/42) \, R _{ ; (a b c d)} - (3/10) \, m^2 \, R _{
 (a b; c d)} - (3/10) \, (\xi
-1/6) \, R R _{
 (a b; c d)} \nonumber \\
&  & \qquad  -(1/2) \, (\xi-1/6) \, R_{;(a}R_{bc;d)}  -(1/3) \,
(\xi-3/20) \, R_{;(ab}R_{cd)} + (1/60) \, R _{ (a b}\Box R _{ c d)}
- (1/12) \, m^2 \, R _{ (a b}R _{ c d)}\nonumber \\
&  & \qquad   -(3/35) \, R^\rho_{\phantom{ \rho}  (a } R_{ |\rho|
b;c d) }+ (1/105) \, R^{ \rho  }_{\phantom{ \rho}  (a } R_{ b c;
|\rho| d) } - (11/210) \, R^{ \rho  }_{\phantom{ \rho} (a;b } R_{
|\rho|   c;d) } - (3/70) \, R^{ \rho }_{\phantom{ \rho} (a ; b }
R_{ c d );   \rho  } \nonumber \\
&  & \qquad   + (17/840) \, R_{ (a b }^{ \phantom{(a b } ; \rho }R_{
c d ); \rho  }  + (2/105) \, R^{ \rho  }_{\phantom{ \rho}
 \sigma } R^{\sigma}_{\phantom{\sigma} (a |\rho | b; c d)}
 + (1/105) \, R^{ \rho  }_{\phantom{ \rho}
 (a ; |\sigma |} R^{\sigma}_{\phantom{\sigma} b |\rho | c; d)}
+ (1/30) \, R^{ \rho  }_{\phantom{ \rho}
 \sigma ;(a } R^{\sigma}_{\phantom{\sigma} b |\rho | c;
 d)} \nonumber \\
&  & \qquad  - (4/175) \, R^{ \rho  }_{\phantom{ \rho}
 (a ; |\sigma | b} R^{\sigma}_{\phantom{\sigma} c |\rho |
 d)}  + (11/525) \, R_{(a b \phantom{;\rho}
\sigma}^{\phantom{(a b } ; \rho  } R^{\sigma}_{\phantom{\sigma} c
|\rho | d)}    + (11/525) \, R^{ \rho  }_{\phantom{ \rho}
 \sigma ;(a b} R^{\sigma}_{\phantom{\sigma} c |\rho | d)}
 + (4/525) \, R^{\rho}_{\phantom{ \rho} (a |\sigma| b}
 \Box R^{\sigma}_{\phantom{\sigma} c |\rho | d)} \nonumber \\
&  & \qquad   - (1/15) \, m^2 \,
 R^{\rho}_{\phantom{ \rho} (a |\sigma| b}
  R^{\sigma}_{\phantom{\sigma} c |\rho | d)} + (4/105) \,
 R^{\rho \sigma \tau}_{\phantom{\rho \sigma \tau} (a} R_{|\rho \sigma \tau| b; c d)} + (1/140) \,
 R^{\rho \phantom{(a |\sigma| b}    ;\tau}_{\phantom{ \rho} (a |\sigma| b}
R^{\sigma}_{\phantom{\sigma} c |\rho | d) ; \tau}  \nonumber \\
&  & \qquad  + (1/28) \,
 R^{\rho \sigma \tau}_{\phantom{\rho \sigma \tau} (a;b} R_{|\rho \sigma \tau|  c; d)}
- (1/12) \, (\xi -1/6) \, R R _{(a b}R _{ c d)} - (1/45) \, R^{ \rho
}_{\phantom{ \rho}  (a  } R_{ |\rho| b}R_{c d) }  + (1/315) \, R^{
\rho  }_{\phantom{ \rho}
 (a } R_{|\sigma| b}
 R^{\sigma}_{\phantom{\sigma} c |\rho | d)} \nonumber \\
& & \qquad    +  (1/90) \, R^{\rho
 \sigma}R_{(a b}R_{|\rho| c
 |\sigma| d)}
- (1/15) \, (\xi-1/6) \,  R R^{\rho}_{\phantom{\rho} (a |\sigma | b}
 R^{\sigma}_{\phantom{\sigma} c |\rho | d)} + (1/90) \, R_{(a b} R^{ \rho  \sigma
\tau}_{\phantom{ \rho \sigma \tau} c} R_{|\rho \sigma \tau|   d)} \nonumber \\
&  & \qquad   + (26/1575) \, R^{ \rho }_{\phantom{ \rho}
 \sigma } R^{\sigma}_{\phantom{\sigma} (a |\tau | b}
 R^{\tau}_{\phantom{\tau} c |\rho | d)} + (2/63) \, R^{ \rho  }_{\phantom{ \rho}
 (a } R^{\sigma \phantom{b} \tau}_{\phantom{\sigma} b \phantom{ \tau } c}
 R_{|\rho \sigma \tau| d)} + (4/1575) \, R^{\rho \sigma \tau
 \kappa} R_{ \rho  (a |\tau|
 b} R_{|\sigma| c |\kappa|
 d)} \nonumber \\
&  & \qquad   + (4/525) \, R^{\rho  \kappa \tau }_{\phantom{\rho
\kappa \tau } (a} R_{|\rho \tau| \phantom{\sigma}
 b}^{\phantom{ |\rho \tau|} \sigma} R_{|\sigma| c |\kappa|
 d)}  + (16/1575) \, R^{\rho  \kappa \tau }_{\phantom{\rho
\kappa \tau } (a} R_{|\rho|  \phantom{\sigma}
 |\tau| b}^{\phantom{ |\rho| } \sigma} R_{|\sigma| c |\kappa|
 d)}  + (8/1575) \, R^{\rho  \tau \kappa
}_{\phantom{\rho
 \tau  \kappa} (a} R_{|\rho \tau| \phantom{\sigma}
 b}^{\phantom{ |\rho \tau|} \sigma} R_{|\sigma| c |\kappa|
 d)}
\end{eqnarray}
\end{subequations}
and
\begin{subequations}\label{CTSExpHAD_d5_5}
\begin{eqnarray}
& & u_2= -(1/2) \, m^4 +(1/6) \, (\xi -1/5) \, \Box R - (\xi -1/6)
\, m^2 \, R \nonumber \\
& & \quad - (1/2) \, (\xi -1/6)^2 \, R^2 + (1/180) \, R_{\rho
\sigma}R^{\rho \sigma}
-  (1/180) \, R_{\rho \sigma \tau \kappa}R^{\rho \sigma  \tau \kappa}\\
& & u_{2 \,\, a}= (1/12) \, (\xi -1/5) \, (\Box R)_{;a}
-(1/2) \, (\xi -1/6) \, m^2 \, R_{;a} \nonumber \\
& & \quad  - (1/2) \, (\xi -1/6)^2 \, RR_{;a} + (1/180) \, R_{\rho
\sigma}R^{\rho \sigma}_{\phantom{\rho \sigma};a} -  (1/180) \,
R_{\rho \sigma \tau \kappa}R^{\rho \sigma  \tau
\kappa}_{\phantom{\rho \sigma \tau \kappa};a} \\
& & u_{2 \,\, a b} = -(1/840)  \,  \Box \Box R_{a b} +(1/20) \,
(\xi-4/21) \, (\Box R)_{;ab} + (1/60) \, m^2 \, \Box R_{a b}
 - (1/3) \, (\xi-3/20) \, m^2 \, R_{;a b} \nonumber \\
& & \quad - (1/12) \, m^4 \,  R_{a b}
   - (1/3) \, (\xi -1/6)(\xi-3/20) \,  R R_{;a b}
   +(1/90) \, (\xi -1/7)
\, R_{;\rho (a} R^\rho_{\phantom{\rho}b)}
 +(1/36) \,(\xi -1/5) \,  (\Box R) R_{a b}  \nonumber \\
& & \quad - (1/4) (\xi -1/6)^2 \, R_{;a} R_{;b} -(1/30) \, (\xi
-3/14) \, R_{;\rho } R^\rho_{\phantom{\rho}(a;b)} + (1/30) \, (\xi
-17/84) \, R_{;\rho } R_{ab}^{\phantom{ab}; \rho} -(1/6)\, (\xi-1/6)
\, m^2 \,  R R_{a b} \nonumber \\
& & \quad  +(1/60)\,(\xi -1/6) \,  R \Box R_{a b} +(1/252) \,
R_{\rho (a} \Box R^\rho_{\phantom{\rho}b)} - (1/45) \, m^2 R_{\rho
a} R^\rho_{\phantom{\rho}b} +(11/3150) \, R^{\rho \sigma} R_{\rho
\sigma;(a b)} \nonumber \\
& & \quad + (1/360) \, R^{\rho \sigma}_{\phantom{\rho
\sigma};a}R_{\rho \sigma;b}  +(1/1050) \, R^{\rho \sigma} R_{\rho (a
; b)\sigma} - (2/1575) \, R^{\rho \sigma} R_{ab; \rho \sigma}  -
(1/1260) \, R^\rho_{\phantom{\rho} a;\sigma} R_{\rho
b}^{\phantom{\rho b};\sigma} + (1/252) \, R^\rho_{\phantom{\rho}
a;\sigma}
R^\sigma_{\phantom{\sigma} b;\rho } \nonumber \\
& & \quad + (1/45) \, (\xi -3/14) \, R^{;\rho \sigma}R_{\rho a
\sigma b}- (1/630) \, (\Box R^{\rho \sigma})R_{\rho a \sigma b} +
(1/90) \, m^2 \, R^{\rho \sigma} R_{\rho a \sigma b} - (1/630) \,
R^{\rho \sigma;\tau} R_{\tau \sigma \rho (a;b)}  \nonumber \\
& & \quad - (1/900) \, R^{\rho \sigma}\Box R_{\rho a \sigma b} -
(1/420) \, R^{\rho \sigma;\tau} R_{\rho a \sigma b;\tau}  + (2/1575)
\, R^{\rho \sigma;\tau}_{\phantom{\rho \sigma;\tau} (a} R_{|\tau
\sigma \rho| b)} - (23/6300) \, R^{\rho \phantom{(a}; \sigma
\tau}_{\phantom{\rho } (a} R_{|\tau \sigma \rho| b)} \nonumber \\
& & \quad + (1/225) \, R^{\rho \phantom{(a}; \sigma
\tau}_{\phantom{\rho } (a} R_{|\rho \sigma \tau | b)} +(1/350) \,
R^{\rho \sigma \tau \kappa} R_{\rho \sigma \tau (a;b) \kappa}-
(4/1575) \, R^{\rho \sigma \tau}_{\phantom{\rho \sigma \tau}a} \Box
R_{\rho \sigma \tau b} + (1/90) \, m^2 \, R^{\rho \sigma
\tau}_{\phantom{\rho \sigma \tau}a} R_{\rho \sigma \tau b}  \nonumber \\
& &  \quad - (29/6300) \, R^{\rho \sigma \tau \kappa} R_{\rho \sigma
\tau \kappa ; (a b) } - (1/420) \, R^{\rho \sigma
\tau}_{\phantom{\rho \sigma \tau}a;\kappa}R_{\rho \sigma \tau
b}^{\phantom{\rho \sigma \tau b};\kappa} - (1/336) \, R^{\rho \sigma
\tau \kappa}_{\phantom{\rho \sigma \tau \kappa};a} R_{\rho \sigma
\tau \kappa ; b } - (1/12) \, (\xi -1/6)^2 \, R^2 R_{ab}\nonumber \\
& &  \quad  - (1/45) \, (\xi -1/6) \, R R_{\rho a}
R^\rho_{\phantom{\rho}b}  +(1/1080) \, R^{\rho \sigma}R_{\rho
\sigma}R_{ab} - (1/945) \, R^{\rho \sigma}R_{\rho a}R_{\sigma b}  +
(1/90) \, (\xi -1/6) \, R R^{\rho \sigma}R_{\rho a \sigma b} \nonumber \\
& &  \quad +(1/1890) \, R^{\rho \tau}R^\sigma_{\phantom{\sigma}
\tau}R_{\rho a \sigma b}  - (8/4725) \, R^{\rho
\sigma}R^\tau_{\phantom{\tau} (a}R_{|\tau \sigma \rho| b)} -(1/9450)
\, R_{\rho \sigma}R^{\rho \kappa \sigma \lambda}R_{\kappa a \lambda
b} +(1/90) \, (\xi -1/6) \, R R^{\rho
\sigma \tau}_{\phantom{\rho \sigma \tau}a }R_{\rho \sigma \tau b} \nonumber \\
& &  \quad - (1/1080) \, R_{ab}R^{\rho \sigma \tau \kappa} R_{\rho
\sigma \tau \kappa }   - (31/18900) \, R_{\rho \sigma}R^{\rho \kappa
\lambda}_{\phantom{\rho \kappa \lambda}a}R^\sigma_{\phantom{\sigma}
\kappa \lambda b} + (1/300) \, R_{\rho \sigma}R^{\rho \kappa
\lambda}_{\phantom{\rho \kappa \lambda}a}R^\sigma_{\phantom{\sigma}
\lambda \kappa b}  \nonumber \\
& &  \quad - (17/18900) \, R^{\rho \sigma}R^{\kappa
\lambda}_{\phantom{\kappa \lambda} \rho a}R_{\kappa \lambda \sigma
b} + (17/7560) \, R^\kappa_{\phantom{\kappa} (a}R^{\rho \sigma
\tau}_{\phantom{\rho \sigma \tau} |\kappa| }R_{|\rho \sigma \tau|
b)} +(17/9450) \, R^{\rho \sigma \tau}_{\phantom{\rho \sigma \tau}
\lambda } R_{\rho \sigma \tau \kappa}R^{\lambda \phantom{a}
\kappa}_{\phantom{\lambda} a  \phantom{\kappa} b} \nonumber \\
& & \quad  - (1/189) \, R^{\rho \kappa \sigma
\lambda}R^\tau_{\phantom{\tau} \rho \sigma a}R_{\tau \kappa \lambda
b} + (1/450) \, R^{\rho \kappa \sigma \lambda}R_{\rho \sigma \tau a
}R_{\kappa \lambda \phantom{\tau} b}^{\phantom{\kappa \lambda} \tau}
- (19/4725) \, R^{\rho \sigma \kappa \lambda}R_{\rho \sigma \tau a
}R_{\kappa \lambda \phantom{\tau} b}^{\phantom{\kappa \lambda} \tau}
\end{eqnarray}
\end{subequations}
and
\begin{eqnarray}\label{CTSExpHAD_d5_7}
& & u_3= -(1/18) \, m^6 -(1/180)  \, (\xi -3/14) \, \Box \Box R +
(1/18) \, (\xi-1/5) \, m^2 \, \Box R - (1/6) \, (\xi -1/6)
 \, m^4 \, R \nonumber \\
& & \qquad + (1/18) \, (\xi-1/6) \,(\xi-1/5) \, R\Box R + (1/36) \,
[\xi^2- (2/5) \, \xi
+17/420] \, R_{;\rho}R^{;\rho} -(1/6) \, (\xi-1/6)^2 \, m^2 \, R^2 \nonumber \\
& & \qquad - (1/270) \, (\xi -3/14) \, R_{;\rho \sigma } R^{\rho
\sigma}   -(1/1890)\, R_{\rho \sigma } \Box R^{\rho \sigma}
+(1/540)\, m^2 \,  R_{\rho \sigma }  R^{\rho \sigma}  -(1/7560)\,
R_{\rho \sigma ;\tau} R^{\rho \sigma ;\tau} \nonumber \\
& & \qquad -(1/3780)\, R_{\rho \tau ;\sigma} R^{\sigma \tau ;\rho }
+ (1/1260) \, R_{\rho \sigma \tau \kappa} \Box R^{\rho \sigma \tau
\kappa} - (1/540) \, m^2 \, R_{\rho \sigma \tau \kappa} R^{\rho
\sigma \tau \kappa} + (1/1680) \, R_{\rho \sigma \tau \kappa
;\lambda} R^{\rho \sigma \tau \kappa ;\lambda} \nonumber \\
& & \qquad  - (1/18) \, (\xi-1/6)^3 \, R^3 + (1/540) \, (\xi-1/6) \,
RR_{\rho \sigma } R^{\rho \sigma} + (1/17010) \, R_{\rho \sigma }
R^{\rho}_{\phantom{\rho} \tau}R^{\sigma \tau} -(1/5670) \, R_{\rho
\sigma }R_{\kappa \lambda }R^{\rho \kappa \sigma \lambda}\nonumber \\
& & \qquad  -(1/540) \, (\xi-1/6) \, RR_{\rho \sigma \tau \kappa}
R^{\rho \sigma \tau \kappa} -(1/2835) \, R_{\kappa \lambda}R^{\kappa
\rho \sigma \tau}R^\lambda_{\phantom{\lambda} \rho \sigma \tau}+
(1/1701) \, R^{\rho \kappa \sigma \lambda}R_{\rho \alpha \sigma
\beta}R_{\kappa \phantom{\alpha} \lambda
\phantom{\beta}}^{\phantom{\kappa} \alpha \phantom{\lambda} \beta} \nonumber \\
& & \qquad  + (11/34020) \, R^{\rho \sigma \kappa \lambda}R_{\rho
\sigma \alpha \beta}R_{\kappa \lambda}^{\phantom{\kappa \lambda}
\alpha \beta}.
\end{eqnarray}
\end{widetext}

For $d=6$, we have
\begin{eqnarray}
& & U_1 =u_1 - u_{1 \,\, a} \sigma^{;a}+\frac{1}{2!} u_{1 \,\, a b}
\sigma^{;a}\sigma^{;b} -\frac{1}{3!} u_{1 \,\, a b c}
\sigma^{;a}\sigma^{;b}\sigma^{;c} \nonumber \\
& & \qquad  \qquad  + \frac{1}{4!} u_{1 \,\, a b c d}
\sigma^{;a}\sigma^{;b}\sigma^{;c}\sigma^{;d} + O \left(\sigma^{5/2}
\right)  \label{CTSExpHAD_d6_2}\\
& & V_0 =v_0 - v_{0 \,\, a} \sigma^{;a}+\frac{1}{2!} v_{0 \,\, a b}
\sigma^{;a}\sigma^{;b}  + O \left(\sigma^{3/2} \right)
\label{CTSExpHAD_d6_4}\\
& & V_1 = v_1  + O \left(\sigma^{1/2} \right) \label{CTSExpHAD_d6_6}
\end{eqnarray}
with
\begin{widetext}
\begin{subequations}\label{CTSExpHAD_d6_3}
\begin{eqnarray}
& & u_1= -(1/2) \, m^2 - (1/2) \,(\xi -1/6) \, R  \\
& & u_{1 \,\, a}= -(1/4)\, (\xi -1/6) \, R_{;a} \\
& & u_{1 \,\, a b} = (1/120)  \,  \Box R_{a b} -(1/6) \,
(\xi-3/20) \, R_{;ab}  -(1/12)\,m^2 \,  R_{a b} \nonumber \\
& & \qquad     -(1/12)\,(\xi -1/6) \, R R_{a b} -(1/90)
\,R^\rho_{\phantom{\rho}a} R_{\rho b} + (1/180) \, R^{\rho
\sigma}R_{\rho a \sigma b}
+ (1/180)  \, R^{\rho \sigma \tau}_{\phantom{\rho \sigma \tau} a }R_{\rho \sigma \tau b} \\
& & u_{1 \,\, a b c} =  -(1/8 ) \, (\xi -2/15) \, R _{ ; (a b c)} +
(1/80) \, ( \Box R_{(a b} )_{;c)}  - (1/8) \, m^2 \, R_{
(a b; c)}  \nonumber \\
& & \qquad     - (1/8) \, (\xi -1/6) \, R R_{ (a b; c)} - (1/8) \,
(\xi -1/6) \, R_{;(a } R_{ bc)} -
(1/30) \, R^{ \rho }_{\phantom{ \rho} (a } R_{ |\rho| b;c) } \nonumber \\
& & \qquad   + (1/120) \, R^{ \rho  }_{\phantom{ \rho}  \sigma}
R^{\sigma}_{\phantom{\sigma}   (a |\rho | b;  c)}
  + (1/120) \, R^{ \rho  }_{\phantom{ \rho}
 \sigma ;(a } R^{\sigma}_{\phantom{\sigma} b |\rho | c)}
 + (1/60) \, R^{
\rho  \sigma \tau}_{\phantom{ \rho \sigma \tau} (a}
R_{|\rho \sigma \tau|   b;  c)}  \\
& & u_{1 \,\, a b c d} = (1/70) \, ( \Box R_{(a b})_{;c d)} - (1/10)
\, (\xi-5/42) \, R _{ ; (a b c d)} - (3/20) \, m^2 \, R _{
 (a b; c d)} - (3/20) \, (\xi
-1/6) \, R R _{
 (a b; c d)} \nonumber \\
&  & \qquad  -(1/4) \, (\xi-1/6) \, R_{;(a}R_{bc;d)}  -(1/6) \,
(\xi-3/20) \, R_{;(ab}R_{cd)} + (1/120) \, R _{ (a b}\Box R _{ c d)}
- (1/24) \, m^2 \, R _{ (a b}R _{ c d)}\nonumber \\
&  & \qquad   -(3/70) \, R^\rho_{\phantom{ \rho}  (a } R_{ |\rho|
b;c d) } + (1/210) \, R^{ \rho  }_{\phantom{ \rho}  (a } R_{ b c;
|\rho| d) } - (11/420) \, R^{ \rho  }_{\phantom{ \rho} (a;b } R_{
|\rho| c;d) } - (3/140) \, R^{ \rho }_{\phantom{ \rho} (a ; b }
R_{ c d );   \rho  } \nonumber \\
&  & \qquad   +(17/1680) \, R_{ (a b }^{ \phantom{(a b } ; \rho }R_{
c d ); \rho  }  + (1/105) \, R^{ \rho  }_{\phantom{ \rho}
 \sigma } R^{\sigma}_{\phantom{\sigma} (a |\rho | b; c d)}
 + (1/210) \, R^{ \rho  }_{\phantom{ \rho}
 (a ; |\sigma |} R^{\sigma}_{\phantom{\sigma} b |\rho | c; d)}
+ (1/60) \, R^{ \rho  }_{\phantom{ \rho}
 \sigma ;(a } R^{\sigma}_{\phantom{\sigma} b |\rho | c;
 d)} \nonumber \\
&  & \qquad  - (2/175) \, R^{ \rho  }_{\phantom{ \rho}
 (a ; |\sigma | b} R^{\sigma}_{\phantom{\sigma} c |\rho |
 d)}  + (11/1050) \, R_{(a b \phantom{;\rho}
\sigma}^{\phantom{(a b } ; \rho  } R^{\sigma}_{\phantom{\sigma} c
|\rho | d)}    + (11/1050) \, R^{ \rho  }_{\phantom{ \rho}
 \sigma ;(a b} R^{\sigma}_{\phantom{\sigma} c |\rho | d)}
 + (2/525) \, R^{\rho}_{\phantom{ \rho} (a |\sigma| b}
 \Box R^{\sigma}_{\phantom{\sigma} c |\rho | d)} \nonumber \\
&  & \qquad   -(1/30) \, m^2 \,
 R^{\rho}_{\phantom{ \rho} (a |\sigma| b}
  R^{\sigma}_{\phantom{\sigma} c |\rho | d)} + (2/105) \,
 R^{\rho \sigma \tau}_{\phantom{\rho \sigma \tau} (a} R_{|\rho \sigma \tau| b; c d)}
 + (1/280) \,
 R^{\rho \phantom{(a |\sigma| b}    ;\tau}_{\phantom{ \rho} (a |\sigma| b}
R^{\sigma}_{\phantom{\sigma} c |\rho | d) ; \tau}  \nonumber \\
&  & \qquad  + (1/56) \,
 R^{\rho \sigma \tau}_{\phantom{\rho \sigma \tau} (a;b} R_{|\rho \sigma \tau|  c; d)}
- (1/24) \, (\xi -1/6) \, R R _{(a b}R _{ c d)} - (1/90) \, R^{ \rho
}_{\phantom{ \rho}  (a  } R_{ |\rho| b}R_{c d) }  + (1/630) \, R^{
\rho }_{\phantom{ \rho}
 (a } R_{|\sigma| b}
 R^{\sigma}_{\phantom{\sigma} c |\rho | d)} \nonumber \\
& & \qquad    +  (1/180) \, R^{\rho
 \sigma}R_{(a b}R_{|\rho| c
 |\sigma| d)}
- (1/30) \, (\xi-1/6) \,  R R^{\rho}_{\phantom{\rho} (a |\sigma | b}
 R^{\sigma}_{\phantom{\sigma} c |\rho | d)} +(1/180) \, R_{(a b} R^{ \rho  \sigma
\tau}_{\phantom{ \rho \sigma \tau} c} R_{|\rho \sigma \tau|   d)} \nonumber \\
&  & \qquad   + (13/1575) \, R^{ \rho }_{\phantom{ \rho}
 \sigma } R^{\sigma}_{\phantom{\sigma} (a |\tau | b}
 R^{\tau}_{\phantom{\tau} c |\rho | d)} + (1/63) \, R^{ \rho  }_{\phantom{ \rho}
 (a } R^{\sigma \phantom{b} \tau}_{\phantom{\sigma} b \phantom{ \tau } c}
 R_{|\rho \sigma \tau| d)} + (2/1575) \, R^{\rho \sigma \tau
 \kappa} R_{ \rho  (a |\tau|
 b} R_{|\sigma| c |\kappa|
 d)} \nonumber \\
&  & \qquad   + (2/525) \, R^{\rho  \kappa \tau }_{\phantom{\rho
\kappa \tau } (a} R_{|\rho \tau| \phantom{\sigma}
 b}^{\phantom{ |\rho \tau|} \sigma} R_{|\sigma| c |\kappa|
 d)} + (8/1575) \, R^{\rho  \kappa \tau }_{\phantom{\rho
\kappa \tau } (a} R_{|\rho|  \phantom{\sigma}
 |\tau| b}^{\phantom{ |\rho| } \sigma} R_{|\sigma| c |\kappa|
 d)}  +(4/1575) \, R^{\rho  \tau \kappa
}_{\phantom{\rho
 \tau  \kappa} (a} R_{|\rho \tau| \phantom{\sigma}
 b}^{\phantom{ |\rho \tau|} \sigma} R_{|\sigma| c |\kappa|
 d)}
\end{eqnarray}
\end{subequations}
and
\begin{subequations}\label{CTSExpHAD_d6_5}
\begin{eqnarray}
& & v_0= -(1/8) \, m^4 +(1/24) \, (\xi -1/5) \, \Box R -
(1/4) \, (\xi -1/6) \, m^2 \, R \nonumber \\
& & \qquad - (1/8) \, (\xi -1/6)^2 \, R^2 + (1/720) \, R_{\rho
\sigma}R^{\rho \sigma}
-  (1/720) \, R_{\rho \sigma \tau \kappa}R^{\rho \sigma  \tau \kappa}\\
& & v_{0 \,\, a}= (1/48) \, (\xi -1/5) \, (\Box R)_{;a}
-(1/8) \, (\xi -1/6) \, m^2 \, R_{;a} \nonumber \\
& & \qquad  - (1/8) \, (\xi -1/6)^2 \, RR_{;a} + (1/720) \, R_{\rho
\sigma}R^{\rho \sigma}_{\phantom{\rho \sigma};a} -  (1/720) \,
R_{\rho \sigma \tau \kappa}R^{\rho \sigma  \tau
\kappa}_{\phantom{\rho \sigma \tau \kappa};a}
\end{eqnarray}
and
\begin{eqnarray}
& & v_{0 \,\, a b} = -(1/3360)  \,  \Box \Box R_{a b} +(1/80) \,
(\xi-4/21) \, (\Box R)_{;ab} + (1/240) \, m^2 \, \Box R_{a b}
 -(1/12) \, (\xi-3/20) \, m^2 \, R_{;a b}  \nonumber \\
& & \quad - (1/48) \, m^4 \,  R_{a b}
  - (1/12) \, (\xi -1/6)(\xi-3/20) \,  R R_{;a b}
   +(1/360) \, (\xi -1/7)
\, R_{;\rho (a} R^\rho_{\phantom{\rho}b)}
 +(1/144) \,(\xi -1/5) \,  (\Box R) R_{a b}  \nonumber \\
& & \quad - (1/16) (\xi -1/6)^2 \, R_{;a} R_{;b} -(1/120) \, (\xi
-3/14) \, R_{;\rho } R^\rho_{\phantom{\rho}(a;b)} + (1/120) \, (\xi
-17/84) \, R_{;\rho } R_{ab}^{\phantom{ab}; \rho} \nonumber \\
& & \quad  -(1/24)\, (\xi-1/6) \, m^2 \,  R R_{a b}  +(1/240)\,(\xi
-1/6) \,  R \Box R_{a b} +(1/1008) \, R_{\rho (a} \Box
R^\rho_{\phantom{\rho}b)} - (1/180) \, m^2 R_{\rho a}
R^\rho_{\phantom{\rho}b} \nonumber \\
& & \quad  +(11/12600) \, R^{\rho \sigma} R_{\rho \sigma;(a b)} +
(1/1440) \, R^{\rho \sigma}_{\phantom{\rho \sigma};a}R_{\rho
\sigma;b} +(1/4200) \, R^{\rho \sigma} R_{\rho (a ; b)\sigma} -
(1/3150) \, R^{\rho \sigma} R_{ab; \rho \sigma} \nonumber \\
& & \quad  - (1/5040) \, R^\rho_{\phantom{\rho} a;\sigma} R_{\rho
b}^{\phantom{\rho b};\sigma} + (1/1008) \, R^\rho_{\phantom{\rho}
a;\sigma} R^\sigma_{\phantom{\sigma} b;\rho }  + (1/180) \, (\xi
-3/14) \, R^{;\rho \sigma}R_{\rho a \sigma b} - (1/2520) \, (\Box
R^{\rho \sigma})R_{\rho a \sigma b} \nonumber \\
& & \quad + (1/360) \, m^2 \, R^{\rho \sigma} R_{\rho a \sigma b} -
(1/2520) \, R^{\rho \sigma;\tau} R_{\tau \sigma \rho (a;b)}  -
(1/3600) \, R^{\rho \sigma}\Box R_{\rho a \sigma b} - (1/1680) \,
R^{\rho \sigma;\tau} R_{\rho a \sigma b;\tau} \nonumber \\
& & \quad +(1/3150) \, R^{\rho \sigma;\tau}_{\phantom{\rho
\sigma;\tau} (a} R_{|\tau \sigma \rho| b)} - (23/25200) \, R^{\rho
\phantom{(a}; \sigma \tau}_{\phantom{\rho } (a} R_{|\tau \sigma
\rho| b)} + (1/900) \, R^{\rho \phantom{(a}; \sigma
\tau}_{\phantom{\rho } (a} R_{|\rho \sigma \tau | b)} + (1/1400) \,
R^{\rho \sigma \tau \kappa} R_{\rho \sigma \tau (a;b) \kappa} \nonumber \\
& & \quad - (1/1575) \, R^{\rho \sigma \tau}_{\phantom{\rho \sigma
\tau}a} \Box R_{\rho \sigma \tau b} + (1/360) \, m^2 \, R^{\rho
\sigma \tau}_{\phantom{\rho \sigma \tau}a} R_{\rho \sigma \tau b} -
(29/25200) \, R^{\rho \sigma \tau \kappa} R_{\rho \sigma \tau \kappa
; (a b) } - (1/1680) \, R^{\rho \sigma \tau}_{\phantom{\rho \sigma
\tau}a;\kappa}R_{\rho \sigma \tau b}^{\phantom{\rho \sigma \tau
b};\kappa} \nonumber \\
& &  \quad - (1/1344) \, R^{\rho \sigma \tau \kappa}_{\phantom{\rho
\sigma \tau \kappa};a} R_{\rho \sigma \tau \kappa ; b } - (1/48) \,
(\xi -1/6)^2 \, R^2 R_{ab} - (1/180) \, (\xi -1/6) \, R R_{\rho a}
R^\rho_{\phantom{\rho}b}  +(1/4320) \, R^{\rho \sigma}R_{\rho
\sigma}R_{ab} \nonumber \\
& &  \quad - (1/3780) \, R^{\rho \sigma}R_{\rho a}R_{\sigma b}  +
(1/360) \, (\xi -1/6) \, R R^{\rho \sigma}R_{\rho a \sigma b}
+(1/7560) \, R^{\rho \tau}R^\sigma_{\phantom{\sigma} \tau}R_{\rho a
\sigma b}  - (2/4725) \, R^{\rho \sigma}R^\tau_{\phantom{\tau}
(a}R_{|\tau \sigma \rho| b)} \nonumber \\
& &  \quad -(1/37800) \, R_{\rho \sigma}R^{\rho \kappa \sigma
\lambda}R_{\kappa a \lambda b} +(1/360) \, (\xi -1/6) \, R R^{\rho
\sigma \tau}_{\phantom{\rho \sigma \tau}a }R_{\rho \sigma \tau b} -
(1/4320) \, R_{ab}R^{\rho \sigma \tau \kappa} R_{\rho \sigma \tau
\kappa }   \nonumber \\
& &  \quad - (31/75600) \, R_{\rho \sigma}R^{\rho \kappa
\lambda}_{\phantom{\rho \kappa \lambda}a}R^\sigma_{\phantom{\sigma}
\kappa \lambda b} + (1/1200) \, R_{\rho \sigma}R^{\rho \kappa
\lambda}_{\phantom{\rho \kappa \lambda}a}R^\sigma_{\phantom{\sigma}
\lambda \kappa b}   - (17/75600) \, R^{\rho \sigma}R^{\kappa
\lambda}_{\phantom{\kappa \lambda} \rho a}R_{\kappa \lambda \sigma
b} \nonumber \\
& &  \quad + (17/30240) \, R^\kappa_{\phantom{\kappa} (a}R^{\rho
\sigma \tau}_{\phantom{\rho \sigma \tau} |\kappa| }R_{|\rho \sigma
\tau| b)} +(17/37800) \, R^{\rho \sigma \tau}_{\phantom{\rho \sigma
\tau} \lambda } R_{\rho \sigma \tau \kappa}R^{\lambda \phantom{a}
\kappa}_{\phantom{\lambda} a  \phantom{\kappa} b}  - (1/756) \,
R^{\rho \kappa \sigma \lambda}R^\tau_{\phantom{\tau} \rho \sigma
a}R_{\tau \kappa \lambda b} \nonumber \\
& & \quad + (1/1800) \, R^{\rho \kappa \sigma \lambda}R_{\rho \sigma
\tau a }R_{\kappa \lambda \phantom{\tau} b}^{\phantom{\kappa
\lambda} \tau} - (19/18900) \, R^{\rho \sigma \kappa \lambda}R_{\rho
\sigma \tau a }R_{\kappa \lambda \phantom{\tau} b}^{\phantom{\kappa
\lambda} \tau}
\end{eqnarray}
\end{subequations}
and
\begin{eqnarray}\label{CTSExpHAD_d6_7}
& & v_1= -(1/48) \, m^6 -(1/480)  \, (\xi -3/14) \, \Box \Box R +
(1/48) \, (\xi-1/5) \, m^2 \, \Box R - (1/16) \, (\xi -1/6)
 \, m^4 \, R \nonumber \\
& & \qquad + (1/48) \, (\xi-1/6) \,(\xi-1/5) \, R\Box R + (1/96) \,
[\xi^2- (2/5) \, \xi
+17/420] \, R_{;\rho}R^{;\rho} -(1/16) \, (\xi-1/6)^2 \, m^2 \, R^2 \nonumber \\
& & \qquad - (1/720) \, (\xi -3/14) \, R_{;\rho \sigma } R^{\rho
\sigma}   -(1/5040)\, R_{\rho \sigma } \Box R^{\rho \sigma}
+(1/1440)\, m^2 \,  R_{\rho \sigma }  R^{\rho \sigma}  -(1/20160)\,
R_{\rho \sigma ;\tau} R^{\rho \sigma ;\tau} \nonumber \\
& & \qquad -(1/10080)\, R_{\rho \tau ;\sigma} R^{\sigma \tau ;\rho }
+ (1/3360) \, R_{\rho \sigma \tau \kappa} \Box R^{\rho \sigma \tau
\kappa} - (1/1440) \, m^2 \, R_{\rho \sigma \tau \kappa} R^{\rho
\sigma \tau \kappa} + (1/4480) \, R_{\rho \sigma \tau \kappa
;\lambda} R^{\rho \sigma \tau \kappa ;\lambda} \nonumber \\
& & \qquad  - (1/48) \, (\xi-1/6)^3 \, R^3 + (1/1440) \, (\xi-1/6)
\, RR_{\rho \sigma } R^{\rho \sigma} + (1/45360) \, R_{\rho \sigma }
R^{\rho}_{\phantom{\rho} \tau}R^{\sigma \tau} -(1/15120) \, R_{\rho
\sigma }R_{\kappa \lambda }R^{\rho \kappa \sigma \lambda}\nonumber \\
& & \qquad  -(1/1440) \, (\xi-1/6) \, RR_{\rho \sigma \tau \kappa}
R^{\rho \sigma \tau \kappa} -(1/7560) \, R_{\kappa \lambda}R^{\kappa
\rho \sigma \tau}R^\lambda_{\phantom{\lambda} \rho \sigma \tau}+
(1/4536) \, R^{\rho \kappa \sigma \lambda}R_{\rho \alpha \sigma
\beta}R_{\kappa \phantom{\alpha} \lambda
\phantom{\beta}}^{\phantom{\kappa} \alpha \phantom{\lambda} \beta} \nonumber \\
& & \qquad  + (11/90720) \, R^{\rho \sigma \kappa \lambda}R_{\rho
\sigma \alpha \beta}R_{\kappa \lambda}^{\phantom{\kappa \lambda}
\alpha \beta}.
\end{eqnarray}
\end{widetext}

\section{Conclusion and perspectives}

In this article, we have considered for a massive scalar field
theory defined on an arbitrary curved spacetime the DeWitt-Schwinger
and Hadamard representations of the associated Feynman propagator
$G^{\mathrm{F}}(x,x')$. By combining the old covariant recursive
method invented by DeWitt \cite{DeWittBrehme,DeWitt65} with the
modern covariant non-recursive techniques introduced and developed
by Avramidi (see Refs.~\cite{Avramidi_PhD,AvramidiLNP2000} and
references therein), we have obtained the covariant Taylor series
expansions of the DeWitt coefficients $A_0(x,x')$, $A_1(x,x')$,
$A_2(x,x')$ and $A_3(x,x')$ up to orders $\sigma^{3}$, $\sigma^{2}$,
$\sigma^{1}$ and $\sigma^{0}$ respectively. We have then constructed
the corresponding geometrical Hadamard coefficients for the
dimensions $d=3,4,5$ and $6$ of spacetime. It should be noted that
the DeWitt and Hadamard coefficients do not formally depend on the
signature of the manifold on which the field theory is defined. As a
consequence, all our results remain valid, {\it mutatis mutandis},
in the Riemannian framework, i.e. when the metric of the
gravitational background is a Riemannian one.

As an immediate first application of the results obtained in this
article, we intend now to develop the Hadamard regularization of the
stress-energy tensor for a quantized scalar field in a general
spacetime of arbitrary dimension \cite{DecaniniFolacci2005b},
emphasizing more particularly the cases corresponding to the
dimensions $d=3,5,6$ of spacetime which have not been treated
explicitly till now.

Our results could be also immediately used in stochastic
semiclassical gravity. Indeed, as we have noted in Sec.~IV, the
results obtained by Phillips and Hu in Ref.~\cite{PhillipsHu03}
which concern the covariant Taylor series expansion of the
four-dimensional Hadamard representation are incorrect. As a
consequence, our own results could be useful to test some of the
conclusions of Ref.~\cite{PhillipsHu03} concerning the behavior of
the noise kernel in the Schwarzschild spacetime and to emphasize
those which remain valid and those which are wrong.

Keeping in mind the various applications in classical and quantum
gravitational physics mentioned in Sec.~I, it seems to us also
interesting to extend the present work i) by going beyond the orders
reached here for the scalar field theory, ii) for the graviton field
propagating on a curved vacuum spacetime and iii) for more general
field theories, i.e., for tensorial field theories coupled to
external gauge fields. Of course, it is obvious that we shall not be
able to realize such a program in the technical framework developed
in this article, i.e., by partially using the old covariant
recursive method of DeWitt. This method has permitted us to go
beyond existing results but at the cost of odious calculations. Even
if it presents the advantage to provide, at each step of the work,
explicit results which can be controlled, it has certainly reached
its limits here. In fact, it seems to us that the program we have
proposed could be certainly realized by fully working in the
framework of the covariant non-recursive approach of Avramidi or by
using the treatment developed in Ref.~\cite{Gilkey84} by Gilkey
which is based on the pseudo-differential operator theory (see
Ref.~\cite{FullingKennedy88} for a covariant version).

\begin{acknowledgments}
We are grateful to Bruce Jensen for help with the English as well as
for various discussions concerning quantum field theory in curved
spacetime during the last eighteen years.
\end{acknowledgments}

\appendix

\section{Hadamard form of the DeWitt-Schwinger representation}

In this Appendix, we shall prove that the DeWitt-Schwinger
representation $G^{\mathrm{F}}_{\mathrm{DS}}(x,x')$ of the Feynman
propagator given by Eqs.~(\ref{DSrep1})-(\ref{DSrep2c}) is a
particular case of Hadamard representation. We shall use the method
developed by Christensen in Refs.~\cite{Christensen1,Christensen2}
for the four-dimensional theory and by DeWitt in
Ref.~\cite{DeWitt03} for the $d$-dimensional one in order to obtain
the first divergent terms of the DeWitt-Schwinger representation.
This method can be used to obtain the full expansion of
$G^{\mathrm{F}}_{\mathrm{DS}}(x,x')$ and to show that it is of the
Hadamard form. This was done in the four-dimensional context by
Brown and Ottewill in Ref.~\cite{BrownOttewill83} and we shall here
extend the proceeding for an arbitrary dimension.

\begin{widetext}
We first substitute (\ref{DSrep2c}) into (\ref{DSrep1}). Then, by
assuming that it is possible to exchange the summation and
integration in the resulting expression we find that
\begin{equation}\label{DSformH1}
G^{\mathrm{F}}_{\mathrm{DS}}(x,x') = - (4 \pi)^{-d/2}
\sum_{n=0}^{+\infty} A_n(x,x')  \int_0^{+\infty} ds \, (is)^{-d/2+n}
\, e^{(i/2s)[ \sigma(x,x')+i\epsilon] -i m^2 s}.
\end{equation}
If we assume that $x'$ is in the light cone of $x$, i.e. that
$\sigma(x,x') <0$, we can express the integral in
Eq.~(\ref{DSformH1}) in term of the Hankel function of the second
kind. By using Eqs.~8.421.7 and 8.476.8 of Ref.~\cite{GR5ed}, we
obtain
\begin{equation}\label{DSformH2}
G^{\mathrm{F}}_{\mathrm{DS}}(x,x') = - \pi (4 \pi)^{-d/2} \, i^{-d}
\sum_{n=0}^{+\infty} (-1)^n A_n(x,x')
\left(\frac{z(x,x')}{2m^2}\right)^{-d/2+1+n}
H^{(2)}_{d/2-1-n}(z(x,x')) \end{equation} with
\begin{equation}\label{DSformH3}
z(x,x') = \left( -2m^2[ \sigma(x,x')+i\epsilon]  \right)^{1/2}.
\end{equation}
It should be noted that Eq.~(\ref{DSformH2}) remains valid when $x'$
is outside the light cone of $x$, i.e. when $\sigma(x,x') > 0$,
provided we consider its analytic continuation from the fourth
quadrant of the complex $z(x,x')$ plane to the first one.

Let us now assume $d$ even. From $H^{(2)}_{-n}(z)=(-1)^n
H^{(2)}_n(z)$ which is valid for $n\in \mathbb{N}$ (see Eq.~8.484.2
of Ref.~\cite{GR5ed}) we can write (\ref{DSformH2}) in the form
\begin{eqnarray}\label{DSformH4}
&  & G^{\mathrm{F}}_{\mathrm{DS}}(x,x') = \pi (4 \pi)^{-d/2}
\sum_{n=1}^{d/2-1} (-1)^n {(m^2)}^n A_{d/2-1-n}(x,x')
 \left(\frac{z(x,x')}{2}\right)^{-n}
H^{(2)}_n(z(x,x'))  \nonumber \\
& & \qquad \qquad \qquad + \pi (4 \pi)^{-d/2} \sum_{n=0}^{+\infty}
\frac{1}{{(m^2)}^n} A_{d/2-1+n}(x,x')
 \left(\frac{z(x,x')}{2}\right)^n
H^{(2)}_n(z(x,x')).
\end{eqnarray}
It is then possible to insert into Eq.~(\ref{DSformH4}) the series
expansion
\begin{eqnarray}\label{DSformH5}
&  & H^{(2)}_n(z)=\left[ 1-\left(2i/\pi\right) \ln
\left(\frac{z}{2}\right) \right] \left(\frac{z}{2}\right)^n
\sum_{k=0}^{+\infty} (-1)^k \frac{(z/2)^{2k}}{k! (n+k)!} \nonumber \\
&  & \quad +(i/\pi) (1-\delta_{n0}) \sum_{k=0}^{n-1}
\frac{(n-k-1)!}{k!}\left(\frac{z}{2}\right)^{-n+2k}+(i/\pi)
\sum_{k=0}^{+\infty}(-1)^k
\frac{\psi(k+1)+\psi(n+k+1)}{k!(n+k)!}\left(\frac{z}{2}\right)^{n+2k}
\end{eqnarray}
which is valid for $n\in \mathbb{N}$ and $|\arg z|<\pi$ (see
Eqs.~8.402 and 8.403 of Ref.~\cite{GR5ed}). Here $\psi$ denotes the
logarithm derivative of the gamma function and it is given by (see
Eq.~8.362.1 of Ref.~\cite{GR5ed})
\begin{equation}\label{DSformH6}
\psi (z)= -\gamma -\sum_{\ell=0}^{+\infty}
\left(\frac{1}{z+\ell}-\frac{1}{1+\ell} \right)
\end{equation}
where $\gamma$ is the Euler constant. A tedious calculation permits
us to prove that $G^{\mathrm{F}}_{\mathrm{DS}}(x,x')$ has the
Hadamard form (\ref{HevRep1})-(\ref{HevRep2}) with the Hadamard
coefficients $U_n(x,x')$ and $V_n(x,x')$ respectively given by
(\ref{RelUetAp2a}) and (\ref{RelUetAp2b}) and with the Hadamard
coefficients $W_n(x,x')$ given by
\begin{eqnarray}\label{DSformH7}
&  & W_n(x,x') = \ln (m^2/2) V_n(x,x') - \left[\psi(n+1) +
\psi(n+d/2) \right]V_n(x,x')  \nonumber \\
& & \quad  -\frac{(-1)^n}{2^{n+d/2-1}n! (d/2-2)!}
\left[\sum_{k=0}^{n+d/2-2} \frac{(-1)^k {(m^2)}^k}{k!} \left(
\sum_{\ell=k+1}^{n+d/2-1} \frac{1}{\ell} \right) A_{n+d/2-1-k}(x,x')
- \sum_{k=0}^{+\infty} \frac{k!}{{(m^2)}^{k+1}} A_{n+d/2+k}(x,x')
\right]. \nonumber \\
\end{eqnarray}
Another tedious calculation using Eqs.~(\ref{DSrep3}),
(\ref{HevRep5B}), (\ref{RelUetAp2b}) and (\ref{DSformH6}) permits us
to verify that these coefficients satisfy the recursion relations
(\ref{HevRep6}). It is finally interesting to note the pathological
behavior of the Hadamard coefficients $W_n(x,x')$ for $m^2 \to 0$
(infra-red divergence).

Let us now assume $d$ odd. From $H^{(2)}_{-n-1/2}(z)=-i(-1)^n
H^{(2)}_{n+1/2}(z)$ valid for $n\in \mathbb{N}$ (see Eq.~8.484.2 of
Ref.~\cite{GR5ed}) we can write (\ref{DSformH2}) in the form
\begin{eqnarray}\label{DSformH8}
&  & G^{\mathrm{F}}_{\mathrm{DS}}(x,x') = -i\pi (4 \pi)^{-d/2}
\sum_{n=0}^{d/2-3/2} (-1)^n {(m^2)}^{n+1/2} A_{d/2-3/2-n}(x,x')
 \left(\frac{2}{z(x,x')}\right)^{n+1/2}
H^{(2)}_{n+1/2}(z(x,x'))  \nonumber \\
& & \qquad \qquad \qquad + \pi (4 \pi)^{-d/2} \sum_{n=0}^{+\infty}
\frac{1}{{(m^2)}^{n+1/2}} A_{d/2-1/2+n}(x,x')
 \left(\frac{z(x,x')}{2}\right)^{n+1/2}
H^{(2)}_{n+1/2}(z(x,x')).
\end{eqnarray}
It is then possible to insert into Eq.~(\ref{DSformH8}) the series
expansion
\begin{eqnarray}\label{DSformH9}
&  & H^{(2)}_{n+1/2}(z)=(-1)^n i \left(\frac{2}{z}\right)^{n+1/2}
\left[ \sum_{k=0}^{+\infty}  \frac{1}{k! \Gamma (k-n+1/2)}
\left(\frac{i z}{2}\right)^{2k} -  \sum_{k=n}^{+\infty}
\frac{1}{(k-n)! \Gamma (k+3/2)} \left(\frac{i z}{2}\right)^{2k+1}
\right]
\end{eqnarray}
which is valid for $n\in \mathbb{N}$ and $|\arg z|<\pi$. We have not
found this useful expansion in the literature. We have constructed
it from
\begin{equation}\label{DSformH10}
H^{(2)}_{n+1/2}(z)=\sqrt{\frac{2}{\pi z}} \, i^{n+1} e^{-iz}
\sum_{k=0}^{n} \frac{(n+k)!}{k! (n-k)!}
 \frac{1}{(2iz)^k}
\end{equation}
which is valid for $n\in \mathbb{N}$ and $|\arg z|<\pi$ (see
Eqs.~8.466.2 of Ref.~\cite{GR5ed}) by replacing $e^{-iz}$ by its
series expansion. Then, an easy calculation permits us to prove that
$G^{\mathrm{F}}_{\mathrm{DS}}(x,x')$ has the Hadamard form
(\ref{HodRep1})-(\ref{HodRep2}) with the Hadamard coefficients
$U_n(x,x')$ given by (\ref{RelUetAo2}) and with the Hadamard
coefficients $W_n(x,x')$ given by
\begin{eqnarray}\label{DSformH11}
& & W_n(x,x') =   -\frac{(-1)^n }{2^{n+d/2-1}n! \Gamma(d/2-1)}
\left[\sum_{k=0}^{n+d/2-3/2} \frac{(-1)^k
{(m^2)}^{k+1/2}}{\Gamma(k+3/2)}  \, \pi \, A_{n+d/2-3/2-k}(x,x') \right. \nonumber \\
&& \left. \qquad \qquad\qquad \phantom{\sum_{k=0}^{n+d/2-3/2}}-
\sum_{k=0}^{+\infty} \frac{\Gamma(k+1/2)}{{(m^2)}^{k+1/2}}
A_{n+d/2-1/2+k}(x,x')\right].
\end{eqnarray}
Using Eq.~(\ref{DSrep3}), it is easy to verify that these
coefficients satisfy the recursion relations (\ref{HodRep4}). Here
again, it is interesting to note the pathological behavior of the
Hadamard coefficients $W_n(x,x')$ for $m^2 \to 0$ (infra-red
divergence).
\end{widetext}

\section{Covariant Taylor series expansions of the bitensors $\Lambda^{\mu \nu}=\sigma^{;\mu \nu}$
and $H^{\mu \nu}=g^\nu_{~ \, \nu'}\sigma^{;\mu \nu'}$}

In this Appendix, we shall provide, up to order $\sigma^{9/2}$, the
covariant Taylor series expansions of the bitensors $\Lambda(x,x')$
and $H(x,x')$ which are both tensors of type $(1,1)$ in $x$ and
scalars in $x'$ and of which components are respectively given by
$\Lambda^{\mu}_{\phantom{\mu}
\nu}=\sigma^{;\mu}_{\phantom{;\mu}\nu}$ and $H^{\mu}_{\phantom{\mu}
\nu}=g_{\nu \nu'}\sigma^{;\mu \nu'}$. The notations used in this
Appendix as well as the results obtained will be extensively used in
Appendixes C,D and E.

With Avramidi \cite{Avramidi_PhD,AvramidiLNP2000}, we first
introduce the bitensors $K_{(p)}(x,x')$ and $\gamma_{(p)}(x,x')$
with $p \ge 2$ which are all tensors of type $(1,1)$ in $x$ and
scalars in $x'$. The bitensors $K_{(p)}(x,x')$ are defined from
their components
\begin{subequations} \label{AppK1}
\begin{eqnarray}\label{AppK1a}
&  & K_{(p) \phantom{\mu} \nu}^{\phantom{(p)} \mu}(x,x')=
K^{\mu}_{\phantom{\mu}\nu \,\, a_1  \dots a_p}(x)
\sigma^{;a_1}(x,x')
\dots \sigma^{;a_p}(x,x') \nonumber \\
& &
\end{eqnarray}
with
\begin{equation}\label{AppK1b}
K^{\mu}_{\phantom{\mu}\nu \,\, a_1 a_2  a_3 \dots
a_p}=R^{\mu}_{\phantom{\mu}(a_1 |\nu | a_2; a_3 \dots a_p)}.
\end{equation}
\end{subequations}
The bitensors $\gamma_{(p)}(x,x')$ are constructed from the
bitensors $K_{(p)}(x,x')$. They are defined by the relation
\begin{widetext}
\begin{eqnarray}\label{AppGamma2}
{\gamma_{(p)}} &=& \sum_{1 \le  k \le  [p/2]} (-1)^{k+1} (2k)! \left( \begin{array}{c}   p     \\
2k \end{array} \right)
\sum_{\begin{subarray}{l} p_1,\dots,p_k \ge 2 \\
p_1+\dots+p_k=p \end{subarray}}  \left( \begin{array}{ccc}  & p-2k &   \\
p_1-2, & \dots  &, p_k-2 \end{array} \right) \nonumber \\
& & \times \frac{K_{(p_k)}}{p(p + 1)}\frac{K_{(p_{k-1})}}{(p_1+\dots
+p_{k-1})(p_1 +\dots +p_{k-1}+ 1)} \dots
\frac{K_{(p_2)}}{(p_1+p_2)(p_1+p_2 + 1)}\frac{K_{(p_1)}}{p_1(p_1 +
1)}   \nonumber \\
\end{eqnarray}
where
\begin{eqnarray}\label{AppCoeffsBIandMULT}
\left( \begin{array}{c}    p    \\
k   \end{array} \right) = \frac{p!}{k! \, (p-k)!} \quad \mathrm{and} \quad \left( \begin{array}{ccc}  & p &   \\
p_1, & \dots  &, p_k \end{array} \right) = \frac{p!}{p_1! \dots
p_k!} \quad \mathrm{if} \quad p_1 + \dots + p_k = p.
\end{eqnarray}
The components of the ${\gamma_{(p)}}(x,x')$ are therefore of the
form
\begin{equation}\label{AppGamma1b}
\gamma_{(p) \phantom{\mu} \nu}^{\phantom{(p)} \mu}(x,x') =
\gamma^\mu_{\phantom{\mu}\nu \,\, a_1 \dots a_p}(x)
\sigma^{;a_1}(x,x') \dots \sigma^{;a_p}(x,x').
\end{equation}
We have obtained the expressions of the bitensors $\gamma_{(p)}$ for
$p=2, \dots, 11$. The results are
\begin{subequations}\label{AppGamma3}
\begin{eqnarray}
& & \gamma_{(2)} =   (1/3)  \,  K_{(2)}  \label{AppGamma3_2}\\
& & \gamma_{(3)} = (1/2)  \,   K_{(3)}  \\
& &\gamma_{(4)} = (3/5)  \,  K_{(4)}  - (1/5)  \, { K_{(2)} }^2 \\
& &\gamma_{(5)} =  (2/3)  \,   K_{(5)}  -(1/3)  \,  K_{(2)}
 K_{(3)}
-(2/3)  \, K_{(3)}  K_{(2)} \\
& &\gamma_{(6)} = (5/7)  \,   K_{(6)} -(3/7)  \,  K_{(2)} K_{(4)}
-(10/7)  \,   { K_{(3)}}^2  -(10/7)  \,   K_{(4)} K_{(2)}  +  (1/7)  \,   { K_{(2)}}^3 \\
& &\gamma_{(7)}= (3/4)  \,    K_{(7)}- (1/2) \,  K_{(2)} K_{(5)}-
(9/4)  \,    K_{(3)} K_{(4)} - (15/4)  \,    K_{(4)} K_{(3)}-
(5/2) \,  K_{(5)} K_{(2)} \nonumber \\
& & \qquad + (1/4) \, { K_{(2)}}^2 K_{(3)} + (1/2) \,  K_{(2)} K_{(3)} K_{(2)}+ (3/4) \,    K_{(3)} { K_{(2)}}^2 \\
& & \gamma_{(8)} = (7/9)  \,  K_{(8)} - (5/9) \, K_{(2)}K_{(6)}
-(28/9) \,  K_{(3)}K_{(5)} -7 \, {K_{(4)}}^2  -(70/9)  \,
K_{(5)}K_{(3)} \nonumber \\
&  & \qquad - (35/9) \, K_{(6)}K_{(2)} + (1/3) \,
{K_{(2)}}^2{K_{(4)}} + (10/9) \, {K_{(2)}}{K_{(3)}}^2  + (10/9) \,
{K_{(2)}}{K_{(4)}}{K_{(2)}} \nonumber \\
&  & \qquad  + (14/9) \, {K_{(3)}}{K_{(2)}}{K_{(3)}} + (28/9) \,
{K_{(3)}}^2{K_{(2)}} + (7/3) \, {K_{(4)}}{K_{(2)}}^2 -(1/9)  \,
{K_{(2)}}^4 \\
& &\gamma_{(9)} =  (4/5)  \, K_{(9)} -(3/5) \,  K_{(2)} K_{(7)} -
4  \,  K_{(3)}K_{(6)} -(56/5) \, K_{(4)}K_{(5)} -(84/5) \, K_{(5)}K_{(4)} \nonumber \\
&  & \qquad  - 14  \,  K_{(6)}K_{(3)} -(28/5) \,  K_{(7)}K_{(2)} +
(2/5) \, {K_{(2)}}^2 K_{(5)}+ (9/5) \, K_{(2)}
K_{(3)}K_{(4)}  \nonumber \\
&  & \qquad + 3 \,   K_{(2)} K_{(4)}K_{(3)} + 2 \,   K_{(2)}
K_{(5)}K_{(2)} +  (12/5)  \,   K_{(3)} K_{(2)}K_{(4)} + 8 \,
{K_{(3)}}^3 + 8 \,   K_{(3)} K_{(4)}K_{(2)} \nonumber \\
&  & \qquad   + (28/5) \,   K_{(4)} K_{(2)}K_{(3)} + (56/5) \,
K_{(4)} K_{(3)}K_{(2)} + (28/5) \,
K_{(5)} {K_{(2)}}^2 - (1/5) \, {K_{(2)}}^3 K_{(3)}\nonumber \\
&  & \qquad  - (2/5) \, {K_{(2)}}^2 K_{(3)}{K_{(2)}}- (3/5) \,
K_{(2)}K_{(3)}{K_{(2)}}^2-(4/5) \, K_{(3)} {K_{(2)}}^3
\end{eqnarray}
and
\begin{eqnarray}
& & \gamma_{(10)} = (9/11)  \, K_{(10)} - (7/11) \, K_{(2)}K_{(8)}
-(54/11)  \,  K_{(3)}K_{(7)} -(180/11) \, K_{(4)}K_{(6)} -(336/11)
\,   {K_{(5)}}^2 \nonumber \\
& &  \qquad -(378/11) \, K_{(6)}K_{(4)} -(252/11)  \, K_{(7)}K_{(3)}
- (84/11) \, K_{(8)}K_{(2)} +(5/11)  \,
{K_{(2)}}^2K_{(6)} \nonumber \\
& &  \qquad + (28/11)  \,   K_{(2)} K_{(3)} K_{(5)} +(63/11) \,
K_{(2)}{K_{(4)}}^2  + (70/11)  \,   K_{(2)}  K_{(5)}K_{(3)} +
(35/11)  \, K_{(2)}  K_{(6)}K_{(2)} \nonumber \\
& &  \qquad + (36/11)  \,   K_{(3)} K_{(2)}K_{(5)} + (162/11)  \,
{K_{(3)}}^2K_{(4)} + (270/11)  \, K_{(3)} K_{(4)} K_{(3)}  +(180/11)
\,   K_{(3)} K_{(5)} K_{(2)} \nonumber \\
& &  \qquad +(108/11)  \, K_{(4)} K_{(2)} K_{(4)} + (360/11)  \,
K_{(4)}{K_{(3)}}^2  + (360/11) \, {K_{(4)}}^2K_{(2)} + (168/11) \,
K_{(5)}
K_{(2)} K_{(3)} \nonumber \\
& &  \qquad  + (336/11)  \,   K_{(5)} K_{(3)} K_{(2)}    + (126/11)
\, K_{(6)}{K_{(2)}}^2 - (3/11)  \, {K_{(2)}}^3K_{(4)}   - (10/11) \,
{K_{(2)}}^2{K_{(3)}}^2 \nonumber \\
& &  \qquad - (10/11) \, {K_{(2)}}^2{K_{(4)}}{K_{(2)}}   - (14/11)
\, K_{(2)} K_{(3)} K_{(2)} K_{(3)} - (28/11) \,   K_{(2)}
{K_{(3)}}^2 K_{(2)} \nonumber \\
& &  \qquad - (21/11)  \,   K_{(2)} K_{(4)}{K_{(2)}}^2   - (18/11)
\, K_{(3)} {K_{(2)}}^2 K_{(3)}  - (36/11)  \, K_{(3)}K_{(2)} K_{(3)}
K_{(2)} \nonumber \\
& &  \qquad -(54/11) \, {K_{(3)}}^2{K_{(2)}}^2
 -(36/11) \, {K_{(4)}}{K_{(2)}}^3 + (1/11) \, {K_{(2)}}^5
\label{AppGamma3_10} \\
& & \gamma_{(11)} = (5/6)  \,  K_{(11)} - (2/3) \, K_{(2)}K_{(9)} -
(35/6)  \,  K_{(3)}K_{(8)}  - (45/2) \, K_{(4)}K_{(7)} -50 \,
K_{(5)}K_{(6)}  \nonumber \\
& & \qquad -70 \, K_{(6)}K_{(5)}   -63 \, K_{(7)}K_{(4)} -35 \,
K_{(8)}K_{(3)} -10 \, K_{(9)}K_{(2)}+ (1/2)  \,   {K_{(2)}}^2
K_{(7)} \nonumber \\
& & \qquad + (10/3) \, K_{(2)} K_{(3)} K_{(6)}   + (28/3) \, K_{(2)}
K_{(4)} K_{(5)} + 14
 \,   K_{(2)} K_{(5)} K_{(4)} + (35/3)  \,
K_{(2)} K_{(6)} K_{(3)} \nonumber \\
&  & \qquad + (14/3)  \, K_{(2)} K_{(7)} K_{(2)} + (25/6)  \,
K_{(3)} K_{(2)} K_{(6)} + (70/3)  \, {K_{(3)}}^2 K_{(5)}  + (105/2)
\,   K_{(3)} {K_{(4)}}^2   \nonumber \\
&  & \qquad+ (175/3) \, K_{(3)} K_{(5)} K_{(3)} + (175/6) \, K_{(3)}
K_{(6)} K_{(2)} + 15 \, K_{(4)} K_{(2)} K_{(5)} + (135/2) \, K_{(4)}
K_{(3)} K_{(4)} \nonumber \\
&  & \qquad  + (225/2)  \, {K_{(4)}}^2 K_{(3)}  + 75 \, K_{(4)}
K_{(5)} K_{(2)} + 30 \, K_{(5)} K_{(2)} K_{(4)} + 100  \,K_{(5)}
{K_{(3)}}^2 + 100 \, K_{(5)} K_{(4)} K_{(2)} \nonumber \\
&  & \qquad + 35  \, K_{(6)} K_{(2)} K_{(3)}  + 70  \, K_{(6)}
K_{(3)} K_{(2)}  + 21 K_{(7)} {K_{(2)}}^2 -(1/3) \, {K_{(2)}}^3
K_{(5)} -(3/2) \, {K_{(2)}}^2 K_{(3)}K_{(4)} \nonumber \\
&  & \qquad - (5/2)  \, {K_{(2)}}^2 K_{(4)}K_{(3)} - (5/3)  \,
{K_{(2)}}^2 K_{(5)}K_{(2)} - 2 \, K_{(2)} K_{(3)} K_{(2)} K_{(4)} -
(20/3)  \, K_{(2)} {K_{(3)}}^3 \nonumber \\
&  & \qquad - (20/3) \, K_{(2)} K_{(3)} K_{(4)} K_{(2)}  - (14/3) \,
K_{(2)} K_{(4)} K_{(2)} K_{(3)}     - (28/3) \, K_{(2)} K_{(4)}
K_{(3)}K_{(2)} \nonumber \\
&  & \qquad - (14/3) \, K_{(2)} K_{(5)} {K_{(2)}}^2  - (5/2) \,
K_{(3)} {K_{(2)}}^2 K_{(4)}    - (25/3) \, K_{(3)} K_{(2)}
{K_{(3)}}^2 - (25/3) \, K_{(3)}K_{(2)} K_{(4)} K_{(2)} \nonumber \\
&  & \qquad  - (35/3) \, {K_{(3)}}^2 K_{(2)} {K_{(3)}}  - (70/3) \,
{K_{(3)}}^3 K_{(2)}  - (35/2)  \, K_{(3)} K_{(4)} {K_{(2)}}^2 -
(15/2)  \,
K_{(4)} {K_{(2)}}^2 K_{(3)} \nonumber \\
&  & \qquad -15 \, K_{(4)}K_{(2)} K_{(3)}K_{(2)}  - (45/2)  \,
K_{(4)}K_{(3)} {K_{(2)}}^2  -10 \, K_{(5)}{K_{(2)}}^3 + (1/6)  \,
  {K_{(2)}}^4 K_{(3)} + (1/3)  \,
  {K_{(2)}}^3 K_{(3)} {K_{(2)}} \nonumber \\
&  & \qquad  + (1/2)  \,
  {K_{(2)}}^2 K_{(3)} {K_{(2)}}^2 + (2/3)  \,
  {K_{(2)}} K_{(3)} {K_{(2)}}^3 + (5/6)  \,
  K_{(3)} {K_{(2)}}^4. \label{AppGamma3_11}
\end{eqnarray}
\end{subequations}
\end{widetext}
In the present Appendix, we shall need only the $\gamma_{(p)}$ with
$p=2, \dots, 9$ in order to construct the expansions of $\Lambda$
and $H$ up to order $\sigma^{9/2}$, but in Appendix C, we shall need
also their expressions for $p=10,11$ in order to obtain the
expansion of $\Delta ^{1/2}$ up to order $\sigma^{11/2}$. It should
be noted that Eqs.~(\ref{AppGamma3_2})-(\ref{AppGamma3_11}) provide
compact expressions of the bitensors $\gamma_{(p)}$ with $p=2,
\dots, 11$. By using (\ref{AppGamma1b}) and (\ref{AppK1}) into
(\ref{AppGamma3}), it is also possible to reexpress these relations
at the level of the components $\gamma^\mu_{\phantom{\mu}\nu \,\,
a_1 \dots a_p}$ of the bitensors $\gamma_{(p)}$. Of course, the
results are much more heavy. For example, the components of the
lowest order bitensors $\gamma_{(p)}$ take the form
\begin{widetext}
\begin{subequations}\label{AppGamma4}
\begin{eqnarray}
& & \gamma^\mu_{\phantom{\mu}\nu \,\, a_1 a_2} =
(1/3)  \,  R^{\mu}_{\phantom{\mu}(a_1 |\nu |  a_2)}  \\
& & \gamma^\mu_{\phantom{\mu}\nu \,\, a_1 a_2 a_3} =
(1/2)  \,   R^{\mu}_{\phantom{\mu}(a_1 |\nu |  a_2; a_3)}  \\
& &\gamma^\mu_{\phantom{\mu}\nu \,\, a_1 a_2 a_3 a_4} = (3/5)  \,
R^{\mu}_{\phantom{\mu}(a_1 |\nu |  a_2; a_3 a_4)}
- (1/5)  \, R^{\mu}_{\phantom{\mu}(a_1 |\rho |  a_2}R^{\rho}_{\phantom{\rho}a_3 |\nu |  a_4)} \\
& &\gamma^\mu_{\phantom{\mu}\nu \,\, a_1 a_2 a_3 a_4 a_5} =  (2/3)
\, R^{\mu}_{\phantom{\mu}(a_1 |\nu |  a_2; a_3 a_4 a_5)} -(1/3) \,
R^{\mu}_{\phantom{\mu}(a_1 |\rho | a_2}R^{\rho}_{\phantom{\rho}a_3
|\nu |  a_4; a_5)}  -(2/3)  \, R^{\mu}_{\phantom{\mu}(a_1 |\rho |
 a_2; a_3}
R^{\rho}_{\phantom{\rho}a_4 |\nu |  a_5)} \\
& &\gamma^\mu_{\phantom{\mu}\nu \,\,  a_1 a_2 a_3 a_4 a_5 a_6} =
(5/7) \, R^{\mu}_{\phantom{\mu}(a_1 |\nu |  a_2; a_3 a_4 a_5 a_6)}
-(3/7) \,  R^{\mu}_{\phantom{\mu}(a_1 |\rho |
a_2}R^{\rho}_{\phantom{\rho}a_3 |\nu |  a_4; a_5 a_6)}\nonumber
\\
& & \quad  -(10/7) \, R^{\mu}_{\phantom{\mu}(a_1 |\rho | a_2 ; a_3
}R^{\rho}_{\phantom{\rho}a_4 |\nu |  a_5; a_6)} -(10/7) \,
R^{\mu}_{\phantom{\mu}(a_1 |\rho |  a_2; a_3 a_4
}R^{\rho}_{\phantom{\rho}a_5 |\nu | \, a_6)}   +  (1/7) \,
R^{\mu}_{\phantom{\mu}(a_1 |\rho | a_2}R^{\rho}_{\phantom{\rho}a_3
|\tau |  a_4 }R^{\tau}_{\phantom{\tau}a_5 |\nu |  a_6)}.
\end{eqnarray}
\end{subequations}
\end{widetext}

The bitensors $\gamma_{(p)}(x,x')$ permit us to construct the
covariant Taylor series expansions of the bitensor $H(x,x')$ and of
its inverse denoted by $\Gamma(x,x')$. Of course, the later
satisfies
\begin{equation}\label{ETA1}
H \Gamma = \Gamma H =  1
\end{equation}
and is also a tensor of type $(1,1)$ in $x$ and a scalar in $x'$. We
have \cite{Avramidi_PhD,AvramidiLNP2000}
\begin{equation}\label{AppGamma1}
\Gamma (x,x') = -1 + \sum_{p=2}^{+\infty} \frac{ (-1)^p}{p!}
\gamma_{(p)}(x,x')
\end{equation}
and
\begin{equation}\label{ETA0sc}
H (x,x')=-1+\sum_{p=2}^{+\infty} \frac{(-1)^p}{p!} \eta_{(p)}(x,x')
\end{equation}
with
\begin{eqnarray}\label{ETA2}
& & \eta_{(p)}=  - \sum_{1 \le  k \le  [p/2]}
\sum_{\begin{subarray}{l} p_1,\dots,p_k \ge 2 \\
p_1+\dots+p_k=p \end{subarray}}  \left( \begin{array}{ccc}  & p &   \\
p_1,  & \dots , & p_k \end{array} \right)   \nonumber \\
& & \qquad \qquad \qquad \qquad \qquad \qquad \times \gamma_{(p_k)}
\dots \gamma_{(p_1)}.
\end{eqnarray}
The components of the bitensors $\eta_{(p)}(x,x')$ are therefore of
the form
\begin{eqnarray}\label{ETA2bis}
& & \eta_{(p) \phantom{\mu} \nu}^{\phantom{(p)} \mu} (x,x')=
\eta^{\mu}_{\phantom{\mu} \nu \,\, a_1 \dots a_p}(x)
\sigma^{;a_1}(x,x') \dots \sigma^{;a_p}(x,x'), \nonumber \\
& &
\end{eqnarray}
with this result being a direct consequence of (\ref{AppGamma1b})
and (\ref{ETA2}). It is possible to express the bitensors
$\eta_{(p)}$ in terms of the bitensors $K_{(p)}$. From (\ref{ETA2})
and (\ref{AppGamma3}), we obtain the $\eta_{(p)}$ for $p=2,\dots,
9$. We have:
\begin{widetext}
\begin{subequations}\label{ETA4}
\begin{eqnarray}
& &\eta_{(2)}  = -(1/3) K_{(2)} \\
& &\eta_{(3)} = - (1/2) K_{(3)} \\
& &\eta_{(4)} = -(3/5)\,
K_{(4)} - (7/15)\, {K_{(2)}}^2 \\
&& \eta_{(5)} = -(2/3)\, K_{(5)} - (4/3)\, K_{(2)} K_{(3)} - K_{(3)}
K_{(2)} \\ && \eta_{(6)} = -(5/7)\, K_{(6)} - (18/7)\, K_{(2)}
K_{(4)} - (25/7)\, {K_{(3)}}^2 -
(11/7)\, K_{(4)} K_{(2)} - (31/21)\, {K_{(2)}}^3  \\
& & \eta_{(7)} = -(3/4)\, K_{(7)} - (25/6)\, K_{(2)}K_{(5)} -
(33/4)\, K_{(3)} K_{(4)} - (27/4)\, K_{(4)} K_{(3)} - (13/6)\,
K_{(5)} K_{(2)} \nonumber \\
& & \qquad - (73/12)\, {K_{(2)}}^2 K_{(3)} - (31/6)\,
K_{(2)} K_{(3)} K_{(2)} - (17/4)\, K_{(3)} {K_{(2)}}^2 \label{ETA4_7} \\
& & \eta_{(8)} = -(7/9)\, K_{(8)} - (55/9)\, K_{(2)} K_{(6)} -
(140/9)\, K_{(3)}K_{(5)} - (91/5)\, {K_{(4)}}^2   - (98/9)\,
K_{(5)} K_{(3)} \nonumber \\
& & \qquad - (25/9)\, K_{(6)} K_{(2)} - (239/15)\, {K_{(2)}}^2
K_{(4)} - (226/9)\, K_{(2)} {K_{(3)}}^2 - (106/9)\, K_{(2)}
K_{(4)} K_{(2)} \nonumber \\
& & \qquad - (182/9)\, K_{(3)} K_{(2)} K_{(3)} - (160/9)\,
{K_{(3)}}^2 K_{(2)} - (43/5)\, K_{(4)} {K_{(2)}}^2 - (127/15)\,
{K_{(2)}}^4  \label{ETA4_8} \\
& & \eta_{(9)} = -(4/5)\, K_{(9)} - (42/5)\, K_{(2)}K_{(7)} - 26\,
K_{(3)} K_{(6)} - (196/5)\, K_{(4)}K_{(5)} - (168/5)\, K_{(5)}
K_{(4)} \nonumber \\
& & \qquad - 16\, K_{(6)} K_{(3)} - (17/5)\, K_{(7)} K_{(2)} -
(168/5)\, {K_{(2)}}^2 K_{(5)} - (378/5)\, K_{(2)} K_{(3)} K_{(4)} \nonumber \\
& & \qquad - 66\, K_{(2)} K_{(4)} K_{(3)}  - 22\, K_{(2)} K_{(5)}
K_{(2)} - 60\, K_{(3)} K_{(2)} K_{(4)} - 98\, {K_{(3)}}^3 - 47\,
K_{(3)} K_{(4)} K_{(2)} \nonumber \\
& & \qquad - (232/5)\, K_{(4)} K_{(2)} K_{(3)} - (209/5)\, K_{(4)}
K_{(3)} K_{(2)}  - (74/5)\, K_{(5)} {K_{(2)}}^2 - (226/5)\,
{K_{(2)}}^3 K_{(3)} \nonumber \\
& & \qquad - (197/5)\, {K_{(2)}}^2 K_{(3)} K_{(2)} - (184/5)\,
K_{(2)} K_{(3)} {K_{(2)}}^2  -
  31\, K_{(3)} {K_{(2)}}^3. \label{ETA4_9}
\end{eqnarray}
\end{subequations}
By using (\ref{ETA2bis}) and (\ref{AppK1}) into (\ref{ETA4}), we can
also obtain the expressions of the components
$\eta^{\mu}_{\phantom{\mu} \nu \,\, a_1 \dots a_p}$ of the bitensors
$\eta_{(p)}$. The components of the lowest order bitensors
$\eta_{(p)}$ take the form
\begin{subequations}\label{ETA5}
\begin{eqnarray}
& & \eta^{\mu}_{\phantom{\mu} \nu \,\, a_1 a_2}   =
-(1/3) R^{\mu}_{\phantom{\mu}(a_1 |\nu |   a_2)} \\
& & \eta^{\mu}_{\phantom{\mu} \nu \,\, a_1 a_2 a_3} =
- (1/2) R^{\mu}_{\phantom{\mu}(a_1 |\nu |   a_2; a_3)} \\
& & \eta^{\mu}_{\phantom{\mu} \nu \,\, a_1 a_2 a_3 a_4} = -(3/5)\,
R^{\mu}_{\phantom{\mu}(a_1 |\nu |   a_2; a_3 a_4)}
- (7/15) \, R^{\mu}_{\phantom{\mu}(a_1 |\rho |   a_2 }R^{\rho}_{\phantom{\rho} a_3 |\nu |  a_4)}  \\
&& \eta^{\mu}_{\phantom{\mu} \nu \,\, a_1 a_2 a_3 a_4 a_5} =
-(2/3)\, R^{\mu}_{\phantom{\mu}(a_1 |\nu |   a_2; a_3 a_4 a_5)} -
(4/3)\, R^{\mu}_{\phantom{\mu}(a_1 |\rho |   a_2
}R^{\rho}_{\phantom{\rho} a_3 |\nu |  a_4; a_5)} -
R^{\mu}_{\phantom{\mu}(a_1 |\rho |   a_2; a_3 }
R^{\rho}_{\phantom{\rho} a_4 |\nu |  a_5)}  \\
&& \eta^{\mu}_{\phantom{\mu} \nu \,\, a_1 a_2 a_3 a_4 a_5 a_6} =
-(5/7)\, R^{\mu}_{\phantom{\mu}(a_1 |\nu |   a_2; a_3 a_4 a_5 a_6)}
- (18/7)\, R^{\mu}_{\phantom{\mu}(a_1 |\rho |   a_2
}R^{\rho}_{\phantom{\rho} a_3 |\nu |  a_4; a_5 a_6)} \nonumber \\
& & \qquad - (25/7)\, R^{\mu}_{\phantom{\mu}(a_1 |\rho | a_2; a_3
}R^{\rho}_{\phantom{\rho} a_4 |\nu |  a_5; a_6)}  - (11/7)\,
R^{\mu}_{\phantom{\mu}(a_1 |\rho |   a_2; a_3 a_4
}R^{\rho}_{\phantom{\rho} a_5 |\nu |  a_6)}  - (31/21)\,
R^{\mu}_{\phantom{\mu}(a_1 |\rho | a_2}R^{\rho}_{\phantom{\rho}a_3
|\tau |  a_4 }R^{\tau}_{\phantom{\tau}a_5 |\nu |  a_6)}. \nonumber
\\
\end{eqnarray}
\end{subequations}
\end{widetext}

Finally, we can now construct the covariant Taylor series expansion
of the bitensor $\Lambda(x,x')$ from the covariant Taylor series
expansions of the bitensors $H(x,x')$ and $\Gamma (x,x')$. Indeed,
by differentiating the identity (\ref{DSrep4}) in $x$ and in $x'$,
we obtain the relation $\sigma^{;\nu
\rho'}=\sigma_{;\mu}^{\phantom{;\mu} \nu} \sigma^{;\mu \rho'}+
\sigma_{;\mu} \sigma^{;\mu \nu \rho'}$. We then multiply this result
by $g_{\rho \rho'}$ and taking into account (\ref{Geodetic1a}), we
can write
\begin{equation}\label{AppGetHetL1}
g_{\rho \rho'} \sigma^{;\nu \rho'}=\sigma_{;\mu}^{\phantom{;\mu}
\nu} \left( g_{\rho \rho'}\sigma^{;\mu \rho'} \right) +
\sigma^{;\mu} \nabla_\mu \left(g_{\rho \rho'} \sigma^{;\nu
\rho'}\right).
\end{equation}
This relation links the components of the bitensors $H$ and
$\Lambda$. It can be rewritten in the form
\begin{equation}\label{AppGetHetL2}
 H = \Lambda H + DH
\end{equation}
where we have introduced the differential operator
\begin{equation}\label{AppOPDIFF_D}
D=\sigma^{;\mu} \nabla_\mu .
\end{equation}
As a consequence, we have
\begin{equation}\label{AppGetHetL3}
\Lambda = 1 - (DH) \Gamma.
\end{equation}
The covariant Taylor series expansion of $\Lambda (x,x')$ is
therefore given by
\begin{equation}\label{AppGetHetL4a}
\Lambda (x,x')=1+\sum_{p=2}^{+\infty} \frac{(-1)^p}{p!}
\lambda_{(p)}(x,x')
\end{equation}
where the ${\lambda_{(p)}}(x,x')$ are also tensors of type $(1,1)$
in $x$ and scalars in $x'$ of which components are of the form
\begin{eqnarray}\label{AppGetHetL4b}
& & \lambda_{(p) \phantom{\mu} \nu}^{\phantom{(p)} \mu} (x,x')=
\lambda^\mu_{\phantom{\mu}\nu \,\, a_1 \dots a_p}(x)
\sigma^{;a_1}(x,x') \dots \sigma^{;a_p}(x,x') \nonumber \\
& &
\end{eqnarray}
and it can be constructed from the covariant Taylor series
expansions of $H(x,x')$ and $\Gamma(x,x')$. By noting the identities
\begin{widetext}
\begin{subequations}
\begin{eqnarray}\label{AppDappK}
& & D \left[ K_{(p)}\right]=p \,  K_{(p)} +   K_{(p+1)}   \\
& & D \left[ K_{(p)}K_{(q)}\right]=(p+q) \,
 K_{(p)}K_{(q)} +  K_{(p+1)}K_{(q)}
+  K_{(p)}K_{(q+1)}  \\
& & D \left[ K_{(p)}K_{(q)}K_{(r)}\right]=(p+q+r) \,
K_{(p)}K_{(q)}K_{(r)} + K_{(p+1)}K_{(q)}K_{(r)}
 \nonumber \\
& & \qquad \qquad +  K_{(p)}K_{(q+1)}K_{(r)}
+  K_{(p)}K_{(q)}K_{(r+1)} \\
& & D\left[ K_{(p)}K_{(q)}K_{(r)}K_{(s)}\right]=(p+q+r+s) \,
K_{(p)}K_{(q)}K_{(r)}K_{(s)} + K_{(p+1)}K_{(q)}K_{(r)}K_{(s)}
 \nonumber \\
& & \qquad \qquad +  K_{(p)}K_{(q+1)}K_{(r)}K_{(s)} +
K_{(p)}K_{(q)}K_{(r+1)}K_{(s)} + K_{(p)}K_{(q)}K_{(r)}K_{(s+1)}
\end{eqnarray}
\end{subequations}
which follow from (\ref{AppK1}) and (\ref{DSrep4}), a tedious
calculation using (\ref{AppGetHetL3}), (\ref{AppGetHetL4a}) as well
as (\ref{ETA0sc}), (\ref{ETA4}), (\ref{AppGamma1}) and
(\ref{AppGamma3}) permits us to obtain the $\lambda_{(p)}$ for
$p=2,\dots,9$. We have
\begin{subequations}\label{lambda5}
\begin{eqnarray}
&  & \lambda_{(2)} = -(2/3) \, {K_{(2)}} \label{lambda5_2} \\
&  & \lambda_{(3)} = -(1/2) \, {K_{(3)}} \label{lambda5_3} \\
&  & \lambda_{(4)} = -\left[ (2/5) \, {K_{(4)}} + (8/15) \, {K_{(2)}}^2 \right] \label{lambda5_4}\\
&  & \lambda_{(5)} = -\left[(1/3) \, {K_{(5)}} + {K_{(2)}} {K_{(3)}}
+ {K_{(3)}} {K_{(2)}} \right] \label{lambda5_5} \\
&  & \lambda_{(6)} = -\left[(2/7) \, {K_{(6)}} + (10/7) \, {K_{(2)}}
{K_{(4)}} + (17/7) \, {K_{(3)}}^2 + (10/7) \, {K_{(4)}}
{K_{(2)}} + (32/21) \, {K_{(2)}}^3 \right] \label{lambda5_6} \\
&  & \lambda_{(7)} = -\left[(1/4) \, {K_{(7)}} + (11/6) \, {K_{(2)}}
{K_{(5)}} + (17/4) \, {K_{(3)}}  {K_{(4)}} + (17/4) \, {K_{(4)}}
{K_{(3)}}  \right. \nonumber \\ & & \qquad \left. + (11/6) \,
{K_{(5)}} {K_{(2)}}  + (17/4) \, {K_{(2)}}^2 {K_{(3)}} + (29/6) \,
{K_{(2)}} {K_{(3)}} {K_{(2)}} +
(17/4) \, {K_{(3)}} {K_{(2)}}^2  \right]  \label{lambda5_7} \\
&  & \lambda_{(8)} = -\left[ (2/9) \, {K_{(8)}} + (20/9) \,
{K_{(2)}}{K_{(6)}} + (58/9) \, {K_{(3)}} {K_{(5)}} + (44/5) \,
{K_{(4)}}^2 + (58/9) \, {K_{(5)}}  {K_{(3)}} \right. \nonumber \\
& & \qquad \left. + (20/9) \, {K_{(6)}} {K_{(2)}}  + (42/5) \,
{K_{(2)}}^2 {K_{(4)}} + (146/9) \, {K_{(2)}} {K_{(3)}}^2
  + (92/9) \, {K_{(2)}} {K_{(4)}} {K_{(2)}} \right.
\nonumber \\ & & \qquad \left. + (124/9) \, {K_{(3)}} {K_{(2)}}
{K_{(3)}} + (146/9) \, {K_{(3)}}^2 {K_{(2)}} + (42/5) \, {K_{(4)}}
{K_{(2)}}^2 + (128/15) \, {K_{(2)}}^4  \right] \label{lambda5_8}
\end{eqnarray}
and
\begin{eqnarray}
&  & \lambda_{(9)} = - \left[ (1/5) \, {K_{(9)}} + (13/5) \,
{K_{(2)}} {K_{(7)}} + 9 \, {K_{(3)}} {K_{(6)}} + (77/5) \, {K_{(4)}}
{K_{(5)}} + (77/5) \, {K_{(5)}}  {K_{(4)}} \right. \nonumber \\ & &
\qquad \left. +
  9 \, {K_{(6)}}  {K_{(3)}}  + (13/5) \, {K_{(7)}} {K_{(2)}}
  + (71/5) \, {K_{(2)}}^2 {K_{(5)}} + (187/5) \, {K_{(2)}} {K_{(3)}} {K_{(4)}}
\right. \nonumber \\ & & \qquad \left. + 40 \, {K_{(2)}}  {K_{(4)}}
{K_{(3)}} +
  18 \, {K_{(2)}} {K_{(5)}} {K_{(2)}}  +
  31 \, {K_{(3)}} {K_{(2)}}  {K_{(4)}} +
  62 \, {K_{(3)}}^3 +
  40 \, {K_{(3)}}  {K_{(4)}} {K_{(2)}}
    \right.
\nonumber \\ & & \qquad \left. +
  31 \, {K_{(4)}} {K_{(2)}}  {K_{(3)}} + (187/5) \, {K_{(4)}}{K_{(3)}}{K_{(2)}}
+ (71/5) \, {K_{(5)}} {K_{(2)}}^2 + 31 \, {K_{(2)}}^3  {K_{(3)}}
  \right.
\nonumber \\ & & \qquad \left. + (181/5) \, {K_{(2)}}^2  {K_{(3)}}
{K_{(2)}}
 + (181/5) \, {K_{(2)}}  {K_{(3)}}
{K_{(2)}}^2  +
  31 \, {K_{(3)}} {K_{(2)}}^3 \right]. \label{lambda5_9}
\end{eqnarray}
\end{subequations}
By using (\ref{AppGetHetL4b}) and (\ref{AppK1}) into
(\ref{lambda5}), we can also obtain the expressions of the
components $\gamma^{\mu}_{\phantom{\mu} \nu \,\, a_1 \dots a_p}$ of
the bitensors $\gamma_{(p)}$. The components of the lowest order
bitensors $\lambda_{(p)}$ take the form
\begin{subequations}\label{lambda5bis}
\begin{eqnarray}
& & \lambda^{\mu}_{\phantom{\mu} \nu \,\, a_1 a_2}   =
-(2/3) R^{\mu}_{\phantom{\mu}(a_1 |\nu |   a_2)} \label{lambda5bis_2} \\
& & \lambda^{\mu}_{\phantom{\mu} \nu \,\, a_1 a_2 a_3} =
- (1/2) R^{\mu}_{\phantom{\mu}(a_1 |\nu |   a_2; a_3)}  \label{lambda5bis_3} \\
& & \lambda^{\mu}_{\phantom{\mu} \nu \,\, a_1 a_2 a_3 a_4} = -\left[
(2/5)\, R^{\mu}_{\phantom{\mu}(a_1 |\nu |   a_2; a_3 a_4)}
+ (8/15) \, R^{\mu}_{\phantom{\mu}(a_1 |\rho |   a_2 }
R^{\rho}_{\phantom{\rho} a_3 |\nu |  a_4)}\right]   \label{lambda5bis_4} \\
&& \lambda^{\mu}_{\phantom{\mu} \nu \,\, a_1 a_2 a_3 a_4 a_5} =
-\left[(1/3)\, R^{\mu}_{\phantom{\mu}(a_1 |\nu |   a_2; a_3 a_4
a_5)} + \, R^{\mu}_{\phantom{\mu}(a_1 |\rho |   a_2
}R^{\rho}_{\phantom{\rho} a_3 |\nu |  a_4; a_5)} +
R^{\mu}_{\phantom{\mu}(a_1 |\rho |   a_2; a_3 }
R^{\rho}_{\phantom{\rho} a_4 |\nu |  a_5)}\right] \\
 && \lambda^{\mu}_{\phantom{\mu} \nu \,\, a_1 a_2 a_3 a_4 a_5
a_6} = -\left[ (2/7)\, R^{\mu}_{\phantom{\mu}(a_1 |\nu |   a_2; a_3
a_4 a_5 a_6)} + (10/7)\, R^{\mu}_{\phantom{\mu}(a_1 |\rho |   a_2
}R^{\rho}_{\phantom{\rho} a_3 |\nu |  a_4; a_5 a_6)} \right. \nonumber \\
& & \qquad  \left. + (17/7)\, R^{\mu}_{\phantom{\mu}(a_1 |\rho |
a_2; a_3 }R^{\rho}_{\phantom{\rho} a_4 |\nu |  a_5; a_6)} + (10/7)\,
R^{\mu}_{\phantom{\mu}(a_1 |\rho |   a_2; a_3 a_4
}R^{\rho}_{\phantom{\rho} a_5 |\nu |  a_6)}  + (32/21)\,
R^{\mu}_{\phantom{\mu}(a_1 |\rho | a_2}R^{\rho}_{\phantom{\rho}a_3
|\tau |  a_4 }R^{\tau}_{\phantom{\tau}a_5 |\nu |  a_6)} \right].
\nonumber \\ \label{lambda5bis_6}
\end{eqnarray}
\end{subequations}
\end{widetext}

The relations (\ref{AppGetHetL4a}) and (\ref{lambda5}) provide a
compact form for the covariant Taylor series expansion up to order
$\sigma^{9/2}$ of the bitensor $\Lambda$. Similarly, the relations
(\ref{ETA0sc}) and (\ref{ETA4}) provide a compact form for the
covariant Taylor series expansion up to order $\sigma^{9/2}$ of the
bitensor $H$. It is also possible to provide the covariant Taylor
series expansions of the bitensors $\Lambda$ and $H$ in a more
explicit form, i.e. by working at the level of their components
$\Lambda^{\mu}_{\phantom{\mu}
\nu}=\sigma^{;\mu}_{\phantom{;\mu}\nu}$ and $H^{\mu}_{\phantom{\mu}
\nu}=g_{\nu \nu'}\sigma^{;\mu \nu'}$. Of course, the corresponding
results are much more heavy. From (\ref{AppGetHetL4a}),
(\ref{AppGetHetL4b}) and  (\ref{lambda5bis}) we obtain
\begin{widetext}
\begin{eqnarray}\label{AppLambda6}
& & \sigma_{; \mu \nu} = g_{\mu \nu} - \frac{1}{3}  \, R_{\mu a_1
\nu a_2} \sigma^{;a_1}\sigma^{;a_2}
+\frac{1}{12}  \,   R_{\mu a_1 \nu a_2; a_3}  \sigma^{;a_1}\sigma^{;a_2} \sigma^{;a_3} \nonumber \\
& & \quad - \left[ \frac{1}{60}  \, R_{\mu a_1 \nu a_2; a_3 a_4} +
\frac{1}{45} \, R_{\mu a_1 \rho a_2} R^{\rho}_{\phantom{\rho} a_3
\nu a_4} \right]
\sigma^{;a_1}\sigma^{;a_2} \sigma^{;a_3} \sigma^{;a_4}    \nonumber \\
& & \quad + \left[  \frac{1}{360} \, R_{\mu a_1 \nu a_2; a_3 a_4
a_5} + \frac{1}{120} \, R_{\mu a_1 \rho a_2}
R^{\rho}_{\phantom{\rho} a_3 \nu a_4; a_5} + \frac{1}{120} \, R_{\mu
a_1 \rho a_2; a_3} R^{\rho}_{\phantom{\rho} a_4 \nu a_5}\right]
\sigma^{;a_1}\sigma^{;a_2} \sigma^{;a_3} \sigma^{;a_4} \sigma^{;a_5} \nonumber \\
& & \quad - \left[ \frac{1}{2520} \, R_{\mu a_1 \nu a_2; a_3 a_4 a_5
a_6} + \frac{1}{504} \, R_{\mu a_1 \rho a_2}
R^{\rho}_{\phantom{\rho} a_3 \nu a_4; a_5 a_6} + \frac{17}{5040} \,
R_{\mu a_1 \rho a_2; a_3} R^{\rho}_{\phantom{\rho} a_4 \nu a_5;
a_6}  \right. \nonumber \\
&  & \qquad  \left. + \frac{1}{504} \, R_{\mu a_1 \rho a_2; a_3 a_4}
R^{\rho}_{\phantom{\rho} a_5 \nu a_6}  + \frac{2}{945} \, R_{\mu a_1
\rho a_2}R^{\rho}_{\phantom{\rho} a_3 \tau
a_4}R^{\tau}_{\phantom{\tau} a_5 \nu a_6}
\right]\sigma^{;a_1}\sigma^{;a_2} \sigma^{;a_3} \sigma^{;a_4}
\sigma^{;a_5} \sigma^{;a_6} \nonumber  \\
& & \quad + \, O \left(\sigma^{7/2} \right)
\end{eqnarray}
(here we have not included the term of order $\sigma^{7/2}$
corresponding to (\ref{lambda5_7}), the term of order $\sigma^4$
corresponding to (\ref{lambda5_8})) and the term of order
$\sigma^{9/2}$ corresponding to (\ref{lambda5_9})) while from
(\ref{ETA0sc}), (\ref{ETA2bis}) and (\ref{ETA5}) we obtain
\begin{eqnarray}\label{AppETA5bis} & & g_{\nu \nu'} \sigma_{;
\mu}^{\phantom{;\mu} \nu'} = - g_{\mu \nu} - \frac{1}{6} \, R_{\mu
a_1 \nu a_2} \sigma^{;a_1}\sigma^{;a_2}
+\frac{1}{12}  \,   R_{\mu a_1 \nu a_2; a_3}  \sigma^{;a_1}\sigma^{;a_2} \sigma^{;a_3} \nonumber \\
& & \quad - \left[ \frac{1}{40}  \, R_{\mu a_1 \nu a_2; a_3 a_4} +
\frac{7}{360} \, R_{\mu a_1 \rho a_2} R^{\rho}_{\phantom{\rho} a_3
\nu a_4} \right]
\sigma^{;a_1}\sigma^{;a_2} \sigma^{;a_3} \sigma^{;a_4}    \nonumber \\
& & \quad + \left[  \frac{1}{180} \, R_{\mu a_1 \nu a_2; a_3 a_4
a_5} + \frac{1}{90} \, R_{\mu a_1 \rho a_2} R^{\rho}_{\phantom{\rho}
a_3 \nu a_4; a_5} + \frac{1}{120} \, R_{\mu a_1 \rho a_2; a_3}
R^{\rho}_{\phantom{\rho} a_4 \nu a_5}\right]
\sigma^{;a_1}\sigma^{;a_2} \sigma^{;a_3} \sigma^{;a_4} \sigma^{;a_5} \nonumber \\
& & \quad - \left[ \frac{1}{1008} \, R_{\mu a_1 \nu a_2; a_3 a_4 a_5
a_6} + \frac{1}{280} \, R_{\mu a_1 \rho a_2}
R^{\rho}_{\phantom{\rho} a_3 \nu a_4; a_5 a_6} + \frac{5}{1008} \,
R_{\mu a_1 \rho a_2; a_3} R^{\rho}_{\phantom{\rho} a_4 \nu a_5;
a_6}  \right. \nonumber \\
&  & \qquad  \left. + \frac{11}{5040} \, R_{\mu a_1 \rho a_2; a_3
a_4} R^{\rho}_{\phantom{\rho} a_5 \nu a_6}  + \frac{31}{15120} \,
R_{\mu a_1 \rho a_2}R^{\rho}_{\phantom{\rho} a_3 \tau
a_4}R^{\tau}_{\phantom{\tau} a_5 \nu a_6}
\right]\sigma^{;a_1}\sigma^{;a_2} \sigma^{;a_3} \sigma^{;a_4}
\sigma^{;a_5} \sigma^{;a_6} \nonumber  \\
& & \quad  \, + O \left(\sigma^{7/2} \right)
\end{eqnarray}
(here we have not included the term of order $\sigma^{7/2}$
corresponding to (\ref{ETA4_7}), the term of order $\sigma^4$
corresponding to (\ref{ETA4_8}) and the term of order $\sigma^{9/2}$
corresponding to (\ref{ETA4_9})).
\end{widetext}

It should be noted that the previous expansions were obtained by
DeWitt \cite{DeWittBrehme,DeWitt65} up to order $\sigma$ and by
Christensen \cite{Christensen1,Christensen2} up to order $\sigma^2$.
They have been recently improved by Anderson, Flanagan and Ottewill
\cite{AndersonFlanaganOttewill05} who have obtained the terms
corresponding to the order $\sigma^{5/2}$. In
Ref.~\cite{PhillipsHu03}, Phillips and Hu have calculated the term
of order $\sigma^3$ for the expansion of $\sigma_{;\mu \nu}$ but
their result is incorrect: even if we simplify their equation
(B25d), there remain three terms of which coefficients disagree with
our own results in (\ref{AppLambda6}).

\section{Covariant Taylor series expansions of the biscalars $\Delta ^{1/2} $
and $\Delta ^{-1/2}{\Delta ^{1/2}}_{;\mu} \sigma^{; \mu}$}

In this Appendix, we shall construct the covariant Taylor series
expansions of the biscalars $\Delta ^{1/2} $ and $\Delta
^{-1/2}{\Delta ^{1/2}}_{;\mu} \sigma^{; \mu}$ up to orders
$\sigma^{11/2}$ and $\sigma^{9/2}$ respectively.

We first consider the biscalar $Z$ which is defined as the logarithm
of $\Delta ^{1/2}$. The general form of its covariant Taylor series
expansion has been obtained by Avramidi
\cite{Avramidi_PhD,AvramidiLNP2000}. It is given by
\begin{equation}\label{AppZeta1}
Z(x,x')= \sum_{p=2}^{+\infty} \frac{ (-1)^p}{p!} \zeta_{(p)}(x,x')
\end{equation}
where the $\zeta_{(p)}(x,x')$ are biscalars of the form
\begin{equation}\label{AppZeta1bis}
\zeta_{(p)}(x,x')=  \zeta_{a_1 \dots a_p}(x) \sigma^{;a_1}(x,x')
\dots \sigma^{;a_p}(x,x')
\end{equation}
which can be constructed from the bitensors $\gamma_{(p)}(x,x')$ by
using the relation
\begin{eqnarray}\label{AppZeta2}
& & \zeta_{(p)} = \sum_{1 \le  k \le  [p/2]} \frac{1}{2k}
\sum_{\begin{subarray}{l} p_1,\dots,p_k \ge 2 \\
p_1+\dots+p_k=p \end{subarray}}  \left( \begin{array}{ccc}  & p &   \\
p_1,  & \dots , & p_k \end{array} \right) \nonumber \\
& & \qquad \qquad  \qquad \qquad \qquad \times  {\rm tr} \left(
\gamma_{(p_1)} \dots \gamma_{(p_k)} \right).
\end{eqnarray}
From this relation and from the relations (\ref{AppGamma3}) which
express the $\gamma_{(p)}$ in terms of the $K_{(p)}$, we can obtain
the $\zeta_{(p)}$. After a long calculation, we have found that the
terms corresponding to $p=2,\dots,11$ are given by
\begin{widetext}
\begin{subequations}\label{AppZeta3bis}
\begin{eqnarray}
& & \zeta_{(2)} =   (1/6) \, {\rm tr} \, K_{(2)} \label{AppZeta3bis_2} \\
& & \zeta_{(3)} = (1/4)  \, {\rm tr} \, K_{(3)} \\
& &\zeta_{(4)} = (3/10) \, {\rm tr} \, K_{(4)}+ (1/15)  \, {\rm tr} \, {K_{(2)}}^2\\
& &\zeta_{(5)} =  (1/3)  \, {\rm tr} \, K_{(5)}+ (1/3)  \, {\rm tr} \, K_{(2)}K_{(3)}\\
& &\zeta_{(6)}= (5/14)  \, {\rm tr} \, K_{(6)}+ (4/7) \, {\rm tr} \,
K_{(2)}K_{(4)}
+ (15/28)  \, {\rm tr} \, {K_{(3)}}^2 + (8/63)  \, {\rm tr} \, {K_{(2)}}^3 \label{AppZeta3bis_6} \\
& &\zeta_{(7)}= (3/8)  \, {\rm tr} \, K_{(7)}+ (5/6) \, {\rm tr} \,
K_{(2)}K_{(5)}+ (9/4)  \, {\rm tr} \, K_{(3)}K_{(4)}
+ (4/3)  \, {\rm tr} \, {K_{(2)}}^2 K_{(3)} \label{AppZeta3bis_7} \\
& &\zeta_{(8)} = (7/18)  \, {\rm tr} \, K_{(8)}+ (10/9) \, {\rm tr}
\, K_{(2)}K_{(6)} + (35/9)  \, {\rm tr} \, K_{(3)}K_{(5)} + (14/5)
\, {\rm tr} \, {K_{(4)}}^2 \nonumber \\
&  & \qquad + (136/45)  \, {\rm tr} \, {K_{(2)}}^2 K_{(4)}    +
(50/9)  \, {\rm tr} \, K_{(2)} {K_{(3)}}^2 +
(8/15) \, {\rm tr} \, {K_{(2)}}^4 \\
& &\zeta_{(9)} =  (2/5)  \, {\rm tr} \, K_{(9)}+ (7/5) \, {\rm tr}
\, K_{(2)} K_{(7)} + 6 \, {\rm tr} \, K_{(3)} K_{(6)}+ (56/5) \,
{\rm tr} \, K_{(4)} K_{(5)} \nonumber \\
&  & \qquad + (28/5)  \, {\rm tr} \, {K_{(2)}}^2 K_{(5)}   + (73/5)
\, {\rm tr} \, K_{(2)} K_{(3)} K_{(4)} + (73/5) \, {\rm tr} \,
K_{(2)} K_{(4)}K_{(3)}+ 9 \, {\rm tr} \, {K_{(3)}}^3 \nonumber \\
&  & \qquad + (48/5)
 \, {\rm tr} \, {K_{(2)}}^3 K_{(3)}\\
 & & \zeta_{(10)} = (9/22)  \, {\rm tr} \, K_{(10)}+
(56/33) \, {\rm tr} \, K_{(2)} K_{(8)} + (189/22)  \, {\rm tr} \,
K_{(3)} K_{(7)} + (216/11)  \, {\rm tr} \,
K_{(4)}K_{(6)} \nonumber \\
&  & \qquad + (140/11)  \, {\rm tr} \, {K_{(5)}}^2 + (304/33) \,
{\rm tr} \, {K_{(2)}}^2 K_{(6)} + (1015/33) \, {\rm tr} \, K_{(2)}
K_{(3)} K_{(5)}  \nonumber \\
& &  \qquad + (480/11) \, {\rm tr} \, K_{(2)}{K_{(4)}}^2   +
(1015/33)  \, {\rm tr} \, K_{(2)} K_{(5)} K_{(3)}    + 81  \, {\rm
tr} \, {K_{(3)}}^2 K_{(4)} \nonumber \\
& &  \qquad  + (896/33) \, {\rm tr} \, {K_{(2)}}^3 K_{(4)} + (149/3)
\, {\rm tr} \, {K_{(2)}}^2{K_{(3)}}^2  + (805/33) \, {\rm tr} \,
K_{(2)} K_{(3)} K_{(2)}
K_{(3)} \nonumber \\
& &  \qquad  + (128/33) \, {\rm tr} \, { K_{(2)}}^5
\end{eqnarray}
and
\begin{eqnarray}
& & \zeta_{(11)} = (5/12)  \, {\rm tr} \, K_{(11)}+ 2 \, {\rm tr} \,
K_{(2)} K_{(9)} + (35/3)  \, {\rm tr} \, K_{(3)} K_{(8)}+ (63/2) \,
{\rm tr} \, K_{(4)} K_{(7)} + 50 \, {\rm tr} \,
 K_{(5)} K_{(6)}   \nonumber \\
& & \qquad + 14   \, {\rm tr} \, {K_{(2)}}^2 K_{(7)} + (170/3) \,
{\rm tr} \, K_{(2)} K_{(3)} K_{(6)} + 103 \, {\rm tr} \, K_{(2)}
K_{(4)} K_{(5)} + 103
 \, {\rm tr} \, K_{(2)} K_{(5)} K_{(4)}   \nonumber \\
&  & \qquad + (170/3)  \, {\rm tr} \, K_{(2)} K_{(6)} K_{(3)} +
(575/3) \, {\rm tr} \, { K_{(3)}}^2  K_{(5)} + 273  \, {\rm tr} \,
K_{(3)} {K_{(4)}}^2
+ (184/3)  \, {\rm tr} \, { K_{(2)}}^3  K_{(5)} \nonumber \\
& & \qquad  + (317/2)  \, {\rm tr} \, {K_{(2)}}^2 K_{(3)} K_{(4)} +
(317/2) \, {\rm tr} \, {K_{(2)}}^2 K_{(4)}K_{(3)} + (461/3)
 \, {\rm tr} \, K_{(2)} K_{(3)} K_{(2)}
K_{(4)}
\nonumber \\
&  & \qquad + (860/3)  \, {\rm tr} \, K_{(2)} {K_{(3)}}^3
 + (320/3) \, {\rm tr} \, {K_{(2)}}^4 K_{(3)}.
\end{eqnarray}
\end{subequations}
By using (\ref{AppZeta1bis}) and (\ref{AppK1}) into
(\ref{AppZeta3bis}), we can also obtain the expressions of the
components $\zeta_{a_1 \dots a_p}$ of the biscalars $\zeta_{(p)}$.
The components of the lowest order biscalars $\zeta_{(p)}$ take the
form
\begin{subequations}\label{AppZeta4}
\begin{eqnarray}
& & \zeta_{a_1 a_2} =
(1/6)  \,  R_{a_1 a_2}  \label{AppZeta4_2}\\
& & \zeta_{a_1 a_2 a_3} =
(1/4)  \,   R_{(a_1 a_2; a_3)}  \\
& &\zeta_{a_1 a_2 a_3 a_4} = (3/10)  \, R_{(a_1 a_2; a_3 a_4)} +
(1/15)  \, R^{\rho}_{\phantom{\rho}(a_1 |\tau| a_2}
R^{\tau}_{\phantom{\tau} a_3 |\rho |  a_4)} \\
& &\zeta_{a_1 a_2 a_3 a_4 a_5} =  (1/3) \, R_{(a_1 a_2; a_3 a_4
a_5)} + (1/3) R^{\rho}_{\phantom{\rho}(a_1 |\tau| a_2}
R^{\tau}_{\phantom{\tau} a_3 |\rho |  a_4;a_5)} \\ & &\zeta_{a_1 a_2
a_3 a_4 a_5 a_6} = (5/14) \, R_{(a_1 a_2; a_3 a_4 a_5 a_6)} +(4/7)
\, R^{\rho}_{\phantom{\rho}(a_1 |\tau| a_2}
R^{\tau}_{\phantom{\tau} a_3 |\rho |  a_4;a_5 a_6)} \nonumber \\
 &  & \qquad \qquad + (15/28) \,  R^{\rho}_{\phantom{\rho}(a_1 |\tau| a_2; a_3}
R^{\tau}_{\phantom{\tau} a_4 |\rho |  a_5; a_6)} + (8/63) \,
R^{\rho}_{\phantom{\rho}(a_1 |\tau | a_2 }R^{\tau}_{\phantom{\tau}
a_3 |\sigma | a_4}R^{\sigma}_{\phantom{\sigma} a_5 |\rho | a_6)}
\label{AppZeta4_6}
\\
& &\zeta_{a_1 a_2 a_3 a_4 a_5 a_6 a_7} = (3/8) \, R_{(a_1 a_2; a_3
a_4 a_5 a_6 a_7)} + (5/6) \,  R^{\rho}_{\phantom{\rho}(a_1 |\tau|
a_2} R^{\tau}_{\phantom{\tau} a_3 |\rho |  a_4; a_5 a_6 a_7)}
\nonumber
\\
&  & \qquad \qquad + (9/4) \, R^{\rho}_{\phantom{\rho}(a_1 |\tau|
a_2; a_3} R^{\tau}_{\phantom{\tau} a_4 |\rho |  a_5; a_6 a_7)} +
(4/3) \, R^{\rho}_{\phantom{\rho}(a_1 |\tau | a_2
}R^{\tau}_{\phantom{\tau}
a_3 |\sigma | a_4}R^{\sigma}_{\phantom{\sigma} a_5 |\rho | a_6;a_7)}\\
& &\zeta_{a_1 a_2 a_3 a_4 a_5 a_6 a_7 a_8} = (7/18) \, R_{(a_1 a_2;
a_3 a_4 a_5 a_6 a_7 a_8)} +(10/9) \, R^{\rho}_{\phantom{\rho}(a_1
|\tau| a_2} R^{\tau}_{\phantom{\tau} a_3 |\rho |  a_4 ; a_5 a_6 a_7
a_8)} \nonumber
\\
&  & \qquad \qquad + (35/9) \,  R^{\rho}_{\phantom{\rho}(a_1 |\tau|
a_2; a_3} R^{\tau}_{\phantom{\tau} a_4 |\rho |  a_5 ; a_6 a_7 a_8)}
+ (14/5) \, R^{\rho}_{\phantom{\rho}(a_1 |\tau| a_2; a_3 a_4 }
R^{\tau}_{\phantom{\tau} a_5 |\rho |  a_6 ; a_7 a_8)} \nonumber
\\
&  & \qquad \qquad + (136/45) \, R^{\rho}_{\phantom{\rho}(a_1 |\tau
| a_2 }R^{\tau}_{\phantom{\tau} a_3 |\sigma |
a_4}R^{\sigma}_{\phantom{\sigma} a_5 |\rho | a_6; a_7 a_8)} + (50/9)
\, R^{\rho}_{\phantom{\rho}(a_1 |\tau | a_2
}R^{\tau}_{\phantom{\tau} a_3 |\sigma | a_4; a_5
}R^{\sigma}_{\phantom{\sigma} a_6 |\rho | a_7; a_8)}\nonumber
\\
&  & \qquad \qquad + (8/15)  \, R^{\rho}_{\phantom{\rho}(a_1 |\tau|
a_2 }R^{\tau}_{\phantom{\tau} a_3 |\sigma|
a_4}R^{\sigma}_{\phantom{\sigma} a_5 |\kappa|  a_6
}R^{\kappa}_{\phantom{\kappa} a_7 |\rho|  a_8 )}.
\end{eqnarray}
\end{subequations}
\end{widetext}

The covariant Taylor series expansion of the biscalar $\Delta
^{1/2}$ can now be constructed from (\ref{AppZeta1}) and
(\ref{AppZeta3bis}). By noting that
\begin{equation}\label{AppVVM}
\Delta ^{1/2}=e^Z
\end{equation}
we can write this expansion in the form
\begin{equation}\label{AppSCU01}
\Delta ^{1/2}(x,x')=1+\sum_{p=2}^{+\infty} \frac{(-1)^p}{p!}
{\delta^{1/2}}_{(p)}(x,x')
\end{equation}
where the biscalars ${\delta^{1/2}}_{(p)}(x,x')$ are of the form
\begin{eqnarray}\label{AppSCU02}
& & {\delta^{1/2}}_{(p)}(x,x')={\delta^{1/2}}_{a_1 \dots
a_p}(x)\sigma^{;a_1}(x,x') \dots \sigma^{;a_p}(x,x'). \nonumber \\
\end{eqnarray}
Here, it should be noted that ${\delta^{1/2}}_{(p)}$ is a notation
and nothing else. The symbol $1/2$ simply recalls that
${\delta^{1/2}}_{(p)}$ is a ``component" of the covariant Taylor
series expansion of $\Delta ^{1/2}$: we do not attribute any meaning
to the symbol $\delta_{(p)}$ itself. Using (\ref{AppZeta1}) and
(\ref{AppZeta3bis}) into (\ref{AppVVM}), we obtain after another
long calculation the expressions of the ${\delta^{1/2}}_{(p)}$ for
$p=2,\dots,11$. We have
\begin{widetext}
\begin{subequations} \label{AppSCU04bis}
\begin{eqnarray}
& & {\delta^{1/2}}_{(2)} =  (1/6) \, {\rm tr} \, K_{(2)}  \label{AppSCU04bis_2} \\
& & {\delta^{1/2}}_{(3)} =   (1/4)  \, {\rm tr} \, K_{(3)} \label{AppSCU04bis_3} \\
& & {\delta^{1/2}}_{(4)} =  (3/10) \, {\rm tr} \, K_{(4)} + (1/15)
\, {\rm tr} \, {K_{(2)}}^2 + (1/12) \, \left({ \rm tr} \, K_{(2)}
\right)^2  \label{AppSCU04bis_4} \\
& & {\delta^{1/2}}_{(5)} =   (1/3)  \, {\rm tr} \, K_{(5)}
 + (1/3)  \,{ \rm tr} \, K_{(2)}K_{(3)}
+ (5/12)   \,{ \rm tr} \, K_{(2)} \,{ \rm tr} \, K_{(3)} \label{AppSCU04bis_5} \\
& & {\delta^{1/2}}_{(6)} = (5/14) \, {\rm tr} \, K_{(6)} + (4/7) \,{
\rm tr} \, K_{(2)} K_{(4)} + (15/28) \, {\rm tr} \, {K_{(3)}}^2
+(3/4)
\,{ \rm tr} \, K_{(2)} \,{ \rm tr} \, K_{(4)} \nonumber \\
& & \qquad + (5/8) \, \left({ \rm tr} \, K_{(3)} \right)^2 + (8/63)
\, {\rm tr} \, {K_{(2)}}^3 + (1/6) \, {\rm tr} \, {K_{(2)}} \, {\rm
tr} \, {K_{(2)}}^2 + (5/72) \, \left({ \rm tr} \, K_{(2)} \right)^3
\label{AppSCU04bis_6}
\\
& & {\delta^{1/2}}_{(7)} = (3/8) \, {\rm tr} \, K_{(7)} + (5/6) \,{
\rm tr} \, K_{(2)} K_{(5)} + (9/4) \,{ \rm tr} \, K_{(3)} K_{(4)}
+(7/6) \,{ \rm tr} \, K_{(2)} \,{ \rm tr} \, K_{(5)}  \nonumber \\
& & \qquad +(21/8) \,{ \rm tr} \, K_{(3)} \,{ \rm tr} \, K_{(4)}
 + (4/3) \, {\rm tr} \, {K_{(2)}}^2 K_{(3)}+ (7/6) \,
{\rm tr} \, {K_{(2)}} \, {\rm tr} \, {K_{(2)}} K_{(3)} \nonumber \\
& & \qquad  + (7/12) \, {\rm tr} \,  {K_{(3)}} \, {\rm tr} \,
{K_{(2)}}^2  + (35/48) \, \left({ \rm tr} \, K_{(2)} \right)^2 \,
{\rm
tr} \, K_{(3)} \label{AppSCU04bis_7}\\
& & {\delta^{1/2}}_{(8)} = (7/18) \, {\rm tr} \, K_{(8)}  + (10/9)
\, {\rm tr} \, K_{(2)}K_{(6)} + (35/9) \,{ \rm tr} \, K_{(3)}
K_{(5)} +
(14/5) \,{ \rm tr} \, {K_{(4)}}^2   \nonumber \\
& & \qquad + (5/3) \,{ \rm tr} \, K_{(2)} \,{ \rm tr} \, K_{(6)}
  + (14/3) \,{ \rm tr} \, K_{(3)} \,{ \rm tr} \, K_{(5)}
 + (63/20) \, \left({ \rm tr} \, K_{(4)} \right)^2
 + (136/45)  \, {\rm tr} \, {K_{(2)}}^2 K_{(4)} \nonumber \\
& & \qquad + (50/9)  \, {\rm tr} \, {K_{(2)}}{K_{(3)}}^2 + (8/3) \,
{\rm tr} \, {K_{(2)}} \, {\rm tr} \, {K_{(2)}} K_{(4)} + (5/2) \,
{\rm tr} \, {K_{(2)}} \, {\rm tr} \, {K_{(3)}}^2  \nonumber \\
& & \qquad+ (14/3) \, {\rm tr} \, {K_{(3)}} \, {\rm tr} \,
{K_{(2)}}{K_{(3)}}  + (7/5) \, {\rm tr} \, {K_{(4)}} \, {\rm tr} \,
{K_{(2)}}^2 + (7/4) \, \left({ \rm tr} \, K_{(2)} \right)^2 \, {\rm
tr} \, K_{(4)} \nonumber \\
& & \qquad + (35/12) \, {\rm tr} \, K_{(2)}  \, \left({ \rm tr} \,
K_{(3)} \right)^2 + (8/15) \, {\rm tr} \, {K_{(2)}}^4  + (16/27) \,
{\rm tr} \, {K_{(2)}} \, {\rm tr} \, {K_{(2)}}^3 \nonumber \\
& & \qquad + (7/45) \, \left({\rm tr} \, {K_{(2)}}^2\right)^2  +
(7/18) \, \left({\rm tr} \, {K_{(2)}}\right)^2 \, {\rm tr} \,
{K_{(2)}}^2 + (35/432) \, \left({\rm tr} \, {K_{(2)}} \right)^4
\label{AppSCU04bis_8} \\
& & {\delta^{1/2}}_{(9)} =
  (2/5)\, {\rm tr} \, K_{(9)}
  + (7/5)\, {\rm tr} \, K_{(2)}K_{(7)}
  + 6\, {\rm tr} \, K_{(3)}K_{(6)}
  + (56/5)\, {\rm tr} \, K_{(4)}K_{(5)}   \nonumber \\ & &\qquad
  + (9/4)\, {\rm tr} \, K_{(2)}\, {\rm tr} \, K_{(7)}
  + (15/2)\, {\rm tr} \, K_{(3)}\, {\rm tr} \, K_{(6)}
  + (63/5)\, {\rm tr} \, K_{(4)}\, {\rm tr} \, K_{(5)}
  + (28/5)\, {\rm tr} \, {K_{(2)}}^2 K_{(5)}   \nonumber \\ & &\qquad
  + (73/5)\, {\rm tr} \, K_{(2)}K_{(3)}K_{(4)}
  + (73/5)\, {\rm tr} \, K_{(2)}K_{(4)}K_{(3)}
  + 9\, {\rm tr} \, {K_{(3)}}^3
  + 5\, {\rm tr} \, K_{(2)}\, {\rm tr} \,
K_{(2)}K_{(5)}    \nonumber \\ & &\qquad
  + (27/2)\, {\rm tr} \, K_{(2)}\, {\rm tr} \, K_{(3)}K_{(4)}
  + 12\, {\rm tr} \, K_{(3)}\, {\rm tr} \, K_{(2)}K_{(4)}
  + (45/4)\, {\rm tr} \, K_{(3)}\, {\rm tr} \, {K_{(3)}}^2
  \nonumber \\ & &\qquad + (63/5)\, {\rm tr} \, K_{(4)}\, {\rm tr} \,
K_{(2)}K_{(3)}
  + (14/5)\, {\rm tr} \, K_{(5)}\, {\rm tr} \, {K_{(2)}}^2
  + (7/2) \, \left( {\rm tr} \, K_{(2)}\right)^2\, {\rm tr} \, K_{(5)}
  \nonumber \\ & &\qquad + (63/4)\, {\rm tr} \, K_{(2)}\, {\rm tr} \, K_{(3)}\, {\rm tr} \, K_{(4)}
  + (35/8)\, \left( {\rm tr} \, K_{(3)}\right)^3
  + (48/5)\, {\rm tr} \, {K_{(2)}}^3 K_{(3)}
  \nonumber \\ & &\qquad + 8\, {\rm tr} \, K_{(2)}\, {\rm tr} \, {K_{(2)}}^2 K_{(3)}
  + (8/3)\, {\rm tr} \, K_{(3)}\, {\rm tr} \, {K_{(2)}}^3
  + (14/5)\, {\rm tr} \, {K_{(2)}}^2 \, {\rm
tr} \, K_{(2)}K_{(3)}
  \nonumber \\ & &\qquad + (7/2)\, \left( {\rm tr} \, K_{(2)}\right)^2\, {\rm tr} \,
K_{(2)}K_{(3)}
  + (7/2)\, {\rm tr} \, K_{(2)}\, {\rm tr} \, K_{(3)}\, {\rm tr} \,
{K_{(2)}}^2
  + (35/24) \, \left( {\rm tr} \, K_{(2)}\right)^3\, {\rm tr} \,
 K_{(3)} \label{AppSCU04bis_9}
 \end{eqnarray}
and
\begin{eqnarray}
& & {\delta^{1/2}}_{(10)} =
  (9/22)\, {\rm tr} \, K_{(10)}
  + (56/33)\, {\rm tr} \, K_{(2)}K_{(8)}
  + (189/22)\, {\rm tr} \, K_{(3)}K_{(7)}
  + (216/11)\, {\rm tr} \, K_{(4)}K_{(6)}     \nonumber \\ & &\qquad
  + (140/11)\, {\rm tr} \, {K_{(5)}}^2
  + (35/12)\, {\rm tr} \, K_{(2)}\, {\rm tr} \, K_{(8)}
  + (45/4)\, {\rm tr} \, K_{(3)}\, {\rm tr} \, K_{(7)}
  + (45/2)\, {\rm tr} \, K_{(4)}\, {\rm tr} \,
K_{(6)}    \nonumber \\ & &\qquad
  + 14 \, \left( {\rm tr} \, K_{(5)}\right)^2
  + (304/33)\, {\rm tr} \, {K_{(2)}}^2 K_{(6)}
  + (1015/33)\, {\rm tr} \, K_{(2)}K_{(3)}K_{(5)}
  + (480/11)\, {\rm tr} \, K_{(2)}{K_{(4)}}^2     \nonumber \\ & &\qquad
  + (1015/33)\, {\rm tr} \, K_{(2)}K_{(5)}K_{(3)}
  + 81\, {\rm tr} \, {K_{(3)}}^2 K_{(4)}
  + (25/3)\, {\rm tr} \, K_{(2)}\, {\rm tr} \, K_{(2)}K_{(6)}
  \nonumber \\ & &\qquad + (175/6)\, {\rm tr} \, K_{(2)}\, {\rm tr} \,
K_{(3)}K_{(5)}
  + 21\, {\rm tr} \, K_{(2)}\, {\rm tr} \, {K_{(4)}}^2
  + 25\, {\rm tr} \, K_{(3)}\, {\rm tr} \, K_{(2)}K_{(5)}
  \nonumber \\ & &\qquad + (135/2)\, {\rm tr} \, K_{(3)}\, {\rm tr} \, K_{(3)}K_{(4)}
  + 36\, {\rm tr} \, K_{(4)}\, {\rm tr} \,
K_{(2)}K_{(4)}
  + (135/4)\, {\rm tr} \, K_{(4)}\, {\rm tr} \, {K_{(3)}}^2
  \nonumber \\ & &\qquad + 28\, {\rm tr} \, K_{(5)}\, {\rm tr} \, K_{(2)}K_{(3)}
  + 5\, {\rm tr} \, K_{(6)}\, {\rm tr} \, {K_{(2)}}^2
  + (25/4) \, \left( {\rm tr} \, K_{(2)}\right)^2\, {\rm tr} \,
K_{(6)}    \nonumber \\ & &\qquad
  + 35\, {\rm tr} \, K_{(2)}\, {\rm tr} \, K_{(3)}\, {\rm tr} \, K_{(5)}
  + (189/8)\, {\rm tr} \, K_{(2)}\, \left(  {\rm tr} \, K_{(4)}\right)^2
  + (315/8)\, \left( {\rm tr} \, K_{(3)}\right)^2\, {\rm tr} \, K_{(4)}
  \nonumber \\ & &\qquad + (896/33)\, {\rm tr} \,
{K_{(2)}}^3 K_{(4)}
  + (149/3)\, {\rm tr} \, {K_{(2)}}^2 {K_{(3)}}^2
  + (805/33)\, {\rm tr} \, {K_{(2)}K_{(3)}}{K_{(2)}K_{(3)}}
   \nonumber \\ & &\qquad
   + (68/3)\, {\rm tr} \, K_{(2)}\, {\rm tr} \, {K_{(2)}}^2 K_{(4)}
  + (125/3)\, {\rm tr} \, K_{(2)}\, {\rm tr} \,
K_{(2)}{K_{(3)}}^2
  + 40\, {\rm tr} \, K_{(3)}\, {\rm tr} \, {K_{(2)}}^2 K_{(3)}
  \nonumber \\ & &\qquad
  + 8\, {\rm tr} \, K_{(4)}\, {\rm tr} \, {K_{(2)}}^3
  + 8\, {\rm tr} \, {K_{(2)}}^2 \, {\rm tr} \, K_{(2)}K_{(4)}
  + (15/2)\, {\rm tr} \, {K_{(2)}}^2 \, {\rm tr} \,
{K_{(3)}}^2     \nonumber \\ & &\qquad
  + 14 \, \left( {\rm tr} \, K_{(2)}K_{(3)} \right)^2
  + 10 \, \left( {\rm tr} \, K_{(2)}\right)^2\, {\rm tr} \, K_{(2)}K_{(4)}
  + (75/8) \, \left( {\rm tr} \, K_{(2)}\right)^2\, {\rm tr} \,
{K_{(3)}}^2
   \nonumber \\ & &\qquad + 35\, {\rm tr} \, K_{(2)}\, {\rm tr} \, K_{(3)}\, {\rm tr} \,
K_{(2)}K_{(3)}
  + (21/2)\, {\rm tr} \, K_{(2)}\, {\rm tr} \, K_{(4)}\, {\rm tr} \,
{K_{(2)}}^2
  + (35/4) \, \left(  {\rm tr} \, K_{(3)}\right)^2\, {\rm tr} \,
{K_{(2)}}^2
  \nonumber \\ & &\qquad + (35/8) \, \left(  {\rm tr} \, K_{(2)}\right)^3\, {\rm tr} \, K_{(4)}
  + (175/16) \, \left(  {\rm tr} \,
          K_{(2)}\right)^2 \, \left(  {\rm tr} \,
                K_{(3)}\right)^2
  + (128/33)\, {\rm tr} \, {K_{(2)}}^5
  \nonumber \\ & &\qquad
  + 4\, {\rm tr} \, K_{(2)}\, {\rm tr} \, {K_{(2)}}^4
 + (16/9)\, {\rm tr} \, {K_{(2)}}^2 \, {\rm tr} \, {K_{(2)}}^3
  + (7/6)\, {\rm tr} \, K_{(2)} \, \left(  {\rm tr} \,
                {K_{(2)}}^2 \right)^2  \nonumber \\ & &\qquad
  + (20/9) \, \left(  {\rm tr} \, K_{(2)}\right)^2\, {\rm tr} \,
{K_{(2)}}^3
  + (35/36) \, \left(  {\rm tr} \, K_{(2)}\right)^3\, {\rm tr} \,
{K_{(2)}}^2
  + (35/288) \, \left(  {\rm tr} \, K_{(2)}\right)^5
  \label{AppSCU04bis_10} \\
& & {\delta^{1/2}}_{(11)} =(5/12)\, {\rm tr} \, K_{(11)}
  + 2\, {\rm tr} \, K_{(2)}K_{(9)}
  + (35/3)\, {\rm tr} \, K_{(3)}K_{(8)}
  + (63/2)\, {\rm tr} \, K_{(4)}K_{(7)}    \nonumber \\ & &\qquad
  + 50\, {\rm tr} \, K_{(5)}K_{(6)}
  + (11/3)\, {\rm tr} \, K_{(2)}\, {\rm tr} \, K_{(9)}
  + (385/24)\, {\rm tr} \, K_{(3)}\, {\rm tr} \, K_{(8)}
  + (297/8)\, {\rm tr} \, K_{(4)}\, {\rm tr} \,
K_{(7)}   \nonumber \\ & &\qquad
  + 55\, {\rm tr} \, K_{(5)}\, {\rm tr} \, K_{(6)}
  + 14\, {\rm tr} \, {K_{(2)}}^2 K_{(7)}
  + (170/3)\, {\rm tr} \, K_{(2)}K_{(3)}K_{(6)}
  + 103\, {\rm tr} \, K_{(2)}K_{(4)}K_{(5)}    \nonumber \\ & &\qquad
  + 103\, {\rm tr} \, K_{(2)}K_{(5)}K_{(4)}
  + (170/3)\, {\rm tr} \, K_{(2)}K_{(6)}K_{(3)}
  + (575/3)\, {\rm tr} \, {K_{(3)}}^2 K_{(5)}
  + 273\, {\rm tr} \, K_{(3)} {K_{(4)}}^2    \nonumber \\ & &\qquad
  + (77/6)\, {\rm tr} \, K_{(2)}\, {\rm tr} \, K_{(2)}K_{(7)}
  + 55\, {\rm tr} \, K_{(2)}\, {\rm tr} \, K_{(3)}K_{(6)}
  + (308/3)\, {\rm tr} \, K_{(2)}\, {\rm tr} \, K_{(4)}K_{(5)}
\nonumber \\ & &\qquad  + (275/6)\, {\rm tr} \, K_{(3)}\, {\rm tr}
\, K_{(2)}K_{(6)}
  + (1925/12)\, {\rm tr} \, K_{(3)}\, {\rm tr} \, K_{(3)}K_{(5)}
  + (231/2)\, {\rm tr} \, K_{(3)}\, {\rm tr} \, {K_{(4)}}^2
\nonumber \\ & &\qquad   + (165/2)\, {\rm tr} \, K_{(4)}\, {\rm tr}
\, K_{(2)}K_{(5)}
  + (891/4)\, {\rm tr} \, K_{(4)}\, {\rm tr} \,
K_{(3)}K_{(4)}
  + 88\, {\rm tr} \, K_{(5)}\, {\rm tr} \, K_{(2)}K_{(4)}
 \nonumber \\ & &\qquad + (165/2)\, {\rm tr} \, K_{(5)}\, {\rm tr} \, {K_{(3)}}^2
  + 55\, {\rm tr} \, K_{(6)}\, {\rm tr} \, K_{(2)}K_{(3)}
  + (33/4)\, {\rm tr} \, K_{(7)}\, {\rm tr} \,
{K_{(2)}}^2
 \nonumber \\ & &\qquad + (165/16)\, \left(  {\rm tr} \, K_{(2)}\right)^2\, {\rm tr} \, K_{(7)}
  + (275/4)\, {\rm tr} \, K_{(2)}\, {\rm tr}
\, K_{(3)}\, {\rm tr} \, K_{(6)}
  + (231/2)\, {\rm tr} \, K_{(2)}\, {\rm tr} \, K_{(4)}\, {\rm tr} \, K_{(5)}
  \nonumber \\ & &\qquad + (385/4)\, \left( {\rm tr} \, K_{(3)}\right)^2\, {\rm tr} \,
K_{(5)}
  + (2079/16)\, {\rm tr} \, K_{(3)}\, \left(  {\rm tr} \, K_{(4)}\right)^2
  + (184/3)\, {\rm tr} \, {K_{(2)}}^3 K_{(5)}
  \nonumber \\ & &\qquad  + (317/2)\, {\rm tr} \, {K_{(2)}}^2 K_{(3)}K_{(4)}
 + (317/2)\, {\rm tr} \,
{K_{(2)}}^2 K_{(4)}K_{(3)}
  + (461/3)\, {\rm tr} \, K_{(2)}K_{(3)}K_{(2)}K_{(4)}
  \nonumber \\ & &\qquad + (860/3)\, {\rm tr} \, K_{(2)} {K_{(3)}}^3
+ (154/3)\, {\rm tr} \, K_{(2)}\, {\rm tr} \, {K_{(2)}}^2 K_{(5)}
  + (803/6)\, {\rm tr} \, K_{(2)}\, {\rm tr} \,
K_{(2)}K_{(3)}K_{(4)}
  \nonumber \\ & &\qquad + (803/6)\, {\rm tr} \, K_{(2)}\, {\rm tr} \, K_{(2)}K_{(4)}K_{(3)}
 + (165/2)\, {\rm tr} \, K_{(2)}\, {\rm tr} \, {K_{(3)}}^3
  + (374/3)\, {\rm tr} \, K_{(3)}\, {\rm tr} \, {K_{(2)}}^2 K_{(4)}
  \nonumber \\ & &\qquad + (1375/6)\, {\rm tr} \, K_{(3)}\, {\rm tr} \,
K_{(2)}{K_{(3)}}^2
  + 132\, {\rm tr} \, K_{(4)}\, {\rm tr} \, {K_{(2)}}^2 K_{(3)}
  + (176/9)\, {\rm tr} \, K_{(5)}\, {\rm tr} \, {K_{(2)}}^3
  \nonumber \\ & &\qquad + (55/3)\, {\rm tr} \, {K_{(2)}}^2 \, {\rm tr} \, K_{(2)}K_{(5)}
   + (99/2)\, {\rm tr} \, {K_{(2)}}^2 \, {\rm tr} \,
K_{(3)}K_{(4)}
  + 88\, {\rm tr} \, K_{(2)}K_{(3)}\, {\rm tr} \, K_{(2)}K_{(4)}
  \nonumber \\ & &\qquad + (165/2)\, {\rm tr} \, K_{(2)}K_{(3)}\, {\rm tr} \, {K_{(3)}}^2
  + (275/12) \, \left(  {\rm tr} \, K_{(2)}\right)^2\, {\rm tr} \,
K_{(2)}K_{(5)}
 + (495/8) \, \left( {\rm tr} \, K_{(2)}\right)^2\, {\rm tr} \,
K_{(3)}K_{(4)}
  \nonumber \\ & &\qquad + 110\, {\rm tr} \, K_{(2)}\, {\rm tr} \, K_{(3)}\, {\rm tr} \,
K_{(2)}K_{(4)}   + (825/8)\, {\rm tr} \, K_{(2)}\, {\rm tr} \,
K_{(3)}\, {\rm tr} \, {K_{(3)}}^2
  \nonumber \\ & &\qquad  + (231/2)\, {\rm tr} \, K_{(2)}\, {\rm tr} \, K_{(4)}\, {\rm tr} \,
K_{(2)}K_{(3)}
  + (77/3)\, {\rm tr} \, K_{(2)}\, {\rm tr} \, K_{(5)}\, {\rm tr} \,
{K_{(2)}}^2   \nonumber \\ & &\qquad
  + (385/4) \, \left(  {\rm tr} \, K_{(3)}\right)^2\, {\rm tr} \,
K_{(2)}K_{(3)}
  + (231/4)\, {\rm tr} \, K_{(3)}\, {\rm tr} \, K_{(4)}\, {\rm tr} \,
{K_{(2)}}^2
  + (385/36) \, \left(  {\rm tr} \, K_{(2)}\right)^3\, {\rm tr} \, K_{(5)}
 \nonumber \\ & &\qquad
 + (1155/16) \, \left(  {\rm tr} \,
          K_{(2)}\right)^2\, {\rm tr} \, K_{(3)}\, {\rm tr} \,
K_{(4)}
  + (1925/48)\, {\rm tr} \, K_{(2)} \, \left(  {\rm tr} \, K_{(3)}\right)^3
  + (320/3)\, {\rm tr} \, {K_{(2)}}^4K_{(3)}
\nonumber \\ & &\qquad + 88\, {\rm tr} \, K_{(2)}\, {\rm tr} \,
{K_{(2)}}^3 K_{(3)}
  + 22\, {\rm tr} \, K_{(3)}\, {\rm tr} \,
{K_{(2)}}^4
  + (88/3)\, {\rm tr} \, {K_{(2)}}^2 \, {\rm tr} \, {K_{(2)}}^2 K_{(3)}
 \nonumber \\ & &\qquad
 + (176/9)\, {\rm tr} \, K_{(2)}K_{(3)}\, {\rm tr} \, {K_{(2)}}^3
  + (110/3) \, \left(  {\rm tr} \, K_{(2)}\right)^2\, {\rm tr} \,
{K_{(2)}}^2 K_{(3)}
  + (220/9)\, {\rm tr} \, K_{(2)}\, {\rm tr} \, K_{(3)}\, {\rm tr} \,
{K_{(2)}}^3  \nonumber \\ & &\qquad
  + (77/3)\, {\rm tr} \, K_{(2)}\, {\rm tr} \, {K_{(2)}}^2 \, {\rm tr} \,
K_{(2)}K_{(3)}
  + (77/12)\, {\rm tr} \, K_{(3)} \, \left(  {\rm tr} \,
          {K_{(2)}}^2 \right)^2
  + (385/36) \, \left(  {\rm tr} \, K_{(2)}\right)^3\, {\rm tr} \,
K_{(2)}K_{(3)} \nonumber \\ & &\qquad + (385/24) \, \left(  {\rm tr}
\, K_{(2)}\right)^2\, {\rm tr} \, K_{(3)}\, {\rm tr} \, {K_{(2)}}^2
+ (1925/576) \, \left(  {\rm tr} \, K_{(2)}\right)^4\, {\rm tr} \,
K_{(3)}.     \label{AppSCU04bis_11}
\end{eqnarray}
\end{subequations}
\end{widetext}

By using (\ref{AppSCU02}) and (\ref{AppK1}) into
(\ref{AppSCU04bis}), we can also obtain the expressions of the
components ${\delta^{1/2}}_{a_1 \dots a_p}$ of the biscalars
${\delta^{1/2}}_{(p)}$. The components of the lowest order biscalars
${\delta^{1/2}}_{(p)}$ take the form
\begin{widetext}
\begin{subequations}\label{AppSCU04}
\begin{eqnarray}
& & {\delta^{1/2}}_{a_1 a_2} =
(1/6)  \,  R_{a_1 a_2}  \label{AppSCU04_2}\\
& & {\delta^{1/2}}_{a_1 a_2 a_3} =
(1/4)  \,   R_{(a_1 a_2; a_3)}  \label{AppSCU04_3}\\
& & {\delta^{1/2}}_{a_1 a_2 a_3 a_4} = (3/10)  \, R_{(a_1 a_2; a_3
a_4)} + (1/15)  \, R^{\rho}_{\phantom{\rho}(a_1 |\tau| a_2}
R^{\tau}_{\phantom{\tau} a_3 |\rho |  a_4)}
+ (1/12)  R_{(a_1 a_2} R_{a_3 a_4)} \label{AppSCU04_4} \\
& &{\delta^{1/2}}_{a_1 a_2 a_3 a_4 a_5} =  (1/3) \, R_{(a_1 a_2; a_3
a_4 a_5)} + (1/3) \, R^{\rho}_{\phantom{\rho}(a_1 |\tau| a_2}
R^{\tau}_{\phantom{\tau}
a_3 |\rho |  a_4;a_5)} + (5/12)  R_{(a_1 a_2} R_{a_3 a_4; a_5)}  \label{AppSCU04_5}\\
& &{\delta^{1/2}}_{a_1 a_2 a_3 a_4 a_5 a_6} = (5/14) \, R_{(a_1 a_2;
a_3 a_4 a_5 a_6)} +(4/7) \, R^{\rho}_{\phantom{\rho}(a_1 |\tau| a_2}
R^{\tau}_{\phantom{\tau} a_3 |\rho |  a_4;a_5 a_6)} \nonumber
\\
&  & \qquad \qquad + (15/28) \, R^{\rho}_{\phantom{\rho}(a_1 |\tau|
a_2; a_3} R^{\tau}_{\phantom{\tau} a_4 |\rho |  a_5; a_6)} + (3/4)
\, R_{(a_1 a_2 }R_{a_3 a_4; a_5 a_6)} + (5/8) \, R_{(a_1 a_2 ;
a_3}R_{ a_4 a_5 ; a_6)} \nonumber \\
&  & \qquad \qquad + (8/63) \, R^{\rho}_{\phantom{\rho}(a_1 |\tau |
a_2 }R^{\tau}_{\phantom{\tau} a_3 |\sigma |
a_4}R^{\sigma}_{\phantom{\sigma} a_5 |\rho | a_6)} + (1/6) \,
R_{(a_1 a_2} R^{\rho}_{\phantom{\rho} a_3 |\tau| a_4}
R^{\tau}_{\phantom{\tau} a_5 |\rho |  a_6)}
\nonumber \\
&  & \qquad \qquad + (5/72) \, R_{(a_1 a_2}R_{a_3 a_4}R_{a_5 a_6)} \label{AppSCU04_6}\\
& &{\delta^{1/2}}_{a_1 a_2 a_3 a_4 a_5 a_6 a_7} =(3/8) \, R_{(a_1
a_2; a_3 a_4 a_5 a_6 a_7)} + (5/6) \,  R^{\rho}_{\phantom{\rho}(a_1
|\tau| a_2} R^{\tau}_{\phantom{\tau} a_3 |\rho |  a_4; a_5 a_6 a_7)}
\nonumber
\\
&  & \qquad \qquad + (9/4) \,  R^{\rho}_{\phantom{\rho}(a_1 |\tau|
a_2; a_3} R^{\tau}_{\phantom{\tau} a_4 |\rho |  a_5; a_6 a_7)}
 + (7/6) \,
R_{(a_1 a_2}R_{a_3 a_4; a_5 a_6 a_7)} + (21/8) \, R_{(a_1 a_2;
a_3}R_{a_4 a_5; a_6 a_7)}   \nonumber
\\
&  & \qquad \qquad + (4/3) \, R^{\rho}_{\phantom{\rho}(a_1 |\tau |
a_2 }R^{\tau}_{\phantom{\tau} a_3 |\sigma |
a_4}R^{\sigma}_{\phantom{\sigma} a_5 |\rho | a_6;a_7)} +  (7/6) \,
R_{(a_1 a_2} R^{\rho}_{\phantom{\rho} a_3 |\tau| a_4}
R^{\tau}_{\phantom{\tau} a_5 |\rho |  a_6; a_7)}\nonumber
\\
&  & \qquad \qquad +  (7/12) \, R_{(a_1 a_2; a_3}
R^{\rho}_{\phantom{\rho} a_4 |\tau| a_5} R^{\tau}_{\phantom{\tau}
a_6 |\rho |  a_7)} + (35/48) \, R_{(a_1
a_2 }R_{a_3 a_4}R_{a_5  a_6;a_7)}  \\
& &{\delta^{1/2}}_{a_1 a_2 a_3 a_4 a_5 a_6 a_7 a_8} =(7/18) \,
R_{(a_1 a_2; a_3 a_4 a_5 a_6 a_7 a_8)}+(10/9) \,
R^{\rho}_{\phantom{\rho}(a_1 |\tau| a_2} R^{\tau}_{\phantom{\tau}
a_3 |\rho |  a_4 ; a_5 a_6 a_7 a_8)} \nonumber
\\
&  & \qquad \qquad + (35/9) \,  R^{\rho}_{\phantom{\rho}(a_1 |\tau|
a_2; a_3} R^{\tau}_{\phantom{\tau} a_4 |\rho |  a_5 ; a_6 a_7 a_8)}
+ (14/5) \, R^{\rho}_{\phantom{\rho}(a_1 |\tau| a_2; a_3 a_4 }
R^{\tau}_{\phantom{\tau} a_5 |\rho |  a_6 ; a_7 a_8)} \nonumber \\
&  & \qquad \qquad + (5/3) \, R_{(a_1 a_2} R_{a_3 a_4; a_5 a_6 a_7
a_8)} + (14/3) \, R_{(a_1 a_2; a_3} R_{a_4 a_5; a_6 a_7 a_8)}
\nonumber \\
&  & \qquad \qquad + (63/20) \, R_{(a_1 a_2; a_3 a_4} R_{a_5 a_6;
a_7 a_8)}  + (136/45) \, R^{\rho}_{\phantom{\rho}(a_1 |\tau | a_2
}R^{\tau}_{\phantom{\tau} a_3 |\sigma |
a_4}R^{\sigma}_{\phantom{\sigma} a_5 |\rho | a_6; a_7 a_8)}
\nonumber \\
&  & \qquad \qquad + (50/9) \, R^{\rho}_{\phantom{\rho}(a_1 |\tau |
a_2 }R^{\tau}_{\phantom{\tau} a_3 |\sigma | a_4; a_5
}R^{\sigma}_{\phantom{\sigma} a_6 |\rho | a_7; a_8)}+ (8/3) \,
R_{(a_1 a_2} R^{\rho}_{\phantom{\rho} a_3 |\tau |
a_4}R^{\tau}_{\phantom{\tau} a_5 |\rho | a_6; a_7 a_8)}
\nonumber \\
&  & \qquad \qquad + (5/2)  \, R_{(a_1 a_2} R^{\rho}_{\phantom{\rho}
a_3 |\tau | a_4;a_5} R^{\tau}_{\phantom{\tau} a_6 |\rho | a_7; a_8)}
+ (14/3)  \, R_{(a_1 a_2; a_3} R^{\rho}_{\phantom{\rho} a_4 |\tau |
a_5} R^{\tau}_{\phantom{\tau} a_6 |\rho | a_7; a_8)}
\nonumber \\
&  & \qquad \qquad + (7/5)  \, R_{(a_1 a_2; a_3 a_4}
R^{\rho}_{\phantom{\rho} a_5 |\tau | a_6} R^{\tau}_{\phantom{\tau}
a_7 |\rho | a_8)} + (7/4) \, R_{(a_1 a_2} R_{a_3 a_4}R_{a_5 a_6; a_7
a_8)}
\nonumber \\
&  & \qquad \qquad + (35/12) R_{(a_1 a_2} R_{a_3 a_4; a_5}R_{a_6
a_7; a_8)} + (8/15)  \, R^{\rho}_{\phantom{\rho}(a_1 |\tau| a_2
}R^{\tau}_{\phantom{\tau} a_3 |\sigma|
a_4}R^{\sigma}_{\phantom{\sigma} a_5 |\kappa|  a_6
}R^{\kappa}_{\phantom{\kappa} a_7 |\rho|  a_8 )}
\nonumber \\
&  & \qquad \qquad + (16/27) R_{(a_1 a_2} R^{\rho}_{\phantom{\rho}
a_3 |\tau| a_4 }R^{\tau}_{\phantom{\tau} a_5 |\sigma|
a_6}R^{\sigma}_{\phantom{\sigma} a_7 |\rho|  a_8)} + (7/18) \,
R_{(a_1 a_2}R_{a_3 a_4} R^{\rho}_{\phantom{\rho} a_5 |\tau| a_6
}R^{\tau}_{\phantom{\tau} a_7 |\rho| a_8)}
\nonumber \\
&  & \qquad \qquad + (7/45) \, R^{\rho}_{\phantom{\rho}(a_1 |\tau|
a_2 }R^{\tau}_{\phantom{\tau} a_3 |\rho| a_4}
R^{\kappa}_{\phantom{\kappa} a_5 |\lambda| a_6
}R^{\lambda}_{\phantom{\lambda} a_7 |\kappa| a_8)} + (35/432) \,
R_{(a_1 a_2}R_{a_3 a_4}R_{a_5 a_6}R_{a_7 a_8)}.
\end{eqnarray}
\end{subequations}
\end{widetext}

The relations (\ref{AppSCU01}) and (\ref{AppSCU04bis}) provide a
compact form for the covariant Taylor series expansion of the
biscalar $\Delta ^{1/2}$. It is also possible to give this expansion
in a more explicit form by using (\ref{AppSCU01}), (\ref{AppSCU02})
and (\ref{AppSCU04}). Of course, the corresponding result is very
heavy. For this explicit expansion up to order $\sigma^4$, we have
\begin{widetext}
\begin{eqnarray}\label{AppSCU05}
& & \Delta ^{1/2} = 1 + \frac{1}{12}  \,  R_{a_1 a_2}
\sigma^{;a_1}\sigma^{;a_2}
-\frac{1}{24}  \,   R_{a_1 a_2; a_3}  \sigma^{;a_1}\sigma^{;a_2} \sigma^{;a_3} \nonumber \\
& & \quad + \left[ \frac{1}{80}  \, R_{a_1 a_2; a_3 a_4} +
\frac{1}{360} \, R^{\rho}_{\phantom{\rho} a_1 \tau a_2}
R^{\tau}_{\phantom{\tau} a_3  \rho  a_4} + \frac{1}{288} R_{a_1 a_2}
R_{a_3 a_4} \right]
\sigma^{;a_1}\sigma^{;a_2} \sigma^{;a_3} \sigma^{;a_4}    \nonumber \\
& & \quad - \left[ \frac{1}{360} \, R_{a_1 a_2; a_3 a_4 a_5} +
\frac{1}{360} \, R^{\rho}_{\phantom{\rho} a_1 \tau a_2}
R^{\tau}_{\phantom{\tau} a_3  \rho a_4;a_5 } + \frac{1}{288} R_{a_1
a_2} R_{a_3 a_4; a_5} \right]
\sigma^{;a_1}\sigma^{;a_2} \sigma^{;a_3} \sigma^{;a_4} \sigma^{;a_5} \nonumber \\
& & \quad + \left[ \frac{1}{2016} \, R_{a_1 a_2; a_3 a_4 a_5 a_6} +
\frac{1}{1260} \, R^{\rho}_{\phantom{\rho} a_1 \tau a_2}
R^{\tau}_{\phantom{\tau} a_3 \rho  a_4;a_5 a_6 } + \frac{1}{1344} \,
R^{\rho}_{\phantom{\rho} a_1 \tau a_2; a_3} R^{\tau}_{\phantom{\tau}
a_4 \rho  a_5; a_6 } \right. \nonumber
\\
&  & \qquad  \left. + \frac{1}{960} \, R_{a_1 a_2 }R_{a_3 a_4; a_5
a_6} + \frac{1}{1152} \, R_{a_1 a_2 ; a_3}R_{ a_4 a_5 ; a_6} +
\frac{1}{5670} \, R^{\rho}_{\phantom{\rho}a_1 \tau a_2
}R^{\tau}_{\phantom{\tau}
a_3 \sigma a_4}R^{\sigma}_{\phantom{\sigma} a_5 \rho  a_6}  \right. \nonumber \\
&  & \qquad  \left. + \frac{1}{4320} \, R_{a_1 a_2}
R^{\rho}_{\phantom{\rho} a_3 \tau a_4} R^{\tau}_{\phantom{\tau} a_5
\rho a_6 }
 + \frac{1}{10368} \, R_{a_1 a_2}R_{a_3 a_4}R_{a_5 a_6} \right]
 \sigma^{;a_1}\sigma^{;a_2} \sigma^{;a_3} \sigma^{;a_4}
 \sigma^{;a_5}\sigma^{;a_6}
\nonumber \\
& & \quad - \left[ \frac{1}{13440} \, R_{ a_1 a_2; a_3 a_4 a_5 a_6
a_7 } + \frac{1}{6048} \,  R^{\rho}_{\phantom{\rho} a_1 \tau a_2}
R^{\tau}_{\phantom{\tau} a_3 \rho  a_4; a_5 a_6 a_7 } \right.
\nonumber
\\
&  & \qquad  \left. + \frac{1}{2240} \, R^{\rho}_{\phantom{\rho} a_1
\tau  a_2; a_3} R^{\tau}_{\phantom{\tau} a_4 \rho  a_5; a_6 a_7 }
 + \frac{1}{4320}
\, R_{a_1 a_2}R_{a_3 a_4; a_5 a_6 a_7} + \frac{1}{1920} \, R_{a_1
a_2; a_3}R_{a_4 a_5; a_6 a_7} \right. \nonumber
\\
&  & \qquad  \left. + \frac{1}{3780} \, R^{\rho}_{\phantom{\rho} a_1
\tau  a_2 }R^{\tau}_{\phantom{\tau} a_3 \sigma
a_4}R^{\sigma}_{\phantom{\sigma} a_5 \rho a_6;a_7} + \frac{1}{4320}
\, R_{a_1 a_2} R^{\rho}_{\phantom{\rho} a_3  \tau a_4}
R^{\tau}_{\phantom{\tau} a_5 \rho  a_6; a_7 }   \right. \nonumber
\\
&  & \qquad  + \left. \frac{1}{8640} \, R_{a_1 a_2; a_3}
R^{\rho}_{\phantom{\rho} a_4 \tau a_5} R^{\tau}_{\phantom{\tau} a_6
\rho  a_7 } + \frac{1}{6912} \, R_{a_1 a_2 }R_{a_3 a_4}R_{a_5
a_6;a_7} \right]
 \sigma^{;a_1}\sigma^{;a_2} \sigma^{;a_3} \sigma^{;a_4}
 \sigma^{;a_5}\sigma^{;a_6}\sigma^{;a_7} \nonumber \\
& & \quad + \left[ \frac{1}{103680} \, R_{ a_1 a_2; a_3 a_4 a_5 a_6
a_7 a_8 }+ \frac{1}{36288} \, R^{\rho}_{\phantom{\rho}a_1
 \tau  a_2} R^{\tau}_{\phantom{\tau} a_3 \rho  a_4 ; a_5 a_6 a_7
a_8 } \right. \nonumber
\\
&  & \qquad  \left. + \frac{1}{10368} \, R^{\rho}_{\phantom{\rho}
a_1 \tau  a_2; a_3} R^{\tau}_{\phantom{\tau} a_4 \rho  a_5 ; a_6 a_7
a_8 } + \frac{1}{14400} \, R^{\rho}_{\phantom{\rho} a_1 \tau a_2;
a_3 a_4 } R^{\tau}_{\phantom{\tau} a_5 \rho  a_6 ; a_7 a_8 } \right.
\nonumber
\\
&  & \qquad  \left. + \frac{1}{24192} \, R_{ a_1 a_2} R_{a_3 a_4;
a_5 a_6 a_7 a_8 } + \frac{1}{8640} \, R_{ a_1 a_2; a_3} R_{a_4 a_5;
a_6 a_7 a_8 } \right. \nonumber
\\
&  & \qquad  \left. + \frac{1}{12800} \, R_{ a_1 a_2; a_3 a_4}
R_{a_5 a_6; a_7 a_8 }  + \frac{17}{226800} \,
R^{\rho}_{\phantom{\rho} a_1 \tau  a_2 }R^{\tau}_{\phantom{\tau} a_3
\sigma a_4}R^{\sigma}_{\phantom{\sigma} a_5 \rho a_6; a_7 a_8 }
\right. \nonumber
\\
&  & \qquad  \left. + \frac{5}{36288} \, R^{\rho}_{\phantom{\rho}
a_1 \tau  a_2 }R^{\tau}_{\phantom{\tau} a_3 \sigma a_4; a_5
}R^{\sigma}_{\phantom{\sigma} a_6 \rho  a_7; a_8 }+ \frac{1}{15120}
\, R_{ a_1 a_2} R^{\rho}_{\phantom{\rho} a_3 \tau
a_4}R^{\tau}_{\phantom{\tau} a_5 \rho a_6; a_7 a_8 } \right.
\nonumber
\\
&  & \qquad  \left. + \frac{1}{16128}  \, R_{ a_1 a_2}
R^{\rho}_{\phantom{\rho} a_3 \tau  a_4;a_5} R^{\tau}_{\phantom{\tau}
a_6 \rho  a_7; a_8 } + \frac{1}{8640} \, R_{ a_1 a_2; a_3}
R^{\rho}_{\phantom{\rho} a_4 \tau  a_5} R^{\tau}_{\phantom{\tau} a_6
\rho  a_7; a_8 } \right. \nonumber
\\
&  & \qquad  \left. + \frac{1}{28800}  \, R_{ a_1 a_2; a_3 a_4}
R^{\rho}_{\phantom{\rho} a_5 \tau  a_6} R^{\tau}_{\phantom{\tau} a_7
\rho  a_8 } + \frac{1}{23040} \, R_{ a_1 a_2} R_{a_3 a_4}R_{a_5 a_6;
a_7 a_8 } \right. \nonumber
\\
&  & \qquad  \left. + \frac{1}{13824}\, R_{ a_1 a_2} R_{a_3 a_4;
a_5}R_{a_6 a_7; a_8 } + \frac{1}{75600}  \, R^{\rho}_{\phantom{\rho}
a_1 \tau a_2 }R^{\tau}_{\phantom{\tau} a_3  \sigma
a_4}R^{\sigma}_{\phantom{\sigma} a_5 \kappa  a_6
}R^{\kappa}_{\phantom{\kappa} a_7 \rho  a_8 } \right. \nonumber
\\
&  & \qquad  \left. + \frac{1}{68040} R_{ a_1 a_2}
R^{\rho}_{\phantom{\rho} a_3 \tau a_4 }R^{\tau}_{\phantom{\tau} a_5
\sigma  a_6}R^{\sigma}_{\phantom{\sigma} a_7 \rho  a_8 } +
\frac{1}{103680} \, R_{ a_1 a_2}R_{a_3 a_4} R^{\rho}_{\phantom{\rho}
a_5 \tau  a_6 }R^{\tau}_{\phantom{\tau} a_7 \rho a_8 } \right.
\nonumber
\\
&  & \qquad  \left. + \frac{1}{259200} \, R^{\rho}_{\phantom{\rho}
a_1 \tau a_2 }R^{\tau}_{\phantom{\tau} a_3 \rho a_4}
R^{\kappa}_{\phantom{\kappa} a_5 \lambda a_6
}R^{\lambda}_{\phantom{\lambda} a_7 \kappa a_8}  \right. \nonumber
\\
&  & \qquad  \left. + \frac{1}{497664} \, R_{a_1 a_2}R_{a_3
a_4}R_{a_5 a_6}R_{a_7 a_8} \right]
 \sigma^{;a_1}\sigma^{;a_2} \sigma^{;a_3} \sigma^{;a_4}
 \sigma^{;a_5}\sigma^{;a_6}\sigma^{;a_7} \sigma^{;a_8} + O \left(\sigma^{9/2} \right)
\end{eqnarray}
(here we have not included the enormous terms given in
(\ref{AppSCU04bis_9}), (\ref{AppSCU04bis_10}) and
(\ref{AppSCU04bis_11}) which correspond respectively to the orders
$\sigma^{9/2}$, $\sigma^5$ and $\sigma^{11/2}$ of the expansion).
\end{widetext}

It should be noted that this expansion was obtained up to order
$\sigma$ by DeWitt \cite{DeWittBrehme,DeWitt65}, up to order
$\sigma^2$ by Christensen \cite{Christensen1,Christensen2} and up to
order $\sigma^{5/2}$ by Brown and Ottewill \cite{BrownOttewill86}.
The term of order $\sigma^3$ disagrees with the recent calculation
of Phillips and Hu in Ref.~\cite{PhillipsHu03}: the signs of the
last three terms in their equation (C12c) are incorrect.

Let us now consider the covariant Taylor series expansion of the
biscalar $T:=\Delta ^{-1/2}{\Delta ^{1/2}}_{;\mu} \sigma^{; \mu}$.
By noting that we have
\begin{equation}\label{AppU0p}
T =Z_{;\mu}\sigma^{; \mu} =DZ
\end{equation}
(this is an immediate consequence of (\ref{AppVVM})), it is obvious
that this expansion is necessarily given by
\begin{equation}\label{AppSTU0p1}
 T (x,x')= \sum_{p=2}^{+\infty} \frac{(-1)^p}{p!} \tau_{(p)} (x,x')
\end{equation}
where the $\tau_{(p)} (x,x')$ are biscalars of the form
\begin{eqnarray}\label{AppSTU0p2}
& & \tau_{(p)} (x,x')=\tau_{a_1 \dots a_p}(x)\sigma^{;
a_1}(x,x') \dots \sigma^{; a_p}(x,x'). \nonumber \\
& &
\end{eqnarray}
By noting the identities
\begin{widetext}
\begin{subequations}\label{AppDapptrK}
\begin{eqnarray}
& & D \left[{\rm tr} \, K_{(p)}\right] =p \, {\rm
tr} \, K_{(p)} + {\rm tr} \, K_{(p+1)}   \\
& & D \left[{\rm tr} \, K_{(p)}K_{(q)}\right] =(p+q) \, {\rm tr} \,
K_{(p)}K_{(q)} + {\rm tr} \, K_{(p+1)}K_{(q)}
+ {\rm tr} \, K_{(p)}K_{(q+1)}  \\
& & D \left[{\rm tr} \, K_{(p)}K_{(q)}K_{(r)}\right]=(p+q+r) \, {\rm
tr} \, K_{(p)}K_{(q)}K_{(r)} + {\rm tr} \, K_{(p+1)}K_{(q)}K_{(r)}
 \nonumber \\
& & \qquad \qquad + {\rm tr} \, K_{(p)}K_{(q+1)}K_{(r)}
+ {\rm tr} \, K_{(p)}K_{(q)}K_{(r+1)} \\
& & D \left[{\rm tr} \,
K_{(p)}K_{(q)}K_{(r)}K_{(s)}\right]=(p+q+r+s) \, {\rm tr} \,
K_{(p)}K_{(q)}K_{(r)}K_{(s)} + {\rm tr} \,
K_{(p+1)}K_{(q)}K_{(r)}K_{(s)}
 \nonumber \\
& & \qquad \qquad + {\rm tr} \, K_{(p)}K_{(q+1)}K_{(r)}K_{(s)} +
{\rm tr} \, K_{(p)}K_{(q)}K_{(r+1)}K_{(s)} + {\rm tr} \,
K_{(p)}K_{(q)}K_{(r)}K_{(s+1)}
\end{eqnarray}
\end{subequations}
which follow from (\ref{AppK1}) and (\ref{DSrep4}), another tedious
calculation using (\ref{AppU0p}), (\ref{AppSTU0p1}) as well as
(\ref{AppZeta1}) and (\ref{AppZeta3bis}) permits us to obtain the
$\tau_{(p)}$ for $p=2,\dots,9$. We have
\begin{subequations}\label{AppSTU0p3}
\begin{eqnarray}
& & \tau_{(2)} =  (1/3) \, {\rm tr} \, K_{(2)} \\
& & \tau_{(3)} =  (1/4)  \, {\rm tr} \, K_{(3)} \\
& & \tau_{(4)} = (1/5) \, {\rm tr} \, K_{(4)} + (4/15) \,
{\rm tr} \, {K_{(2)}}^2  \\
& & \tau_{(5)} =   (1/6)  \, {\rm tr} \, K_{(5)} +
\, {\rm tr} \, K_{(2)}K_{(3)}   \\
& & \tau_{(6)}= (1/7)  \, {\rm tr} \, K_{(6)}+ (10/7) \, {\rm tr} \,
K_{(2)}K_{(4)} + (17/14)  \, {\rm tr} \, {K_{(3)}}^2
+ (16/21)  \, {\rm tr} \, {K_{(2)}}^3  \\
& & \tau_{(7)}=  (1/8)  \, {\rm tr} \, K_{(7)} + (11/6) \, {\rm tr}
\, K_{(2)}K_{(5)} + (17/4)  \, {\rm tr} \, K_{(3)}K_{(4)}
+ (20/3)  \, {\rm tr} \, {K_{(2)}}^2 K_{(3)}  \label{AppSTU0p3_7} \\
& & \tau_{(8)} = (1/9)  \, {\rm tr} \, K_{(8)}+ (20/9) \, {\rm tr}
\, K_{(2)}K_{(6)} + (58/9)  \,
{\rm tr} \, K_{(3)}K_{(5)} + (22/5) \, {\rm tr} \, {K_{(4)}}^2   \nonumber \\
&  & \qquad + (608/45)  \, {\rm tr} \, {K_{(2)}}^2 K_{(4)} + (208/9)
\, {\rm tr} \, K_{(2)} {K_{(3)}}^2 + (64/15) \, {\rm tr} \,
{K_{(2)}}^4   \label{AppSTU0p3_8} \\
& & \tau_{(9)} =(1/10)  \, {\rm tr} \, K_{(9)} + (13/5) \, {\rm tr}
\, K_{(2)} K_{(7)} + 9 \, {\rm tr} \, K_{(3)} K_{(6)} + (77/5) \,
{\rm tr} \, K_{(4)} K_{(5)} \nonumber \\
&  & \qquad + (116/5)  \, {\rm tr} \, {K_{(2)}}^2 K_{(5)}     +
(271/5) \, {\rm tr} \, K_{(2)} K_{(3)} K_{(4)} + (271/5) \, {\rm tr}
\, K_{(2)} K_{(4)}K_{(3)} \nonumber \\
&  & \qquad + 31 \, {\rm tr} \, {K_{(3)}}^3 +(336/5)
 \, {\rm tr} \, {K_{(2)}}^3 K_{(3)}.
\label{AppSTU0p3_9}
\end{eqnarray}
\end{subequations}
By using (\ref{AppSTU0p2}) and (\ref{AppK1}) into (\ref{AppSTU0p3}),
we can also obtain the expressions of the components $\tau_{a_1
\dots a_p}$ of the biscalars $\tau_{(p)}$. The components of the
lowest order biscalars $\tau_{(p)}$ take the form
\begin{subequations}\label{AppSTU0p3bis}
\begin{eqnarray}
& & \tau_{a_1 a_2} =
(1/3)  \,  R_{a_1 a_2}  \\
& & \tau_{a_1 a_2 a_3} =
(1/4)  \,  R_{(a_1 a_2; a_3)}  \\
& & \tau_{a_1 a_2 a_3 a_4} = (1/5)  \,  R_{(a_1 a_2; a_3 a_4)} +
(4/15)  \, R^{\rho}_{\phantom{\rho}(a_1 |\tau| a_2}
R^{\tau}_{\phantom{\tau}
a_3 |\rho | a_4 )} \\
& & \tau_{a_1 a_2 a_3 a_4 a_5} = (1/6)  \,  R_{(a_1 a_2; a_3 a_4
a_5)} +    \, R^{\rho}_{\phantom{\rho}(a_1 |\tau| a_2}
R^{\tau}_{\phantom{\tau}
a_3 |\rho | a_4 ; a_5)} \\
& & \tau_{a_1 a_2 a_3 a_4 a_5 a_6} =  (1/7)  \, R_{(a_1 a_2; a_3 a_4
a_5 a_6)} + (10/7) \, R^{\rho}_{\phantom{\rho}(a_1 |\tau| a_2}
R^{\tau}_{\phantom{\tau}
a_3 |\rho | a_4 ; a_5 a_6)} \nonumber \\
&  &  \qquad + (17/14) \, R^{\rho}_{\phantom{\rho}(a_1 |\tau| a_2;
a_3 } R^{\tau}_{\phantom{\tau} a_4 |\rho | a_5 ; a_6)} + (16/21) \,
R^{\rho}_{\phantom{\rho}(a_1 |\tau | a_2 }R^{\tau}_{\phantom{\tau}
a_3 |\sigma | a_4}R^{\sigma}_{\phantom{\sigma} a_5 |\rho | a_6)}.
\end{eqnarray}
\end{subequations}

The relations (\ref{AppSTU0p1}) and (\ref{AppSTU0p3}) provide a
compact form for the covariant Taylor series expansion of the
biscalar $T = \Delta ^{-1/2}{\Delta ^{1/2}}_{;\mu} \sigma^{; \mu}$.
Of course, it is also possible to give that covariant Taylor series
expansion in a more explicit form by replacing into
(\ref{AppSTU0p1}) and (\ref{AppSTU0p3}) the $K_{(p)}$ by their
expressions (\ref{AppK1}). We then have
\begin{eqnarray}\label{AppSTU0p4}
& & T = \Delta ^{-1/2}{\Delta ^{1/2}}_{;\mu} \sigma^{; \mu} =
\frac{1}{6}  \,  R_{a_1 a_2} \sigma^{;a_1}\sigma^{;a_2}
-\frac{1}{24}  \,   R_{a_1 a_2; a_3}  \sigma^{;a_1}\sigma^{;a_2} \sigma^{;a_3} \nonumber \\
& & \quad + \left[ \frac{1}{120}  \, R_{a_1 a_2; a_3 a_4} +
\frac{1}{90} \, R^{\rho}_{\phantom{\rho} a_1  \tau  a_2}
R^{\tau}_{\phantom{\tau} a_3  \rho   a_4 } \right]
\sigma^{;a_1}\sigma^{;a_2} \sigma^{;a_3} \sigma^{;a_4}    \nonumber \\
& & \quad - \left[ \frac{1}{720} \, R_{a_1 a_2; a_3 a_4 a_5} +
\frac{1}{120} \, R^{\rho}_{\phantom{\rho} a_1  \tau  a_2}
R^{\tau}_{\phantom{\tau} a_3  \rho   a_4 ; a_5} \right]
\sigma^{;a_1}\sigma^{;a_2} \sigma^{;a_3} \sigma^{;a_4} \sigma^{;a_5} \nonumber \\
& & \quad + \left[ \frac{1}{5040} \, R_{a_1 a_2; a_3 a_4 a_5 a_6} +
\frac{1}{504} \, R^{\rho}_{\phantom{\rho} a_1  \tau  a_2}
R^{\tau}_{\phantom{\tau} a_3  \rho   a_4 ; a_5 a_6}
 \right. \nonumber
\\
&  & \qquad  \left. + \frac{17}{10080} \, R^{\rho}_{\phantom{\rho}
a_1  \tau  a_2; a_3} R^{\tau}_{\phantom{\tau} a_4  \rho   a_5 ; a_6}
+ \frac{1}{945} \, R^{\rho}_{\phantom{\rho}a_1 \tau a_2
}R^{\tau}_{\phantom{\tau} a_3 \sigma
a_4}R^{\sigma}_{\phantom{\sigma} a_5 \rho  a_6}  \right]
 \sigma^{;a_1}\sigma^{;a_2} \sigma^{;a_3} \sigma^{;a_4}
 \sigma^{;a_5}\sigma^{;a_6} \nonumber \\
& & \quad + O \left(\sigma^{7/2} \right)
\end{eqnarray}
(here we have not included the terms given in (\ref{AppSTU0p3_7}),
(\ref{AppSTU0p3_8}) and (\ref{AppSTU0p3_9}) which correspond
respectively to the orders $\sigma^{7/2}$, $\sigma^4$ and
$\sigma^{9/2}$ of the expansion).
\end{widetext}

\section{Covariant Taylor series expansions of the bitensors $F$,
$F_{;\mu}$, $F_{;\mu \nu}$ and $\Box F$ when $F$ is a symmetric
biscalar}

In this Appendix, we shall first construct the covariant Taylor
series expansions of the covariant derivative, the second covariant
derivative and the d'Alembertian of an arbitrary biscalar $F(x,x')$
from that of this biscalar. We shall then express the constraints
induced on all these expansions by the symmetry of the biscalar
$F(x,x')$ in the exchange of $x$ and $x'$.

Let us first consider an arbitrary biscalar $F(x,x')$. Its covariant
Taylor series expansion is given by
\begin{equation}\label{CTSExpF1}
F(x,x')=f(x) +\sum_{p=1}^{+\infty} \frac{(-1)^p}{p!} f_{(p)}(x,x')
\end{equation}
where the $f_{(p)}(x,x')$ are biscalars in $x$ and $x'$ which are of
the form
\begin{equation}
f_{(p)}(x,x')=f_{a_1 \dots a_p}(x) \sigma^{;a_1}(x,x') \dots
\sigma^{;a_p}(x,x'). \label{CTSExpF2}
\end{equation}
Its covariant derivative $(\nabla F) (x,x')$, its second covariant
derivative $(\nabla \nabla F) (x,x')$ and its d'Alembertian $(\Box
F) (x,x')$ possess covariant Taylor series expansions given by
\begin{eqnarray}
& & (\nabla F) (x,x')={\overline{f}} (x) +\sum_{p=1}^{+\infty}
\frac{(-1)^p}{p!} {\overline{f}}_{(p)}(x,x') \label{CTSExpNabF1} \\
& & (\nabla \nabla F) (x,x')={\overline{\overline{f}}} (x)
+\sum_{p=1}^{+\infty}
\frac{(-1)^p}{p!} {\overline{\overline{f}}}_{(p)}(x,x') \label{CTSExpNabNabF1} \\
& & (\Box F) (x,x')=f''(x) +\sum_{p=1}^{+\infty} \frac{(-1)^p}{p!}
f''_{(p)}(x,x'). \label{CTSExpBoxF1}
\end{eqnarray}
In Eq.~(\ref{CTSExpNabF1}), ${\overline{f}} (x)$ is a tensor of type
$(0,1)$ in $x$ of which components are of the form
${\overline{f}}_\mu (x)$ while the ${\overline{f}}_{(p)}(x,x')$ with
$p=1,2,\dots $ are tensors of type $(0,1)$ in $x$ and scalars in
$x'$ of which components are of the form
\begin{equation}
{\overline{f}}_{\mu \,\, a_1 \dots a_p}(x) \sigma^{;a_1}(x,x') \dots
\sigma^{;a_p}(x,x'). \label{CTSExpNabF2}
\end{equation}
In Eq.~(\ref{CTSExpNabNabF1}), $\overline{\overline{f}} (x)$ is a
tensor of type $(0,2)$ in $x$ of which components are of the form
${\overline{\overline{f}}}_{\mu \nu} (x)$ while the
${\overline{\overline{f}}}_{(p)}(x,x')$ with $p=1,2,\dots $ are
tensors of type $(0,2)$ in $x$ and scalars in $x'$ of which
components are of the form
\begin{equation}
{\overline{\overline{f}}}_{\mu \nu  \,\, a_1 \dots a_p}(x)
\sigma^{;a_1}(x,x') \dots \sigma^{;a_p}(x,x').
\label{CTSExpNabNabF2}
\end{equation}
In Eq.~(\ref{CTSExpBoxF1}), $f'' (x)$ is a scalar in $x$ while the
$f''_{(p)}(x,x')$ with $p=1,2,\dots $ are biscalars of the form
\begin{equation}
f''_{(p)}(x,x')=f''_{a_1 \dots a_p}(x) \sigma^{;a_1}(x,x') \dots
\sigma^{;a_p}(x,x'). \label{CTSExpBoxF2}
\end{equation}
Of course, we can establish relationships between the components of
the covariant Taylor series expansions of $F (x,x')$, $(\nabla F)
(x,x')$, $(\nabla \nabla F) (x,x')$ and $(\Box F) (x,x')$. By taking
the covariant derivative of (\ref{CTSExpF1}) and by using
(\ref{CTSExpF2}) as well as the covariant Taylor series expansions
of $\sigma^{;\mu}_{\phantom{;\mu} \nu}$ given in
Eqs.~(\ref{AppGetHetL4a}) and (\ref{AppGetHetL4b}), we can link the
components of the covariant Taylor series expansions of $(\nabla F)
(x,x')$ and $F (x,x')$. We have
\begin{widetext}
\begin{subequations}\label{AppNabF_2}
\begin{eqnarray}
&  &  {\overline{f}}_\mu= f_{;\mu} -f_\mu     \\
&  & {\overline{f}}_{ \mu \,\, a_1}=  f_{a_1;\mu}- f_{\mu a_1 }  \\
&  &  {\overline{f}}_{ \mu \,\, a_1 a_2}= f_{a_1 a_2; \mu } - f_{
\mu a_1
a_2} - f_\rho \lambda^\rho_{\phantom{\rho} \mu \,\, a_1 a_2} \\
&  &  {\overline{f}}_{ \mu \,\, a_1 \dots a_p}=  f_{a_1 \dots a_p;
\mu } - f_{\mu a_1 \dots a_p } - f_\rho \lambda^\rho_{\phantom{\rho}
\mu \,\, a_1 \dots a_p}
\nonumber \\
& & \qquad \qquad \qquad \qquad - \sum_{\begin{subarray}{l} r+s=p  \\
r\ge 1, \, s \ge 2 \end{subarray}}    \left( \begin{array}{c}  p    \\
r \,  s \end{array} \right) f_{\rho (a_1 \dots a_r}
\lambda^\rho_{\phantom{\rho} |\mu| \,\, a_{r+1} \dots a_p) } \quad
\mathrm{for} \quad p \ge 3.
\end{eqnarray}
\end{subequations}
\end{widetext}
Similarly, by taking the covariant derivative of (\ref{CTSExpNabF1})
and by using (\ref{CTSExpNabF2}) as well as the covariant Taylor
series expansions of $\sigma^{;\mu}_{\phantom{;\mu} \nu}$ given in
Eqs.~(\ref{AppGetHetL4a}) and (\ref{AppGetHetL4b}), we can link the
components of the covariant Taylor series expansions of $(\nabla
\nabla F) (x,x')$ and $(\nabla F) (x,x')$. We have
\begin{widetext}
\begin{subequations}\label{AppNabNabF_2}
\begin{eqnarray}
&  &  {\overline{\overline{f}}}_{\mu \nu}=  {\overline{f}}_{\mu;\nu}
- {\overline{f}}_{ \mu
\,\, \nu}  \\
&  &  {\overline{\overline{f}}}_{\mu \nu  \,\, a_1}=
{\overline{f}}_{ \mu \,\, a_1 ;\nu} -
{\overline{f}}_{ \mu \,\, \nu a_1} \\
&  &  {\overline{\overline{f}}}_{\mu \nu  \,\, a_1 a_2}=
{\overline{f}}_{\mu \,\, a_1 a_2; \nu } - {\overline{f}}_{ \mu \,\,
\nu a_1 a_2}
 - {\overline{f}}_{ \mu \,\, \rho} \, \lambda^\rho_{\phantom{\rho} \nu \,\, a_1 a_2} \\
&  & {\overline{\overline{f}}}_{\mu \nu \,\, a_1 \dots a_p}=
{\overline{f}}_{\mu \,\, a_1 \dots a_p; \nu } - {\overline{f}}_{\mu
\,\, \nu a_1 \dots a_p } - {\overline{f}}_{ \mu \,\, \rho} \,
\lambda^\rho_{\phantom{\rho} \nu \,\, a_1 \dots a_p}
\nonumber \\
& & \qquad \qquad \qquad \qquad  - \sum_{\begin{subarray}{l} r+s=p  \\
r\ge 1, \, s \ge 2 \end{subarray}}    \left( \begin{array}{c}  p    \\
r \,  s \end{array} \right) {\overline{f}}_{\mu \,\, \rho (a_1 \dots
a_r} \lambda^\rho_{\phantom{\rho} |\nu| \,\, a_{r+1} \dots a_p) }
\quad \mathrm{for} \quad p \ge 3.
\end{eqnarray}
\end{subequations}
\end{widetext}
It is also possible to link the components of the covariant Taylor
series expansions of $(\nabla \nabla F)(x,x')$ and $F(x,x')$. This
can be done by using (\ref{AppNabF_2}) into (\ref{AppNabNabF_2}).
The resulting relations are rather complicated and we do not display
them here. However, it should be noted that they permit us to show
the symmetry of ${\overline{\overline{f}}}_{\mu \nu}$ and of the
${\overline{\overline{f}}}_{\mu \nu \,\, a_1 \dots a_n}$ in the
exchange of the indices $\mu$ and $\nu$, a result which does not
explicitly appear in Eq.~(\ref{AppNabNabF_2}). It is also possible
to link the components of the covariant Taylor series expansion of
$(\Box F) (x,x')$ and $F (x,x')$. This can be done from the previous
results by noting that
\begin{eqnarray}
& & f''= g^{\mu \nu}{\overline{\overline{f}}}_{\mu \nu} \, \, \,
\mathrm{and} \, \, \,  f''_{a_1 \dots a_p}=g^{\mu
\nu}{\overline{\overline{f}}}_{\mu \nu \,\, a_1 \dots a_p} \, \, \,
\mathrm{for} \,\,\, p \ge 1. \nonumber \\
\label{AppBoxF_1}
\end{eqnarray}

We now assume that the biscalar $F(x,x')$ is symmetric in the
exchange of $x$ and $x'$. It is well-known that this property
induces constraints on the coefficients $f_{a_1 \dots a_p}(x)$ of
the covariant Taylor series expansion of $F(x,x')$ (see, for
example, Ref.~\cite{BrownOttewill83}). These constraints are of the
form
\begin{eqnarray}\label{AppConstrF_1}
& & f_{a_1 \dots a_p}=(-1)^p f_{a_1 \dots a_p} \nonumber \\
& & \qquad + \sum_{k=0}^{p-1} (-1)^k \left( \begin{array}{c}    p    \\
k   \end{array} \right) f_{(a_1 \dots a_k;a_{k+1} \dots a_p)}\, \,
\mathrm{for} \,\, p \ge 1. \nonumber \\
& &
\end{eqnarray}
They permit us to determine the odd coefficients of the covariant
Taylor series expansion of $F(x,x')$ in terms of the even ones. We
have for the odd coefficients of lowest-orders
\begin{subequations} \label{AppConstrF_2}
\begin{eqnarray}
& & f_{a_1}=(1/2) \, f_{; a_1} \label{AppConstrF_2a} \\
& & f_{a_1 a_2 a_3}= (3/2) \, f_{(a_1 a_2; a_3)} - (1/4) \, f_{; (a_1 a_2 a_3)} \label{AppConstrF_2b}\\
& & f_{a_1 a_2 a_3 a_4 a_5}= (5/2) \, f_{(a_1 a_2 a_3 a_4 ;
a_5)}-(5/2) \, f_{(a_1 a_2; a_3 a_4 a_5)}
\nonumber \\
& & \quad + (1/2) \, f_{; (a_1 a_2 a_3 a_4 a_5)} \label{AppConstrF_2c}\\
& & f_{a_1 a_2 a_3 a_4 a_5 a_6 a_7}= (7/2) \, f_{(a_1 a_2 a_3 a_4
a_5 a_6; a_7)} \nonumber \\
& & \quad -(35/4) \, f_{(a_1 a_2 a_3 a_4; a_5 a_6 a_7)}
+ (21/2) \, f_{(a_1 a_2; a_3 a_4 a_5 a_6 a_7)} \nonumber \\
& & \quad - (17/8) \, f_{;(a_1 a_2 a_3 a_4 a_5 a_6 a_7)} \label{AppConstrF_2d}\\
& & f_{a_1 a_2 a_3 a_4 a_5 a_6 a_7 a_8 a_9}= (9/2) \, f_{(a_1 a_2
a_3 a_4 a_5 a_6 a_7 a_8; a_9)}\nonumber \\
& & \quad -21 \, f_{(a_1 a_2 a_3 a_4 a_5 a_6; a_7 a_8 a_9)}+ 63 \,
f_{(a_1 a_2 a_3 a_4; a_5 a_6 a_7 a_8 a_9)}\nonumber \\
& & \quad -(153/2) \, f_{(a_1 a_2; a_3 a_4 a_5 a_6 a_7 a_8 a_9)} \nonumber \\
& & \quad +(31/2) \, f_{;(a_1 a_2 a_3 a_4 a_5 a_6 a_7 a_8 a_9)}.
\label{AppConstrF_2e}
\end{eqnarray}
\end{subequations}

Of course, the constraints (\ref{AppConstrF_2}) permit us to
``simplify" the covariant Taylor series expansions of $(\nabla F)
(x,x')$, $(\nabla \nabla F) (x,x')$ and $(\Box F) (x,x')$. As far as
the later is concerned, it is given by (\ref{CTSExpBoxF1}) and by
using (\ref{AppBoxF_1}), (\ref{AppNabNabF_2}) and (\ref{AppNabF_2})
as well as (\ref{AppConstrF_2a}) and (\ref{AppConstrF_2b}) we obtain
for its coefficients of lowest orders:
\begin{widetext}
\begin{subequations}\label{AppBoxFsymm}
\begin{eqnarray}
& & f''=  f^\rho_{\phantom{\rho} \rho} \\
& & f''_{a_1}=\left[\frac{1}{4} \, (\Box f)_{;a_1} +  \frac{1}{2}
\, f^\rho_{\phantom{\rho} \rho; a_1} - f^\rho_{\phantom{\rho}
a_1;\rho} + \frac{1}{2} \, R^\rho_{\phantom{\rho}
a_1}f_{;\rho}\right]  \\
 & & f''_{a_1 a_2}=  \left[ f^\rho_{\phantom{\rho} \rho a_1
a_2} + \frac{1}{2} \, (\Box f)_{;a_1 a_2} -2 \,
f^\rho_{\phantom{\rho} a_1;\rho a_2} + \frac{1}{2} \, R_{\rho
a_1;a_2}f^{;\rho} + \frac{1}{12} \, R_{a_1 a_2;\rho}f^{;\rho}
\right.
\nonumber \\
& & \qquad \qquad\qquad\qquad\qquad\qquad \left. - \frac{4}{3} \,
R^\rho_{\phantom{\rho} a_1}f_{\rho a_2} + R^\rho_{\phantom{\rho}
a_1}f_{; \rho a_2} + \frac{2}{3} \, R_{\rho a_1 \sigma a_2}f^{\rho
\sigma} \right].
\end{eqnarray}
\end{subequations}
\end{widetext}

Finally, it is interesting to construct the covariant Taylor series
expansion of the biscalar $F_{;\mu} \sigma^{;\mu}$. It can be
obtained from (\ref{CTSExpNabF1}), (\ref{CTSExpNabF2}),
(\ref{AppNabF_2}) and (\ref{AppConstrF_2}) by noting that
$\lambda^\rho_{\phantom{\rho} \mu \,\, a_1 \dots a_p} \sigma^{;\mu}
\sigma^{;a_1} \dots \sigma^{;a_p}=0$ (this relation being a direct
consequence of the symmetries of the Riemann tensor). We have
\begin{eqnarray}\label{AppFS_cse}
& & F_{;\mu} \sigma^{;\mu} = (1/2) \, f_{;a_1}\sigma^{;a_1} +
 \left[f_{a_1 a_2}- (1/2) \, f_{;a_1 a_2}  \right]
\sigma^{;a_1}\sigma^{;a_2} \nonumber \\
& & \quad - \left[ (1/4) \, f_{a_1 a_2;a_3}- (1/8) \,f_{;a_1 a_2
a_3} \right]
 \sigma^{;a_1}\sigma^{;a_2} \sigma^{;a_3} \nonumber \\
& & \quad + \left[(1/6) \, f_{a_1 a_2 a_3 a_4} - (1/4) \, f_{a_1
a_2;a_3 a_4} \right. \nonumber \\
& & \qquad    \left. + (1/24) \, f_{;a_1 a_2 a_3 a_4} \right]
\sigma^{;a_1}\sigma^{;a_2} \sigma^{;a_3} \sigma^{;a_4}  + O
\left(\sigma^{5/2} \right). \nonumber \\
\end{eqnarray}

\section{Covariant Taylor series expansions of the biscalar $\Box \Delta ^{1/2} $.}

In this Appendix, we shall obtain the covariant Taylor series
expansion of the biscalar $\Box \Delta ^{1/2}$ up to order
$\sigma^2$. We shall derive these results from three intermediate
long calculations concerning the covariant Taylor series expansions
of $Z_{;\mu}$, $Z_{;\mu \nu}$ and $\Box Z$ up to orders
$\sigma^{5/2}$, $\sigma^2$ and $\sigma^2$ respectively. In those
calculations, we have extensively used the commutation of covariant
derivatives in the form (\ref{CD_NabNabTensor}) as well as the
Bianchi identities
\begin{subequations}\label{AppBianchi_1}
\begin{eqnarray}
& & R_{abcd}+R_{adbc}+R_{acdb}=0 \label{AppBianchi_1a} \\
& & R_{abcd;e}+R_{abec;d}+R_{abde;c}=0 \label{AppBianchi_1b}
\end{eqnarray}
\end{subequations}
and their consequences
\begin{subequations}\label{AppBianchi_2}
\begin{eqnarray}
& & R_{ abcd}^{\phantom{abcd};d}= R_{ac;b}-R_{bc;a}  \label{AppBianchi_2a} \\
& &R_{ ab}^{\phantom{ab};b}= (1/2) \, R_{;a} \label{AppBianchi_2b}
\end{eqnarray}
\end{subequations}
and
\begin{subequations}\label{AppBianchi_3}
\begin{eqnarray}
& & R^{ \rho \sigma \tau}_{\phantom{ \rho \sigma \tau} a}R_{\tau
\sigma \rho b}= (1/2) \, R^{ \rho \sigma \tau}_{\phantom{ \rho
\sigma \tau} a}R_{\rho \sigma \tau b} \\
& & R^{ \rho \sigma \tau}_{\phantom{ \rho \sigma \tau} a}R_{\tau
\sigma \rho b;c}= (1/2) \, R^{ \rho \sigma \tau}_{\phantom{ \rho
\sigma \tau} a}R_{\rho \sigma \tau b;c} \\
& & R^{ \rho \sigma \tau}_{\phantom{ \rho \sigma \tau} a}R_{\tau
\sigma \rho b;cd}= (1/2) \, R^{ \rho \sigma \tau}_{\phantom{ \rho
\sigma \tau} a}R_{\rho \sigma \tau b;cd} \\
& & R^{ \rho \sigma \tau}_{\phantom{ \rho \sigma \tau} a;b}R_{\tau
\sigma \rho c;d}= (1/2) \, R^{ \rho \sigma \tau}_{\phantom{ \rho
\sigma \tau} a;b}R_{\rho \sigma \tau c;d}
\end{eqnarray}
\end{subequations}
and
\begin{subequations}\label{AppBianchi_4}
\begin{eqnarray}
& & R^{ \rho \sigma \tau}_{\phantom{ \rho \sigma \tau} a}R_{\tau b
\sigma c ;\rho}= -(1/2) \, R^{ \rho \sigma \tau}_{\phantom{ \rho
\sigma \tau} a}R_{\rho \sigma \tau b;c} \\
& & R^{ \rho \sigma \tau}_{\phantom{ \rho \sigma \tau} a}R_{\tau b
\sigma c ;\rho d}= -(1/2) \, R^{ \rho \sigma \tau}_{\phantom{ \rho
\sigma \tau} a}R_{\rho \sigma \tau b;cd} \\
& & R^{ \rho \sigma \tau}_{\phantom{ \rho \sigma \tau} a;b}R_{\tau c
\sigma d ;\rho}= -(1/2) \, R^{ \rho \sigma \tau}_{\phantom{ \rho
\sigma \tau} a;b}R_{\rho \sigma \tau c;d}.
\end{eqnarray}
\end{subequations}

Let us first consider the covariant Taylor series expansion of
$Z_{;\mu}$. Up to order $\sigma^{5/2}$, it is of the form
\begin{widetext}
\begin{eqnarray}\label{AppNabZ_1}
& & Z_{;\mu}=-{\overline{\zeta}}_{ \mu \,\,
a_1}\sigma^{;a_1}+\frac{1}{2!}{\overline{\zeta}}_{ \mu \,\, a_1
a_2}\sigma^{;a_1}\sigma^{;a_2}-\frac{1}{3!}{\overline{\zeta}}_{ \mu
\,\,
a_1 a_2 a_3}\sigma^{;a_1}\sigma^{;a_2}\sigma^{;a_3} \nonumber \\
& & \qquad \quad  + \frac{1}{4!}{\overline{\zeta}}_{ \mu \,\, a_1
a_2 a_3
a_4}\sigma^{;a_1}\sigma^{;a_2}\sigma^{;a_3}\sigma^{;a_4}-\frac{1}{5!}{\overline{\zeta}}_{
\mu \,\, a_1 a_2 a_3 a_4
a_5}\sigma^{;a_1}\sigma^{;a_2}\sigma^{;a_3}\sigma^{;a_4}\sigma^{;a_5}+
O \left(\sigma^{3} \right)
\end{eqnarray}
where the coefficients ${\overline{\zeta}}_{ \mu \,\, a_1 \dots
a_p}$ with $p=1,\dots,5$ are obtained from
(\ref{AppZeta4_2})-(\ref{AppZeta4_6}) by using (\ref{AppNabF_2}) and
(\ref{lambda5bis_2})-(\ref{lambda5bis_4}) and are given by
\begin{subequations}\label{AppNabZ_2}
\begin{eqnarray}
&  &  {\overline{\zeta}}_{ \mu \,\, a_1}= -(1/6) \, R_{\mu a_1}    \\
&  &  {\overline{\zeta}}_{ \mu \,\, a_1 a_2}= (1/12) \, R_{a_1 a_2;
\mu}
- (1/6) \, R_{\mu (a_1; a_2)}  \\
&  &  {\overline{\zeta}}_{ \mu \,\, a_1 a_2 a_3}=  (1/10) \, R_{(a_1
a_2; |\mu | a_3)} -(3/20) \, R_{\mu (a_1; a_2 a_3)} -(1/15) \,
R^{\rho}_{\phantom{\rho}  \mu \tau (a_1 } R^{\tau}_{\phantom{\tau}
a_2 |\rho | a_3)} \nonumber \\
&  &  \quad \qquad - (1/60) \, R_{\rho (a_1}
R^{\rho}_{\phantom{\rho} a_2 |\mu| a_3)}
\\
&  &  {\overline{\zeta}}_{ \mu \,\ a_1 a_2 a_3 a_4}=   (1/10) \,
R_{(a_1 a_2; |\mu | a_3 a_4)}-(2/15) \, R_{\mu (a_1; a_2 a_3 a_4)} +
(1/15) \, R^{\rho}_{\phantom{\rho} (a_1 |\tau| a_2}
R^{\tau}_{\phantom{\tau}
a_3  |\rho|  a_4); \mu}  \nonumber \\
&  &  \quad \qquad  - (2/15)    \, R^{\rho}_{\phantom{\rho}  \mu
\tau (a_1 } R^{\tau}_{\phantom{\tau} a_2 |\rho | a_3; a_4)}  -
(2/15) \, R^{\rho}_{\phantom{\rho}   (a_1 |\tau | a_2 }
R^{\tau}_{\phantom{\tau} | \mu \rho |  a_3; a_4)} - (2/15) \,
R_{\rho (a_1; a_2} R^{\rho}_{\phantom{\rho}
a_3  |\mu | a_4)}  \nonumber \\
&  &  \quad \qquad + (1/10) \,  R_{(a_1 a_2 ; |\rho|}
R^{\rho}_{\phantom{\rho} a_3  |\mu | a_4)}
\end{eqnarray}
and
\begin{eqnarray}
 &  &  {\overline{\zeta}}_{ \mu \,\, a_1 a_2 a_3 a_4 a_5}=(2/21) \, R_{(a_1 a_2;
|\mu | a_3 a_4 a_5)}-(5/42) \, R_{\mu (a_1; a_2 a_3 a_4 a_5)} +
(1/7) \, R^{\rho}_{\phantom{\rho} (a_1 |\tau| a_2}
R^{\tau}_{\phantom{\tau}
a_3  |\rho|  a_4; |\mu| a_5)}  \nonumber \\
&  &  \quad  -(4/21)  \, R^{\rho}_{\phantom{\rho}  \mu \tau (a_1 }
R^{\tau}_{\phantom{\tau} a_2 |\rho | a_3; a_4 a_5)}  - (4/21) \,
R^{\rho}_{\phantom{\rho}   (a_1 |\tau | a_2 }
R^{\tau}_{\phantom{\tau} | \mu \rho |  a_3; a_4 a_5)} + (13/84) \,
 R^{\rho}_{\phantom{\rho}   (a_1 |\tau | a_2  ;| \mu |}
R^{\tau}_{\phantom{\tau}   a_3 |\rho| a_4; a_5)}
\nonumber \\
&  &  \quad  -(5/14)  \, R^{\rho}_{\phantom{\rho}  \mu \tau (a_1 ;
a_2} R^{\tau}_{\phantom{\tau} a_3 |\rho | a_4; a_5)}  + (1/42) \,
R_{\rho  (a_1} R^{\rho}_{\phantom{\rho} a_2 | \mu| a_3; a_4 a_5)} -
(3/14) \, R_{\rho  (a_1; a_2} R^{\rho}_{\phantom{\rho}
a_3 | \mu| a_4; a_5)} \nonumber \\
&  &  \quad  + (17/84) \, R_{(a_1 a_2; |\rho|}
R^{\rho}_{\phantom{\rho} a_3 | \mu| a_4; a_5)} - (2/7) \, R_{\rho
(a_1; a_2 a_3} R^{\rho}_{\phantom{\rho} a_4 | \mu| a_5)} + (5/21) \,
R_{(a_1 a_2; |\rho| a_3} R^{\rho}_{\phantom{\rho} a_4 | \mu|
a_5)} \nonumber \\
&  &  \quad  -(1/126) \, R_{\rho (a_1} R^{\rho}_{\phantom{\rho} a_2
|\tau | a_3} R^{\tau}_{\phantom{\tau} a_4 |\mu| a_5)}-(8/63)  \,
R^{\rho}_{\phantom{\rho}  \mu \tau (a_1 } R^{\tau}_{\phantom{\tau}
a_2 |\sigma | a_3}R^{\sigma}_{\phantom{\sigma} a_4 |\rho | a_5)}
\nonumber \\
&  &  \quad -(2/63)  \, R^{\sigma}_{\phantom{\sigma}  \rho \tau (a_1
} R^{\tau}_{\phantom{\tau} a_2 |\sigma |
a_3}R^{\rho}_{\phantom{\rho} a_4 |\mu | a_5)}.
 \end{eqnarray}
\end{subequations}

Let us now consider the covariant Taylor series expansion of
$Z_{;\mu \nu}$. Up to order $\sigma^2$, it is of the form
\begin{eqnarray}\label{AppNabNabZ_1}
& & Z_{;\mu \nu}={\overline{\overline{\zeta}}}_{ \mu
\nu}-{\overline{\overline{\zeta}}}_{ \mu \nu \,\,
a_1}\sigma^{;a_1}+\frac{1}{2!}{\overline{\overline{\zeta}}}_{ \mu\nu
\,\, a_1
a_2}\sigma^{;a_1}\sigma^{;a_2}-\frac{1}{3!}{\overline{\overline{\zeta}}}_{
\mu\nu \,\, a_1 a_2 a_3}\sigma^{;a_1}\sigma^{;a_2}\sigma^{;a_3}
\nonumber \\
& & \qquad \qquad \qquad \qquad +
\frac{1}{4!}{\overline{\overline{\zeta}}}_{ \mu\nu \,\, a_1 a_2 a_3
a_4}\sigma^{;a_1}\sigma^{;a_2}\sigma^{;a_3}\sigma^{;a_4} + O
\left(\sigma^{5/2} \right)
\end{eqnarray}
where the coefficients ${\overline{\overline{\zeta}}}_{ \mu \nu}$
and ${\overline{\overline{\zeta}}}_{ \mu \nu \,\, a_1 \dots a_p}$
with $p=1,\dots,4$ can be obtained from (\ref{AppNabZ_2}) by using
(\ref{AppNabNabF_2}) and (\ref{lambda5bis_2})-(\ref{lambda5bis_4})
and are given by
\begin{subequations}\label{AppNabNabZ_2}
\begin{eqnarray}
&  &  {\overline{\overline{\zeta}}}^{\,  \mu \nu}= (1/6) \, R^{\mu \nu}    \\
&  &  {\overline{\overline{\zeta}}}^{\,  \mu \nu}_{\phantom{ \mu
\nu}   a_1} = (1/12) \, R^{\mu \nu}_{\phantom{\mu \nu} ;a_1}
-(1/6) \, R^{(\mu \phantom{ a_1} ; \nu)}_{\phantom{( \mu}  a_1}    \\
&  &  {\overline{\overline{\zeta}}}^{\,  \mu \nu}_{\phantom{ \mu
\nu}  a_1 a_2}= (1/20) \, R_{a_1 a_2}^{\phantom{a_1 a_2}; (\mu \nu)}
+ (1/20) \, R^{ \mu \nu }_{\phantom{\mu \nu};(a_1 a_2)} -(2/15) \,
R^{(\mu \phantom{(a_1} ;\nu )}_{\phantom{(\mu } (a_1 \phantom{ \mu }
a_2)} + (1/90) \, R^{\rho (\mu} R^{\phantom{\rho (a_1} \nu)}_{\rho
(a_1
\phantom{\nu} a_2)}  \nonumber \\
&  & \quad   + (11/90) \, R_{\rho (a_1} R^{\rho (\mu \nu)
}_{\phantom{\rho (\mu \nu)  } a_2)}   +(1/45) \,  R^{\mu
\phantom{\rho} \nu}_{\phantom{\mu} \rho \phantom{\nu} \tau} R^{\rho
\phantom{(a_1} \tau}_{\phantom{\rho} (a_1 \phantom{\tau}a_2) } +
(1/45) \, R^{\mu }_{\phantom{\mu} \rho \tau (a_1 } R^{\nu \rho \tau
}_{ \phantom{\nu  \rho \tau}  a_2)}
 \nonumber \\
&  & \quad   + (1/45) \,  R^{\mu   }_{\phantom{\mu} \rho \tau (a_1 }
R^{\nu \tau \rho    }_{ \phantom{\nu  \tau \rho }  a_2)} \\
&  &  {\overline{\overline{\zeta}}}^{\,  \mu \nu}_{\phantom{ \mu
\nu}   a_1 a_2 a_3}= (1/30) \, R^{ \mu \nu }_{\phantom{\mu \nu};(a_1
a_2 a_3)} + (1/20) \, R_{(a_1 a_2 \phantom{; (\mu \nu)}
a_3)}^{\phantom{a_1 a_2}; (\mu \nu)} -   (1/10)
\, R_{(a_1  \phantom{ (\mu ; \nu)} a_2 a_3)}^{\phantom{(a_1} (\mu ; \nu)}  \nonumber \\
&  & \quad   + (1/12) \, R_{\rho (a_1} R^{\rho (\mu \nu)
}_{\phantom{\rho (\mu \nu)  } a_2;a_3)}  + (1/60)   \, R_{\rho (a_1}
R^{\rho \phantom{a_2 a_3)} (\mu ;\nu) }_{\phantom{\rho} a_2 a_3)}  +
(1/30)   \, R_{\rho (a_1}^{\phantom{\rho (a_1}; (\mu} R^{\nu)
\phantom{a_2} \rho }_{\phantom{\nu)} a_2 \phantom{\rho}
a_3)} \nonumber \\
&  & \quad   +  (1/15)   \, R^{(\mu}_{\phantom{(\mu} \rho; (a_1}
R^{\nu) \phantom{a_2} \rho }_{\phantom{\nu)} a_2 \phantom{\rho}
a_3)} -  (1/10)   \, R^{(\mu}_{\phantom{(\mu} (a_1; |\rho| } R^{\nu)
\phantom{a_2} \rho }_{\phantom{\nu)} a_2 \phantom{\rho} a_3)} +
(7/30) \, R_{\rho (a_1; a_2} R^{\rho (\mu \nu)
}_{\phantom{\rho (\mu \nu)  } a_3)} \nonumber \\
&  & \quad    -  (1/10) \, R_{(a_1 a_2; |\rho |} R^{\rho (\mu \nu)
}_{\phantom{\rho (\mu \nu)  } a_3)} -  (1/15) \,
R^{\rho}_{\phantom{\rho} (a_1 |\tau| a_2} R^{\tau \phantom{a_3) \rho
} (\mu;\nu)}_{\phantom{\tau} a_3) \rho } -  (1/15) \, R^{\rho
\phantom{(a_1 |\tau| a_2}; (\mu }_{\phantom{\rho} (a_1 |\tau| a_2}
R^{|\tau| \phantom{a_3) \rho } \nu)}_{\phantom{|\tau|} a_3) \rho }  \nonumber \\
&  & \quad     + (1/15) \,  R^{(\mu   }_{\phantom{(\mu} \rho \tau
(a_1 } R^{\nu) \tau \rho    }_{ \phantom{\nu  \tau \rho } a_2;a_3)}
+ (1/15) \, R^{(\mu }_{\phantom{(\mu} \rho \tau (a_1 } R^{\nu) \rho
\tau  }_{ \phantom{\nu) \rho \tau}  a_2;a_3)} +(1/30) \,  R^{\mu
\phantom{\rho} \nu}_{\phantom{\mu} \rho \phantom{\nu} \tau ; (a_1}
R^{\rho \phantom{ a_1} \tau}_{\phantom{\rho}  a_2
\phantom{\tau} a_3) }   \nonumber \\
&  & \quad   +  (1/30) \,  R^{\mu \phantom{\rho} \nu}_{\phantom{\mu}
\rho \phantom{\nu} \tau } R^{\rho \phantom{ (a_1}
\tau}_{\phantom{\rho}  (a_1 \phantom{\tau} a_2;a_3) }
\end{eqnarray}
and
\begin{eqnarray}
&  &  {\overline{\overline{\zeta}}}^{\mu \nu}_{\phantom{  \mu \nu}
 a_1 a_2 a_3 a_4}= (3/70) \, R_{(a_1 a_2 \phantom{(\mu \nu)} a_3
a_4) }^{\phantom{(a_1 a_2}; (\mu \nu)} + (1/42) \, R^{ \mu \nu
}_{\phantom{\mu \nu};(a_1 a_2 a_3 a_4)} -(8/105) \, R^{(\mu
\phantom{(a_1} ;\nu )}_{\phantom{(\mu } (a_1 \phantom{ (\mu } a_2
a_3 a_4 )} \nonumber \\
&  & \quad   -(1/105) \, R^{\rho (\mu} R^{\nu)}_{\phantom{\nu)} (a_1
|\rho| a_2;a_3 a_4)}+ (3/35) \, R^{\rho \phantom{a_1};
(\mu}_{\phantom{\rho} (a_1}R^{\nu)}_{\phantom{\nu)} a_2 |\rho| a_3;
a_4)}+ (3/35) \, R^{\rho (\mu}_{\phantom{\rho (\mu}
;(a_1}R^{\nu)}_{\phantom{\nu)} a_2 |\rho| a_3; a_4)}\nonumber \\
&  & \quad   -(17/105) \, R^{(\mu
\phantom{(a_1};|\rho|}_{\phantom{\mu} (a_1}R^{\nu)}_{\phantom{\nu)}
a_2 |\rho| a_3;a_4)} -  (4/105) \, R^{\rho \phantom{a_1};
(\mu}_{\phantom{\rho} (a_1 \phantom{(\mu} a_2
}R^{\nu)}_{\phantom{\nu)} a_3 |\rho| a_4)} + (11/105) \ R_{(a_1
a_2}^{\phantom{(a_1 a_2} ;(\mu |\rho|}
R^{\nu)}_{\phantom{\nu)} a_3 |\rho| a_4)}\nonumber \\
&  & \quad + (4/35) \, R^{\rho (\mu}_{\phantom{\rho (\mu} ;(a_1 a_2
}R^{\nu)}_{\phantom{\nu)} a_3 |\rho| a_4)} -(4/21) \, R^{(\mu
\phantom{(a_1}; |\rho|}_{\phantom{(\mu} (a_1 \phantom{;|\rho|} a_2
}R^{\nu)}_{\phantom{\nu)} a_3 |\rho| a_4)} + (2/35) \,
R^{\rho}_{\phantom{\rho} (a_1} R^{\phantom{|\rho|} (\mu \nu)
}_{|\rho| \phantom{(\mu \nu)} a_2;a_3 a_4)} \nonumber \\
&  & \quad  + (1/105) \, R^{\rho}_{\phantom{\rho} (a_1} R_{|\rho|
a_2 a_3 \phantom{(\mu ; \nu)} a_4)}^{\phantom{|\rho| a_2 a_3} (\mu;
\nu) } +(53/210) \, R^{\rho}_{\phantom{\rho} (a_1;a_2}
R^{\phantom{|\rho|} (\mu \nu) }_{|\rho| \phantom{(\mu \nu)}
a_3;a_4)}  + (19/210) \, R^{\rho}_{\phantom{\rho} (a_1;a_2}
R_{|\rho| a_3 a_4) \phantom{(\mu ; \nu)}}^{\phantom{|\rho| a_2 a_3}
(\mu; \nu) } \nonumber \\
&  & \quad + (11/35) \, R^{\rho}_{\phantom{\rho} (a_1;a_2 a_3}
R^{\phantom{|\rho|} (\mu \nu) }_{|\rho| \phantom{(\mu \nu)} a_4)} -
(43/420) \, R_{(a_1 a_2}^{\phantom{(a_1
a_2};\rho}R^{\phantom{|\rho|} (\mu \nu) }_{|\rho| \phantom{(\mu
\nu)} a_3;a_4)}-(5/84) \, R_{(a_1 a_2}^{\phantom{(a_1
a_2};\rho}R_{|\rho| a_3 a_4) \phantom{(\mu ; \nu)}}^{\phantom{|\rho|
a_2 a_3} (\mu; \nu) }  \nonumber \\
&  & \quad  -(4/21) \, R_{(a_1 a_2 \phantom{;\rho}
a_3}^{\phantom{(a_1 a_2};\rho}R^{\phantom{|\rho|} (\mu \nu)
}_{|\rho| \phantom{(\mu \nu)} a_4)} + (4/105) \,
R^{\rho}_{\phantom{\rho} (a_1 |\tau| a_2 }R^{\tau \phantom{a_3
|\rho| a_4)};(\mu \nu) }_{\phantom{\tau} a_3 |\rho| a_4)}   + (1/28)
\, R^{\rho \phantom{(a_1 |\tau| a_2};\mu}_{\phantom{\rho} (a_1
|\tau| a_2}R^{\tau \phantom{a_3 |\rho| a_4)};\nu}_{\phantom{\rho}
a_3 |\rho| a_4)} \nonumber \\
&  & \quad  +(4/105) \, R^{\rho \mu \phantom{\tau}
\nu}_{\phantom{\rho \mu} \tau \phantom{\nu}}R^\tau_{\phantom{\tau}
(a_1 |\rho| a_2; a_3 a_4)}+ (4/105) \, R^{\rho \phantom{(a_1 |\tau|
a_2}}_{\phantom{\rho} (a_1 |\tau| a_2} R^{\tau \mu \phantom{\rho}
\nu}_{\phantom{\tau \mu} \rho \phantom{\nu} ; a_3 a_4)} +(1/14) \,
R^{\rho \mu \phantom{\tau} \nu}_{\phantom{\rho \mu} \tau
\phantom{\nu}; (a_1}R^\tau_{\phantom{\tau} a_2 |\rho| a_3; a_4)} \nonumber \\
&  & \quad  - (4/35) \, R^{\rho \phantom{(a_1 |\tau| a_2};
(\mu}_{\phantom{\rho} (a_1 |\tau| a_2  \phantom{;(\mu} a_3}
R^{|\tau| \nu)}_{\phantom{|\tau| \nu)} |\rho| a_4)}  - (4/35) \,
R^{\rho}_{\phantom{\rho} (a_1 |\tau| a_2} R^{\tau  (\mu
\phantom{|\rho| a_3}; \nu)}_{\phantom{\tau (\mu} |\rho| a_3
\phantom{;\nu)} a_4)} - (13/105) \, R^{\rho (\mu \phantom{\tau
(a_1}; \nu)}_{\phantom{\rho (\mu} \tau (a_1 \phantom{;\nu)}}
R^{\tau}_{\phantom{\tau} a_2 |\rho| a_3 ;a_4)} \nonumber \\
&  & \quad - (13/105) \, R^{\rho \phantom{(a_1 |\tau| a_2};
(\mu}_{\phantom{\rho} (a_1 |\tau| a_2 } R^{|\tau|
\nu)}_{\phantom{|\tau| \nu)} |\rho| a_3 ; a_4)}  +(8/105) \, R^{\rho
(\mu \phantom{\tau (a_1} }_{\phantom{\rho (\mu} \tau (a_1 }
R^{|\tau| \nu)}_{\phantom{|\tau| \nu)} |\rho| a_2 ; a_3 a_4)}
+(8/105) \,  R^{(\mu}_{\phantom{(\mu} \rho \tau (a_1}R^{\nu) \rho
\tau}_{\phantom{\nu) \rho \tau} a_2;a_3 a_4)} \nonumber \\
&  & \quad +(1/14) \, R^{\mu}_{\phantom{\mu} \rho \tau
(a_1;a_2}R^{\nu \tau \rho}_{\phantom{\nu \tau \rho } a_3; a_4)}
+(1/14) \,  R^{\mu}_{\phantom{\mu} \rho \tau (a_1;a_2}R^{\nu  \rho
\tau}_{\phantom{\nu \rho \tau } a_3; a_4)} +(17/315) \, R^{\rho
\tau} R^\mu_{\phantom{\mu} (a_1 |\rho| a_2} R^\nu_{\phantom{\nu} a_3
|\tau| a_4)} \nonumber \\
&  & \quad  + (1/315) \, R^{(\mu}_{\phantom{(\mu} \rho}
R^{|\rho|}_{\phantom{|\rho|} (a_1 |\tau| a_2} R^{|\tau|
\phantom{a_3} \nu)}_{\phantom{|\tau|} a_3 \phantom{\nu)} a_4)} -
(68/315) \, R_{\rho (a_1} R^{\rho \phantom{a_2} (\mu
}_{\phantom{\rho} a_2 \phantom{(\mu } |\tau |} R^{|\tau|
\phantom{a_3} \nu)}_{\phantom{|\tau|} a_3
\phantom{\nu)} a_4)}  \nonumber \\
&  & \quad + (46/315) \, R_{\rho (a_1} R^{\rho \phantom{|\tau|} (\mu
}_{\phantom{\rho} |\tau| \phantom{(\mu } a_2} R^{|\tau|
\phantom{a_3} \nu)}_{\phantom{|\tau|} a_3 \phantom{\nu)} a_4)} +
(2/315) \, R_{\rho (a_1} R^{\rho}_{\phantom{\rho} a_2 |\tau|
a_3}R^{\tau (\mu \nu)}_{\phantom{\tau (\mu \nu)} a_4)} \nonumber \\
&  & \quad + (8/315) \, R^{\rho (\mu \phantom{\tau}
\nu)}_{\phantom{\rho (\mu} \tau} R^{\tau}_{\phantom{\tau} (a_1
|\sigma| a_2}R^{\sigma}_{\phantom{\sigma} a_3 |\rho| a_4)} +
(32/315) \, R^\sigma_{\phantom{\sigma} \rho \tau (a_1}
R^{\tau}_{\phantom{\tau} a_2 |\sigma| a_3}R^{\rho (\mu
\nu)}_{\phantom{\rho (\mu \nu)} a_4)} \nonumber \\
&  & \quad  + (4/315) \, R^\rho_{\phantom{\rho} (a_1 |\tau| a_2}
R^{\tau (\mu}_{\phantom{\tau (\mu} |\rho \sigma|} R^{|\sigma|
\phantom{a_3} \nu)}_{\phantom{|\sigma| } a_3 \phantom{\nu)} a_4)}  +
(4/315) \, R^{\rho (\mu}_{\phantom{\rho (\mu} \tau (a_1} R^{|\sigma
| \phantom{a_2} \nu)}_{\phantom{|\sigma | } a_2 \phantom{\nu)} a_3}
R^\tau_{\phantom{\tau} |\sigma \rho| a_4)} \nonumber \\
&  & \quad  + (4/315) \, R^{\rho (\mu}_{\phantom{\rho (\mu} \tau
(a_1} R^{|\sigma | \phantom{a_2} \nu)}_{\phantom{|\sigma | } a_2
\phantom{\nu)} a_3} R^\tau_{\phantom{\tau} a_4) \rho \sigma} +
(16/315) \, R^{\rho (\mu}_{\phantom{\rho (\mu} \tau (a_1} R^{|\tau |
\nu)}_{\phantom{|\tau | \nu)} |\sigma| a_2}
R^\sigma_{\phantom{\sigma} a_3 |\rho| a_4)}
\nonumber \\
&  & \quad + (8/315) \, R^\mu_{\phantom{\mu} \rho \tau (a_1}R^{\nu
\phantom{|\sigma|} \tau}_{\phantom{\nu} |\sigma| \phantom{\tau} a_2}
R^{\rho \phantom{a_3} \sigma}_{\phantom{\rho} a_3 \phantom{\sigma}
a_4)} + (8/315) \, R^\mu_{\phantom{\mu} \rho \tau (a_1} R^{\nu \rho}
_{\phantom{\nu \rho} |\sigma| a_2} R^{\sigma \phantom{a_3}
\tau}_{\phantom{\sigma} a_3 \phantom{\tau} a_4)}.
\end{eqnarray}
\end{subequations}
Here, the symmetry of all these coefficients in the exchange of the
indices $\mu$ et $\nu$ should be noted.

The covariant Taylor series expansion of $\Box Z$ up to order
$\sigma^2$ can be now obtained. It is of the form
\begin{eqnarray} \label{AppBoxZ_1}
& & \Box Z=  \zeta''
-\zeta''_{~a_1}\sigma^{;a_1}+\frac{1}{2!}\zeta''_{~a_1
a_2}\sigma^{;a_1}\sigma^{;a_2}-\frac{1}{3!}\zeta''_{~a_1 a_2
a_3}\sigma^{;a_1}\sigma^{;a_2}\sigma^{;a_3} +
\frac{1}{4!}\zeta''_{~a_1 a_2 a_3
a_4}\sigma^{;a_1}\sigma^{;a_2}\sigma^{;a_3}\sigma^{;a_4} + O
\left(\sigma^{5/2} \right)
\end{eqnarray}
where the coefficients $\zeta''$ and $\zeta''_{~a_1 \dots a_p}$ with
$p=1,\dots,4$ can be obtained from (\ref{AppNabNabZ_2}) by using
(\ref{AppBoxF_1}) and are given by
\begin{subequations}\label{AppBoxZ_2}
\begin{eqnarray}
&  &  \zeta''= (1/6) \, R     \\
&  &   \zeta''_{~a_1} = 0   \\
&  &   \zeta''_{~a_1 a_2}= (1/20) \, \Box R_{a_1 a_2}  - (1/60) \, R
_{ ;a_1 a_2} -(11/90) \, R^{ \rho  }_{\phantom{ \rho}  a_1  } R_{
\rho   a_2 }
  \nonumber \\
&  & \qquad   +(1/30) \,  R_{ \rho   \tau} R^{\rho \phantom{ a_1}
\tau}_{\phantom{\rho}  a_1 \phantom{\tau}a_2  }
 + (1/30) \,  R^{\rho \sigma \tau    }_{\phantom{\rho \sigma \tau} a_1 }
 R_{ \rho \sigma \tau  a_2 } \\
&  &   \zeta''_{~a_1 a_2 a_3}= - (1/60 ) \, R _{ ; (a_1 a_2 a_3)} +
(1/20) \,  (\Box R_{(a_1 a_2})_{;a_3)} -(3/10) \, R^{ \rho
}_{\phantom{ \rho}  (a_1 } R_{ |\rho|   a_2;a_3) }
  + (1/12) \, R^{ \rho  }_{\phantom{ \rho}  (a_1  }
R_{a_2 a_3);  \rho}  \nonumber \\
&  & \qquad + (1/30) \, R^{ \rho  }_{\phantom{ \rho}
 \sigma ;(a_1 } R^{\sigma}_{\phantom{\sigma} a_2 |\rho | a_3)}
 + (1/30) \, R^{ \rho  }_{\phantom{ \rho}  \sigma}
R^{\sigma}_{\phantom{\sigma}   (a_1 |\rho | a_2;  a_3)}  + (1/15) \,
R^{ \rho  \sigma \tau}_{\phantom{ \rho \sigma \tau} (a_1}
R_{|\rho \sigma \tau|   a_2;  a_3)} \\
&  &   \zeta''_{~a_1 a_2 a_3 a_4}=(3/70) \, ( \Box R_{(a_1 a_2}
)_{;a_3 a_4)} - (1/70) \, R _{ ; (a_1 a_2 a_3 a_4)}
-(38/105) \, R^{ \rho  }_{\phantom{ \rho}  (a_1  } R_{ |\rho|   a_2;a_3 a_4) }\nonumber \\
&  & \qquad + (19/105) \, R^{ \rho  }_{\phantom{ \rho}  (a_1  } R_{
a_2 a_3;  |\rho| a_4) } - (17/105) \, R^{ \rho  }_{\phantom{ \rho}
(a_1;a_2  } R_{ |\rho|   a_3;a_4) } - (1/21) \, R^{ \rho
}_{\phantom{ \rho}  (a_1 ; a_2 } R_{ a_3 a_4 );   \rho  } \nonumber \\
&  & \qquad + (5/84) \, R_{ (a_1 a_2 }^{ \phantom{(a_1 a_2 } ; \rho
}R_{ a_3 a_4 ); \rho  } + (1/35) \, R^{ \rho  }_{\phantom{ \rho}
 \sigma } R^{\sigma}_{\phantom{\sigma} (a_1 |\rho | a_2; a_3 a_4)}
+ (1/21) \, R^{ \rho  }_{\phantom{ \rho}
 (a_1 ; |\sigma |} R^{\sigma}_{\phantom{\sigma} a_2 |\rho | a_3; a_4)}
\nonumber \\
&  & \qquad + (1/30) \, R^{ \rho  }_{\phantom{ \rho}
 \sigma ;(a_1 } R^{\sigma}_{\phantom{\sigma} a_2 |\rho | a_3;
 a_4)} - (4/35) \, R^{ \rho  }_{\phantom{ \rho}
 (a_1 ; |\sigma | a_2} R^{\sigma}_{\phantom{\sigma} a_3 |\rho |
 a_4)}+ (11/105) \, R_{(a_1 a_2 \phantom{;\rho} \sigma}^{\phantom{(a_1 a_2 } ;
\rho  } R^{\sigma}_{\phantom{\sigma} a_3 |\rho | a_4)} \nonumber \\
&  & \qquad + (4/105) \, R^{ \rho  }_{\phantom{ \rho}
 \sigma ;(a_1 a_2} R^{\sigma}_{\phantom{\sigma} a_3 |\rho | a_4)}
 + (4/105) \, R^{\rho}_{\phantom{ \rho} (a_1 |\sigma| a_2}
 \Box R^{\sigma}_{\phantom{\sigma} a_3 |\rho | a_4)} + (2/35) \,
 R^{\rho \sigma \tau}_{\phantom{\rho \sigma \tau} (a_1} R_{|\rho \sigma \tau| a_2; a_3 a_4)}\nonumber \\
&  & \qquad+ (1/28) \,
 R^{\rho \phantom{(a_1 |\sigma| a_2}    ;\tau}_{\phantom{ \rho} (a_1 |\sigma| a_2}
R^{\sigma}_{\phantom{\sigma} a_3 |\rho | a_4) ; \tau} + (19/420) \,
 R^{\rho \sigma \tau}_{\phantom{\rho \sigma \tau} (a_1;a_2} R_{|\rho \sigma \tau|  a_3; a_4)}
 - (2/315) \, R^{ \rho  }_{\phantom{ \rho}
 (a_1 } R_{|\sigma| a_2}
 R^{\sigma}_{\phantom{\sigma} a_3 |\rho | a_4)}
\nonumber \\
&  & \qquad + (26/315) \, R^{ \rho  }_{\phantom{ \rho}
 \sigma } R^{\sigma}_{\phantom{\sigma} (a_1 |\tau | a_2}
 R^{\tau}_{\phantom{\tau} a_3 |\rho | a_4)} + (26/105) \, R^{ \rho  }_{\phantom{ \rho}
 (a_1 } R^{\sigma \phantom{a_2} \tau}_{\phantom{\sigma} a_2 \phantom{ \tau } a_3}
 R_{|\rho \sigma \tau| a_4)} \nonumber \\
&  & \qquad  + (4/315) \, R^{\rho \sigma \tau
 \kappa} R_{ \rho  (a_1 |\tau|
 a_2} R_{|\sigma| a_3 |\kappa|
 a_4)} + (4/105) \, R^{\rho  \kappa \tau }_{\phantom{\rho
\kappa \tau } (a_1} R_{|\rho \tau| \phantom{\sigma}
 a_2}^{\phantom{ |\rho \tau|} \sigma} R_{|\sigma| a_3 |\kappa|
 a_4)}\nonumber \\
&  & \qquad  + (16/315) \, R^{\rho  \kappa \tau }_{\phantom{\rho
\kappa \tau } (a_1} R_{|\rho  \phantom{\sigma}
 \tau| a_2}^{\phantom{ |\rho| } \sigma} R_{|\sigma| a_3 |\kappa|
 a_4)}  + (8/315) \, R^{\rho  \tau \kappa }_{\phantom{\rho
 \tau  \kappa} (a_1} R_{|\rho \tau| \phantom{\sigma}
 a_2}^{\phantom{ |\rho \tau|} \sigma} R_{|\sigma| a_3 |\kappa|
 a_4)}.
\end{eqnarray}
\end{subequations}

Finally, by noting that
\begin{equation}
\Box \Delta^{1/2}= (\Box Z +Z_{; \mu}Z^{; \mu})\Delta^{1/2}
\label{AppVBoxVVMfinal_1}
\end{equation}
we can constructed the covariant Taylor series expansion of $\Box
\Delta^{1/2}$ up to order $\sigma^2$. We have
\begin{eqnarray}\label{AppVBoxVVMfinal_2}
& & \Box \Delta^{1/2} ={\delta^{1/2 \, ''}} - {\delta^{1/2 \,
''}}_{a_1}\sigma^{;a_1}+\frac{1}{2!}{\delta^{1/2 \, ''}}_{a_1
a_2}\sigma^{;a_1}\sigma^{;a_2} \nonumber \\
& & \qquad \qquad   -\frac{1}{3!} {\delta^{1/2 \,
''}}_{a_1 a_2 a_3}\sigma^{;a_1}\sigma^{;a_2}\sigma^{;a_3} +
\frac{1}{4!} {\delta^{1/2 \, ''}}_{a_1 a_2 a_3
a_4}\sigma^{;a_1}\sigma^{;a_2}\sigma^{;a_3}\sigma^{;a_4} + O
\left(\sigma^{5/2} \right)
\end{eqnarray}
where the coefficients ${\delta^{1/2 \, ''}}$ and ${\delta^{1/2 \,
''}}_{a_1 \dots a_p}$ with $p=1,\dots,4$ can be obtained from
(\ref{AppVBoxVVMfinal_1}) by using the expansions of $\Delta^{1/2}$
(see Eq.~(\ref{AppSCU05})), $Z_{;\mu}$ (see
Eqs.~(\ref{AppNabZ_1})-(\ref{AppNabZ_2})) and $\Box Z$ (see
Eqs.~(\ref{AppBoxZ_1}) and (\ref{AppBoxZ_2})).  They are given by
\begin{subequations}\label{AppBoxVVMfinal_3}
\begin{eqnarray}
&  &  {\delta^{1/2 \, ''}}= (1/6) \, R     \label{AppBoxVVMfinal_3a}\\
&  &   {\delta^{1/2 \,
''}}_{a_1} = 0   \label{AppBoxVVMfinal_3b} \\
&  &   {\delta^{1/2 \, ''}}_{a_1 a_2}= (1/20) \, \Box R_{a_1 a_2}  -
(1/60) \, R _{ ;a_1 a_2} + (1/36) \, R R_{a_1 a_2} -(1/15) \, R^{
\rho  }_{\phantom{ \rho} a_1 } R_{ \rho   a_2 }
\nonumber \\
&  & \qquad   +(1/30) \,  R_{ \rho   \tau} R^{\rho \phantom{ a_1}
\tau}_{\phantom{\rho}  a_1 \phantom{\tau}a_2  }
 + (1/30) \,  R^{\rho \sigma \tau    }_{\phantom{\rho \sigma \tau} a_1 }
 R_{ \rho \sigma \tau  a_2 }  \label{AppBoxVVMfinal_3c} \\
&  &   {\delta^{1/2 \, ''}}_{a_1 a_2 a_3} = - (1/60 ) \, R _{ ; (a_1
a_2 a_3)} + (1/20) \, ( \Box R_{(a_1 a_2} )_{;a_3)} + (1/24 ) \, R R
_{  (a_1 a_2; a_3)} \nonumber \\ &  & \qquad -(2/15) \, R^{ \rho  }_{\phantom{ \rho} (a_1 }
R_{ |\rho| a_2;a_3) }
+ (1/30) \, R^{ \rho  }_{\phantom{ \rho}
 \sigma ;(a_1 } R^{\sigma}_{\phantom{\sigma} a_2 |\rho | a_3)}
 \nonumber \\ &  & \qquad  + (1/30) \, R^{ \rho  }_{\phantom{ \rho}  \sigma}
R^{\sigma}_{\phantom{\sigma}   (a_1 |\rho | a_2;  a_3)} + (1/15) \,
R^{ \rho  \sigma \tau}_{\phantom{ \rho \sigma \tau} (a_1}
R_{|\rho \sigma \tau|   a_2;  a_3)} \label{AppBoxVVMfinal_3d} \\
&  &   {\delta^{1/2 \, ''}}_{a_1 a_2 a_3 a_4}=(3/70) \, ( \Box
R_{(a_1 a_2})_{;a_3 a_4)} - (1/70) \, R _{ ; (a_1 a_2 a_3
a_4)} - (1/60) \, R _{ ; (a_1 a_2}R _{ a_3 a_4)} \nonumber \\
&  & \qquad +  (1/20) \, R R _{
 (a_1 a_2; a_3 a_4)} + (1/20) \, R _{  (a_1 a_2}\Box R _{ a_3 a_4)}
-(17/105) \, R^{ \rho  }_{\phantom{ \rho}  (a_1  } R_{ |\rho|   a_2;a_3 a_4) }\nonumber \\
&  & \qquad + (1/21) \, R^{ \rho  }_{\phantom{ \rho}  (a_1  } R_{
a_2 a_3;  |\rho| a_4) } + (1/210) \, R^{ \rho  }_{\phantom{ \rho}
(a_1;a_2  } R_{ |\rho|   a_3;a_4) } - (3/14) \, R^{ \rho
}_{\phantom{ \rho}  (a_1 ; a_2 } R_{ a_3 a_4 );   \rho  } \nonumber \\
&  & \qquad + (17/168) \, R_{ (a_1 a_2 }^{ \phantom{(a_1 a_2 } ;
\rho }R_{ a_3 a_4 ); \rho  } + (1/35) \, R^{ \rho  }_{\phantom{
\rho}
 \sigma } R^{\sigma}_{\phantom{\sigma} (a_1 |\rho | a_2; a_3 a_4)}
+ (1/21) \, R^{ \rho  }_{\phantom{ \rho}
 (a_1 ; |\sigma |} R^{\sigma}_{\phantom{\sigma} a_2 |\rho | a_3; a_4)}
\nonumber \\
&  & \qquad + (1/30) \, R^{ \rho  }_{\phantom{ \rho}
 \sigma ;(a_1 } R^{\sigma}_{\phantom{\sigma} a_2 |\rho | a_3;
 a_4)} - (4/35) \, R^{ \rho  }_{\phantom{ \rho}
 (a_1 ; |\sigma | a_2} R^{\sigma}_{\phantom{\sigma} a_3 |\rho |
 a_4)}+ (11/105) \, R_{(a_1 a_2 \phantom{;\rho} \sigma}^{\phantom{(a_1 a_2 } ;
\rho  } R^{\sigma}_{\phantom{\sigma} a_3 |\rho | a_4)} \nonumber \\
&  & \qquad + (4/105) \, R^{ \rho  }_{\phantom{ \rho}
 \sigma ;(a_1 a_2} R^{\sigma}_{\phantom{\sigma} a_3 |\rho | a_4)}
 + (4/105) \, R^{\rho}_{\phantom{ \rho} (a_1 |\sigma| a_2}
 \Box R^{\sigma}_{\phantom{\sigma} a_3 |\rho | a_4)} + (2/35) \,
 R^{\rho \sigma \tau}_{\phantom{\rho \sigma \tau} (a_1} R_{|\rho \sigma \tau| a_2; a_3 a_4)}\nonumber \\
&  & \qquad+ (1/28) \,
 R^{\rho \phantom{(a_1 |\sigma| a_2}    ;\tau}_{\phantom{ \rho} (a_1 |\sigma| a_2}
R^{\sigma}_{\phantom{\sigma} a_3 |\rho | a_4) ; \tau} + (19/420) \,
 R^{\rho \sigma \tau}_{\phantom{\rho \sigma \tau} (a_1;a_2} R_{|\rho \sigma \tau|  a_3; a_4)}
+ (1/72) \, R R _{(a_1 a_2}R _{ a_3 a_4)} \nonumber \\
& & \qquad - (1/15) \, R^{ \rho }_{\phantom{ \rho}  (a_1  } R_{
|\rho| a_2}R_{a_3 a_4) } + (1/63) \, R^{ \rho  }_{\phantom{ \rho}
 (a_1 } R_{|\sigma| a_2}
 R^{\sigma}_{\phantom{\sigma} a_3 |\rho | a_4)} +  (1/30) \, R^{\rho
 \sigma}R_{(a_1 a_2}R_{|\rho| a_3
 |\sigma| a_4)}
\nonumber \\
&  & \qquad + (1/90) \, R R^{\rho}_{\phantom{\rho} (a_1 |\sigma |
a_2}
 R^{\sigma}_{\phantom{\sigma} a_3 |\rho | a_4)} + (1/30) \,
R_{(a_1 a_2} R^{ \rho  \sigma \tau}_{\phantom{ \rho \sigma \tau}
a_3} R_{|\rho \sigma \tau|   a_4)} \nonumber \\ &  & \qquad + (26/315) \, R^{ \rho
}_{\phantom{ \rho}
 \sigma } R^{\sigma}_{\phantom{\sigma} (a_1 |\tau | a_2}
 R^{\tau}_{\phantom{\tau} a_3 |\rho | a_4)}
+ (10/63) \, R^{ \rho  }_{\phantom{ \rho}
 (a_1 } R^{\sigma \phantom{a_2} \tau}_{\phantom{\sigma} a_2 \phantom{ \tau } a_3}
 R_{|\rho \sigma \tau| a_4)}
\nonumber \\ &  & \qquad
+ (4/315) \, R^{\rho \sigma \tau
 \kappa} R_{ \rho  (a_1 |\tau|
 a_2} R_{|\sigma| a_3 |\kappa|
 a_4)} + (4/105) \, R^{\rho  \kappa \tau }_{\phantom{\rho
\kappa \tau } (a_1} R_{|\rho \tau| \phantom{\sigma}
 a_2}^{\phantom{ |\rho \tau|} \sigma} R_{|\sigma| a_3 |\kappa|
 a_4)} \nonumber \\
&  & \qquad + (16/315) \, R^{\rho  \kappa \tau }_{\phantom{\rho
\kappa \tau } (a_1} R_{|\rho|  \phantom{\sigma}
 |\tau| a_2}^{\phantom{ |\rho| } \sigma} R_{|\sigma| a_3 |\kappa|
 a_4)}  + (8/315) \, R^{\rho  \tau \kappa }_{\phantom{\rho
 \tau  \kappa} (a_1} R_{|\rho \tau| \phantom{\sigma}
 a_2}^{\phantom{ |\rho \tau|} \sigma} R_{|\sigma| a_3 |\kappa|
 a_4)}. \label{AppBoxVVMfinal_3e}
\end{eqnarray}
\end{subequations}
\end{widetext}
The coefficients (\ref{AppBoxVVMfinal_3a})-(\ref{AppBoxVVMfinal_3c})
were calculated by Christensen \cite{Christensen1,Christensen2}. In
Ref.~\cite{BrownOttewill83}, Brown and Ottewill have obtained some
of the terms of the coefficient (\ref{AppBoxVVMfinal_3d}) and their
result has been corrected in the recent article by Anderson,
Flanagan and Ottewill \cite{AndersonFlanaganOttewill05}. To our
knowledge, the expression of the coefficient
(\ref{AppBoxVVMfinal_3e}) is new. Finally, from
(\ref{AppVBoxVVMfinal_2}) and (\ref{AppBoxVVMfinal_3}) we can write
\begin{widetext}
\begin{eqnarray}\label{AppVBoxVVMfinal_4}
& & \Box \Delta^{1/2} = \frac{1}{6} \, R
\nonumber \\
&  & \quad    + \left[\frac{1}{40} \, \Box R_{a_1 a_2}  -
\frac{1}{120} \, R _{ ;a_1 a_2} + \frac{1}{72} \, R R_{a_1 a_2}
-\frac{1}{30} \, R^{ \rho }_{\phantom{ \rho} a_1 } R_{ \rho a_2 }
\right. \nonumber \\ &  & \qquad \quad  \left.
 +
\frac{1}{60} \,  R^{ \rho   \tau} R_{\rho a_1 \tau a_2  }
 + \frac{1}{60} \,  R^{\rho \sigma \tau }_{\phantom{\rho \sigma \tau} a_1 }
 R_{ \rho \sigma \tau  a_2 }
\right]\sigma^{;a_1}\sigma^{;a_2} \nonumber \\
& & \quad -  \left[  - \frac{1}{360} \, R _{ ; a_1 a_2 a_3} +
\frac{1}{120} \, ( \Box R_{a_1 a_2})_{;a_3 } + \frac{1}{144} \, R R
_{  a_1 a_2; a_3 }- \frac{1}{45} \, R^{ \rho }_{\phantom{ \rho} a_1
} R_{ \rho a_2;a_3  }
\right. \nonumber \\ &  & \qquad \quad  \left.
+ \frac{1}{180} \, R^{ \rho  }_{\phantom{
\rho}
 \sigma ;a_1 } R^{\sigma}_{\phantom{\sigma} a_2 \rho  a_3 }
 + \frac{1}{180} \, R^{ \rho  }_{\phantom{ \rho}  \sigma}
R^{\sigma}_{\phantom{\sigma}   a_1 \rho  a_2;  a_3 } + \frac{1}{90}
\, R^{ \rho  \sigma \tau}_{\phantom{ \rho \sigma \tau} a_1}
R_{\rho \sigma \tau   a_2;  a_3 }\right]\sigma^{;a_1}\sigma^{;a_2}\sigma^{;a_3} \nonumber \\
& & \quad +  \left[  \frac{1}{560} \, ( \Box R_{a_1 a_2} )_{;a_3
a_4} - \frac{1}{1680} \, R _{ ; a_1 a_2 a_3 a_4} - \frac{1}{1440} \,
R _{ ; a_1 a_2}R _{ a_3 a_4} +  \frac{1}{480} \, R R _{
 a_1 a_2; a_3 a_4} \right. \nonumber \\
&  & \qquad \quad \left. +  \frac{1}{480} \, R _{  a_1 a_2}\Box R _{
a_3 a_4} -\frac{17}{2520} \, R^{ \rho  }_{\phantom{ \rho}  a_1  }
R_{ \rho   a_2;a_3 a_4 } + \frac{1}{504} \, R^{ \rho  }_{\phantom{
\rho} a_1  } R_{ a_2 a_3;  \rho a_4 } + \frac{1}{5040} \, R^{ \rho
}_{\phantom{ \rho} a_1;a_2  } R_{ \rho   a_3;a_4 } \right. \nonumber \\
&  & \qquad \quad \left. - \frac{1}{112} \, R^{ \rho }_{\phantom{
\rho}  a_1 ; a_2 } R_{ a_3 a_4;   \rho  } + \frac{17}{4032} \, R_{
a_1 a_2 }^{ \phantom{a_1 a_2 } ; \rho }R_{ a_3 a_4 ; \rho  } +
\frac{1}{840} \, R^{ \rho }_{\phantom{ \rho}
 \sigma } R^{\sigma}_{\phantom{\sigma} a_1 \rho  a_2; a_3 a_4}
+ \frac{1}{504} \, R^{ \rho  }_{\phantom{ \rho}
 a_1 ; \sigma } R^{\sigma}_{\phantom{\sigma} a_2 \rho  a_3; a_4}
\right. \nonumber \\
&  & \qquad \quad \left. + \frac{1}{720} \, R^{ \rho  }_{\phantom{
\rho}
 \sigma ;a_1 } R^{\sigma}_{\phantom{\sigma} a_2 \rho  a_3;
 a_4} - \frac{1}{210} \, R^{ \rho  }_{\phantom{ \rho}
 a_1 ; \sigma  a_2} R^{\sigma}_{\phantom{\sigma} a_3 \rho
 a_4}+ \frac{11}{2520} \, R_{a_1 a_2 \phantom{;\rho} \sigma}^{\phantom{a_1 a_2 } ;
\rho  } R^{\sigma}_{\phantom{\sigma} a_3 \rho  a_4}  + \frac{1}{630}
\, R^{ \rho  }_{\phantom{ \rho}
 \sigma ;a_1 a_2} R^{\sigma}_{\phantom{\sigma} a_3 \rho  a_4} \right. \nonumber \\
&  & \qquad \quad \left.
 + \frac{1}{630} \, R^{\rho}_{\phantom{ \rho} a_1 \sigma a_2}
 \Box R^{\sigma}_{\phantom{\sigma} a_3 \rho  a_4} + \frac{1}{420} \,
 R^{\rho \sigma \tau}_{\phantom{\rho \sigma \tau} a_1} R_{\rho \sigma \tau a_2; a_3 a_4}  + \frac{1}{672} \,
 R^{\rho \phantom{a_1 \sigma a_2}    ;\tau}_{\phantom{ \rho} a_1 \sigma a_2}
R^{\sigma}_{\phantom{\sigma} a_3 \rho  a_4 ; \tau} \right. \nonumber \\ &  & \qquad \quad  \left.
 +
\frac{19}{10080} \,
 R^{\rho \sigma \tau}_{\phantom{\rho \sigma \tau} a_1;a_2} R_{\rho \sigma \tau  a_3;
 a_4}
 + \frac{1}{1728} \, R R _{a_1 a_2}R _{ a_3
a_4}  - \frac{1}{360} \, R^{ \rho }_{\phantom{ \rho} a_1  } R_{ \rho
a_2}R_{a_3 a_4 } \right. \nonumber \\ &  & \qquad \quad  \left.
+ \frac{1}{1512} \, R^{ \rho }_{\phantom{ \rho}
 a_1 } R_{\sigma a_2}
 R^{\sigma}_{\phantom{\sigma} a_3 \rho  a_4} +  \frac{1}{720} \, R^{\rho
 \sigma}R_{a_1 a_2}R_{\rho a_3
 \sigma a_4}
+ \frac{1}{2160} \, R
R^{\rho}_{\phantom{\rho} a_1 \sigma
 a_2}
 R^{\sigma}_{\phantom{\sigma} a_3 \rho  a_4} \right. \nonumber \\ &  & \qquad \quad  \left.
 + \frac{1}{720} \,
R_{a_1 a_2} R^{ \rho  \sigma \tau}_{\phantom{ \rho \sigma \tau} a_3}
R_{\rho \sigma \tau   a_4} + \frac{13}{3780} \, R^{ \rho
}_{\phantom{ \rho}
 \sigma } R^{\sigma}_{\phantom{\sigma} a_1 \tau  a_2}
 R^{\tau}_{\phantom{\tau} a_3 \rho  a_4}
+ \frac{5}{756} \, R^{ \rho  }_{\phantom{
\rho}
 a_1 } R^{\sigma \phantom{a_2} \tau}_{\phantom{\sigma} a_2 \phantom{ \tau } a_3}
 R_{\rho \sigma \tau a_4} \right. \nonumber \\&  & \qquad \quad \left.
   + \frac{1}{1890} \, R^{\rho \sigma \tau
 \kappa} R_{ \rho  a_1 \tau
 a_2} R_{\sigma a_3 \kappa
 a_4} + \frac{1}{630} \, R^{\rho  \kappa \tau }_{\phantom{\rho
\kappa \tau } a_1} R_{\rho \tau \phantom{\sigma}
 a_2}^{\phantom{ \rho \tau} \sigma} R_{\sigma a_3 \kappa
 a_4} \right. \nonumber \\
&  & \qquad \quad \left. + \frac{2}{945} \, R^{\rho  \kappa \tau
}_{\phantom{\rho \kappa \tau } a_1} R_{\rho  \phantom{\sigma}
 \tau a_2}^{\phantom{ \rho } \sigma} R_{\sigma a_3 \kappa
 a_4}  + \frac{1}{945} \, R^{\rho  \tau \kappa }_{\phantom{\rho
 \tau  \kappa} a_1} R_{\rho \tau \phantom{\sigma}
 a_2}^{\phantom{ \rho \tau} \sigma} R_{\sigma a_3 \kappa
 a_4} \right]\sigma^{;a_1}\sigma^{;a_2}\sigma^{;a_3}\sigma^{;a_4}  + O \left(\sigma^{5/2}
 \right).
\end{eqnarray}
\end{widetext}

\newpage

\bibliography{DS_H}

\end{document}